\documentclass{aa}  

\usepackage{graphicx}
\usepackage{txfonts}
\usepackage{color}
\usepackage{longtable}

\begin{document} 

   \title{Opening PANDORA's box: APEX observations of CO in PNe}
   
   \author{L. Guzman-Ramirez\inst{1,2}
         \and
          A.~I. G\'omez-Ru\'iz\inst{3}
          \and
          H.~M.~J. Boffin\inst{4}
          \and
          D. Jones\inst{5,6}
          \and
          R. Wesson\inst{7}
          \and
          A.~A. Zijlstra\inst{8,9}
          \and
          C.~L. Smith\inst{10}
          \and
          Lars-$\mathring{\rm A}$ke Nyman\inst{2,10}}

   \institute{Leiden Observatory, Leiden University, Niels Bohrweg 2, 2333 CA Leiden, The Netherlands \\
              \email{guzmanl@strw.leidenuniv.nl}
         \and
             European Southern Observatory, Alonso de C\'ordova 3107, Santiago, Chile
         \and    
             CONACYT Instituto Nacional de Astrof\'isica, \'Optica y Electr\'onica, Luis E. Erro 1, 72840 Tonantzintla, Puebla, M\'exico
         \and    
             European Southern Observatory, Karl-Schwarzschild-str. 2, D-85748 Garching, Germany
          \and 
            Instituto de Astrof\'isica de Canarias, E-38205 La Laguna, Tenerife, Spain
          \and 
            Departamento de Astrof\'isica, Universidad de La Laguna, E-38206 La Laguna, Tenerife, Spain
          \and 
            Department of Physics and Astronomy, University College London, Gower Street, London WC1E 6BT, UK
          \and 
            Jodrell Bank Centre for Astrophysics, University of Manchester, Manchester, UK
          \and 
            Department of Physics \& Laboratory for Space Research, University of Hong Kong, Pok Fu Lam Road, Hong Kong
          \and 
            Centre for Research in Earth and Space Sciences, York University, 4700 Keele Street, Toronto, ON, M3J 1P3, Canada
          \and 
            Joint ALMA Observatory, Alonso de C\'ordova 3107, Vitacura, Santiago, Chile}

   \date{Received xx, 2018; accepted xx, 2018}

  \abstract
   {Observations of molecular gas have played a key role in developing the current understanding of the late stages of stellar evolution.}
   {The survey Planetary nebulae AND their cO Reservoir with APEX (PANDORA) was designed to study the circumstellar shells of evolved stars with the aim to estimate their physical parameters.}
   {Millimetre carbon monoxide (CO) emission is the most useful probe of the warm molecular component ejected by low- to intermediate-mass stars. CO is the second-most abundant molecule in the Universe, and the millimetre transitions are easily excited, thus making it particularly useful to study the mass, structure, and kinematics of the molecular gas. We present a large survey of the CO (J $= 3-2$) line using the Atacama Pathfinder EXperiment (APEX) telescope in a sample of 93 proto-planetary nebulae and planetary nebulae. }
   {CO (J $= 3-2$) was detected in 21 of the 93 objects. Only two objects (IRC+10216 and PN M 2-9) had previous CO (J $= 3-2$) detections, therefore we present the first detection of CO (J $= 3-2$) in the following 19 objects: Frosty Leo, HD 101584, IRAS 19475+3119, PN M1-11, V* V852 Cen, IC 4406, Hen 2-113, Hen 2-133, PN Fg 3, PN Cn 3-1, PN M 2-43, PN M1-63, PN M1-65, BD +30 3639, Hen 2-447, Hen 2-459, PN M3-35, NGC 3132, and NGC 6326.}
   {CO (J $= 3-2$) was detected in all 4 observed pPNe (100\%), 15 of the 75 PNe (20\%), one of the 4 wide binaries (25\%), and in 1 of the 10 close binaries (10\%). Using the CO (J $= 3-2$) line, we estimated the column density and mass of each source. The H$_2$ column density ranges from 1.7x10$^{18}$ to 4.2x10$^{21}$cm$^{-2}$ and the molecular mass ranges from 2.7x10$^{-4}$ to 1.7x10$^{-1}$M$_{\odot}$.}

   \keywords{Line: identification -- Molecular data -- Catalogs -- (ISM:) planetary nebulae: general}
   \maketitle
   
%

\section{Introduction}
Planetary nebulae (PNe) represent the final evolutionary phase of low- and intermediate-mass stars, where the extensive mass lost by the star on the asymptotic giant branch (AGB) is ionised by the emerging white dwarf. The ejecta quickly disperses and merges with the surrounding interstellar medium (ISM), whereby the total mass lost from these stars constitutes approximately half of the Milky Way interstellar gas and dust \citep{leitner11}. 
In the ejecta of AGB stars, the CO molecule locks away the less abundant element, leaving the remaining free oxygen (O) or carbon (C) to drive the chemistry and dust formation.  Dust grains and molecules, predominantly CO given the large quantities of C and O in the ejecta, condense in these AGB winds, forming substantial circumstellar envelopes that are detectable in the infrared and millimetre domains. The evolutionary phase between the end of the AGB and the beginning of the planetary nebula (PN) phase has long been a missing link in understanding the late stages of stellar evolution. Observations of molecular gas have played a key role in developing the current understanding of this intermediate phase. 

Millimetre and sub-millimetre CO line emission is a very useful probe of the warm molecular component of the neutral gas surrounding AGB stars. This molecule therefore is a standard probe of their envelopes: CO is relatively abundant, and the sub-millimetre transitions are easily excited. In Table 1 we present the general parameters for the most commonly observed CO rotational transitions, including their frequency, density, and their excitation temperature (E$_{upper}$/$k_B$). It can therefore be used to study the mass, structure, and kinematics of the bulk of the molecular gas.   CO envelopes surrounding PNe have been detected in many CO surveys, including those of \citet{huggins89}, using the NRAO 12m telescope, and \citet{huggins96}, using the SEST 15m and IRAM 30m telescopes. Several recent lines of investigation, including optical imaging of PNe \citep{sahai98, lopez03}, kinematic studies of proto-planetary nebulae \citep[pPNe][]{bujarrabal01}, and
observations of individual molecular envelopes \citep[e.g. He 3-1475 and AFGL~617; ][]{huggins04,cox03}, have focused the attention on the complex geometry of PN formation.  The origins of these complex geometries are not very well understood, but it is very likely that similar structures are found in the molecular gas surrounding young PNe \citep{huggins05} and as such share a formation mechanism. 

\begin{table}
\caption{Most commonly observed CO transitions with their general parameters}
\label{CO}
\begin{tabular}{cccc}
\hline

Line    &       Frequency       &Density        &E$_{upper}$/$k_B$ \\
        &       (GHz)           &(cm$^{-3}$)    &(K) \\
\hline  
CO (J $= 1-0$)  &115.271                &1.4x10$^3$     &5.5  \\
CO (J $= 2-1$)  &230.518                &2.3X10$^3$     &17  \\
CO (J $= 3-2$)  &345.796                &7.0x10$^4$     &34  \\
\hline
\end{tabular}
\end{table}

The dynamical evolution from the AGB to the PN phase has only been observed for a very small sample of objects, which clearly places limits on our understanding of this regime. \citet{knapp98} studied CO (J $= 2-1$) and CO (J $= 3-2$) in a sample of 45 AGB stars, finding about half of the line profiles to be non-parabolic in shape, with profiles associated with complex winds. Complex line profiles occur when a star is undergoing a change in luminosity, pulsation mode, or chemistry. 

Recent improvements in frequency coverage, frequency resolution, and receiver sensitivity enable homogeneous, unbiased, low-noise observations with high velocity resolution. Motivated by the lack of a large enough sample of observations of PNe in CO, especially the more excited CO (J $= 3-2$) transition, we observed a sample of 93 young and evolved PNe using the Swedish Heterodyne Facility Instrument (SHeFI) instrument at the Atacama Pathfinder EXperiment (APEX) telescope. 
We selected the CO (J $= 3-2$) line rather than lower J transitions because it offers the advantage that higher J transitions have higher critical densities for collisional excitation (see Table \ref{CO}). For this reason, emission from molecular clouds is less extensive and the observations are less strongly affected by contamination of Galactic emission. 

We here present the first data release and a preliminary analysis of all the observations. In Section 2 we present the observations, Section 3 summarises the results, and Section 4 presents a discussion of the main outcomes of the survey and future prospects.

\section{Observations and data reduction}
APEX is ideally suited to surveying the CO (J $= 3-2$) emission from pPNe/PNe in the Southern sky. The atmosphere above Chajnantor has sufficiently transparent conditions (precipitable water vapour, PWV $<$2.5mm 80\% of the time) to allow CO (J $= 3-2$) and nearby transitions to be easily detected. The APEX-2 SHeFI receiver with its eXtended bandwidth Fast Fourier Transform Spectrometer (XFFTS) backend, installed at APEX, offers sufficient sensitivity, bandwidth, and spectral resolution to achieve our primary goal of characterising $^{12}$CO (J $= 3-2$) emission in pPNe and PNe. The bandwidth used was 4 GHz, centring the intermediate frequency (IF) at 344.7 GHz. 

Data were processed using \textsc{Continuum and Line Analysis Single-dish Software} (CLASS), a part of the \textsc{gildas} software\footnote{\textsc{gildas} is a radio astronomy software developed by   IRAM. See {\tt http://www.iram.fr/IRAMFR/GILDAS}}. We subtracted the baselines around the detected CO line to obtain an averaged spectrum of each observation. After this, spectra from all epochs were averaged into a single, final spectrum for each object. To account for the telescope baseline, a polynomial fit of order 3 (in some cases 5) was removed form the averaged spectrum.

APEX observed 93 pPNe and PNe in total. The pPNe sample was extracted from the catalogue of young PNe of \citet{sahai11}, while the PNe sample was taken from two different sources: PNe that have been observed previously in CO (J $=2-1$) by co-author L.-A. Nyman using SEST, and the second from the list of PNe with known binary central stars compiled and maintained by co-author D. Jones (http://drdjones.net/bCSPN.html). Targets at $|b|<3.5\deg$ were excluded in order to avoid confusion with interstellar CO emission \citep{dame01}. Objects with angular dimensions larger than the telescope beam size (18\arcsec) were also excluded (except for IC 4406, NGC 3132, and M2-9).  Each source was observed for a total time of 30\,min (including overheads), corresponding to a typical on-source time of 6\,min, which with nominal T$_{sys}$ for the expected observing conditions and source elevations (PWV$<$2.5mm) provides a sensitivity of 0.04K\,km\,s$^{-1}$. The total telescope time dedicated to the survey was 46.5\,hours.

\section{Results}

In Table \ref{observations} we present the complete list of objects observed, columns 1 and 2 are the RA and DEC, respectively, columns
3 and 4 contain the PN G and common name of each object, respectively, and column 5 shows the rms of the observations for each target in mK. Column 6 specifies whether the line comes from the object or if it is contamination from the interstellar medium (ISM), and Column 7 refers to the figure in which the final averaged spectrum is shown.

CO (J $= 3-2$) was detected in 35 of the 93 sources. The spectra of another 7 objects show the CO (J $= 3-2$) transition in absorption (negative intensity features); 1 additional object was found to present a line profile with apparent P-Cygni shape. Figures \ref{Fig1} to \ref{Fig11} show the complete spectrum per object. 
For the objects with null-detections, the rms value can be used
to estimate an upper limit on the CO line intensity. 

For the objects where CO (J $= 3-2$) was detected as a negative intensity feature and for the one object with the apparent P-Cygni shape, the literature velocity of the object was compared with the measured velocity of the line. In all the cases, the negative intensity lines were found to be interstellar CO (J $= 3-2$) in the OFF positions. The P-Cygni shaped line profile is likely to be caused by emission from a cloud (or two clouds) in both the ON and OFF positions. The line width is also an indication of interstellar material, CO from the ISM has typical line widths of $\sim$1km/s \citep{hacar16}, while CO from evolved stars shows line widths of at least a few (up to several tens) km/s.

Using the same logic, we found that for six objects, the detected CO (J $= 3-2$) central line velocities do not correspond to the respective velocity of the object and thus are most likely interstellar emission, these were named CO from the ISM.  
These objects are PN M1-6 , IC 4191, PN PC 12, Hen 2-182, PN M3-15, and Hen 2-39. 

Therefore we have 21 true CO (J $= 3-2$) detections. Based on these true detections, we found that CO (J $= 3-2$) has been detected in two objects, IRC+10216 and PN M 2-9 (more detail on their detections is given in the following section). The final number of true CO (J $= 3-2$) detections that have not previously been reported in the literature is 19. This is the first time that CO (J $= 3-2$) has been detected in Frosty Leo, HD 101584, IRAS 19475+3119, PN M1-11, V* V852 Cen, IC 4406, Hen 2-113, Hen 2-133, PN Fg 3, PN Cn 3-1, PN M 2-43, PN M1-63, PN M1-65, BD +30 3639, Hen 2-447, Hen 2-459, PN M3-35, NGC 3132, and NGC 6326. The following PNe M1-11, Fg 3, Cn3-1, M1-65, Hen 2-133, NGC6326, and M3-35 show line widths smaller than 5km/s, therefore these detections should be considered with caution. Each of the detections is discussed in more detail in the following subsections. 

In Table \ref{detections} we present only the objects with CO (J $= 3-2$) line detections. Column 1 contains the name of each object, column 2 lists the PN G name, column 3 the heliocentric velocity in kms$^{-1}$, column 4presents the line width obtained by the fit in km\,s$^{-1}$ (when using several Gaussians, this represents the total width of the line including all components). Column 5 lists the integrated intensity over the velocity range of the line in K\,km\,s$^{-1}$. Columns 6 and 7 show the CO and H$_2$ column density in units of cm$^{-2}$, Column 8 refers to the mass estimate of the H$_2$ component in solar masses, and Column 10 finally refers to the figure where the spectrum is plotted and fitted.
All the line intensities were calculated using either a Gaussian fit (for single-peaked lines) or a shell profile fit (for double-peaked lines). In some objects more than one Gaussian line was needed to fit the line profile. Figures \ref{Fig12} to \ref{Fig14} show the spectra with the detected lines highlighted (the literature values for the heliocentric velocities of the objects are also given). Each line has a fit (either Gaussian or shell) marked in red. Four objects (MPA J1508-6455, Hen 2-447, IRAS 19475+3119, and PN K4-41) have an unknown heliocentric velocity, and it is therefore not indicated in their spectrum.



\subsection{Proto-planetary nebulae}
We detected CO (J $= 3-2$)  in all of the four surveyed pPNe Frosty Leo, IRC+10216, HD 101584, and IRAS 19475+3119. IRC+10216 was known to have CO (J $= 3-2$) emission, for the other sources this was not the case, and these detections therefore represent the first detections of CO (J $= 3-2$) in these objects.

\subsubsection{Frosty Leo}
At first detection of CO (J $= 3-2$) in this object, its line profile shows much structure, although to first order, it can be fitted with a one-component Gaussian.
Previous studies of this object have focussed mainly on CO (J $= 2-1$) and CO (J $= 1-0$) \citep{likkel87,castro05,Sahai2000}.

\subsubsection{IRC+10216}
The CO (J $= 3-2$) line profile of IRC-10216 was fitted with a one-component Gaussian.
IRC+10216 is a very well studied object, the first detection of CO (J $= 3-2$) in this object was made using the James Clerk Maxwell Telescope (\citealp{williams92}). A more detailed study is presented in \cite{patel2010}, where the authors performed a line survey using the Submillimeter Array covering from 293.9 to 354.8 GHz. 

\subsubsection{HD 101584}
At first detection of CO (J $= 3-2$) in this object, it needed more than one Gaussian to fit the line profile. 
Many molecular observations of this object can be found in the literature, where CO (J $= 2-1$) and CO (J $= 1-0$) have been detected. The first detection of CO (J $= 1-0$) has been reported by \cite{loup90}, with further molecular CO studies presented in \cite{trams90,oudmaijer95,olof99,olof15,olof16,olof17}.

\subsubsection{IRAS 19475+3119}
At first detection of CO (J $= 3-2$) in this object, its line profile can be fitted with a one-component Gaussian.
This object has CO (J $= 2-1$) and CO (J $= 1-0$) detections in the literature \citep{likkel87,sanchez06,sahai07,castro10,hsu11}.

\subsection{Planetary nebulae}
In the sample of 75 PNe, CO (J $= 3-2$) was detected in 15: in PN M1-11, V* V852 Cen, IC 4406, Hen 2-113, Hen 2-133, PN M 2-9, PN Fg 3, PN Cn 3-1, PN M 2-43, PN M1-63, PN M1-65, BD +30 3639, Hen 2-447, Hen 2-459, and PN M3-35. 
 
\subsubsection{PN M1-11, PN Fg 3, PN Cn 3-1, and PN M1-65}
At first detection of CO (J $= 3-2$) in these objects, the line profiles can be fitted with a one-component Gaussian. No other molecular lines are detected in any of these objects. For all these sources, the line widths appear to be very narrow. This may indicate that the emission comes from the ISM or might be due to a cloud with the same velocity as the PN. Therefore these detections should be followed-up and confirmed with higher sensitivity instruments). 

\subsubsection{V* V852 Cen}
At first detection of CO (J $= 3-2$) in this object, for its line profile, multiple Gaussian fits were required. This object does not have any other CO lines detected.

\subsubsection{IC 4406}
At first detection of CO (J $= 3-2$) in this object, for its line profile, a two-component Gaussian fit was required. The two-component line profile could be associated with an expanding ring, or a disc in rotation. This object has previously been detected in CO (J $= 2-1$) and CO (J $= 1-0$) \citep{cox91,sahai91,cox92,smith15}.

\subsubsection{Hen 2-113}
At first detection of CO (J $= 3-2$) in this object, for its line profile, a three-component Gaussian fit was required. The line profile is very broad; higher resolution might reveal a high-velocity molecular gas component for this object. There are no other CO detections for this object.

\subsubsection{Hen 2-133}
The heliocentric velocity of PN Hen 2-133 was also unknown, but we found archival observations taken using the Ultraviolet and Visual Echelle Spectrograph \citep[UVES][]{dekker00} instrument at the Very Large Telescope (VLT) (program 095.D-0067(A), PI Garc\'ia-Hernandez). They were reduced using Reflex \citep{freud13}. Emission lines present in the resulting spectrum were fitted using the automated line fitting algorithm \citep[ALFA,][]{wesson16}; they indicate a line-of-sight velocity at time of observation of -27.36km/s, which was then corrected to heliocentric velocity using the \textsc{starlink} RV command \citep{wallace07}, giving a value of -25.05km/s.
Using this information, we detect CO (J $= 3-2$) for the first time in this object. Its line profile can be fitted with a one-component Gaussian. Its line profile is very narrow, which may mean that the line is interstellar rather than nebular, although it shows some broadening at the wings of the line. No other CO lines are detected in this object.

\subsubsection{PN M 2-9}
The CO (J $= 3-2$) line profile was fitted with a two-component Gaussian. The line shows a beautiful double-peaked profile, which could be interpreted as an indication of a disc or a ring close to the star. 
The first detection of CO (J $= 3-2$) in this object was made using the Atacama Large Millimetre/submillimeter Array (ALMA) (\citealp{castro17}). CO (J $= 2-1$) has also been detected in this object using the Plateau de Bure interferometer \citep{castro12}

\subsubsection{PN M 2-43}
At first detection of CO (J $= 3-2$) in this object, its line profile is very complex. It needed a three-component Gaussian for its fit. 
This might be an indication of several components, meaning that this object might have much structure, but a further more detailed investigation is needed. 
CO (J $= 3-2$) was observed in this object using the JCMT, but only upper limits were obtained \citep{gussie95}.
CO (J $= 1-0$) was detected in this object using the 13.7m millimetre-wave telescope of Purple Mountain Observatory \citep{zhang00}.

\subsubsection{PN M1-63}
At first detection of CO (J $= 3-2$) in this object, its line profile is very complex (similar to M2-9), with a three-component Gaussian needed in order to obtain a reasonable fit. 
This might be an indication of several components, meaning that this object might have much structure, but a further more detailed investigation is needed. No other CO lines are detected in this object.

\subsubsection{BD +30 3639}
This is the first detection of CO (J $= 3-2$) in this object. The CO (J $= 3-2$) line profile was fitted using a one-component Gaussian. The molecular observations of CO (J $= 2-1$) and CO (J $= 1-0$) in this object were made using the IRAM 30m telescope \citep{bachiller91}, with further molecular CO studies presented in \cite{bachiller92,shupe98,bachiller00,freeman16}.

\subsubsection{Hen 2-447}
At first detection of CO (J $= 3-2$) in this object, its line profile needed a two-Gaussian component for its fit. 
This might be an indication of a ring or a disc, but a further more detailed investigation is needed. No other CO lines are
detected in this object.

\subsubsection{Hen 2-459}
At first detection of CO (J $= 3-2$) in this object, its line profile needed only a one-component Gaussian for its fit, although the line is very broad ($\sim$100\,km/s).
The width of the line might be an indication of a faster component, but a further more detailed investigation is needed. No other CO lines are detected in this object.

\subsubsection{PN M3-35}
At first detection of CO (J $= 3-2$) in this object, its line profile needed only a one-component Gaussian for its fit. Its line width is smaller than 5km/s, which might be an indication that the emission comes from the ISM rather than the nebula. CO (J $= 3-2$) was observed in this object using the JCMT, but only upper limits were obtained \citep{gussie95}.

\subsection{Planetary nebulae with binary central stars}
Of the initial sample of 93 observed objects, 14 were known to host binary central stars \citep{jones15, jones17}. Of these 14, 4 are wide binaries (orbital periods $\sim$ 10--1000 days) and 10 are close binaries (orbital periods $\sim$ 1 day).
We detected  CO (J $= 3-2$) in emission in only 2 objects: NGC 3132, and NGC~6326.
  
\subsubsection{NGC 3132}
The central star of NGC 3132 is a visual binary, and as such is the widest of all the binary central stars observed here \citep{ciardullo99}. This is the first time CO (J $= 3-2$) has been detected in this object. Its line profile is very complex, it needed a three-component Gaussian for its fit. 
This might be an indication of several components, meaning that this object might have much structure, but a further more detailed investigation is needed. 
This object was previously observed with the 12m National Radioastronomy Observatory (NRAO) antenna at Kitt Peak, but CO (J $= 3-2$) was not detected \citep{phillips92}. Further molecular CO (J $= 2-1$) and CO (J $= 1-0$) studies on this object are presented in \cite{zuckerman90,sahai90}.

\subsubsection{NGC 6326}
The central star of NGC 6326 close binary system with a period less than one day \citep{miza11}. NGC 6326 shows CO (J $= 3-2$). This is the first time CO (J $= 3-2$) has been detected in this object. The line width is smaller than 4km/s, which might be an indication that the emission comes from the ISM rather than the nebula. The line profile could be fitted with a one-component Gaussian.  

\subsection{Sources where the line velocity does not match the PN velocity}
The following PNe show CO (J $= 3-2$) emission lines, but the line is either at a different velocity than the object (therefore might not be associated to the object) or the line is too thin (and therefore associated to interstellar CO (J $= 3-2$). For IC 4191, PN PC 12, PN M3-15, and Hen 2-39 these CO (J $= 3-2$) seems to be more likely to be emission form the ISM. 

\subsubsection{PN M1-6}
At first detection of CO (J $= 3-2$) in this object, its line profile can be fitted with a one-component Gaussian. CO (J $= 3-2$) emission is detected at around 15km/s, while the heliocentric velocity of this object is reported to be 65.5km/s (see references in the table).  

\subsubsection{IC 4191}
At first detection of CO (J $= 3-2$) in this object, its line profile can be fitted with a one-component Gaussian. CO (J $= 3-2$) emission is detected at around 10km/s, while the heliocentric velocity of this object is reported to be -18.3km/s (see references in the table).  Since the line profile of this object is very thin, this emission is likely associated with the ISM. No other molecular lines are detected in this object.

\subsubsection{PN PC 12}
At first detection of CO (J $= 3-2$) in this object, its line profile can be fitted with a one-component Gaussian. CO (J $= 3-2$) emission is detected at around 8km/s, while the heliocentric velocity of this object is reported to be -60.3km/s (see references in the table).  Since the line profile of this object is very thin, this emission is likely associated with the ISM. No other CO lines are detected in this object.

\subsubsection{Hen 2-182}
At first detection of CO (J $= 3-2$) in this object, its line profile can be fitted with a one-component Gaussian. CO (J $= 3-2$) emission is detected at around 10km/s, while the heliocentric velocity of this object is reported to be -89.5km/s (see references in the table).  The line profile does not seem as narrow as the lines associated with emission from the ISM, but further confirmation is required. No other CO lines are detected in
this object.

\subsubsection{PN M3-15}
At first detection of CO (J $= 3-2$) in this object, its line profile can be fitted with a one-component Gaussian. The CO (J $= 3-2$) emission presents at around 10km/s, while the heliocentric velocity of this object is reported to be 97.2km/s (see references in the table).  Since the line profile of this object is very thin, this emission is likely to originate from the ISM. No other CO lines are detected in this object.

\subsubsection{Hen 2-39}
The central star of Hen 2-39 is a barium star, and as such, is a wide binary system \citep[although the exact orbital period is unknown;][]{miza13}. This is the first time CO (J $= 3-2$) has been detected in this object. Its line profile can be fitted with a one-component Gaussian. The line profile on this object is very thin, and even though it is centred at the nebular heliocentric velocity, this line could still be associated with emission from the ISM. Further investigation is needed. No other CO lines are
detected in this object.

\subsection{Column density and masses}
We estimated the column density N(CO) and N(H$_2$) for all the sources with CO (J $= 3-2$) detections (see Table \ref{detections}). Using N(H$_2$), we estimated the molecular mass of each source. \\

The column density calculations assume optically thin emission and local thermal equilibrium (LTE). Under these assumptions, the CO column density, N(CO), is proportional to the line integrated intensity. We followed the standard formulations to obtain CO column densities (e.g. \citet{huggins96}), assuming an excitation temperature of 15 K for all objects. Hence the correlation between N(CO) and the CO (J $= 3-2$) integrated intensity is given by N(CO)$=$ 6.59x10$^{14}$ x $\int$CO (J $= 3-2$) dv. The N(CO) values reported in Table \ref{detections} include the beam filling factor corrections estimated using the source sizes reported in Table \ref{morphology}. The molecular hydrogen column density, N(H$_2$), is then obtained multiplying N(CO) by the abundance ratio X (CO/H$_2=$) 3x10$^{-4}$ (see, e.g. \citet{huggins96}). Finally, to compute the H$_2$ mass, spherical symmetry is assumed and the distances to nebulae listed in Table \ref{morphology} used.\\

We assumed the CO emission of the nebula to be distributed in a uniform sphere and did not take CO photodissociation into account, therefore nebula masses presented here should be taken with a degree of caution. We also assumed optically thin line emission, which means that for those nebulae where this assumption does not hold, the mass estimates should be taken as lower limits.

The N$_{H_2}$ values range from 1.7x10$^{18}$ to 4.2x10$^{21}$cm$^{-2}$. The molecular masses have values in the range from 2.7x10$^{-4}$ to 1.7x10$^{-1}$M$_{\odot}$, which is in the same range as was found by \citet{huggins96}.

\begin{figure}
\includegraphics[width=10cm, height=7cm]{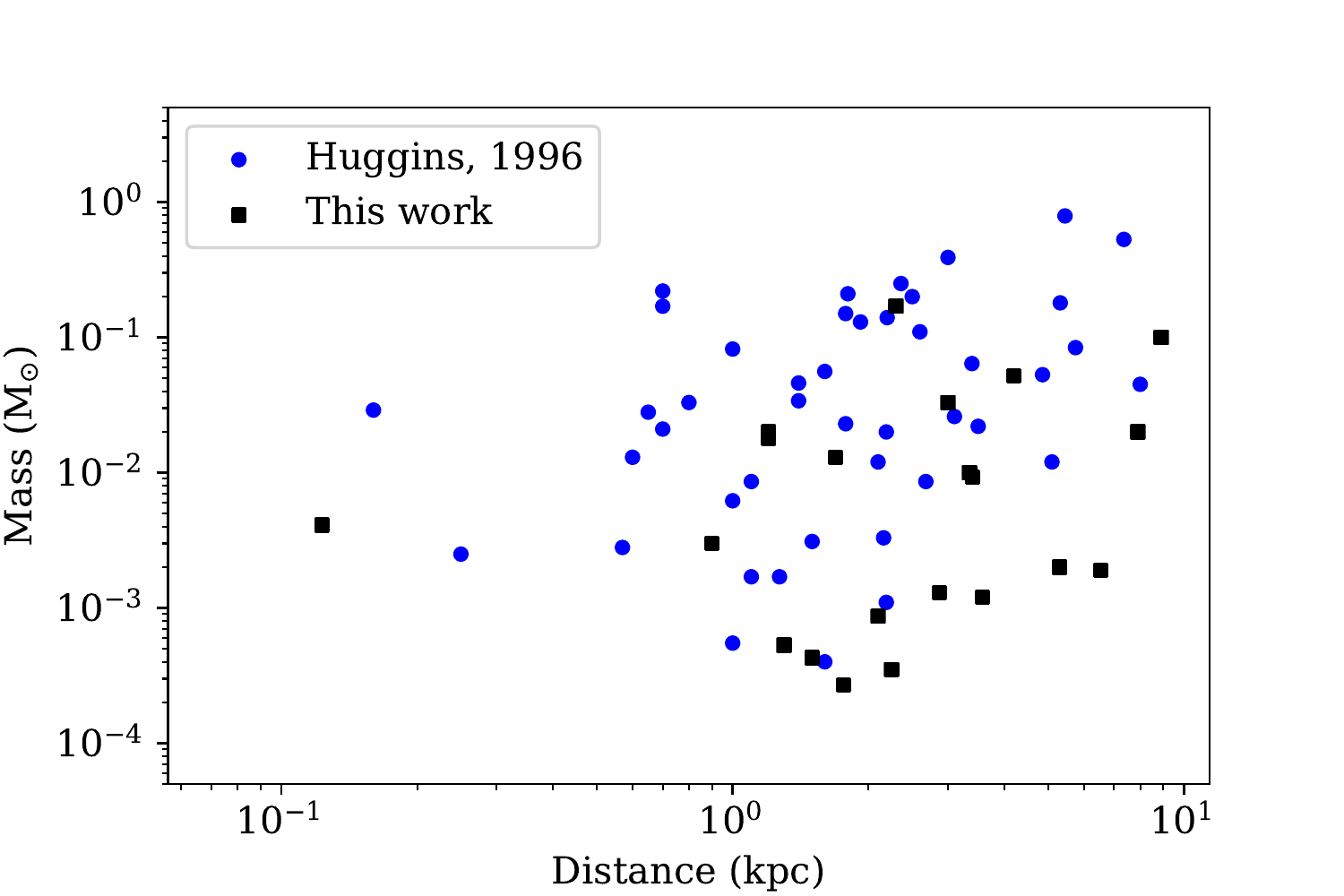}
\caption[]{Distance to the source vs. the molecular mass of its envelope. The plot combines the data from this work (black squares) with the data published by \citet{huggins96} (blue dots).}
\label{M_D}
\end{figure}

Following \citet{huggins96}, we plot the masses versus the distance to each source. In Figure \ref{M_D} we present the distribution of masses estimated with respect to their distance. Black squares are the objects from this paper, and for comparison, we also plot the values taken from \citet{huggins96}, which are represented here as the blue dots. 
This figure shows that the mass is spread over almost four orders of magnitude, and a correlation between mass and distance is
expected, meaning that observationally, we are biased to see the most massive objects at greater
distance. 

It is difficult to draw strong conclusions from the mass estimates because of the significant assumptions made whilst calculating them. However, the mass estimates of these nebulae calculated using the CO (J $= 3-2$) transition are very similar to those determined using CO (J $= 2-1$) or CO (J $= 1-0$), indicating that these transitions trace similar regions of the nebula. It
might have been expected that the CO (J $= 3-2$) would trace a hotter inner region of the nebula (torus/disc), but this seems to not be the case.

\subsection{RADEX diagnostic plot}

We used the offline version of RADEX \citep{Tak07} to generate a diagnostic plot for the CO (J $= 3-2$) to CO (J $= 2-1$) intensity ratio as a function of kinetic temperatures and H$_2$ volume densities for the sources that had complementary CO (J $= 2-1$) observations. Only a few sources are reported in the literature with the appropriate data to analyse them together with our CO (J $= 3-2$) observations. BD$+$30 3639 was the source with the best complementary CO (J $= 2-1$) data, given the size of the source and the beam size of the CO (J $= 2-1$) observations. A discussion of the use of these diagnostic plots is presented for different molecules in the appendix of \citet{Tak07}. Linear molecules, such as CO, are density probes at low densities, while at higher densities, the line ratios are more sensitive to temperature. However, with no other information than a single line ratio (such as our case), it is only possible to provide lower limits to density and temperature. 

\begin{figure}
\centering
\includegraphics[width=6.5cm, height=9cm, angle=-90]{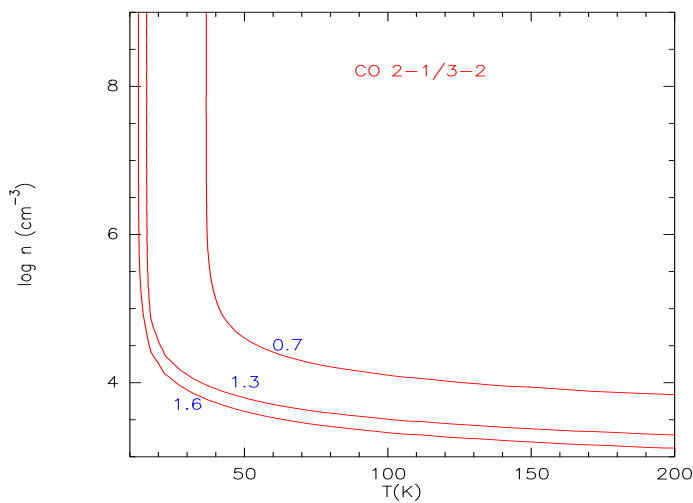}
\caption{Results from the LVG analysis of the CO lines. The curves represent the predicted CO (J $= 3-2$) to CO (J $= 2-1$) intensity ratio for a column density of 1$\times$10$^{16}$ cm$^{-2}$ and line width of 50 km/s. The ratios that match the observed CO line ratio and are indicated with solid red curves.}
\label{CO-LVG}
\end{figure}

\begin{figure*}
\vspace{2cm}
\centering
\hbox{
\centering
\includegraphics[width=6cm, height=5.5cm]{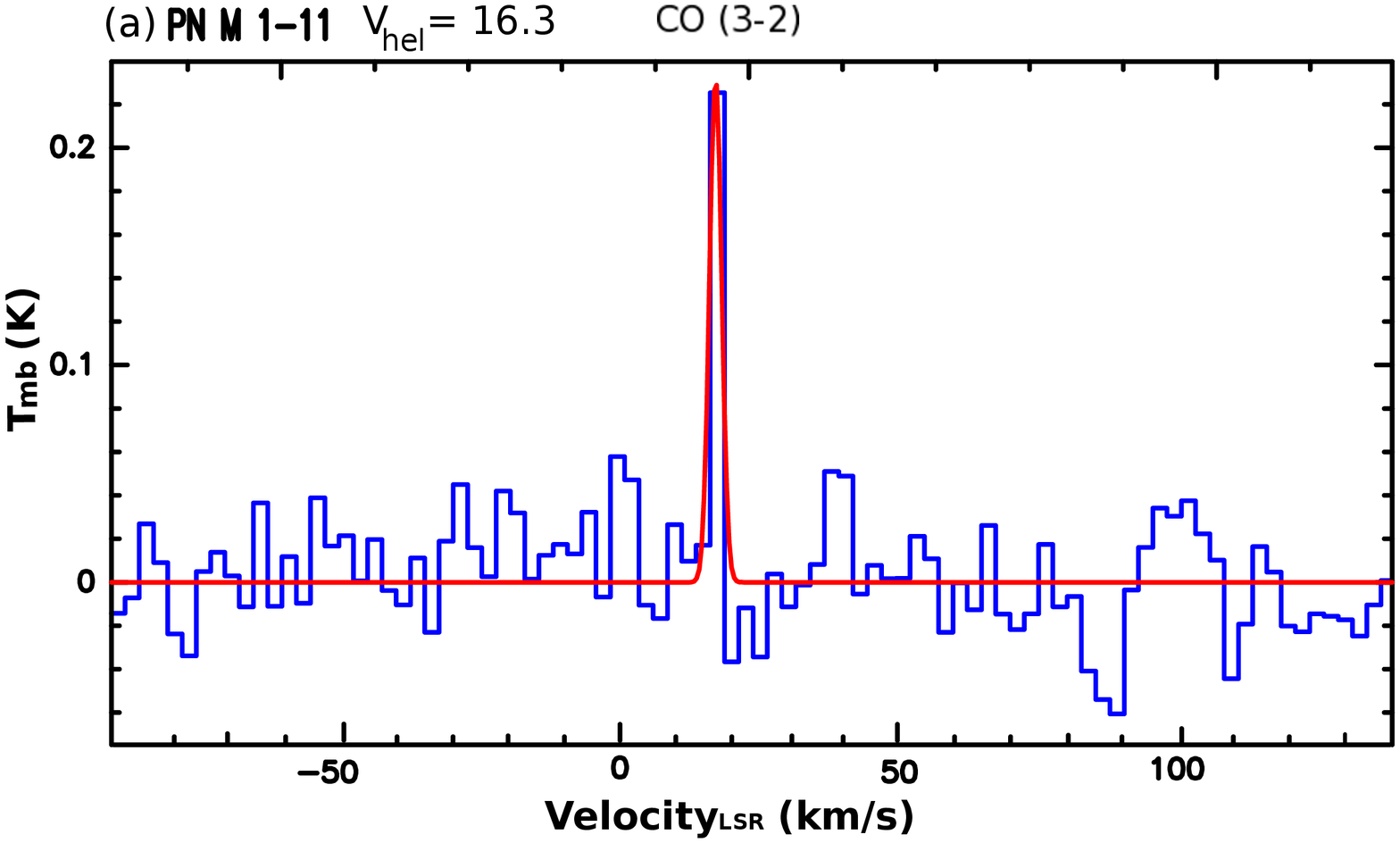}
\includegraphics[width=6cm, height=5.5cm]{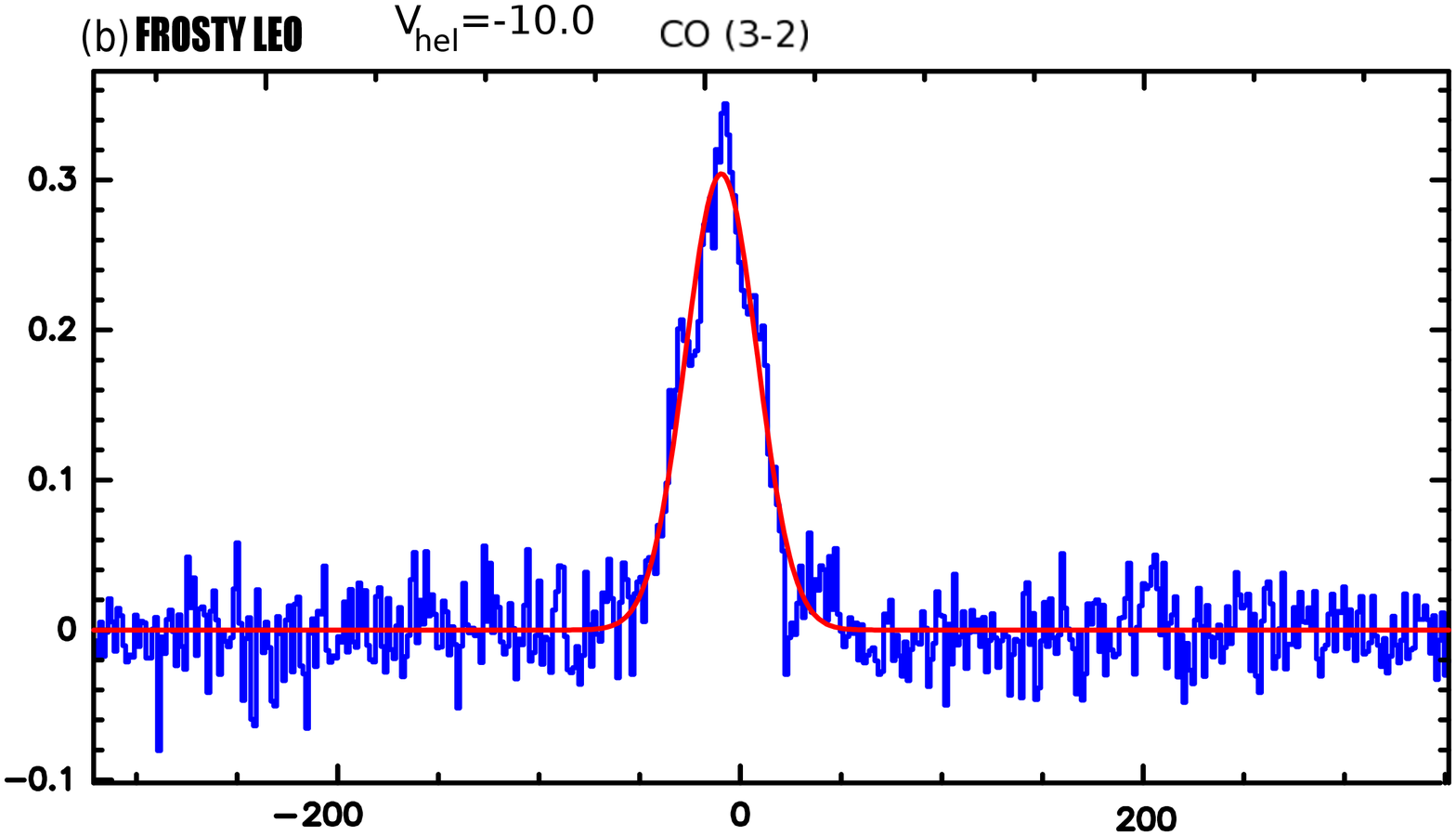}
\includegraphics[width=6cm, height=5.5cm]{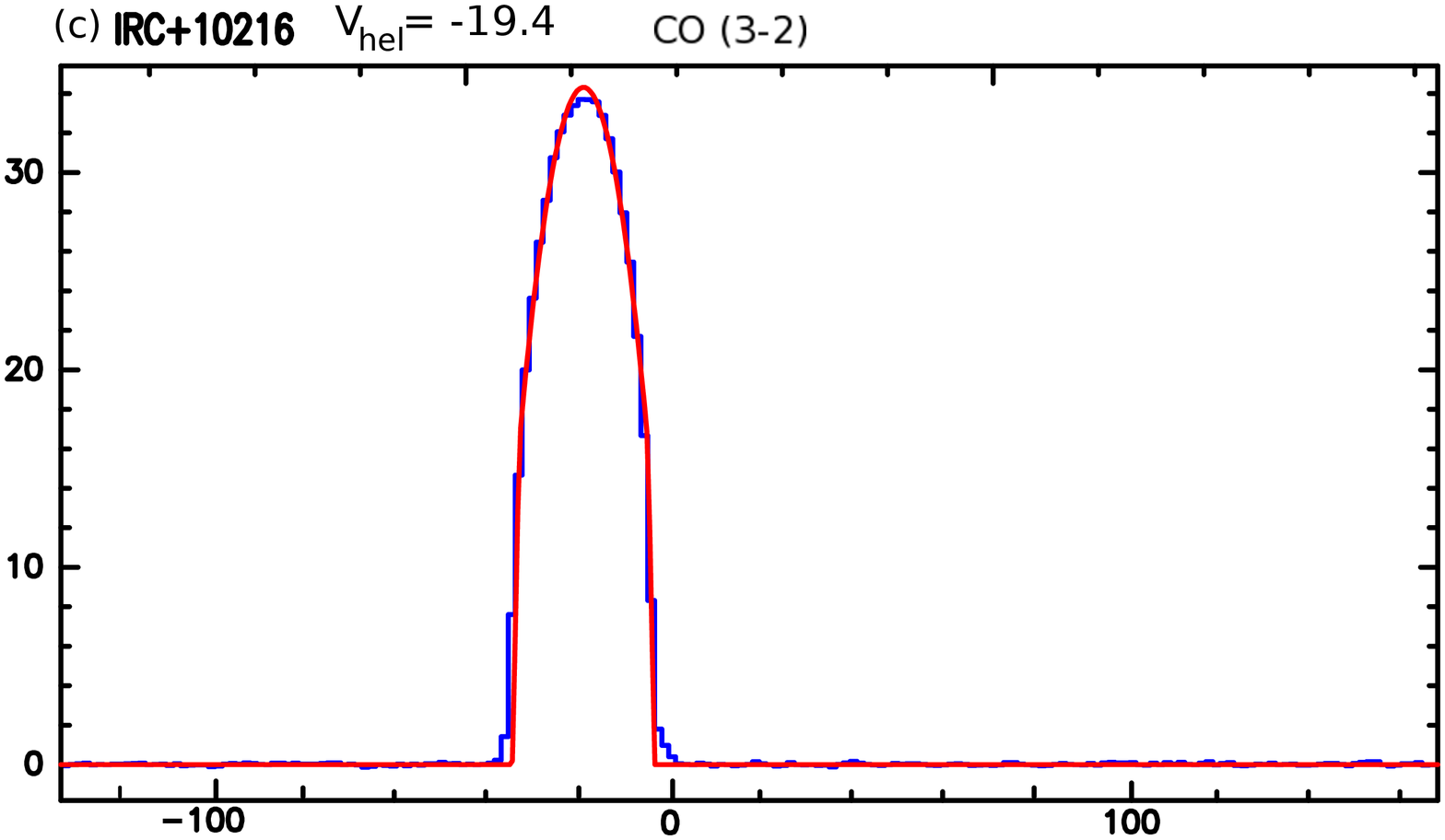}}
\vspace{2cm}
\hbox{
\centering
\includegraphics[width=6cm, height=5.5cm]{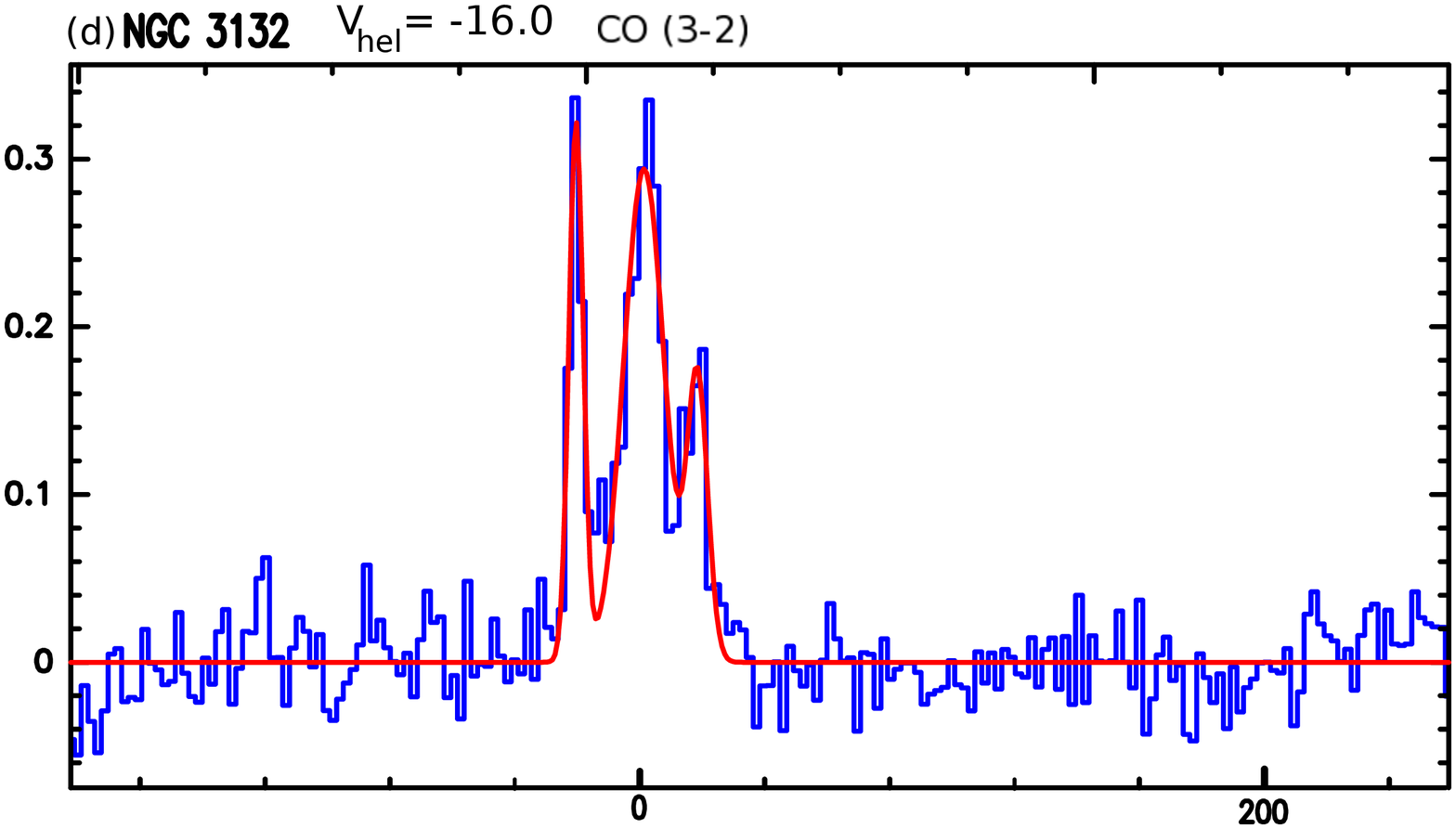}
\includegraphics[width=6cm, height=5.5cm]{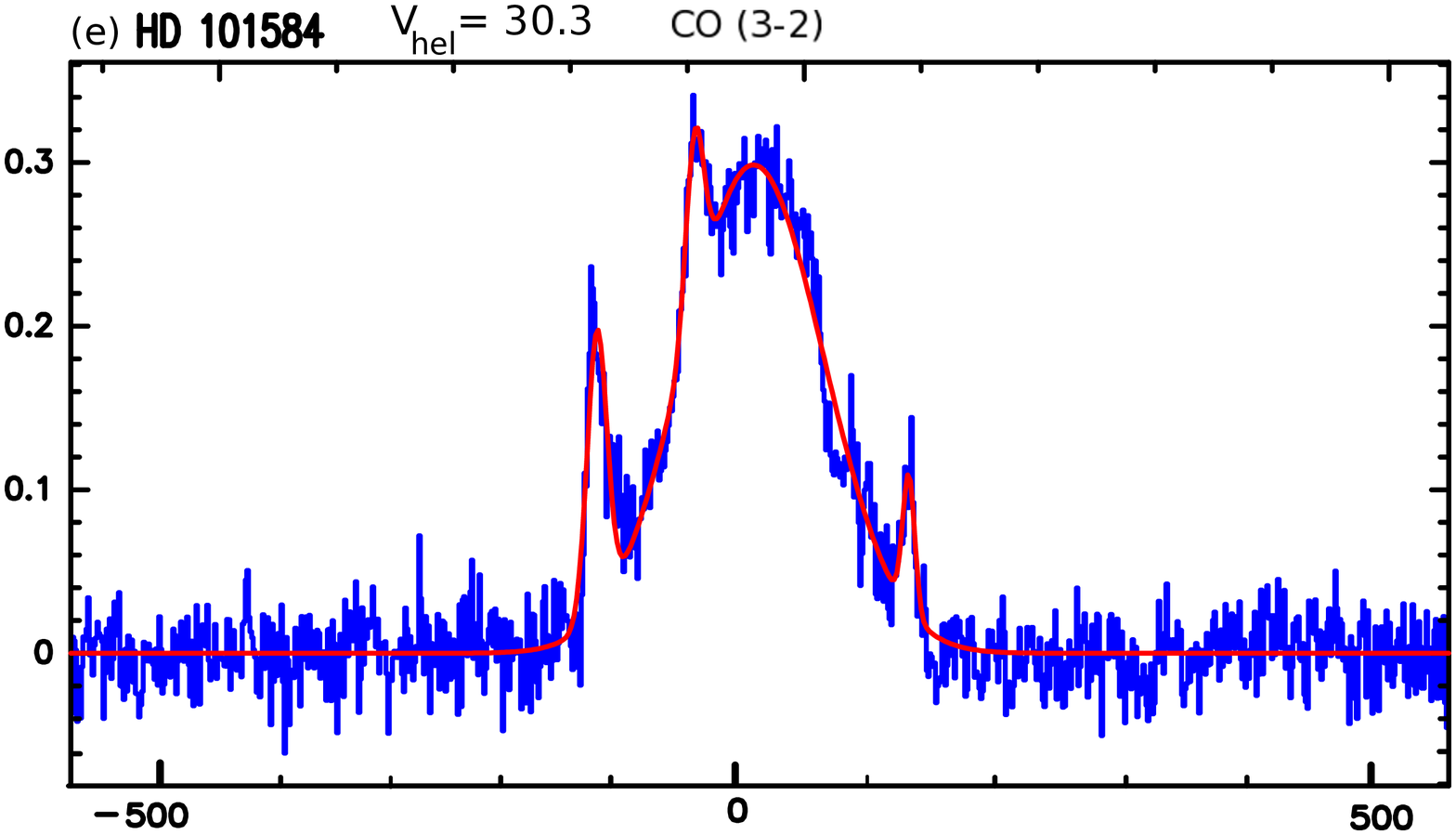}
\includegraphics[width=6cm, height=5.5cm]{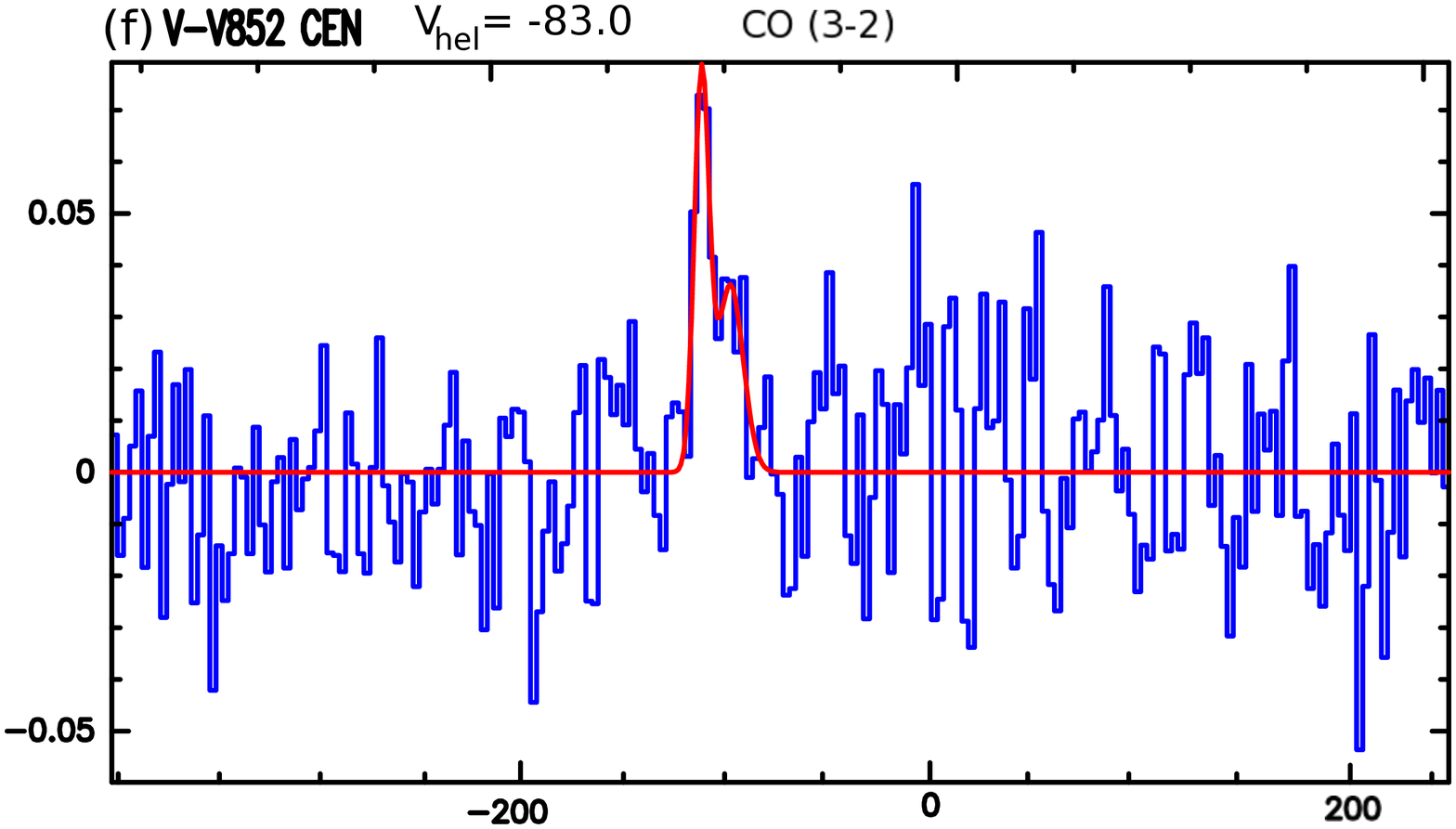}}
\vspace{2cm}
\hbox{
\centering
\vspace{2cm}
\includegraphics[width=6cm, height=5.5cm]{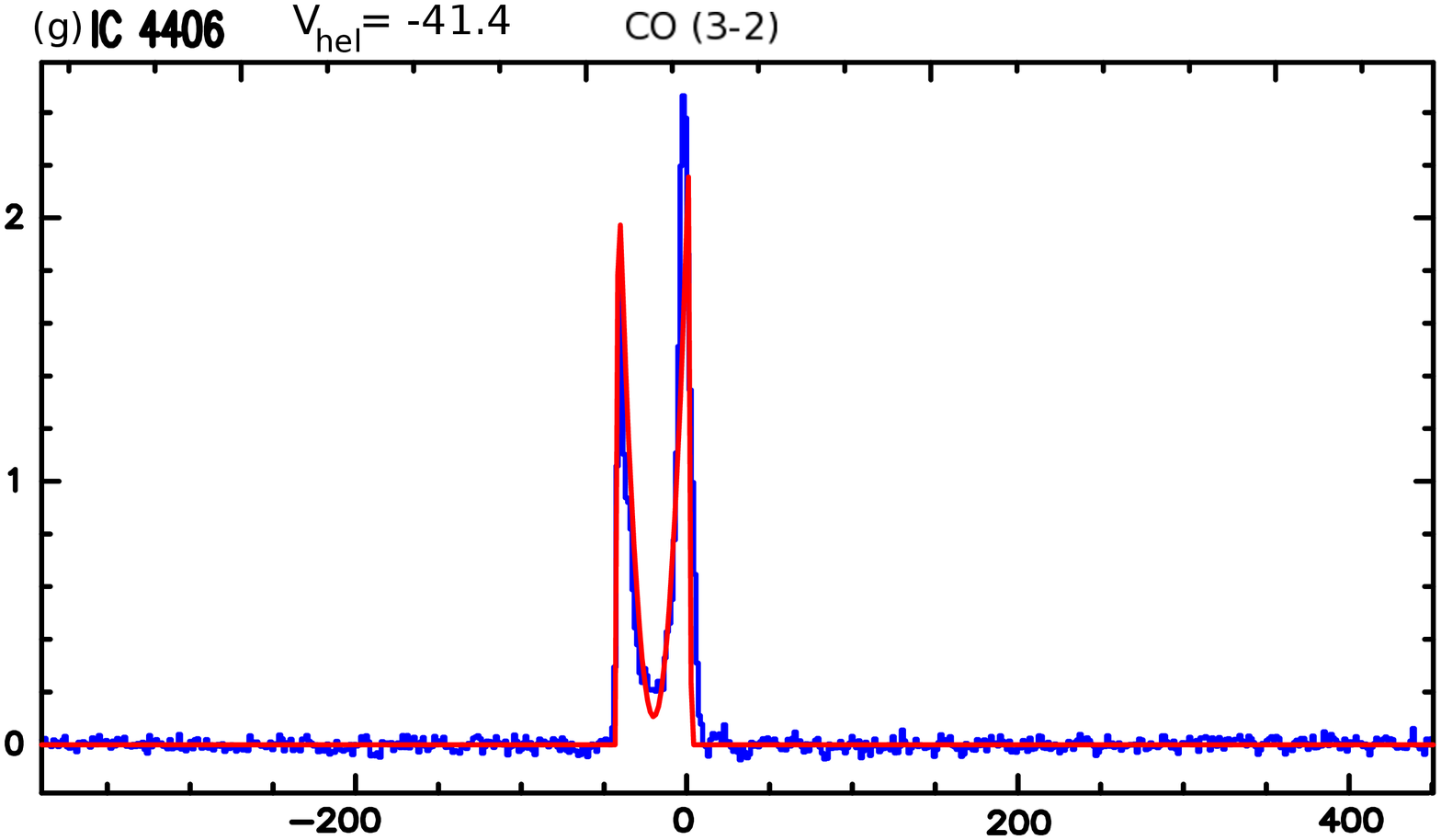}
\includegraphics[width=6cm, height=5.5cm]{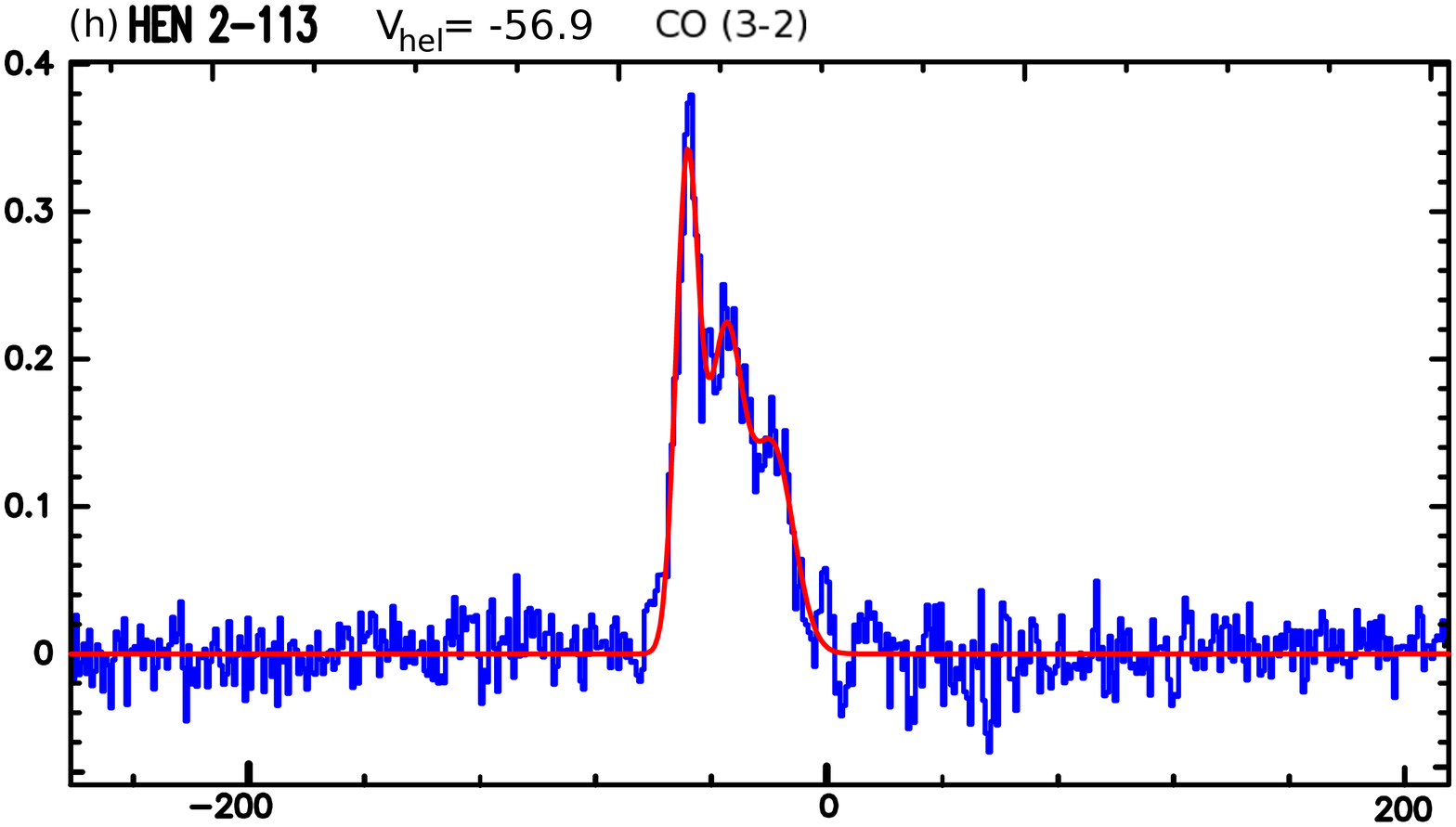}
\includegraphics[width=6cm, height=5.5cm]{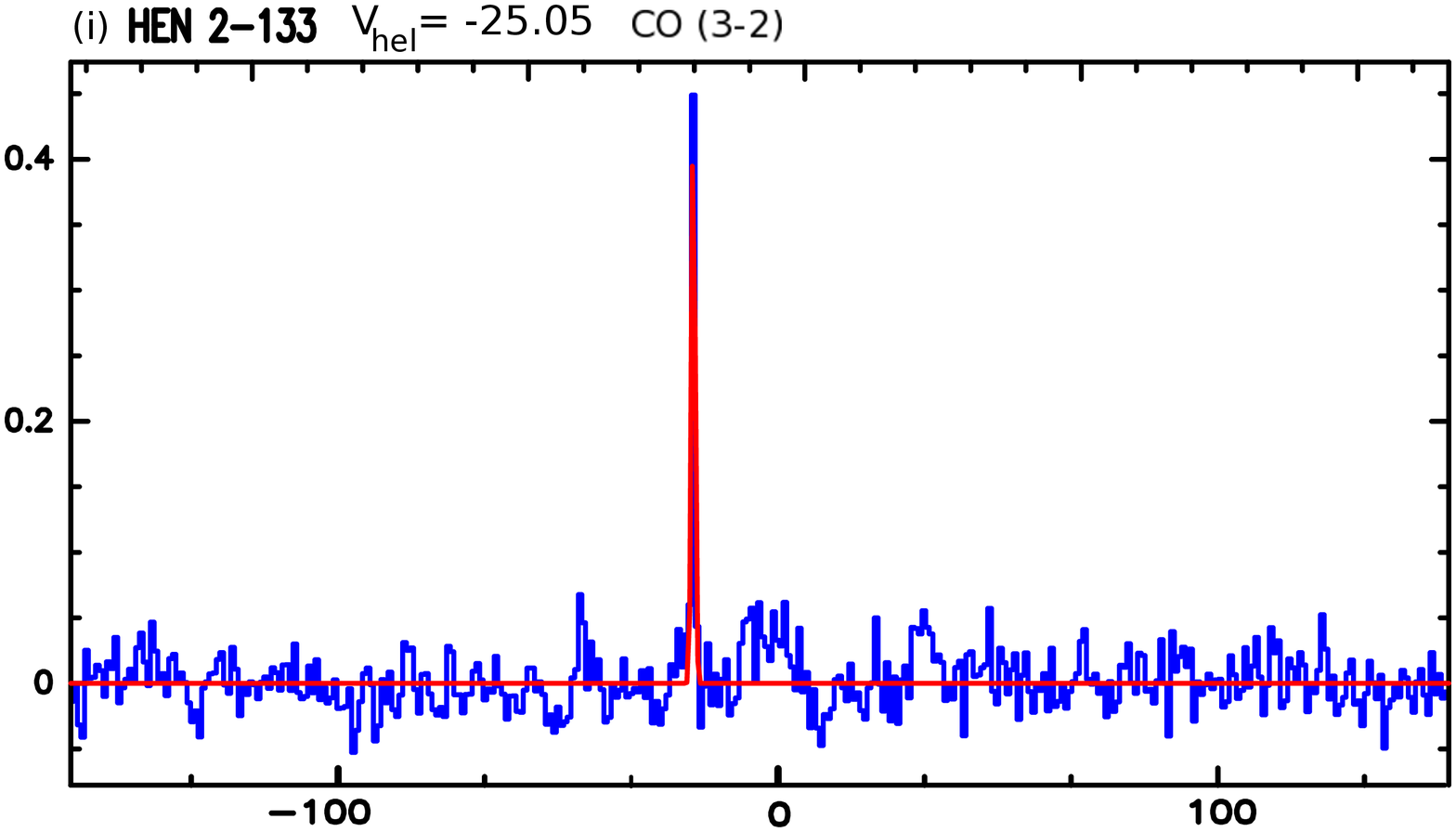}}
\caption[]{APEX CO (3-2) line detections in pPNe and PNe.}
\label{Fig12}
\end{figure*}

\begin{figure*}
\vspace{2cm}
\centering
\hbox{
\centering
\includegraphics[width=6cm, height=5.5cm]{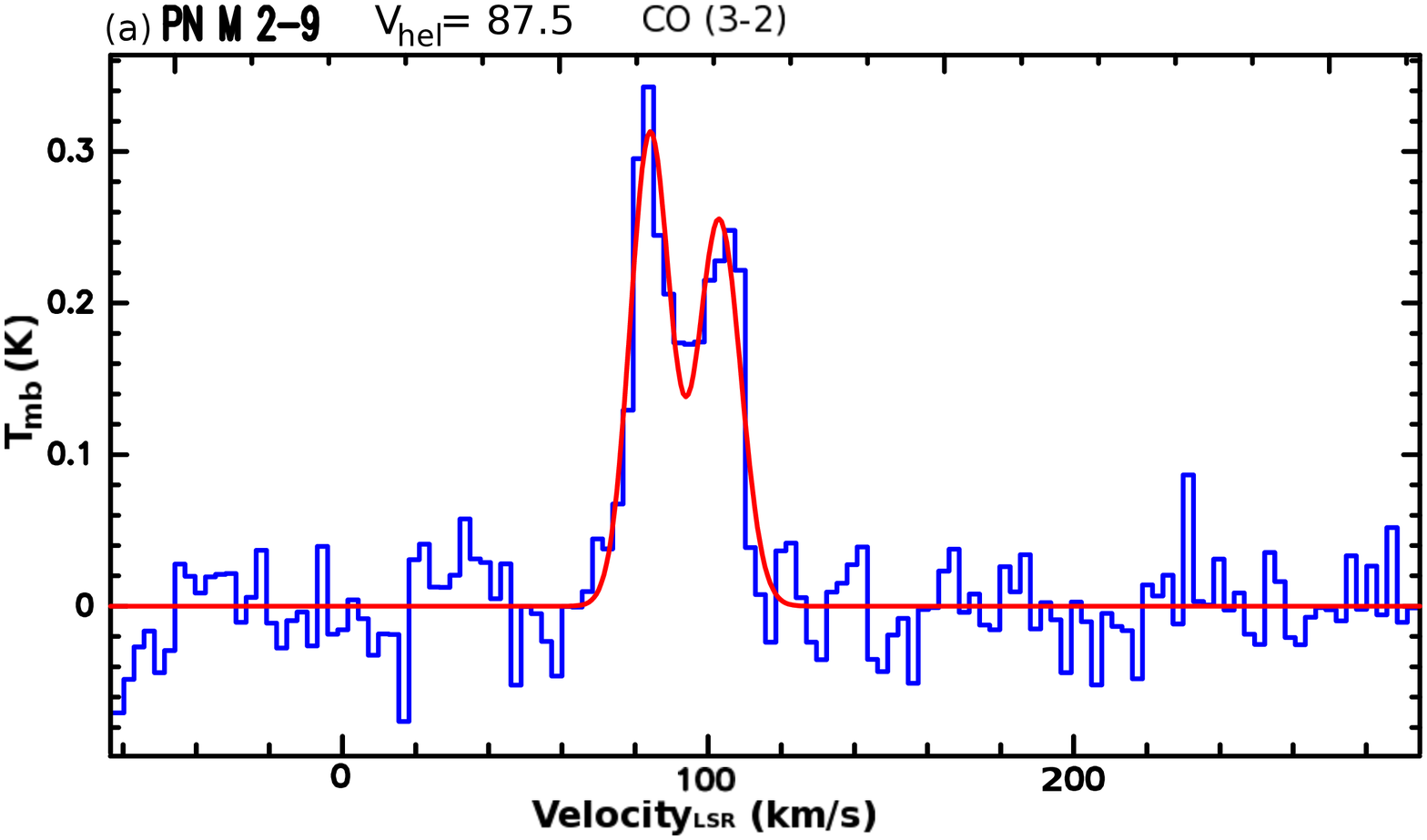}
\includegraphics[width=6cm, height=5.5cm]{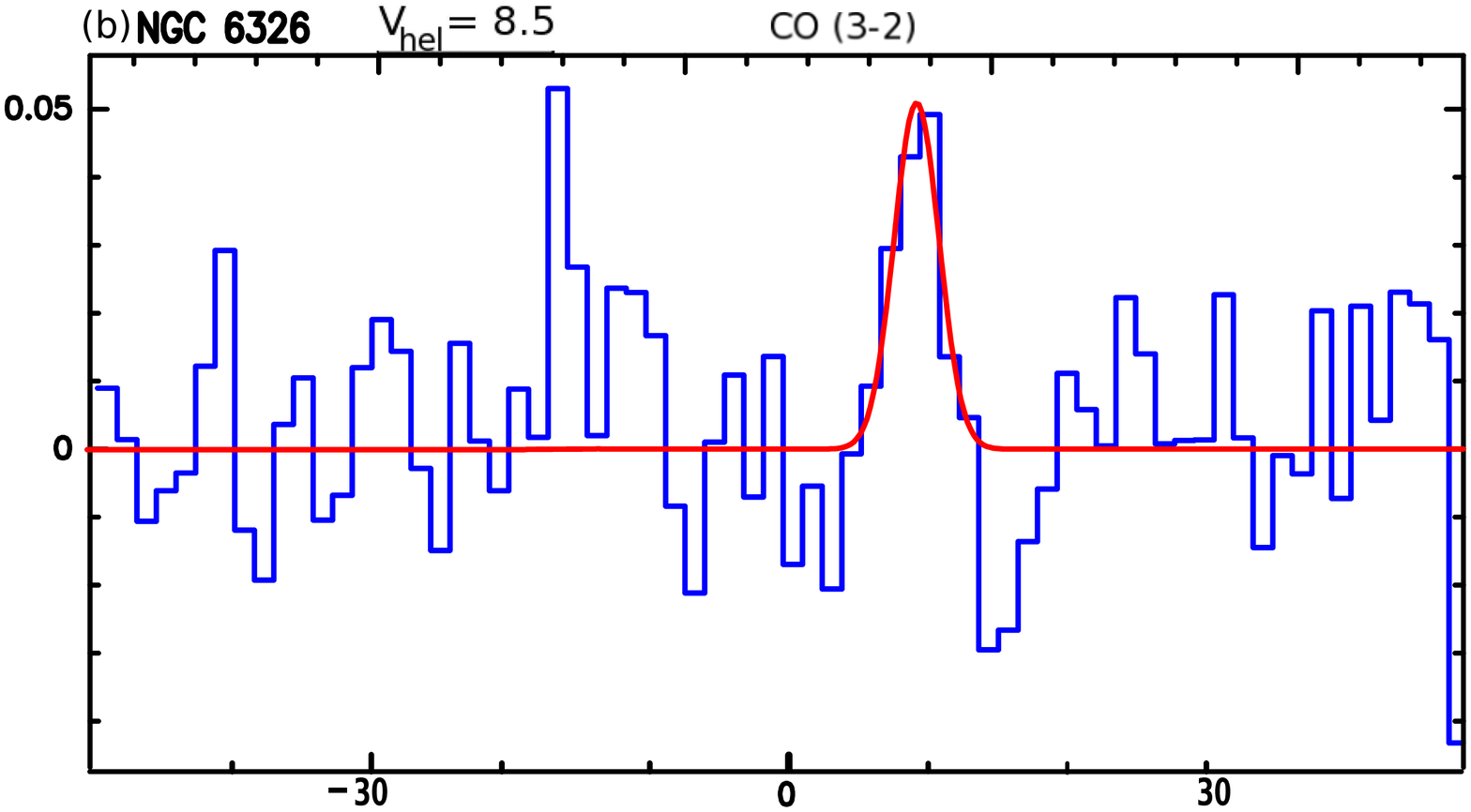}
\includegraphics[width=6cm, height=5.5cm]{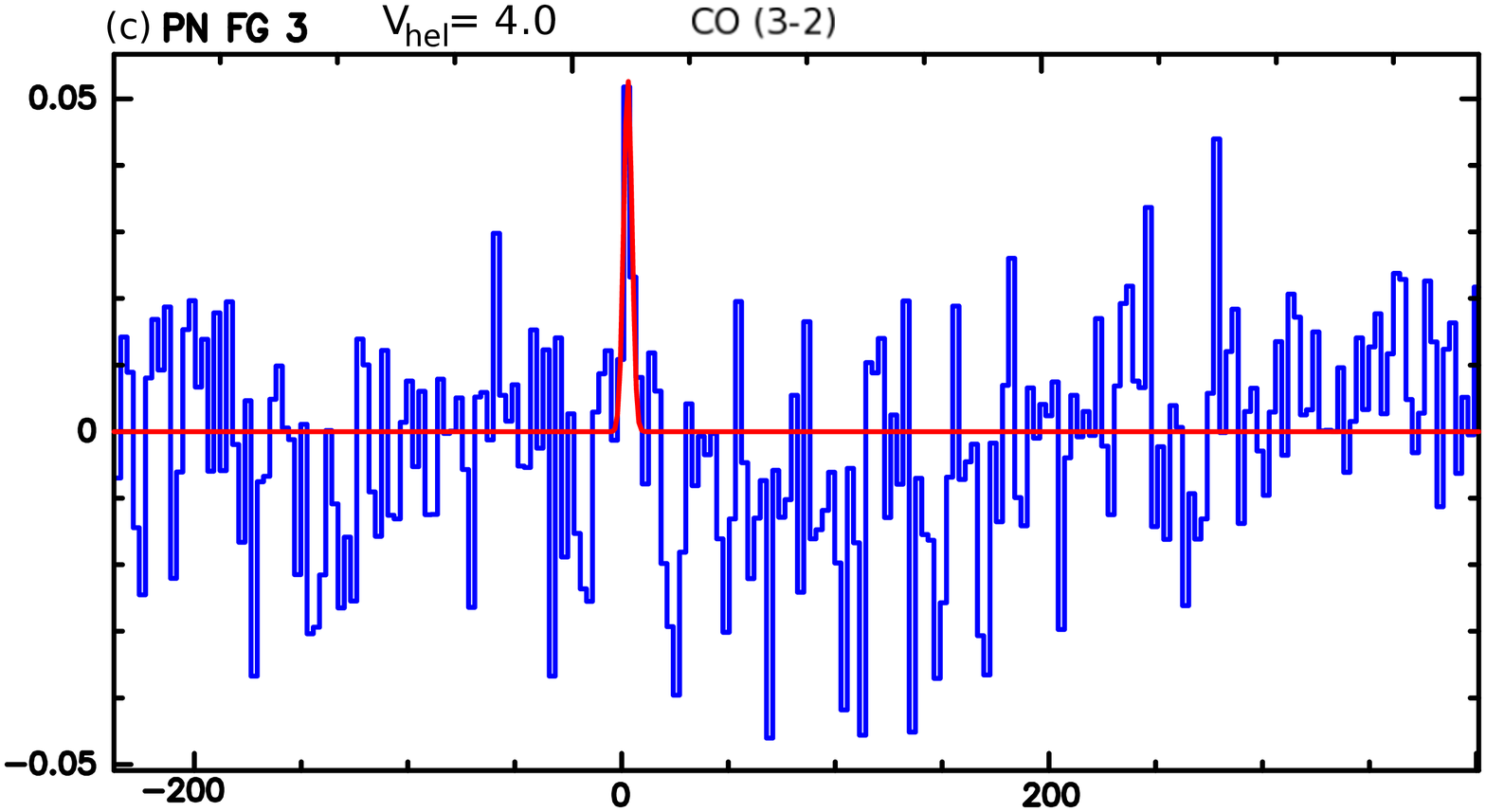}}
\vspace{2cm}
\hbox{
\centering
\vspace{2cm}
\includegraphics[width=6cm, height=5.5cm]{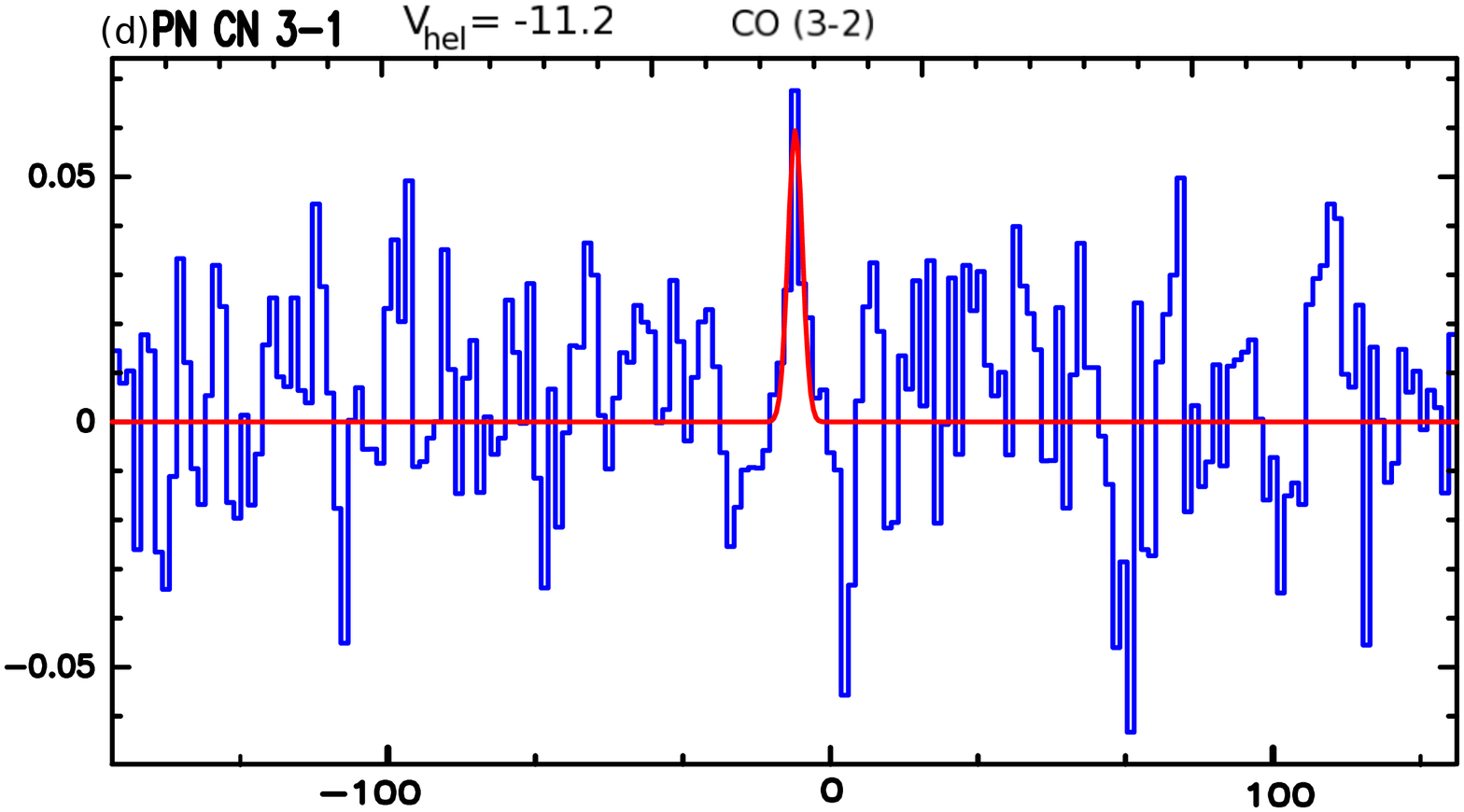}
\includegraphics[width=6cm, height=5.5cm]{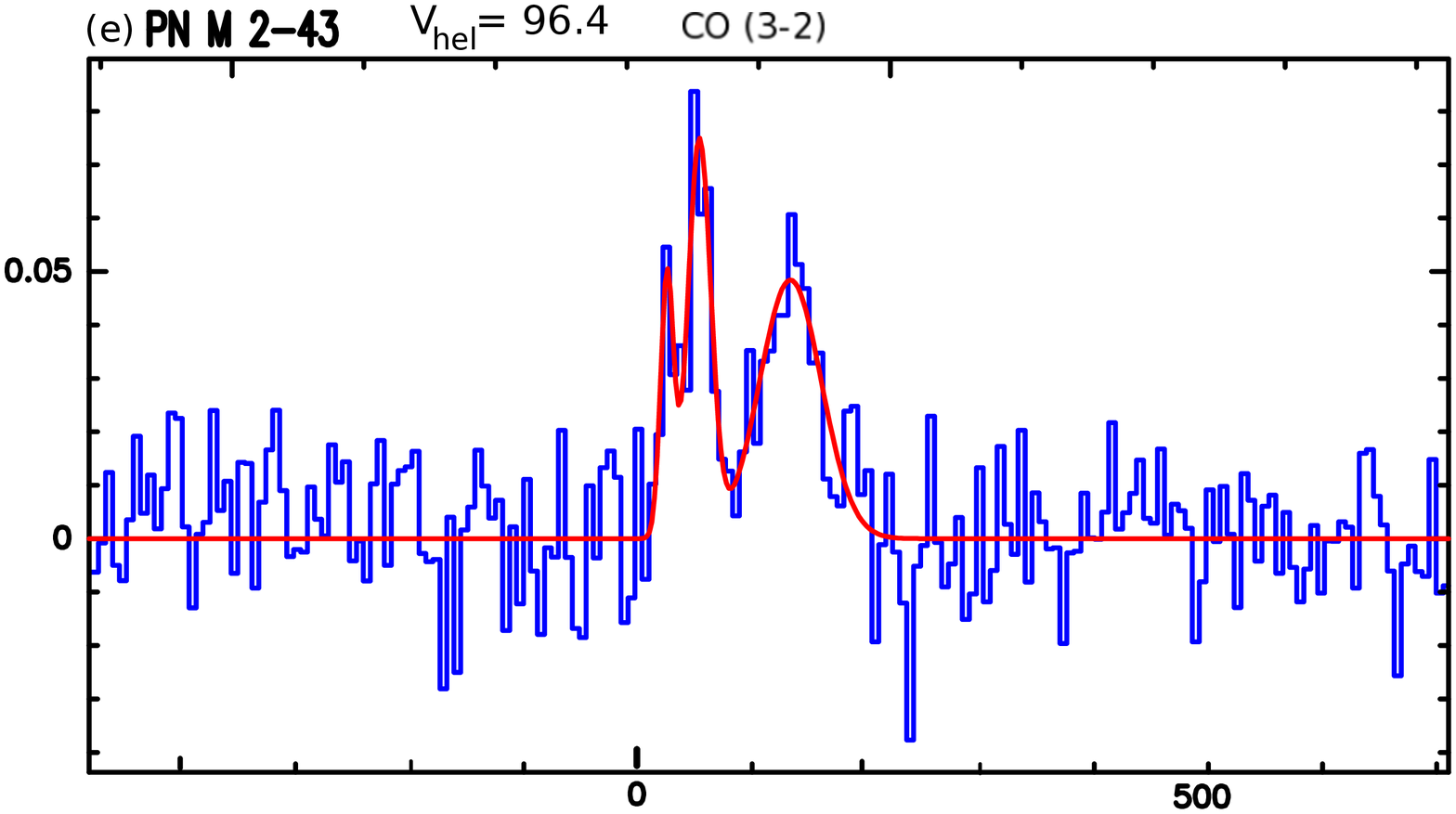}
\includegraphics[width=6cm, height=5.5cm]{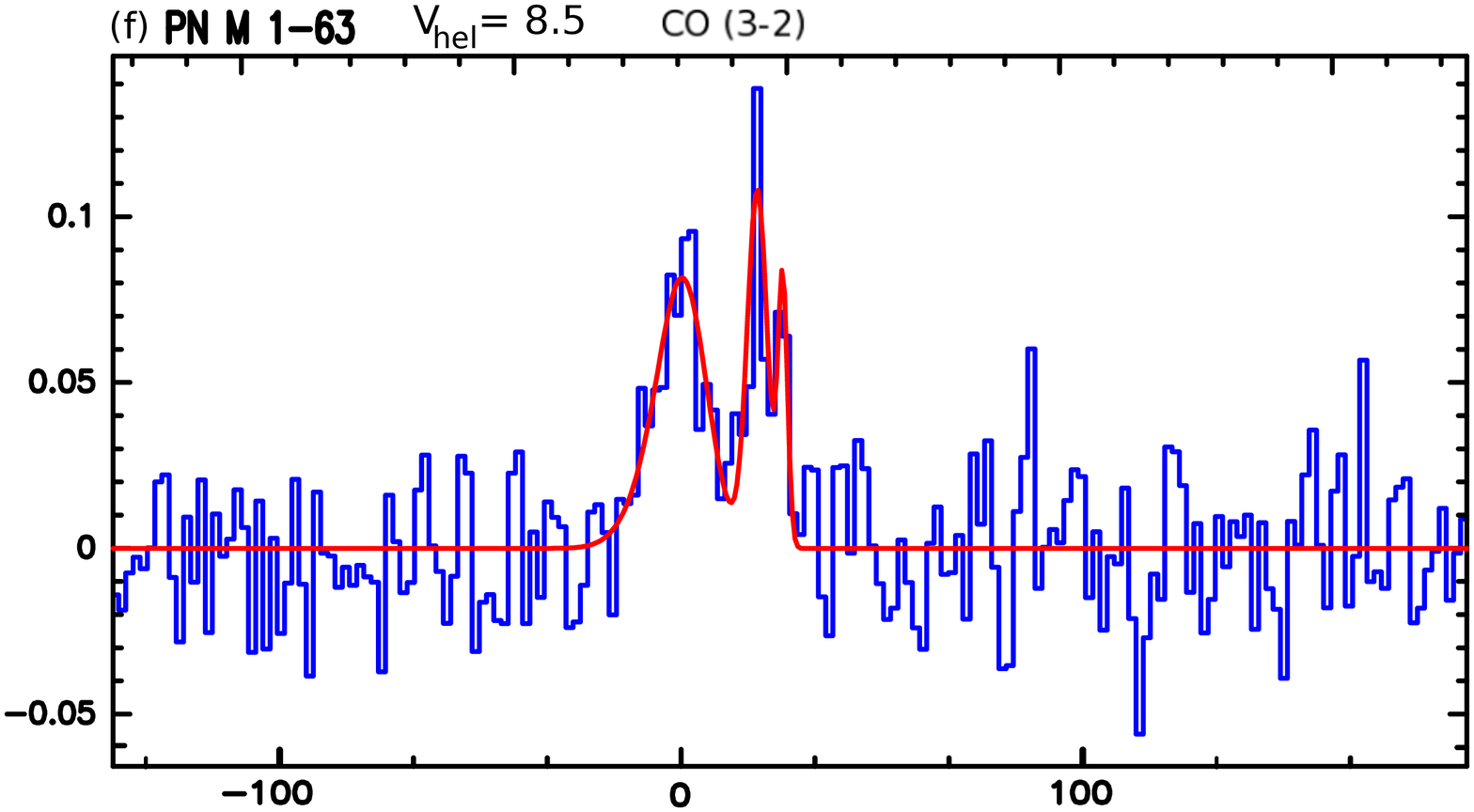}}
\hbox{
\centering
\includegraphics[width=6cm, height=5.5cm]{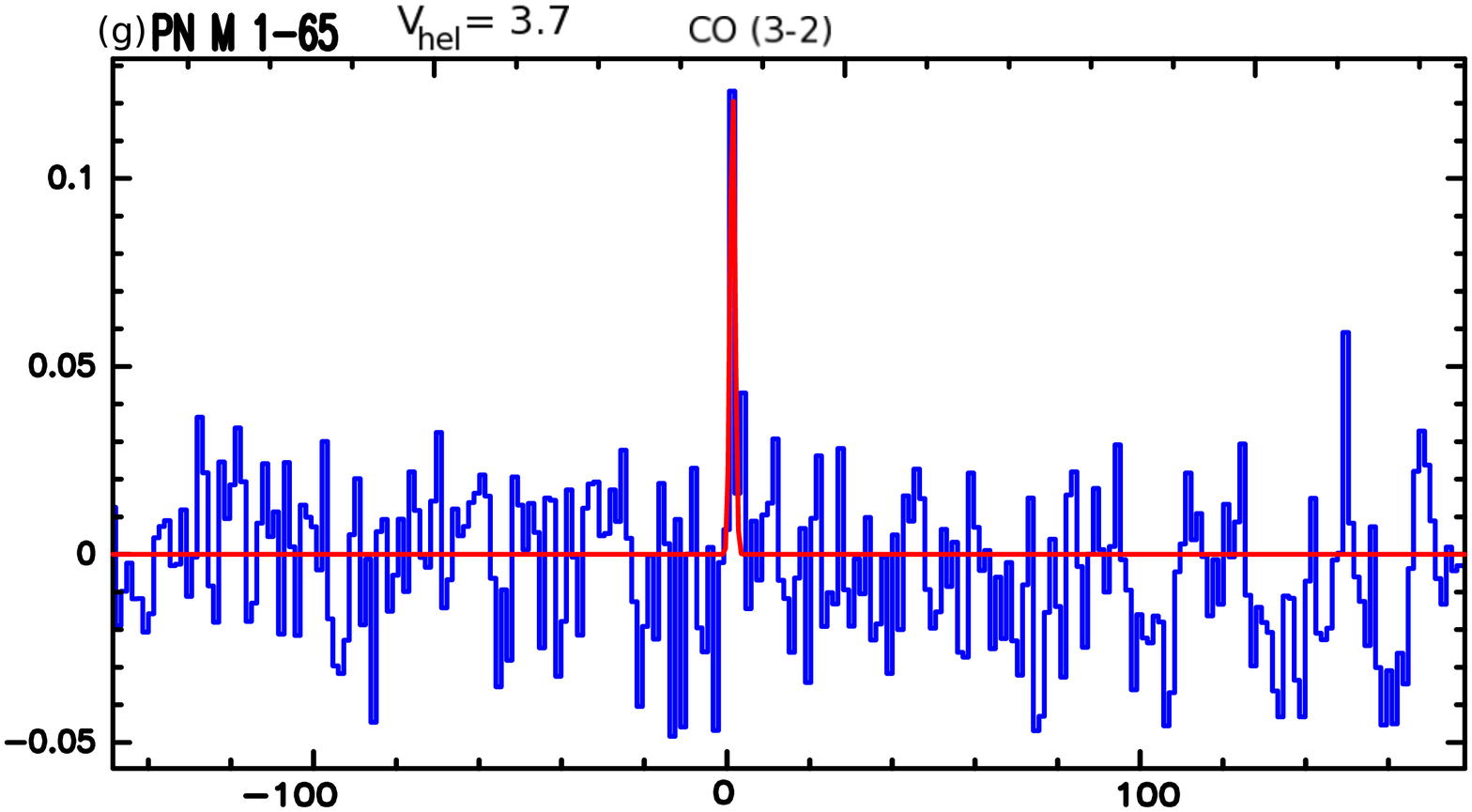}
\includegraphics[width=6cm, height=5.5cm]{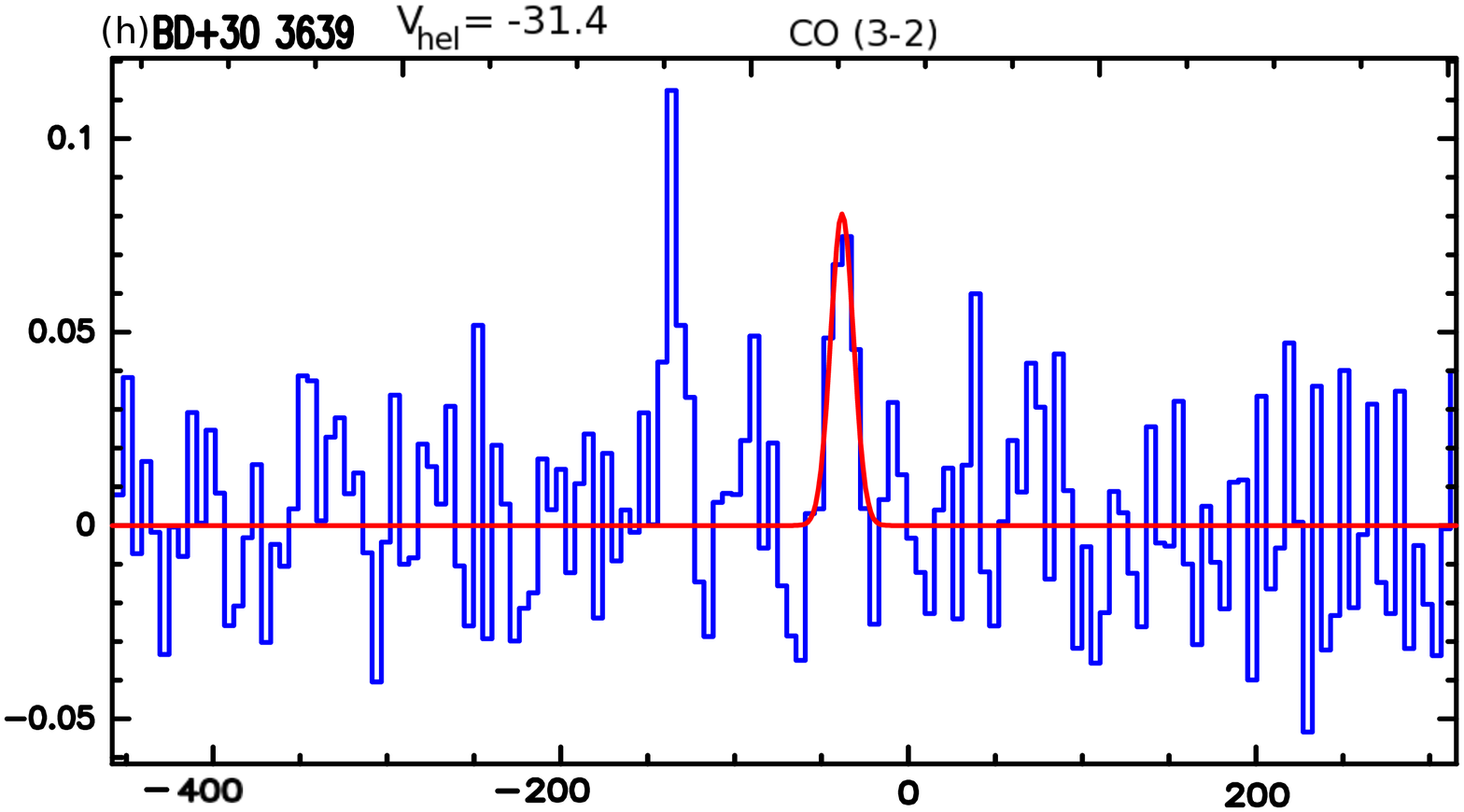}
\includegraphics[width=6cm, height=5.5cm]{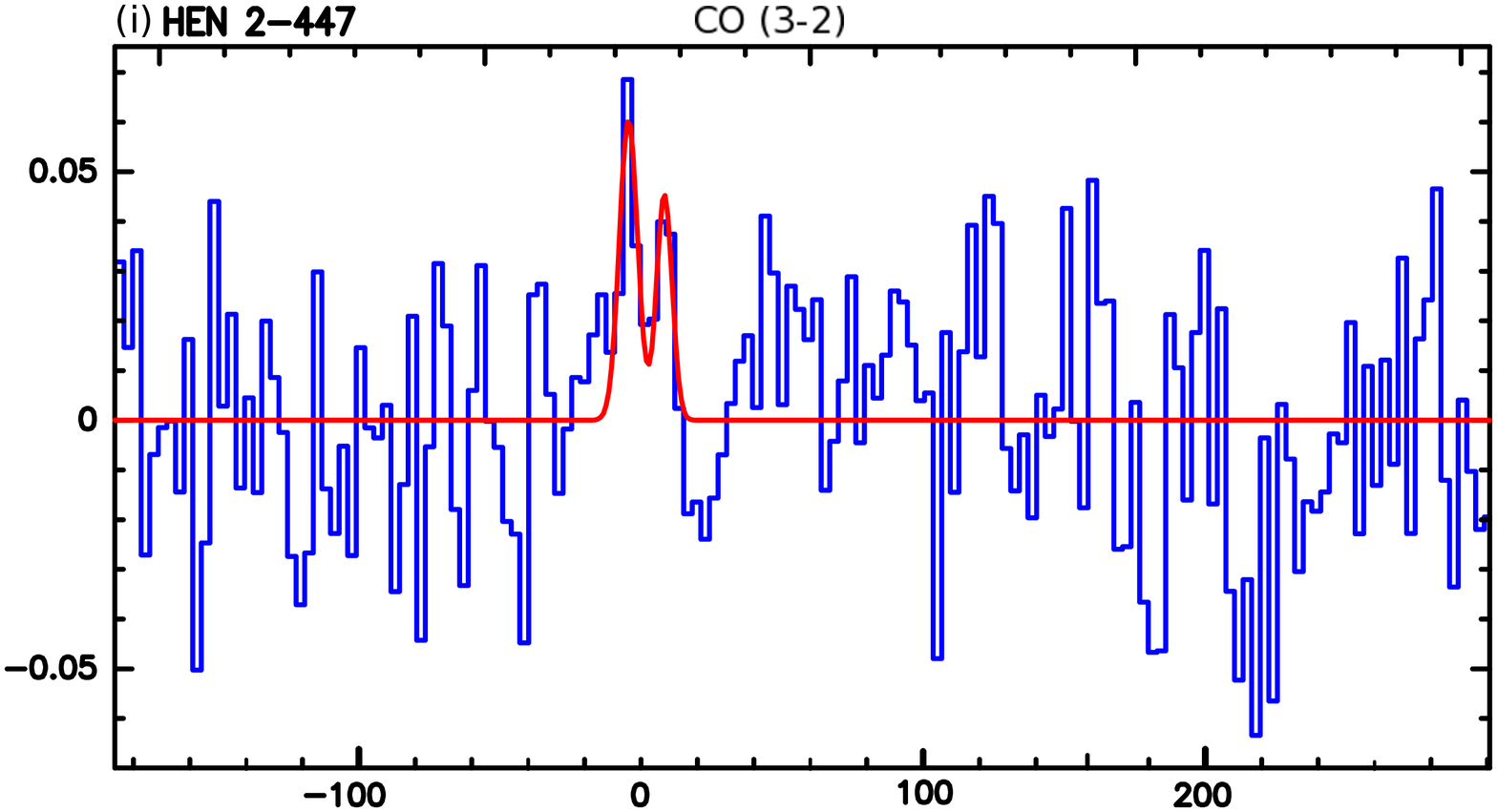}}
\caption[]{APEX CO (3-2) and other line detections in pPNe and PNe.}
\label{Fig13}
\end{figure*}

\begin{figure*}
\vspace{2cm}
\centering
\hbox{
\centering
\includegraphics[width=6cm, height=5.5cm]{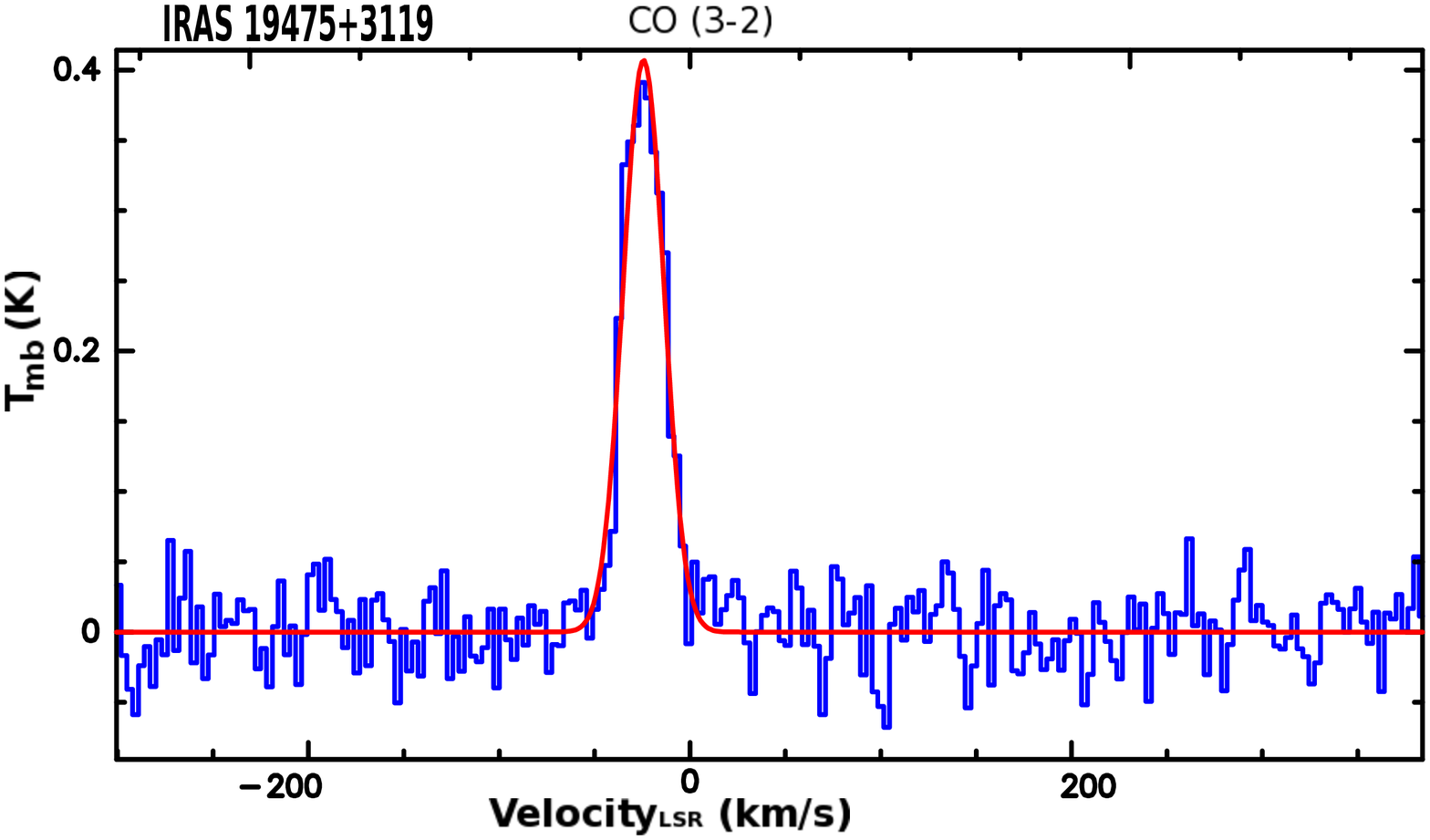}
\includegraphics[width=6cm, height=5.5cm]{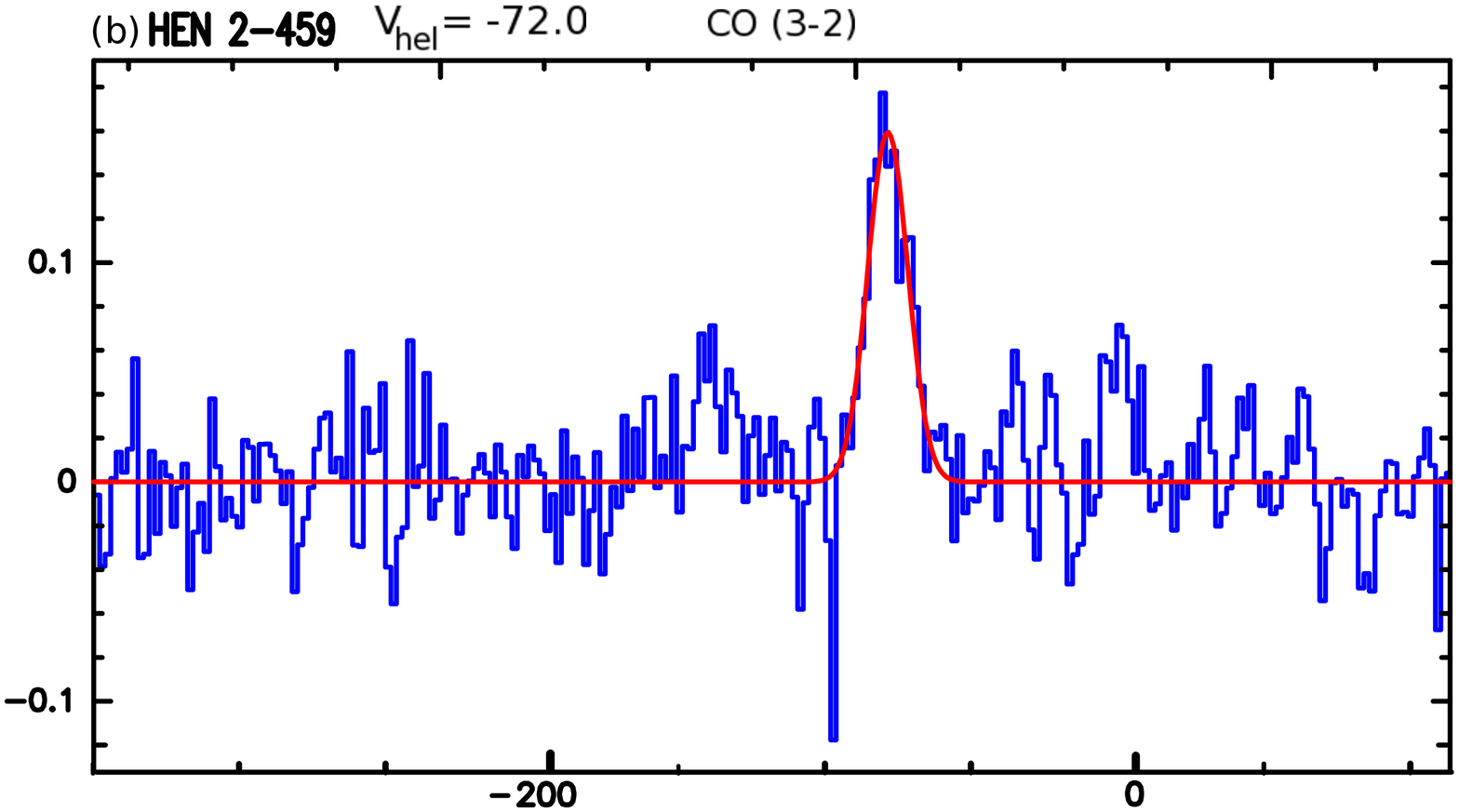}
\includegraphics[width=6cm, height=5.5cm]{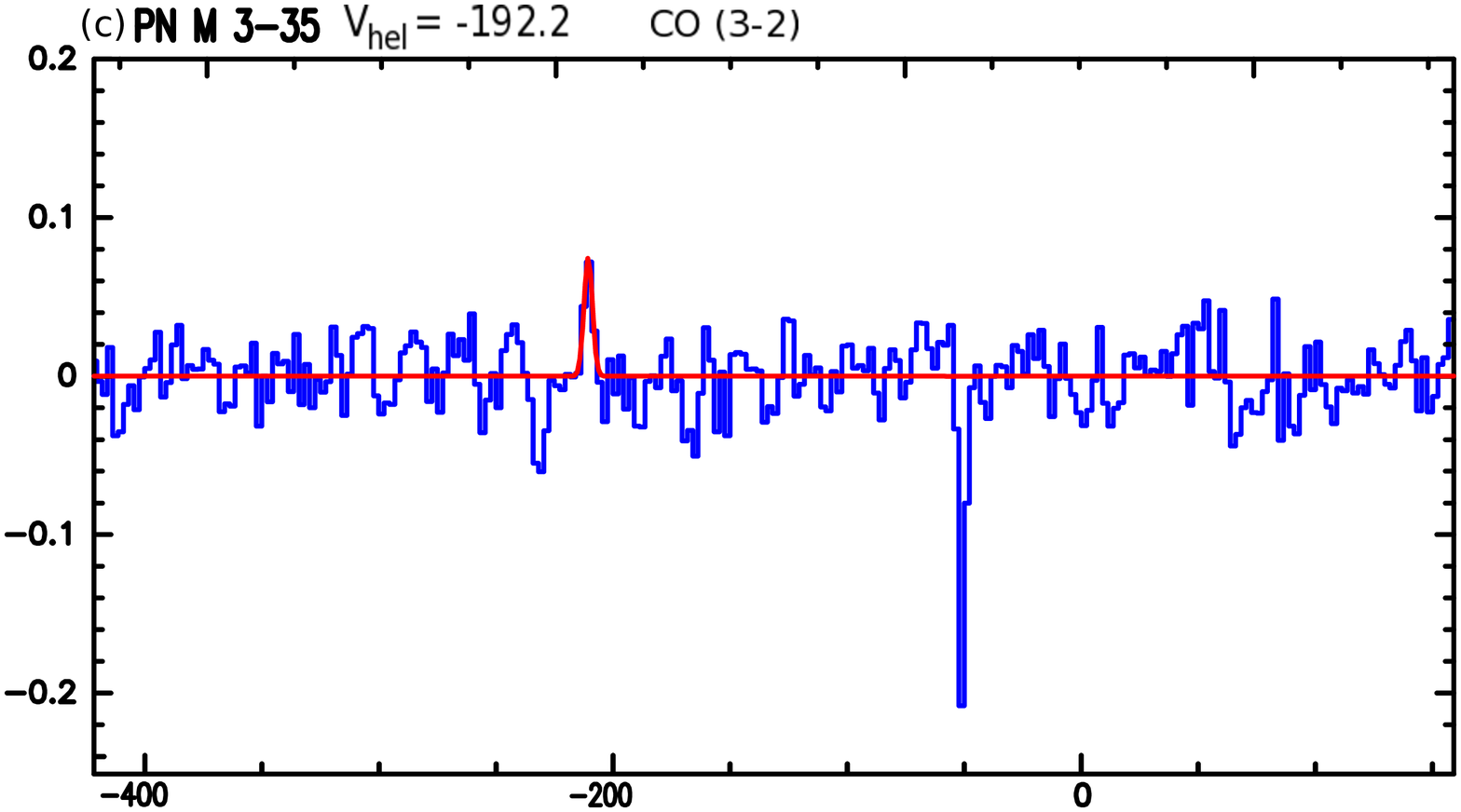}}
\vspace{2cm}
\caption[]{APEX CO (3-2) and other line detections in pPNe and PNe.}
\label{Fig14}
\end{figure*}

In Fig. \ref{CO-LVG} we show the diagnostic plot for the CO (J $= 3-2$) to CO (J $= 2-1$) intensity ratio, assuming 2.73 K as the background temperature and a line width of 50 km s$^{-1}$. The plot shows the results for a column density of 1$\times$10$^{16}$ cm$^{-2}$, but we find that the CO line ratio is not very sensitive to column density within the range 3*10$^{15}$ cm$^{-2}$ to $\sim$ 4*10$^{16}$ cm$^{-2}$. The plotted ratios corresponded to the observed values for BD$+$30 3639 (1.6), NGC 3132 (1.3), IC 4406, and M2-9 (0.7, for the last two). However, for the latter three sources, the measured ratios are only lower limits because of different CO (J $= 2-1$) and CO (J $= 3-2$) beam sizes (the former larger than the latter) and the source size (larger than both beam sizes).  
The CO intensity ratio constrains a lower limit to the H$_2$ volume density, n$_{\rm H_2} >$ 10$^{3}$--10$^{3.7}$ cm$^{-3}$, and kinetic temperature, T$_{\rm kin} >$ 10--35 K. The highest lower limit values are given for IC 4406 and M2-9.  

\subsection{Correlation with H$_2$}
We would like to probe the relationship between the detection of CO and H$_2$.
We compared our sample with the sample analysed by \cite{kastner96}, whose study was made by imaging the molecular hydrogen emission ($\lambda$ =  2.122 $\mu$m) using the near-infrared cryogenic optical bench mounted at Kitt Peak National Observatory, from the National Optical Astronomy Observatories. They observed 60 PNe and detected H$_2$ in 23 of them.
When compared with out sample, 9 objects are in common.
Of these 9 objects, 4 were detected in H$_2$ , and all are also detected in CO as part of our survey. The objects are NGC 3132, IC 4406, PN M2-9, and BD+30 3639. The remaining objects (five) were not detected in H$_2$ and are they detected here in CO either. These PNe are PN M3-2, PN MyCn18, NGC 6778, NGC 6790, and HD 193538.
Although the overlap between PANDORA and the H$_2$ survey of \cite{kastner96} is small, we find a clear correlation between the detection of CO and the detection of H$_2$. 

\subsection{Morphology}
We investigated the morphological classification for the all the PNe that were observed.
The morphology of the sources is included in Tables \ref{observations} and \ref{morphology} for reference. Table \ref{observations} lists all the sources, whilst Table \ref{morphology} lists only those with CO detections. The different morphologies are B (bipolar), E (elliptical), R (round), I (irregular), and S (symmetric) according to the Hong Kong/AAO/Strasbourg H$\alpha$ planetary nebula database \citep[HASH;][]{parker16}. The HASH planetary nebula database is an online
research platform providing free and easy access to the largest and most comprehensive
catalogue of known Galactic PNe and a repository of observational data (imaging and spectroscopy) for these and related astronomical objects. The number of sources with the same morphology are 6 with no morphology, 12 S (symmetric), 7 R (round), 1 I (irregular), 26 E (elliptical), and 41 B (bipolar). 

We find that CO (J $= 3-2$) was detected in the 6 sources that have no morphology classification, in 9 out of the 26 ellipticals, and 9 out of the 41 PNe with bipolar morphology. CO (J $= 3-2$) was not detected in any of the symmetric, round, or irregular PNe. 
This bipolar fraction amongst the PNe with CO emission is a lower limit, since the ellipticals could also be bipolar, depending on the viewing angle \citep{jones12}. We can conclude that CO (J $= 3-2$) is more likely to be found in PNe with bipolar or elliptical morphologies than in symmetric, round, or irregular PNe.

\section{Conclusions}
We present the largest survey of the CO (J $= 3-2$) line ever taken using the Atacama Pathfinder EXperiment (APEX) telescope in a sample of 93 pPNe and PNe. This survey was designed to study the circumstellar shells of evolved stars with the aim to investigate their physical parameters. Of the 93 objects, we detected CO (J $= 3-2$) in 35, and another 14 observations show spurious detections associated with the ISM.  

CO (J $= 3-2$) has been detected in two objects, IRC+10216 and PN M 2-9. The final number of true CO (J $= 3-2$) first detections is 19; 3 pPNe and 16 PNe (including one known to host wide binaries and one known to host a close binary central star).  

Evaluating the success rates of detection for different object types within our sample, CO (J $= 3-2$) was detected in all 4 pPNe (100\% success rate), 15 of the 74 PNe (20\% success rate), one of the 4 wide binaries (25\% success rate), and 1 of the 10 close binaries (10\% success rate). 

Using the CO (J $= 3-2$) line, we estimated the column density and mass of each source. The H$_2$ column density ranges from 1.7x10$^{18}$ to 4.2x10$^{21}$cm$^{-2}$ , and the molecular mass ranges from 2.7x10$^{-4}$ to 1.7x10$^{-1}$M$_{\odot}$

Although the overlap between PANDORA and the H$_2$ survey of \cite{kastner96} is small, we find a clear correlation between the detection of CO and the detection of H$_2$, all the sources that show H$_2$ also present CO (J $= 3-2$) and also vice versa all the sources that do not present H$_2$ also do not present CO (J $= 3-2$).

An interesting and perhaps somewhat unexpected correlation was found with morphology in that CO (J $= 3-2$) will be preferentially found in PN with bipolar or elliptical morphologies (the uncertainty in the final fraction is due to the ambiguity of morphology with viewing angle). 

Data from the literature, CO (J $= 2-1$) for three sources and the RADEX software were used to constrain a lower limit to the H$_2$ volume density, n$_{\rm H_2} >$ 10$^{3}$--10$^{3.7}$ cm$^{-3}$, and kinetic temperature, T$_{\rm kin} >$ 10--35 K.

The results presented here represent an extremely valuable
catalogue for use in selecting the samples for future follow-up observations
with the Atacama Large Millimetre and sub-millimetre
Array (ALMA), particularly in sampling the spatial distribution of
the molecular gas in these objects. 


\section*{Acknowledgements}
We all thank the anonymous referee for all the comments and suggestions. LGR thanks the time allocation committee for all the telescope time awarded at APEX(SHeFI). LGR is co-funded under the Marie Curie Actions of the European Commission (FP7-COFUND). RW is supported by European Research Grant SNDUST. DJ acknowledges the support by the Spanish Ministry of Economy and Competitiveness (MINECO) under the grant AYA2017-83383-P.

\bibliographystyle{aa}
\bibliography{reference} 

\begin{thebibliography}{71}
\expandafter\ifx\csname natexlab\endcsname\relax\def\natexlab#1{#1}\fi

\bibitem[{{Acker} {et~al.}(1992){Acker}, {Cuisinier}, {Stenholm}, \&
  {Terzan}}]{acker92}
{Acker}, A., {Cuisinier}, F., {Stenholm}, B., \& {Terzan}, A. 1992, \aap, 264,
  217

\bibitem[{{Akras} \& {Steffen}(2012)}]{akras12}
{Akras}, S. \& {Steffen}, W. 2012, \mnras, 423, 925

\bibitem[{{Bachiller} {et~al.}(2000){Bachiller}, {Forveille}, {Huggins}, {Cox},
  \& {Maillard}}]{bachiller00}
{Bachiller}, R., {Forveille}, T., {Huggins}, P.~J., {Cox}, P., \& {Maillard},
  J.~P. 2000, \aap, 353, L5

\bibitem[{{Bachiller} {et~al.}(1991){Bachiller}, {Huggins}, {Cox}, \&
  {Forveille}}]{bachiller91}
{Bachiller}, R., {Huggins}, P.~J., {Cox}, P., \& {Forveille}, T. 1991, \aap,
  247, 525

\bibitem[{{Bachiller} {et~al.}(1992){Bachiller}, {Huggins}, {Martin-Pintado},
  \& {Cox}}]{bachiller92}
{Bachiller}, R., {Huggins}, P.~J., {Martin-Pintado}, J., \& {Cox}, P. 1992,
  \aap, 256, 231

\bibitem[{{Bujarrabal} {et~al.}(2001){Bujarrabal}, {Castro-Carrizo}, {Alcolea},
  \& {S{\'a}nchez Contreras}}]{bujarrabal01}
{Bujarrabal}, V., {Castro-Carrizo}, A., {Alcolea}, J., \& {S{\'a}nchez
  Contreras}, C. 2001, \aap, 377, 868

\bibitem[{{Cahn} {et~al.}(1992){Cahn}, {Kaler}, \& {Stanghellini}}]{cahn92}
{Cahn}, J.~H., {Kaler}, J.~B., \& {Stanghellini}, L. 1992, \aaps, 94, 399

\bibitem[{{Castro-Carrizo} {et~al.}(2017){Castro-Carrizo}, {Bujarrabal},
  {Neri}, {Alcolea}, {S{\'a}nchez Contreras}, {Santander-Garc{\'{\i}}a}, \&
  {Nyman}}]{castro17}
{Castro-Carrizo}, A., {Bujarrabal}, V., {Neri}, R., {et~al.} 2017, \aap, 600,
  A4

\bibitem[{{Castro-Carrizo} {et~al.}(2005){Castro-Carrizo}, {Bujarrabal},
  {S{\'a}nchez Contreras}, {Sahai}, \& {Alcolea}}]{castro05}
{Castro-Carrizo}, A., {Bujarrabal}, V., {S{\'a}nchez Contreras}, C., {Sahai},
  R., \& {Alcolea}, J. 2005, \aap, 431, 979

\bibitem[{{Castro-Carrizo} {et~al.}(2012){Castro-Carrizo}, {Neri},
  {Bujarrabal}, {Chesneau}, {Cox}, \& {Bachiller}}]{castro12}
{Castro-Carrizo}, A., {Neri}, R., {Bujarrabal}, V., {et~al.} 2012, \aap, 545,
  A1

\bibitem[{{Castro-Carrizo} {et~al.}(2010){Castro-Carrizo}, {Quintana-Lacaci},
  {Neri}, {Bujarrabal}, {Sch{\"o}ier}, {Winters}, {Olofsson}, {Lindqvist},
  {Alcolea}, {Lucas}, \& {Grewing}}]{castro10}
{Castro-Carrizo}, A., {Quintana-Lacaci}, G., {Neri}, R., {et~al.} 2010, \aap,
  523, A59

\bibitem[{{Ciardullo} {et~al.}(1999){Ciardullo}, {Bond}, {Sipior}, {Fullton},
  {Zhang}, \& {Schaefer}}]{ciardullo99}
{Ciardullo}, R., {Bond}, H.~E., {Sipior}, M.~S., {et~al.} 1999, \aj, 118, 488

\bibitem[{{Cox} {et~al.}(1991){Cox}, {Huggins}, {Bachiller}, \&
  {Forveille}}]{cox91}
{Cox}, P., {Huggins}, P.~J., {Bachiller}, R., \& {Forveille}, T. 1991, \aap,
  250, 533

\bibitem[{{Cox} {et~al.}(2003){Cox}, {Huggins}, {Maillard}, {Muthu},
  {Bachiller}, \& {Forveille}}]{cox03}
{Cox}, P., {Huggins}, P.~J., {Maillard}, J.-P., {et~al.} 2003, \apjl, 586, L87

\bibitem[{{Cox} {et~al.}(1992){Cox}, {Omont}, {Huggins}, {Bachiller}, \&
  {Forveille}}]{cox92}
{Cox}, P., {Omont}, A., {Huggins}, P.~J., {Bachiller}, R., \& {Forveille}, T.
  1992, \aap, 266, 420

\bibitem[{{Dame} {et~al.}(2001){Dame}, {Hartmann}, \& {Thaddeus}}]{dame01}
{Dame}, T.~M., {Hartmann}, D., \& {Thaddeus}, P. 2001, \apj, 547, 792

\bibitem[{{Dekker} {et~al.}(2000){Dekker}, {D'Odorico}, {Kaufer}, {Delabre}, \&
  {Kotzlowski}}]{dekker00}
{Dekker}, H., {D'Odorico}, S., {Kaufer}, A., {Delabre}, B., \& {Kotzlowski}, H.
  2000, in \procspie, Vol. 4008, Optical and IR Telescope Instrumentation and
  Detectors, ed. M.~{Iye} \& A.~F. {Moorwood}, 534--545

\bibitem[{{Dougados} {et~al.}(1992){Dougados}, {Rouan}, \& {Lena}}]{dougados92}
{Dougados}, C., {Rouan}, D., \& {Lena}, P. 1992, \aap, 253, 464

\bibitem[{{Durand} {et~al.}(1998){Durand}, {Acker}, \& {Zijlstra}}]{durand98}
{Durand}, S., {Acker}, A., \& {Zijlstra}, A. 1998, \aaps, 132, 13

\bibitem[{{Freeman} \& {Kastner}(2016)}]{freeman16}
{Freeman}, M.~J. \& {Kastner}, J.~H. 2016, \apjs, 226, 15

\bibitem[{{Freudling} {et~al.}(2013){Freudling}, {Romaniello}, {Bramich},
  {Ballester}, {Forchi}, {Garc{\'{\i}}a-Dabl{\'o}}, {Moehler}, \&
  {Neeser}}]{freud13}
{Freudling}, W., {Romaniello}, M., {Bramich}, D.~M., {et~al.} 2013, \aap, 559,
  A96

\bibitem[{{Frew}(2008)}]{frew08}
{Frew}, D.~J. 2008, PhD thesis, Department of Physics, Macquarie University,
  NSW 2109, Australia

\bibitem[{{Groenewegen} {et~al.}(2012){Groenewegen}, {Barlow}, {Blommaert},
  {Cernicharo}, {Decin}, {Gomez}, {Hargrave}, {Kerschbaum}, {Ladjal}, {Lim},
  {Matsuura}, {Olofsson}, {Sibthorpe}, {Swinyard}, {Ueta}, \&
  {Yates}}]{groen12}
{Groenewegen}, M.~A.~T., {Barlow}, M.~J., {Blommaert}, J.~A.~D.~L., {et~al.}
  2012, \aap, 543, L8

\bibitem[{{Gussie} \& {Taylor}(1995)}]{gussie95}
{Gussie}, G.~T. \& {Taylor}, A.~R. 1995, \mnras, 273, 801

\bibitem[{{Guzm{\'a}n} {et~al.}(2006){Guzm{\'a}n}, {G{\'o}mez}, \&
  {Rodr{\'{\i}}guez}}]{me06}
{Guzm{\'a}n}, L., {G{\'o}mez}, Y., \& {Rodr{\'{\i}}guez}, L.~F. 2006, \rmxaa,
  42, 127

\bibitem[{{Hacar} {et~al.}(2016){Hacar}, {Alves}, {Burkert}, \&
  {Goldsmith}}]{hacar16}
{Hacar}, A., {Alves}, J., {Burkert}, A., \& {Goldsmith}, P. 2016, \aap, 591,
  A104

\bibitem[{{Hsu} \& {Lee}(2011)}]{hsu11}
{Hsu}, M.-C. \& {Lee}, C.-F. 2011, \apj, 736, 30

\bibitem[{{Huggins} {et~al.}(1996){Huggins}, {Bachiller}, {Cox}, \&
  {Forveille}}]{huggins96}
{Huggins}, P.~J., {Bachiller}, R., {Cox}, P., \& {Forveille}, T. 1996, \aap,
  315, 284

\bibitem[{Huggins {et~al.}(2005)Huggins, Bachiller, Planesas, Forveille, \&
  Cox}]{huggins05}
Huggins, P.~J., Bachiller, R., Planesas, P., Forveille, T., \& Cox, P. 2005,
  The Astrophysical Journal Supplement Series, 160, 272

\bibitem[{{Huggins} \& {Healy}(1989)}]{huggins89}
{Huggins}, P.~J. \& {Healy}, A.~P. 1989, \apj, 346, 201

\bibitem[{{Huggins} {et~al.}(2004){Huggins}, {Muthu}, {Bachiller}, {Forveille},
  \& {Cox}}]{huggins04}
{Huggins}, P.~J., {Muthu}, C., {Bachiller}, R., {Forveille}, T., \& {Cox}, P.
  2004, \aap, 414, 581

\bibitem[{{Jones} \& {Boffin}(2017)}]{jones17}
{Jones}, D. \& {Boffin}, H.~M.~J. 2017, Nature Astronomy, 1, 0117

\bibitem[{{Jones} {et~al.}(2015){Jones}, {Boffin}, {Rodr{\'{\i}}guez-Gil},
  {Wesson}, {Corradi}, {Miszalski}, \& {Mohamed}}]{jones15}
{Jones}, D., {Boffin}, H.~M.~J., {Rodr{\'{\i}}guez-Gil}, P., {et~al.} 2015,
  \aap, 580, A19

\bibitem[{{Jones} {et~al.}(2012){Jones}, {Mitchell}, {Lloyd}, {Pollacco},
  {O'Brien}, {Meaburn}, \& {Vaytet}}]{jones12}
{Jones}, D., {Mitchell}, D.~L., {Lloyd}, M., {et~al.} 2012, \mnras, 420, 2271

\bibitem[{{Kastner} {et~al.}(1996){Kastner}, {Weintraub}, {Gatley}, {Merrill},
  \& {Probst}}]{kastner96}
{Kastner}, J.~H., {Weintraub}, D.~A., {Gatley}, I., {Merrill}, K.~M., \&
  {Probst}, R.~G. 1996, \apj, 462, 777

\bibitem[{{Kharchenko} {et~al.}(2007){Kharchenko}, {Scholz}, {Piskunov},
  {R{\"o}ser}, \& {Schilbach}}]{kha07}
{Kharchenko}, N.~V., {Scholz}, R.-D., {Piskunov}, A.~E., {R{\"o}ser}, S., \&
  {Schilbach}, E. 2007, Astronomische Nachrichten, 328, 889

\bibitem[{{Knapp} {et~al.}(1998){Knapp}, {Young}, {Lee}, \&
  {Jorissen}}]{knapp98}
{Knapp}, G.~R., {Young}, K., {Lee}, E., \& {Jorissen}, A. 1998, \apjs, 117, 209

\bibitem[{{Leitner} \& {Kravtsov}(2011)}]{leitner11}
{Leitner}, S.~N. \& {Kravtsov}, A.~V. 2011, \apj, 734, 48

\bibitem[{{Likkel} {et~al.}(1987){Likkel}, {Morris}, {Omont}, \&
  {Forveille}}]{likkel87}
{Likkel}, L., {Morris}, M., {Omont}, A., \& {Forveille}, T. 1987, \aap, 173,
  L11

\bibitem[{{L{\'o}pez}(2003)}]{lopez03}
{L{\'o}pez}, J.~A. 2003, in IAU Symposium, Vol. 209, Planetary Nebulae: Their
  Evolution and Role in the Universe, ed. S.~{Kwok}, M.~{Dopita}, \&
  R.~{Sutherland}, 483

\bibitem[{{Loup} {et~al.}(1990){Loup}, {Forveille}, {Omont}, \&
  {Nyman}}]{loup90}
{Loup}, C., {Forveille}, T., {Omont}, A., \& {Nyman}, L.~A. 1990, \aap, 227,
  L29

\bibitem[{{Miszalski} {et~al.}(2013){Miszalski}, {Boffin}, {Jones}, {Karakas},
  {K{\"o}ppen}, {Tyndall}, {Mohamed}, {Rodr{\'{\i}}guez-Gil}, \&
  {Santander-Garc{\'{\i}}a}}]{miza13}
{Miszalski}, B., {Boffin}, H.~M.~J., {Jones}, D., {et~al.} 2013, \mnras, 436,
  3068

\bibitem[{{Miszalski} {et~al.}(2011){Miszalski}, {Jones},
  {Rodr{\'{\i}}guez-Gil}, {Boffin}, {Corradi}, \&
  {Santander-Garc{\'{\i}}a}}]{miza11}
{Miszalski}, B., {Jones}, D., {Rodr{\'{\i}}guez-Gil}, P., {et~al.} 2011, \aap,
  531, A158

\bibitem[{{Olofsson} \& {Nyman}(1999)}]{olof99}
{Olofsson}, H. \& {Nyman}, L.-{\AA}. 1999, \aap, 347, 194

\bibitem[{{Olofsson} {et~al.}(2016){Olofsson}, {Vlemmings}, {Maercker},
  {Humphreys}, {Lindqvist}, {Nyman}, \& {Ramstedt}}]{olof16}
{Olofsson}, H., {Vlemmings}, W., {Maercker}, M., {et~al.} 2016, in Journal of
  Physics Conference Series, Vol. 728, Journal of Physics Conference Series,
  042005

\bibitem[{{Olofsson} {et~al.}(2017){Olofsson}, {Vlemmings}, {Bergman},
  {Humphreys}, {Lindqvist}, {Maercker}, {Nyman}, {Ramstedt}, \&
  {Tafoya}}]{olof17}
{Olofsson}, H., {Vlemmings}, W.~H.~T., {Bergman}, P., {et~al.} 2017, \aap, 603,
  L2

\bibitem[{{Olofsson} {et~al.}(2015){Olofsson}, {Vlemmings}, {Maercker},
  {Humphreys}, {Lindqvist}, {Nyman}, \& {Ramstedt}}]{olof15}
{Olofsson}, H., {Vlemmings}, W.~H.~T., {Maercker}, M., {et~al.} 2015, \aap,
  576, L15

\bibitem[{{Oudmaijer} {et~al.}(1995){Oudmaijer}, {Waters}, {van der Veen}, \&
  {Geballe}}]{oudmaijer95}
{Oudmaijer}, R.~D., {Waters}, L.~B.~F.~M., {van der Veen}, W.~E.~C.~J., \&
  {Geballe}, T.~R. 1995, \aap, 299, 69

\bibitem[{{Parker} {et~al.}(2016){Parker}, {Boji{\v c}i{\'c}}, \&
  {Frew}}]{parker16}
{Parker}, Q.~A., {Boji{\v c}i{\'c}}, I.~S., \& {Frew}, D.~J. 2016, in Journal
  of Physics Conference Series, Vol. 728, Journal of Physics Conference Series,
  032008

\bibitem[{{Patel} {et~al.}(2010){Patel}, {Young}, {Wilson}, {Thaddeus},
  {Menten}, {McCarthy}, {Bruenken}, {He}, {Trung}, {Reid}, {Keto}, \&
  {Gottlieb}}]{patel2010}
{Patel}, N.~A., {Young}, K.~H., {Wilson}, R.~W., {et~al.} 2010, in Bulletin of
  the American Astronomical Society, Vol.~42, American Astronomical Society
  Meeting Abstracts \#215, 541

\bibitem[{{Phillips} {et~al.}(1992){Phillips}, {Williams}, {Mampaso}, \&
  {Ukita}}]{phillips92}
{Phillips}, J.~P., {Williams}, P.~G., {Mampaso}, A., \& {Ukita}, N. 1992,
  \apss, 188, 171

\bibitem[{Quintana-Lacaci {et~al.}(2016)Quintana-Lacaci, Cernicharo, Agundez,
  Prieto, Castro-Carrizo, Marcelino, Cabezas, Pena, Alonso, Zuniga, Requena,
  Bastida, Kalugina, Lique, \& GuÃ©lin}]{quintana16}
Quintana-Lacaci, G., Cernicharo, J., Agundez, M., {et~al.} 2016, The
  Astrophysical Journal, 818, 192

\bibitem[{{Robinson} {et~al.}(1992){Robinson}, {Smith}, \&
  {Hyland}}]{robinson92}
{Robinson}, G., {Smith}, R.~G., \& {Hyland}, A.~R. 1992, \mnras, 256, 437

\bibitem[{{Sahai} {et~al.}(2000){Sahai}, {Bujarrabal}, {Castro-Carrizo}, \&
  {Zijlstra}}]{Sahai2000}
{Sahai}, R., {Bujarrabal}, V., {Castro-Carrizo}, A., \& {Zijlstra}, A. 2000,
  \aap, 360, L9

\bibitem[{{Sahai} {et~al.}(2011){Sahai}, {Morris}, \& {Villar}}]{sahai11}
{Sahai}, R., {Morris}, M.~R., \& {Villar}, G.~G. 2011, \aj, 141, 134

\bibitem[{{Sahai} {et~al.}(2007){Sahai}, {S{\'a}nchez Contreras}, {Morris}, \&
  {Claussen}}]{sahai07}
{Sahai}, R., {S{\'a}nchez Contreras}, C., {Morris}, M., \& {Claussen}, M. 2007,
  \apj, 658, 410

\bibitem[{{Sahai} \& {Trauger}(1998)}]{sahai98}
{Sahai}, R. \& {Trauger}, J.~T. 1998, \aj, 116, 1357

\bibitem[{{Sahai} {et~al.}(1990){Sahai}, {Wootten}, \& {Clegg}}]{sahai90}
{Sahai}, R., {Wootten}, A., \& {Clegg}, R.~E.~S. 1990, \aap, 234, L1

\bibitem[{{Sahai} {et~al.}(1991){Sahai}, {Wootten}, {Schwarz}, \&
  {Clegg}}]{sahai91}
{Sahai}, R., {Wootten}, A., {Schwarz}, H.~E., \& {Clegg}, R.~E.~S. 1991, \aap,
  251, 560

\bibitem[{{S{\'a}nchez Contreras} {et~al.}(2006){S{\'a}nchez Contreras},
  {Bujarrabal}, {Castro-Carrizo}, {Alcolea}, \& {Sargent}}]{sanchez06}
{S{\'a}nchez Contreras}, C., {Bujarrabal}, V., {Castro-Carrizo}, A., {Alcolea},
  J., \& {Sargent}, A. 2006, \apj, 643, 945

\bibitem[{{Santander-Garc{\'{\i}}a} {et~al.}(2008){Santander-Garc{\'{\i}}a},
  {Corradi}, {Mampaso}, {Morisset}, {Munari}, {Schirmer}, {Balick}, \&
  {Livio}}]{santander08}
{Santander-Garc{\'{\i}}a}, M., {Corradi}, R.~L.~M., {Mampaso}, A., {et~al.}
  2008, \aap, 485, 117

\bibitem[{{Shupe} {et~al.}(1998){Shupe}, {Larkin}, {Knop}, {Armus}, {Matthews},
  \& {Soifer}}]{shupe98}
{Shupe}, D.~L., {Larkin}, J.~E., {Knop}, R.~A., {et~al.} 1998, \apj, 498, 267

\bibitem[{{Smith} {et~al.}(2015){Smith}, {Zijlstra}, \& {Fuller}}]{smith15}
{Smith}, C.~L., {Zijlstra}, A.~A., \& {Fuller}, G.~A. 2015, \mnras, 454, 177

\bibitem[{{Tajitsu} \& {Tamura}(1998)}]{tajitsu98}
{Tajitsu}, A. \& {Tamura}, S. 1998, \aj, 115, 1989

\bibitem[{{Trams} {et~al.}(1990){Trams}, {Lamers}, {van der Veen}, {Waelkens},
  \& {Waters}}]{trams90}
{Trams}, N.~R., {Lamers}, H.~J.~G.~L.~M., {van der Veen}, W.~E.~C.~J.,
  {Waelkens}, C., \& {Waters}, L.~B.~F.~M. 1990, \aap, 233, 153

\bibitem[{{van der Tak} {et~al.}(2007){van der Tak}, {Black}, {Sch{\"o}ier},
  {Jansen}, \& {van Dishoeck}}]{Tak07}
{van der Tak}, F.~F.~S., {Black}, J.~H., {Sch{\"o}ier}, F.~L., {Jansen}, D.~J.,
  \& {van Dishoeck}, E.~F. 2007, \aap, 468, 627

\bibitem[{{Wallace} \& {Clayton}(2014)}]{wallace07}
{Wallace}, P.~T. \& {Clayton}, C.~A. 2014, {RV: Radial Components of Observer's
  Velocity}, Astrophysics Source Code Library

\bibitem[{{Wesson}(2016)}]{wesson16}
{Wesson}, R. 2016, \mnras, 456, 3774

\bibitem[{{Williams} \& {White}(1992)}]{williams92}
{Williams}, P.~G. \& {White}, G.~J. 1992, \aap, 266, 365

\bibitem[{{Zhang} {et~al.}(2000){Zhang}, {Sun}, \& {Ping}}]{zhang00}
{Zhang}, H.-y., {Sun}, J., \& {Ping}, J.-s. 2000, \caa, 24, 309

\bibitem[{{Zuckerman} {et~al.}(1990){Zuckerman}, {Kastner}, {Balick}, \&
  {Gatley}}]{zuckerman90}
{Zuckerman}, B., {Kastner}, J.~H., {Balick}, B., \& {Gatley}, I. 1990, \apjl,
  356, L59

\end{thebibliography}

\clearpage
\onecolumn
\begin{longtable}{llllccl}
\caption{APEX CO (3-2) observations. Column 1 contains the name of each object, columns 2 and 3 are the RA and DEC, respectively, the Column 4 shows the rms of the observations for each target in mK. Column 5 specifies whether CO (J $= 3-2$) was detected, and Column 6 refers first to the to the morphology of the PN where B (bipolar), E (elliptical), R (round), I (irregular), and S (symmetric) are according to the Hong Kong/AAO/Strasbourg H$\alpha$ planetary nebula database \citep[HASH;][]{parker16}, and then it refers to the figure in which the final averaged spectrum is shown. All these figures are shown in the appendix.}
\label{observations}\\
\hline
 RA   & Dec  & PN G & Name & rms (mK) & CO (J $= 3-2$) & Notes\\
\hline
 05 55 06.7 &  $-$22 54 02 &  228.2-22.1 &  LoTr 1 &  19.70    &   -- & R - Figure \ref{Fig1}(a)  \\ 
   06 35 45.1 &  $-$00 05 37 & 226.7+05.6 & PN M 1-6 &    17.90  &  ISM & B - Figure \ref{Fig1}(b) \\ 
  06 54 13.4 &  $-$10 45 38 &   007.6+06.9 &    PN PM 1-23 & 16.72    &   --&E - Figure \ref{Fig1}(c)  \\ 
 07 02 46.8 &  $-$13 42 33 &  226.4-03.7 &    PN PB 1 &   16.27   &   --&  E - Figure \ref{Fig1}(d)  \\ 
   07 11 16.6 &  $-$19 51 02 &   232.8-04.7 &    PN M 1-11 &26.30    & yes & E - Figure \ref{Fig1}(e) \\ 
   07 14 49.9 &  $-$27 50 23 & 240.3-07.6 &   PN M 3-2 &  14.58   & --&  B - Figure \ref{Fig1}(f) \\ 
  07 19 21.4 &  $-$21 43 55 &  235.3-03.9 &   PN M 1-12 &   16.64    &  -- & B - Figure \ref{Fig1}(g) \\ 
   07 27 56.5 &  $-$20 13 22 &  234.9-01.4 &   PN M 1-14 &   18.39   &  -- & E - Figure \ref{Fig1}(h)\\ 
  07 47 20.1 &  $-$51 15 03 &   264.4-12.7 & Hen 2-5 &  24.24    &  -- & S - Figure \ref{Fig1}(i) \\ 
    08 20 40.2 &  $-$46 22 58 & 263.0-05.5 &  PN PB 2 &    24.72   &  --& S - Figure \ref{Fig2}(a) \\ 
   09 13 52.8 &  $-$55 28 16 &     275.3-04.7 & Hen 2-21 &  24.80   &  --&S - Figure \ref{Fig2}(b) \\ 
     09 39 53.9 &   $+$11 58 52 &  221.9+42.7 &  Frosty Leo &  22.69   & yes  & -- - Figure \ref{Fig2}(c) \\ 
   09 47 57.4  & $+$13 16 43 & 221.4+45.1 &   IRC$+$10216 & 84.96 & yes & -- - Figure \ref{Fig2}(d)\\
    09 52 59.0 &   $+$13 44 35 & 221.6+46.4 &  EGB 6 &   13.29  & --  & E - Figure \ref{Fig2}(e)\\
  10 03 49.2 &  $-$60 43 48 &  283.8-04.2 &  Hen 2-39 &  19.45    &  ISM  & E - Figure \ref{Fig2}(f)\\
   10 07 01.7 &  $-$40 26 11 &  272.1+12.3 &   NGC 3132 & 26.90  & yes  & B - Figure \ref{Fig2}(g)\\  
    10 07 23.6 &  $-$63 54 30 &  286.0-06.5 &  Hen 2-41 &   27.79   &  --&S - Figure \ref{Fig2}(h) \\ 
  11 17 43.1 &  $-$70 49 32 &  295.3-09.3 &   Hen 2-62 &   23.94   &  --& S - Figure \ref{Fig2}(i) \\
    11 24 01.1 &  $-$52 51 19 & 289.8+07.7 &   Hen 2-63 &   21.11   &   -- & R - Figure \ref{Fig3}(a)  \\ 
   11 31 45.4 &  $-$65 58 13 &  294.9-04.3 &   Hen 2-68 &   23.85   &   -- & S - Figure \ref{Fig3}(b) \\ 
  11 39 11.2 &  $-$68 52 08 &    296.4-06.9 &  Hen 2-71 &  26.11   &   -- & B - Figure \ref{Fig3}(c)  \\ 
    11 40 58.8 &  $-$55 34 25 & 293.0+05.9 &   HD 101584 &   18.66   & yes & -- - Figure \ref{Fig3}(d)\\ 
 11 50 17.7 &  $-$57 10 56 &   294.6+04.7 &  NGC 3918 &  18.59   &    -- & B - Figure \ref{Fig3}(e) \\ 
   13 08 47.3 &  $-$67 38 37 & 304.5-04.8 &  IC 4191 &    19.27    &   ISM  & B - Figure \ref{Fig3}(f) \\ 
   13 39 35.1 &  $-$67 22 51 &  307.5-04.9 &   PN MyCn18 &    30.74   &     -- & B - Figure \ref{Fig3}(g)\\ 
  14 11 52.1 &  $-$51 26 24 & 315.4+09.4 &   V* V852 Cen &   22.74     & yes  & -- - Figure \ref{Fig3}(h) \\ 
  14 22 26.2 &  $-$44 09 04      &   319.6+15.7 &  IC 4406 &   21.08    & yes  & B - Figure \ref{Fig3}(i) \\
   14 59 53.4 &  $-$54 18 07 & 321.0+03.9 &   Hen 2-113  &    15.79    &   yes   & B - Figure \ref{Fig4}(a) \\ 
   15 08 06.5 &  $-$64 55 40 &  316.7-05.8 &   MPA J1508-6455 &   18.99   & ISM & E - Figure \ref{Fig4}(b) \\ 
   15 37 11.1 &  $-$71 54 52 & 315.1-13.0 &  Hen 2-131 &  19.54       &    -- & E - Figure \ref{Fig4}(c)  \\ 
   15 41 58.8 &   $-$56 36 25 & 324.8-01.1 &   Hen 2-133 &   19.00    &   yes  &E - Figure \ref{Fig4}(d) \\ 
   16 02 13.1 &  $-$41 33 36 &  336.9+08.3 &  PN StWr 4-10 &  26.54       &   ISM &S - Figure \ref{Fig4}(e)\\ 
   16 13 28.1 &  $-$34 35 39 & 343.4+11.9 &   PN H 1-1 &    26.85       &    --  &E - Figure \ref{Fig4}(f) \\ 
   16 15 42.3 &  $-$59 54 01 & 326.0-06.5 &   Hen 2-151 &  22.97       &    -- &S - Figure \ref{Fig4}(g)  \\ 
   16 42 33.4 &  $-$38 54 40 & 344.2+04.7 &  Hen 2-178 &   18.56      &    --&E - Figure \ref{Fig4}(h) \\ 
   16 43 53.8 &  $-$18 57 12 & 000.1+17.2 &  PN PC 12    &   19.77    &  ISM & B - Figure \ref{Fig4}(i) \\ 
   16 45 00.2 &  $-$51 12 20 &   335.2-03.6 &  PN HaTr 4 &   21.71      &  ISM &B - Figure \ref{Fig5}(a)\\ 
  16 53 37.1 &  $-$31 40 33 &  351.3+07.6 &   PN H 1-4 &   24.82      &    -- &E - Figure \ref{Fig5}(b) \\ 
   16 54 35.2 &  $-$64 14 28 & 325.8-12.8 &   Hen 2-182 &   21.00  &   ISM  & S - Figure \ref{Fig5}(c) \\ 
  17 03 02.8&  $-$53 55 53 &  334.8-07.4 &   Hen 3-1312 &  19.65      &    -- &B - Figure \ref{Fig5}(d) \\ 
    17 05 37.9 &  $-$10 08 34 &  010.8+18.0 &    PN M 2-9 &  22.02   &  yes& -- - Figure \ref{Fig5}(e) \\ 
   17 11 27.4 &  $-$47 25 01 & 340.9-04.6 &  PN Sa 1-5   &    20.07      & ISM& S - Figure \ref{Fig5}(f)\\ 
    17 16 21.1 &  $-$59 29 23 & 331.3-12.1 &   Hen 3-1357 &  18.97       &    -- &B - Figure \ref{Fig5}(g) \\ 
   17 20 46.3 &  $-$51 45 15 & 338.1-08.3&  NGC 6326 &   18.38    &  yes & B - Figure \ref{Fig5}(h) \\                 
  17 28 57.6 &  $-$19 15 53 &  006.1+08.3 &  PN M 1-20 &   26.07   &   -- &E - Figure \ref{Fig5}(i) \\   
   17 29 02.0 &  $-$15 13 04 &  009.6+10.5 &   PN A66 41 &  20.61    &    -- & B - Figure \ref{Fig6}(a) \\        
   17 34 26.8 &  $-$22 53 19 & 003.8+05.3 &   PN H 2-15 &   18.58    &   -- & B - Figure \ref{Fig6}(b) \\ 
   17 35 41.9 &  $-$40 11 27 & 349.3-04.2 &   PN Lo 16 &    14.27       &    -- &B -  Figure \ref{Fig6}(c) \\     
   17 38 30.3 &  $-$22 08 38 & 004.9+04.9 &   PN M 1-25 &   21.95    &    -- &E - Figure \ref{Fig6}(d) \\         
  17 38 57.1 &  $-$18 17 35 &  008.2+06.8 &  Hen 2-260 &   18.96    &  -- & B - Figure \ref{Fig6}(e)   \\        
    17 43 28.7 &  $-$21 09 51 & 006.3+04.4 &   PN H 2-18 &  24.79    &  -- &B -  Figure \ref{Fig6}(f) \\  
    17 45 31.7 &  $-$20 58 01 & 006.8+04.1 &   PN M 3-15 &  24.36    &  ISM  &B - Figure \ref{Fig6}(g) \\         
    17 54 21.1 &  $-$15 55 52 &012.2+04.9 &   PN HuBi 1 &  17.23     &  -- & R - Figure \ref{Fig6}(h)\\  
  17 54 22.9 &   $+$27 59 58 & 053.3+24.0 &   PN Vy 1-2 & 20.34   &    -- & E - Figure \ref{Fig6}(i) \\ 
  17 56 33.7 &  $-$43 03 19 &348.8-09.0 &   Hen 2-306 &    18.11  &    -- &S - Figure \ref{Fig7}(a)\\    
   17 58 58.8 &  $-$15 32 14& 013.1+04.1 &   PN M 1-33 &   16.32    &   -- & B - Figure \ref{Fig7}(b)  \\ 
 18 00 11.8 &  $-$38 49 52 &    352.9-07.5 &  PN Fg 3 &  17.93    & yes  &B - Figure \ref{Fig7}(c) \\  
  18 17 34.1 &   $+$10 09 03 &  038.2+12.0 &   PN Cn 3-1 &    19.33   &   yes  & E - Figure \ref{Fig7}(d)\\  
  18 20 08.8 &  $-$24 15 05 &   007.8-04.4 &    PN H 1-65 &  16.32    &   -- &E - Figure \ref{Fig7}(e) \\    
   18 21 23.8 &  $-$06 01 55 &   024.1+03.8 &    PN M 2-40 &  14.81   &  -- &B -  Figure \ref{Fig7}(f) \\    
  18 24 44.5 &   $+$02 29 28 &032.1+07.0&   PN PC 19 &    14.88    &  -- & B - Figure \ref{Fig7}(g) \\  
  18 26 40.0 &  $-$02 42 57 & 027.6+04.2 &   PN M 2-43 &   14.93    &   yes  & E - Figure \ref{Fig7}(h)\\        
  18 32 34.7 &  $-$25 07 44&  008.3-07.3 &  NGC 6644 &  16.15    &   -- & B - Figure \ref{Fig7}(i) \\         
     18 33 54.6 &  $-$22 38 41 & 010.7-06.4 & IC 4732 &   15.57    &  -- &E - Figure \ref{Fig8}(a) \\ 
    18 36 32.2 &   $-$19 19 29 &  014.0-05.5 & PN V-V 3-5 &   15.99   &  -- & E - Figure \ref{Fig8}(b) \\    
    18 37 11.1 &  $-$19 02 21 &  014.3-05.5 & PN V-V 3 &  16.56    &    -- &E - Figure \ref{Fig8}(c)  \\        
     18 45 55.1 &   $-$14 27 37 &  019.4-05.3 & PN M 1-61 &  16.55   & -- &R - Figure \ref{Fig8}(d) \\       
  18 49 47.5 &   $+$20 50 39 &     051.4+09.6 & PN Hu 2-1 &  14.48   &-- &B - Figure \ref{Fig8}(e) \\ 
  18 51 30.9 &  $-$13 10 36 &  021.1-05.9 & PN M 1-63 &    23.08    &  yes  &E - Figure \ref{Fig8}(f) \\         
    18 54 20.0 &  $-$08 47 33 &  025.3-04.6 & PN K 4-8 &   15.30   & -- & E - Figure \ref{Fig8}(g) \\    
    18 56 33.6 &   $+$10 52 09 & 043.1+03.8 &  PN M 1-65 &    18.44   & yes & E - Figure \ref{Fig8}(h) \\    
  19 03 10.1 &   $+$14 06 59 &    046.8+03.8 & Sh 2-78 &  13.70    & -- & I - Figure \ref{Fig8}(i) \\   
     19 04 51.5 &   $+$15 47 39 &  048.5+04.2 &  PN K 4-16       &  14.01   &-- & S - Figure \ref{Fig9}(a) \\    
   19 16 28.2 &  $-$09 02 36 &  027.6-09.6 & IC 4846 &   14.91  & -- & B - Figure \ref{Fig9}(b)  \\    
     19 17 05.7 &   $+$25 37 33 & 058.6+06.1 & PN A66 57 &    24.59   &   -- & B - Figure \ref{Fig9}(c) \\ 
     19 18 24.9 &  $-$01 35 47 &  034.5-06.7 & NGC 6778 &   15.35   &   -- & B - Figure \ref{Fig9}(d) \\  
    19 22 56.9 &   $+$01 30 46 &  037.8-06.3 & NGC 6790 &    15.65   & ISM  & B - Figure \ref{Fig9}(e) \\    
     19 32 39.6 &   $+$07 27 52 & 044.3-05.6 & PN K 3-36 &    19.04   & -- & B - Figure \ref{Fig9}(f) \\  
    19 34 33.5 &   $+$05 41 02 & 042.9-06.9 & NGC 6807 &    17.79   & -- &R - Figure \ref{Fig9}(g)  \\    
    19 34 45.2 &   $+$30 30 58 &  064.7+05.0 & BD$+$30 3639 &   32.89   & yes & E - Figure \ref{Fig9}(h)\\    
    19 41 09.3 &   $+$14 56 58 &  051.9-03.8 & PN M 1-73 &   14.04   &   -- &B - Figure \ref{Fig9}(i) \\ 
     19 42 10.4 &   $+$17 05 14 &  053.8-03.0 & PN A66 63 &   14.28   &  -- &B - Figure \ref{Fig10}(a)  \\   
      19 43 59.5 &   $+$17 09 01 & 054.2-03.4 & The Necklace &  14.66   &   -- & B - Figure \ref{Fig10}(b)  \\       
   19 45 22.1 &   $+$21 20 03 &   057.9-01.5 & Hen 2-447 &    24.27   &  yes   &B - Figure \ref{Fig10}(c)\\ 
    19 49 29.5 &   $+$31 27 16 & 067.1+02.7 & IRAS 19475+3119 &    31.64    & yes  &-- - Figure \ref{Fig10}(d)  \\    
    19 54 24.3 &   $+$20 52 47 & 058.6-03.6 & V 458 Vul &    23.37    &  -- &B - Figure \ref{Fig10}(e) \\            
     19 56 34.0 &   $+$32 22 12 & 068.7+01.9 & PN K 4-41         &    23.72   & ISM  & R - Figure \ref{Fig10}(f) \\      
     20 03 11.4 &   $+$30 32 34 &  067.9-00.2 & PN K 3-52        &   23.23   &  -- & B - Figure \ref{Fig10}(g)  \\       
    20 04 58.6 &   $+$25 26 37 &   063.8-03.3 & PN K 3-54        &  22.15    & -- &R - Figure \ref{Fig10}(h)\\   
    20 12 42.8 &   $+$19 59 22 &  060.1-07.7 & NGC 6886 &    21.09   & ISM  &B - Figure \ref{Fig10}(i) \\ 
    20 13 57.8 &   $+$29 33 55 & 068.3-02.7 & Hen 2-459 &    34.72    &  yes  & B - Figure \ref{Fig11}(a) \\         
    20 20 08.7 &   $+$16 43 53 & 058.3-10.9 & HD 193538 &     23.02   &  --& B - Figure \ref{Fig11}(b)  \\   
     20 21 03.7 &   $+$32 29 24 &  071.6-02.3 & PN M 3-35 &  22.79    &  yes and ISM & E - Figure \ref{Fig11}(c)\\ 
\end{longtable}
\clearpage
\twocolumn

\begin{table*}
\footnotesize
\caption{APEX CO (J $= 3-2$) line detections. Column 1 contains the name of each object, column 2 shows the peak of the line in K, column 3 presents the line width obtained by the fit in km\,s$^{-1}$ (when using several Gaussians; this represents the total width of the line including all components). Column 4 shows the integrated intensity over the velocity range of the line in K\,km\,s$^{-1}$, and Column 5 refers to the mass estimate and the figure where the fitted spectrum is shown. References: $^a$\protect\cite{durand98}, $^b$\protect\cite{dougados92}, $^c$\protect\cite{quintana16}, $^d$\protect\cite{kha07}, and $^e$\protect\cite{santander08}}
\label{detections}
\begin{tabular}{llllllllr}
\hline
Name&  PN G & V$_\mathrm{hel}$  &  Line width & $\int$CO($3-2$) dv & $N$(CO) & $N$(H$_2$) & M$_{H_2}$ & Figure\\
 && (kms$^{-1}$) & (kms$^{-1}$) & (K\,kms$^{-1}$) & (cm$^{-2}$) & (cm$^{-2}$)  & (M$_{\odot}$) & \\
\hline
    PN M 1-11 & 232.8-04.7 &  16.3$^a$&  2.646$\pm$0.230 &  0.654$\pm$0.084   & 1.24x$10^{15}$  &  4.10x$10^{18}$   &       8.7x$10^{-4}$     &   \ref{Fig12}(a)\\ 
  Frosty Leo &   \textit{221.9+42.7} &-10.0$^b$&  44.268$\pm$1.015 &  14.331$\pm$0.281  &  3.96x$10^{16}$ & 1.31x$10^{20}$ &       3.3x$10^{-2}$ &   \ref{Fig12}(b)\\ 
    IRC$+$10216 & \textit{221.4+45.1} &-19.4$^c$ &  22.222$\pm$0.015 & 815.992$\pm$4.420 &  1.29x$10^{18}$   & 4.25x$10^{21}$        & 4.1x$10^{-3}$ &   \ref{Fig12}(c)\\
   NGC 3132 &  272.1+12.3 & -16.0$^a$ &   54.418$\pm$1.980 & 11.400$\pm$0.352  &  7.51x$10^{15}$  &  2.47x$10^{19}$  & 2.0x$10^{-2}$  &   \ref{Fig12}(d)\\  
    HD 101584 & \textit{293.0+05.9} & 30.3$^d$ &   188.459$\pm$1.868 & 59.816$\pm$0.594   &  8.35x$10^{17}$  &  2.75x$10^{21}$       & 1.8x$10^{-2}$   &   \ref{Fig12}(e)\\
  V* V852 Cen & 315.4+09.4 &  -83.0$^e$ &   22.572$\pm$4.951 &  1.270$\pm$0.363    &  3.50x$10^{15}$ &   1.16x$10^{19}$              & 2.0x$10^{-2}$  &   \ref{Fig12}(f) \\ 
   IC 4406 &  319.6+15.7 &-41.4$^a$ &    20.494$\pm$0.789        &   42.207$\pm$0.494    &  2.78x$10^{16}$       & 9.18x$10^{19}$        & 1.7x$10^{-1}$   &  \ref{Fig12}(g)\\ 
    Hen 2-113 & 321.0+03.9 &-56.9$^a$  &  57.356$\pm$1.089 &   14.580$\pm$0.107  &  4.03x$10^{16}$   & 1.33x$10^{20}$        & 3.0x$10^{-3}$   &  \ref{Fig12}(h)\\ 
   Hen 2-133 &324.8-01.1 &  -25.05 &    1.724$\pm$0.122 &  0.826$\pm$0.063    &  4.13x$10^{15}$    & 1.36x$10^{19}$        & 4.3x$10^{-4}$   & \ref{Fig12}(i)\\  
   PN M 2-9 &  010.8+18.0 &87.5$^a$ &  17.085$\pm$0.748 &  4.680$\pm$0.240  &  3.08x$10^{15}$      & 1.01x$10^{19}$        & 1.3x$10^{-2}$   &  \ref{Fig13}(a)\\ 
    NGC 6326 &338.1-08.3 & 8.5$^a$ &    3.175$\pm$8.564&  0.253$\pm$0.071   &   5.33x$10^{14}$ &  1.76x$10^{18}$  & 2.0x$10^{-3}$  &  \ref{Fig13}(b) \\ 
   PN Fg 3 & 352.9-07.5 &4.0$^a$ &   4.789$\pm$1.744 &  0.269$\pm$0.092   &  7.44x$10^{14}$     & 2.45x$10^{18}$        & 3.5x$10^{-4}$   &  \ref{Fig13}(c) \\  
  PN Cn 3-1 & 038.2+12.0 &  -11.2$^a$ &   7.232$\pm$4.627 &   0.457$\pm$0.189   &   4.21x$10^{15}$   & 1.39x$10^{19}$        & 1.2x$10^{-3}$  &  \ref{Fig13}(d)\\
    PN M 2-43 & 027.6+04.2 &96.4$^a$ &    146.366$\pm$10.220 &  5.869$\pm$0.369   &    2.94x$10^{16}$        & 9.70x$10^{19}$        & 1.0x$10^{-1}$   & \ref{Fig13}(e)\\ 
   PN M 1-63 & 021.1-05.9 &  8.5$^a$ & 61.477$\pm$5.513 &  3.805$\pm$0.090   &  4.86x$10^{16}$     & 1.60x$10^{20}$        & 9.3x$10^{-3}$   &  \ref{Fig13}(f)\\   
   PN M 1-65 &  043.1+03.8 & 3.7$^a$ &    1.619$\pm$0.957 &   0.216$\pm$0.062  &    3.70x$10^{15}$   & 1.22x$10^{19}$        & 1.9x$10^{-3}$  &  \ref{Fig13}(g)\\      
   BD$+$30 3639 & 064.7+05.0 & -31.4$^a$ & 13.739$\pm$3.506 &   1.314$\pm$0.302  &  5.19x$10^{15}$    & 1.71x$10^{19}$        & 5.3x$10^{-4}$ & \ref{Fig13}(h) \\       
    Hen 2-447 & 057.9-01.5 &-- &    21.103$\pm$6.043 &   0.802$\pm$0.282 & 1.95x$10^{16}$        & 6.45x$10^{19}$        & 1.3x$10^{-3}$  &  \ref{Fig13}(i)\\
   IRAS 19475+3119 &067.1+02.7 & -- &    25.192$\pm$0.895 &  10.918$\pm$0.344    & 2.59x$10^{16}$    & 8.54x$10^{19}$        & 5.2x$10^{-2}$   & \ref{Fig14}(a)  \\    
   Hen 2-459 & 068.3-02.7 & -72.0$^a$ &   26.265$\pm$2.392 &  4.459$\pm$0.368    &   4.11x$10^{16}$  & 1.35x$10^{20}$        & 1.0x$10^{-2}$  &  \ref{Fig14}(b) \\     
   PN M 3-35 & 071.6-02.3 & -192.2$^a$ &   4.965$\pm$1.356 & 0.394$\pm$0.105   &  2.59x$10^{15}$   &  8.57x$10^{18}$         & 2.7x$10^{-4}$ & \ref{Fig14}(c)\\ 
\end{tabular}
\end{table*}

\begin{table*}
\footnotesize
\caption{Size, distance, and morphology of PNe with CO emission. References: $^1$\protect\cite{cahn92}, $^2$\protect\cite{robinson92}, $^3$\protect\cite{groen12}, $^4$\protect\cite{frew08}, $^5$\protect\cite{sahai07}, $^6$\protect\cite{acker92}, $^7$\protect\cite{tajitsu98}, $^8$\protect\cite{me06}, and $^9$\protect\cite{akras12}}
\label{morphology}
\begin{tabular}{lllll}
\hline
Name    & Size & Distance & Morphology           &      Page on HASH \\
                & (arcsec) & (kpc) &     &       \\
\hline 
M 1-11  & 13&2.1$^1$ &  Elliptical      &http://hashpn.space/objectInfoPage.php?id=778\\
Frosty Leo & 10& 3.0$^2$&Bipolar & http://hashpn.space/objectInfoPage.php?id=12777\\
IRC+10216  &15 &0.1$^3$ &Unresolved & -- \\
NGC 3132 &45 & 1.2$^4$& Bipolar &http://hashpn.space/objectInfoPage.php?id=840\\
HD101584 & 4&1.2$^4$& Unresolved & -- \\
V* V852 Cen &10 &7.9$^6$& Unresolved & -- \\
IC 4406 & 35& 2.3$^1$&  Bipolar &http://hashpn.space/objectInfoPage.php?id=990\\
Hen 2-113        & 10& 0.9$^7$& Bipolar &       http://hashpn.space/objectInfoPage.php?id=996\\
Hen 2-133 & 7& 1.5$^1$&     Elliptical    &    http://hashpn.space/objectInfoPage.php?id=1010 \\
M 2-9   & 42& 1.7$^1$&  Bipolar &       http://hashpn.space/objectInfoPage.php?id=232\\
NGC 6326& 12&5.3$^4$ &  Bipolar &http://hashpn.space/objectInfoPage.php?id=1080\\
Fg 3            & 10&2.2$^1$ &  Bipolar &http://hashpn.space/objectInfoPage.php?id=1170\\
Cn 3-1  & 5&3.6$^1$ &   Elliptical              &http://hashpn.space/objectInfoPage.php?id=392\\
M 2-43  & 7& 8.5$^8$&   Elliptical              &http://hashpn.space/objectInfoPage.php?id=339\\
M 1-63  & 4.2& 3.4$^7$& Elliptical              &http://hashpn.space/objectInfoPage.php?id=301\\
M 1-65  & 3.6&6.5$^1$ & Elliptical      &http://hashpn.space/objectInfoPage.php?id=412\\
BD+30 3639      & 8&1.5$^9$ &Elliptical &       http://hashpn.space/objectInfoPage.php?id=512\\
Hen 2-447               &3 & 2.8$^1$&Bipolar    &http://hashpn.space/objectInfoPage.php?id=482\\
IRAS 19475+3119 &11 &4.2$^4$& Unresolved & -- \\
Hen 2-459       & 5 & 3.3$^1$&  Bipolar &http://hashpn.space/objectInfoPage.php?id=523\\
M 3-35  & 6&1.7$^1$ &   Elliptical      &http://hashpn.space/objectInfoPage.php?id=537\\
\end{tabular}
\end{table*}

\clearpage

\clearpage
\begin{appendix}
\section{APEX full spectrum of each source}

\begin{figure}[h!]
\vspace{2cm}
\hbox{
\includegraphics[width=6cm, height=5.5cm]{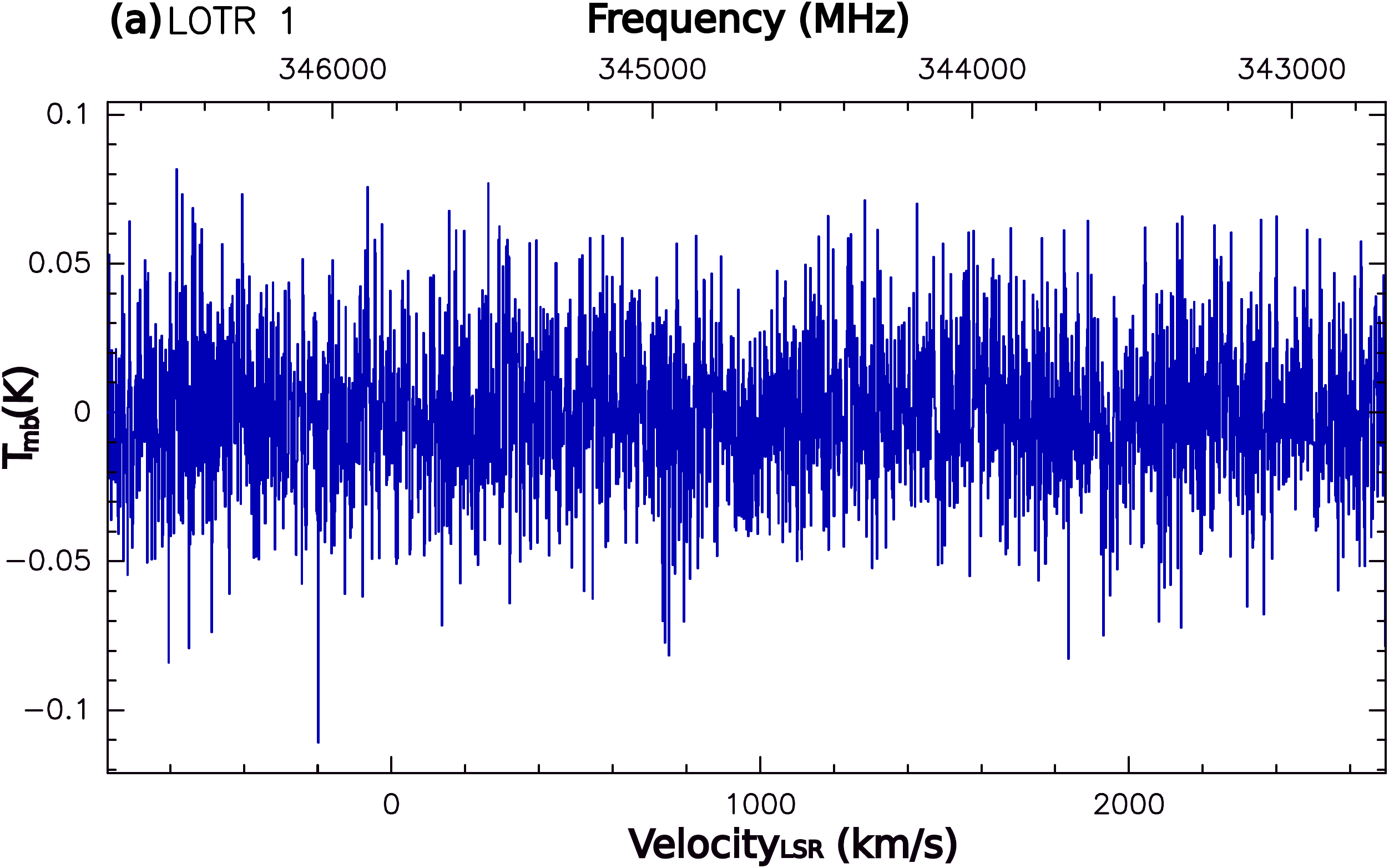}
\includegraphics[width=6cm, height=5.5cm]{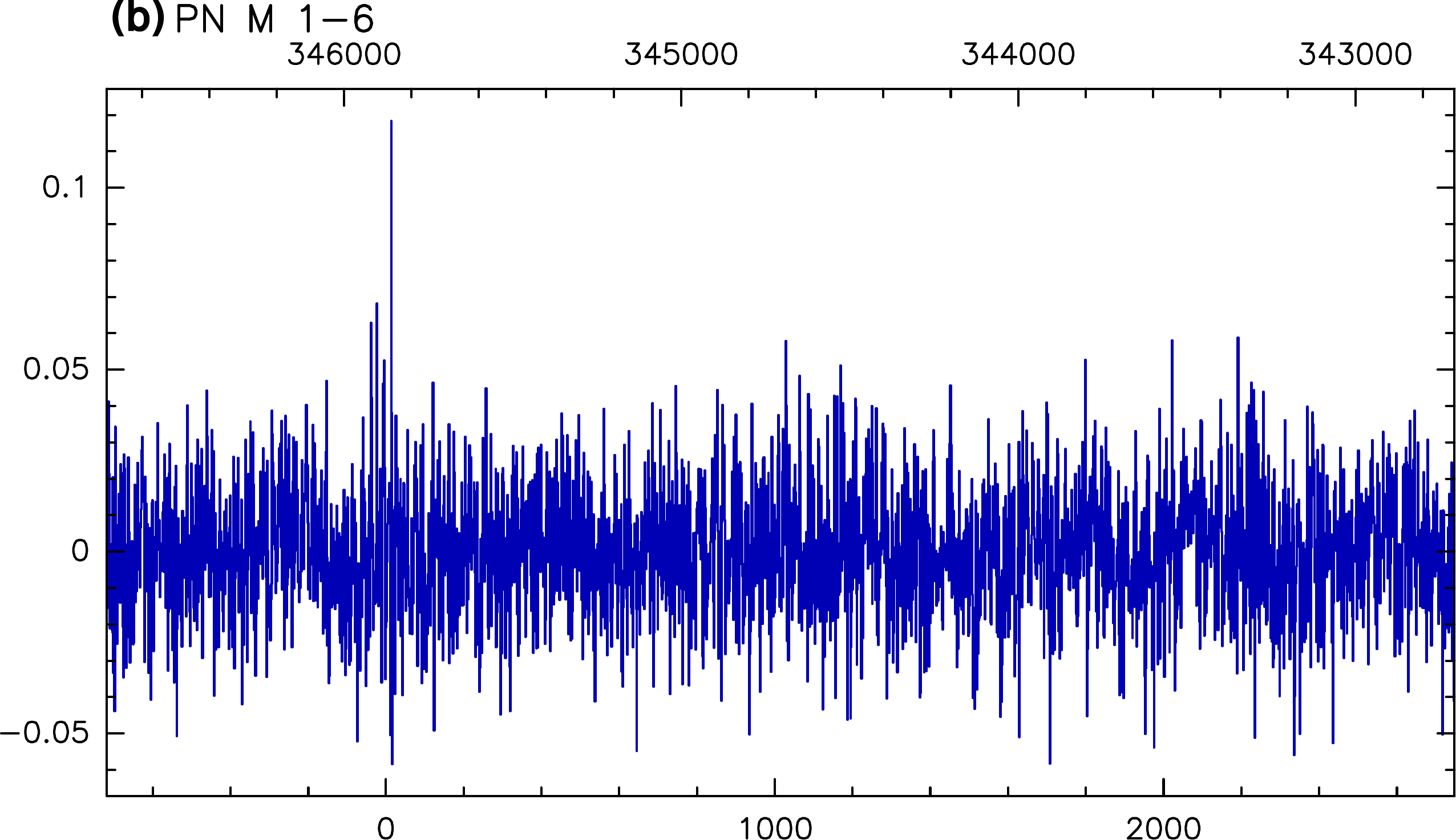}
\includegraphics[width=6cm, height=5.5cm]{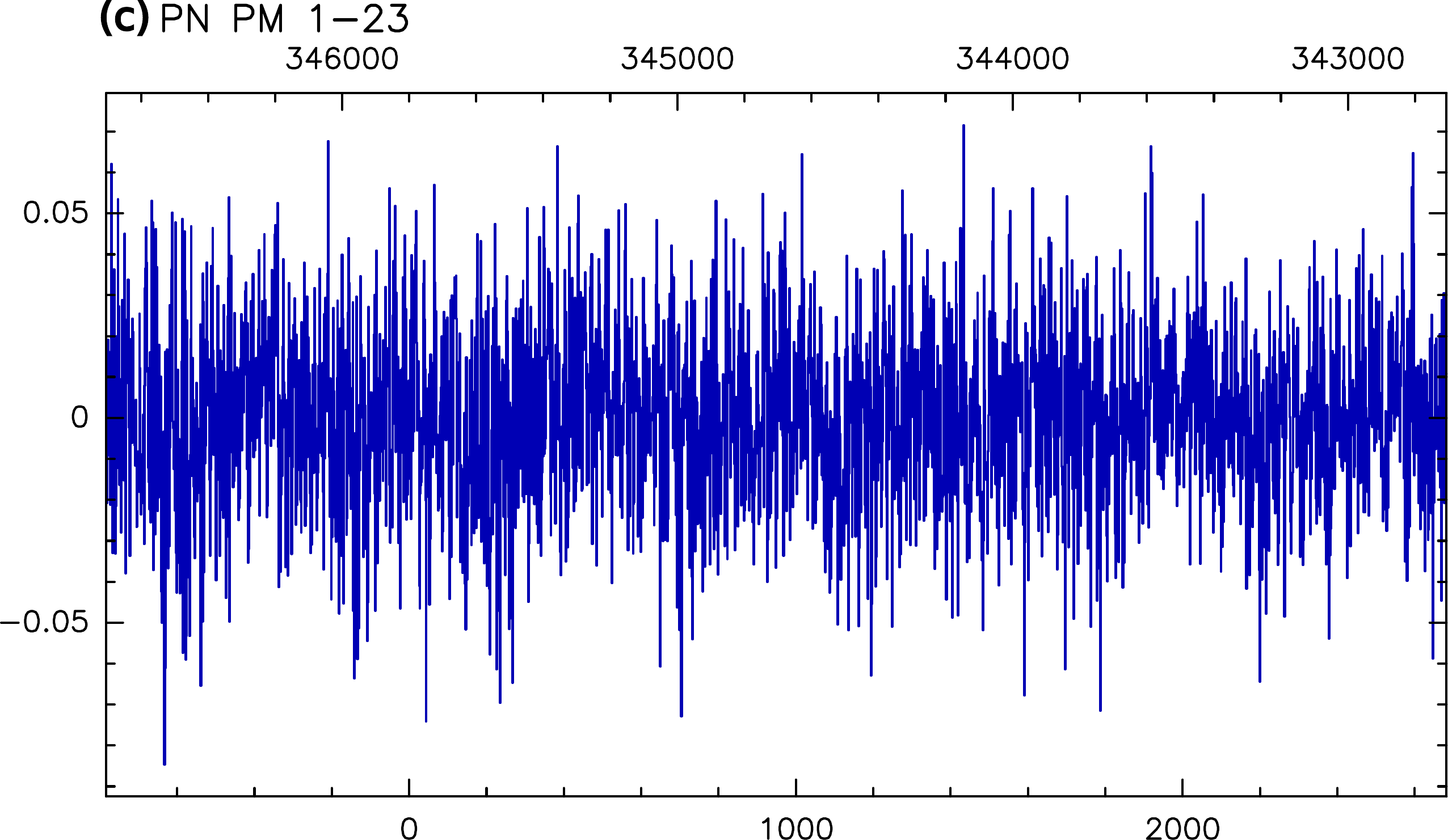}
\vspace{1cm}}
\hbox{
\vspace{1cm}
\includegraphics[width=6cm, height=5.5cm]{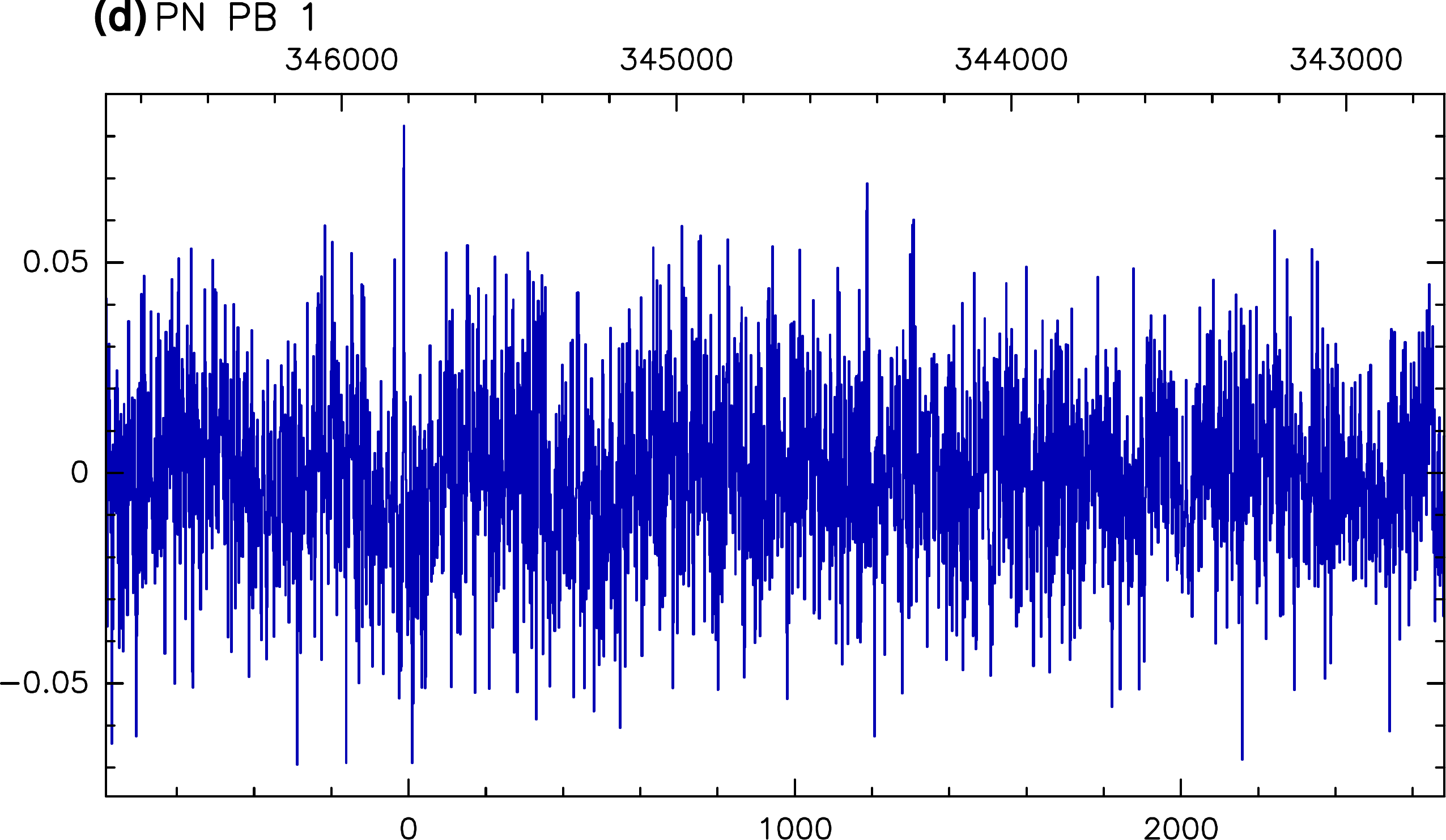}
\includegraphics[width=6cm, height=5.5cm]{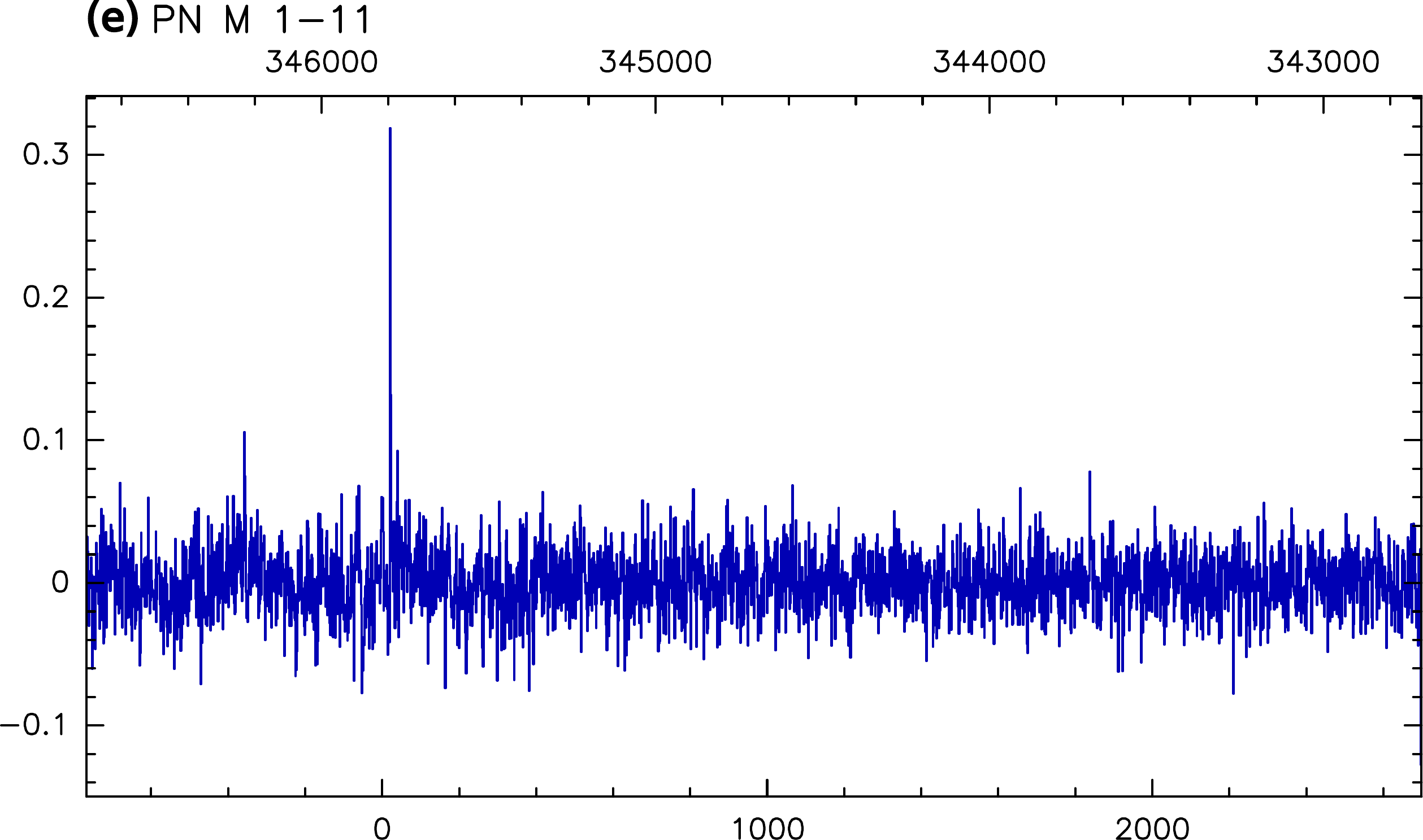}
\includegraphics[width=6cm, height=5.5cm]{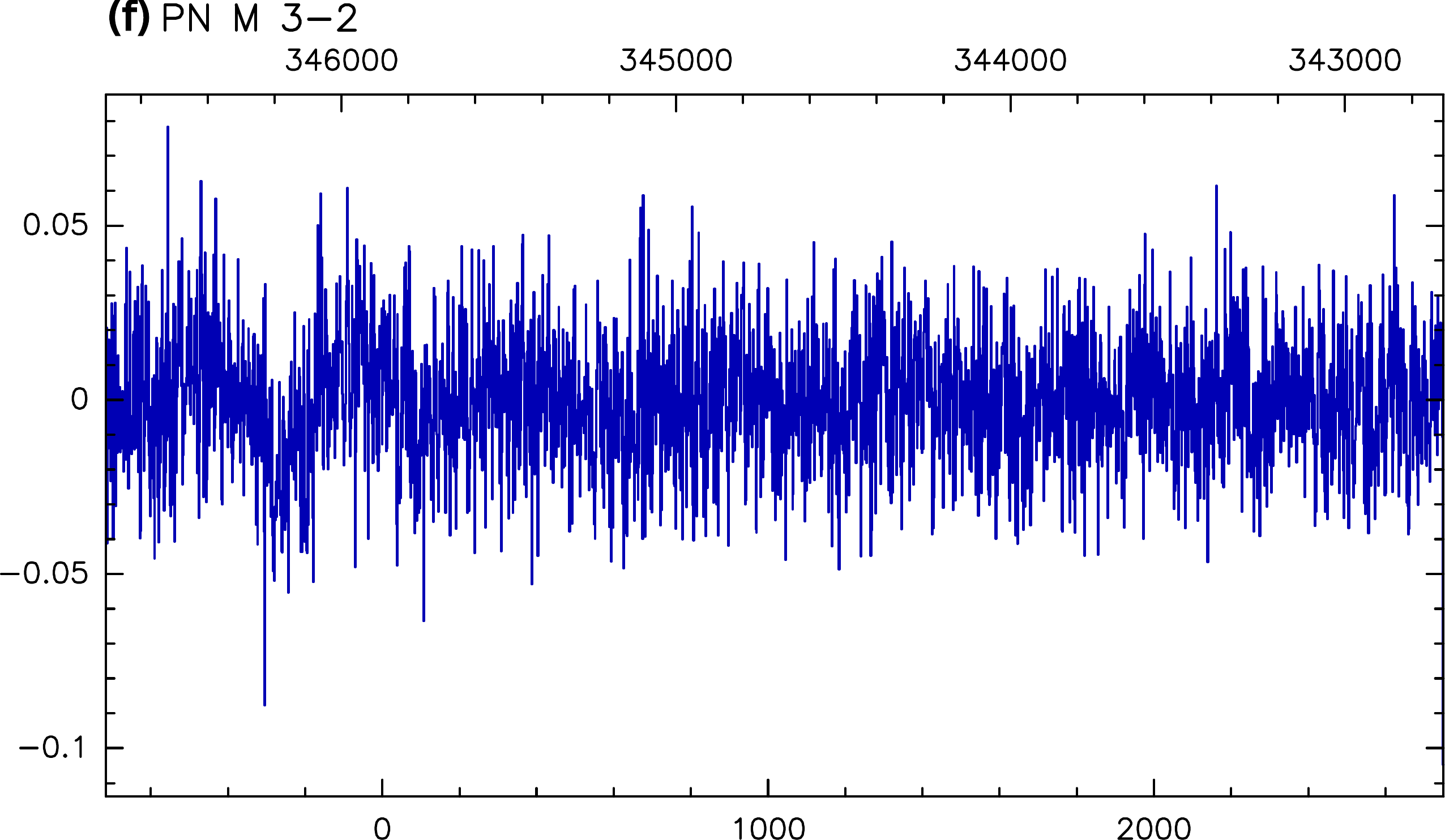}}
\hbox{
\includegraphics[width=6cm, height=5.5cm]{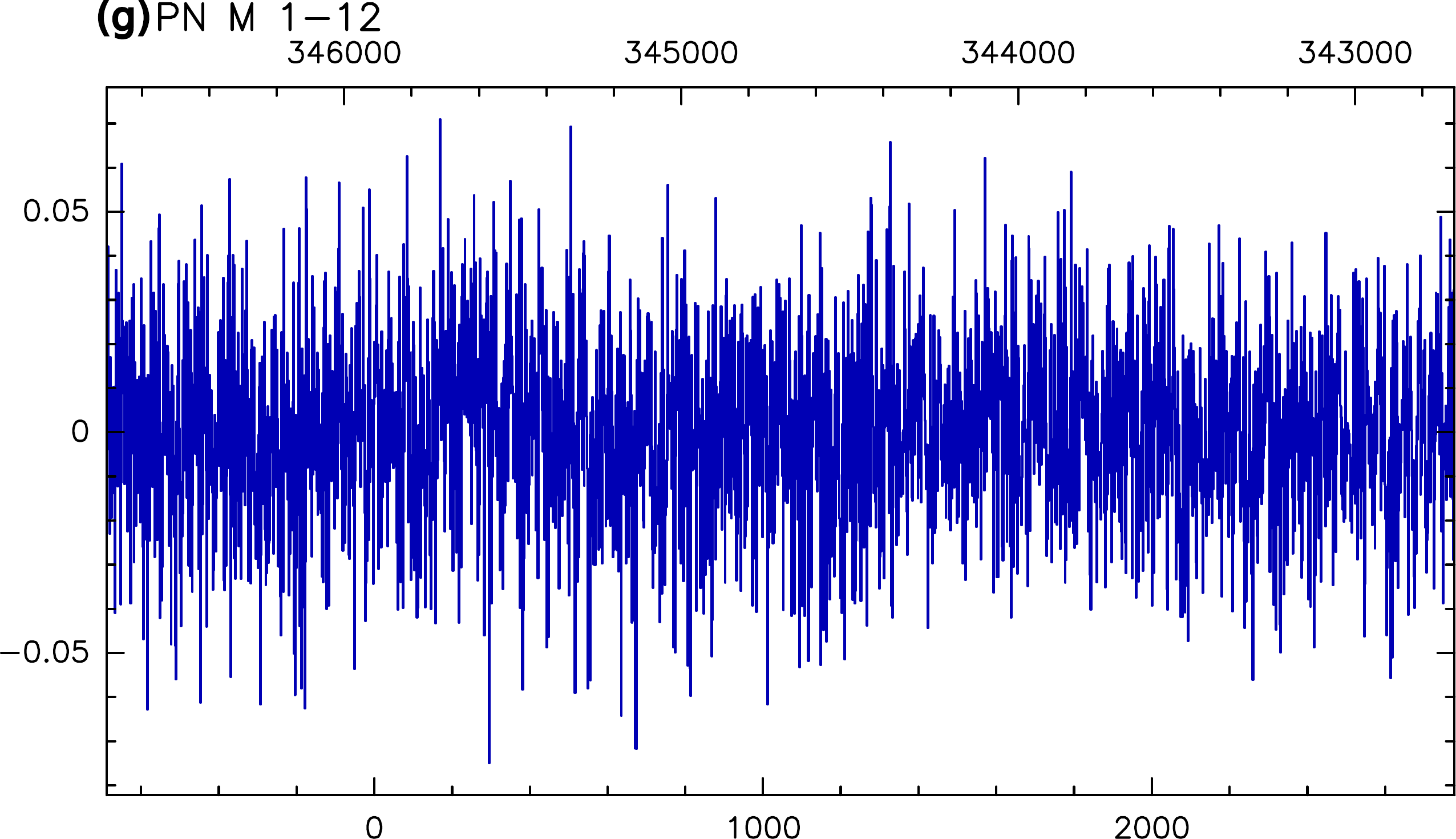}
\includegraphics[width=6cm, height=5.5cm]{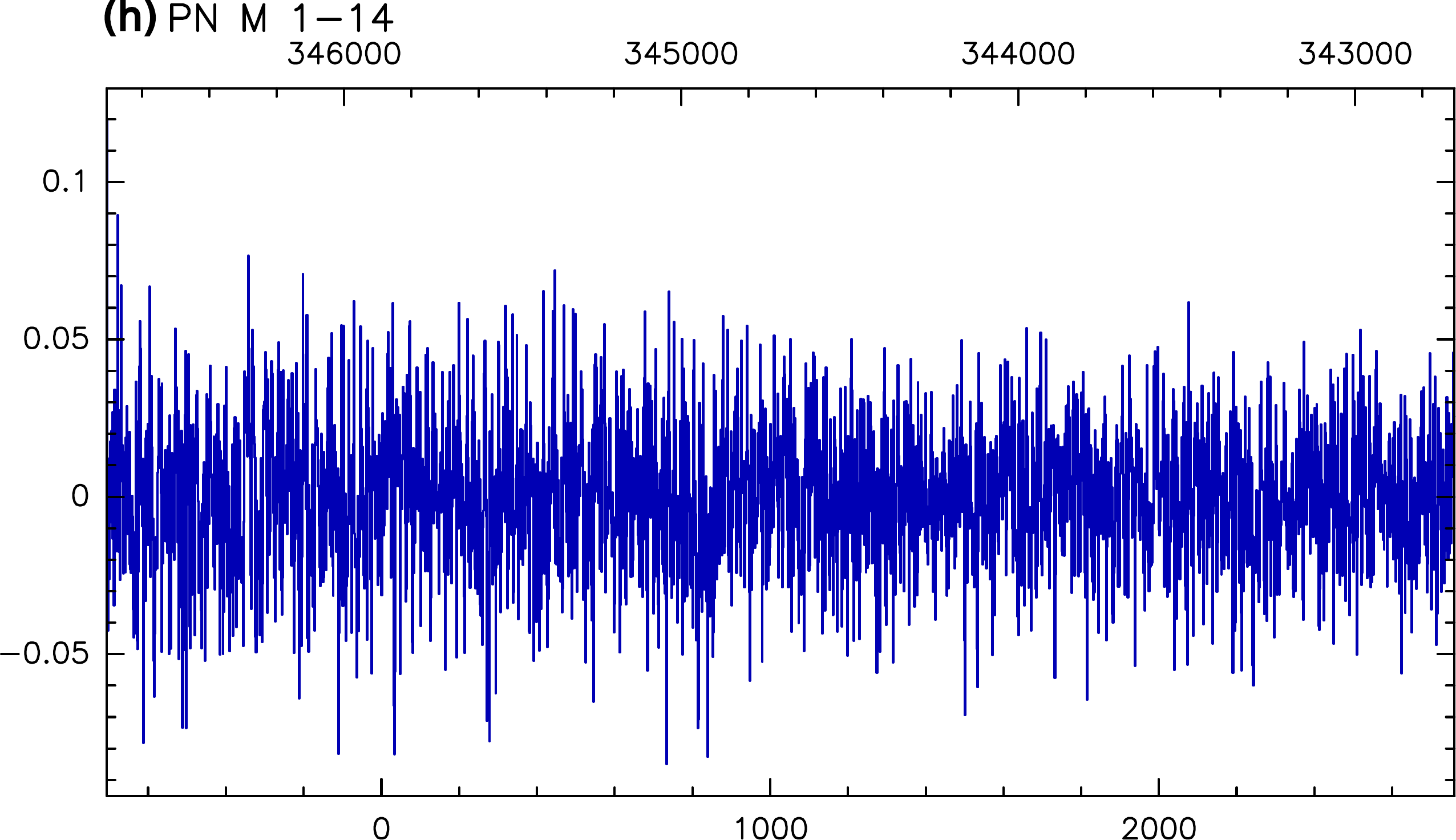}
\includegraphics[width=6cm, height=5.5cm]{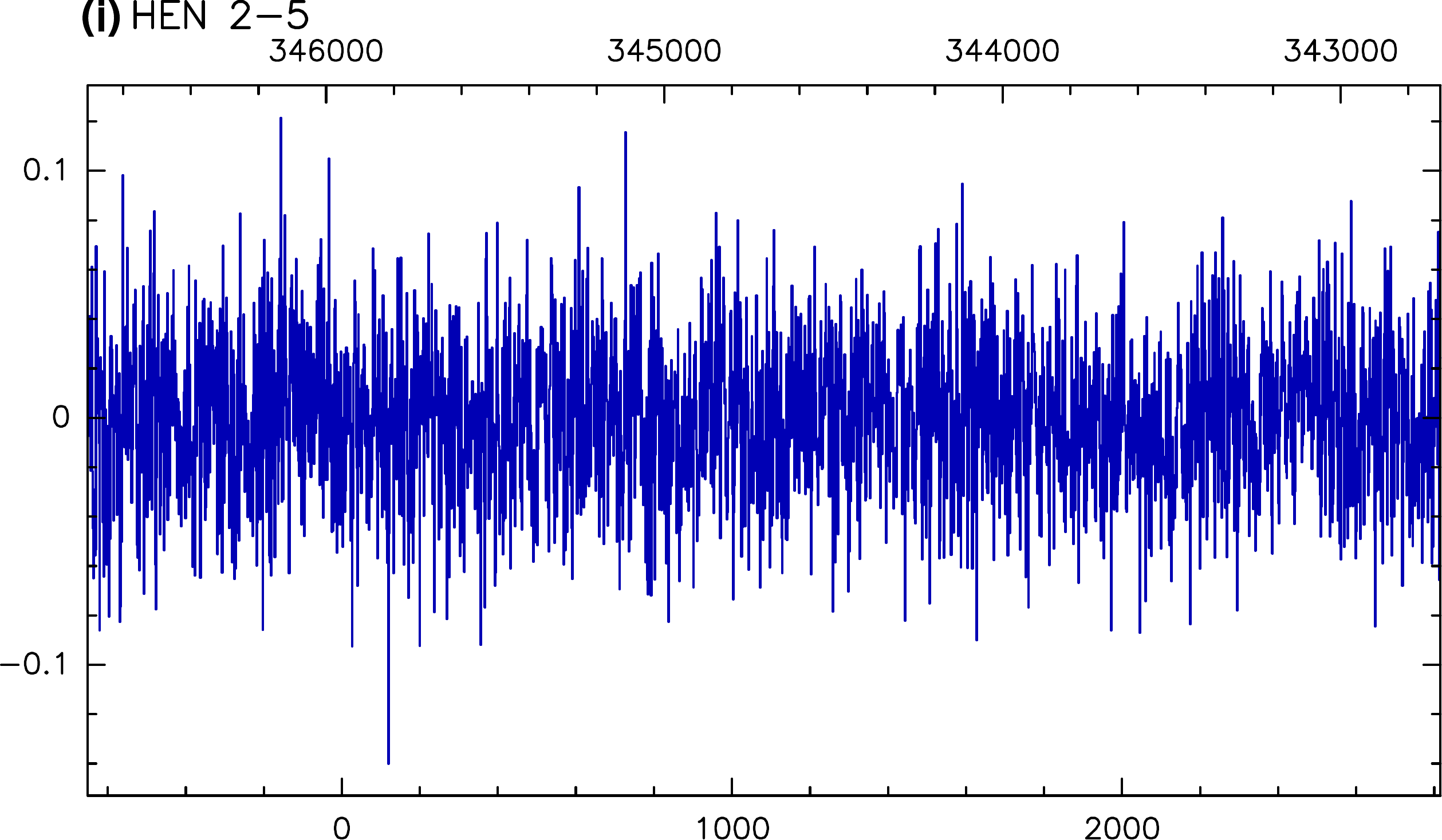}}
\caption[]{pPNe and PNe observations using the APEX telescope.}
\label{Fig1}
\end{figure} 

\begin{figure*}
\vspace{2cm}
\centering
\hbox{
\centering
\includegraphics[width=6cm, height=5.5cm]{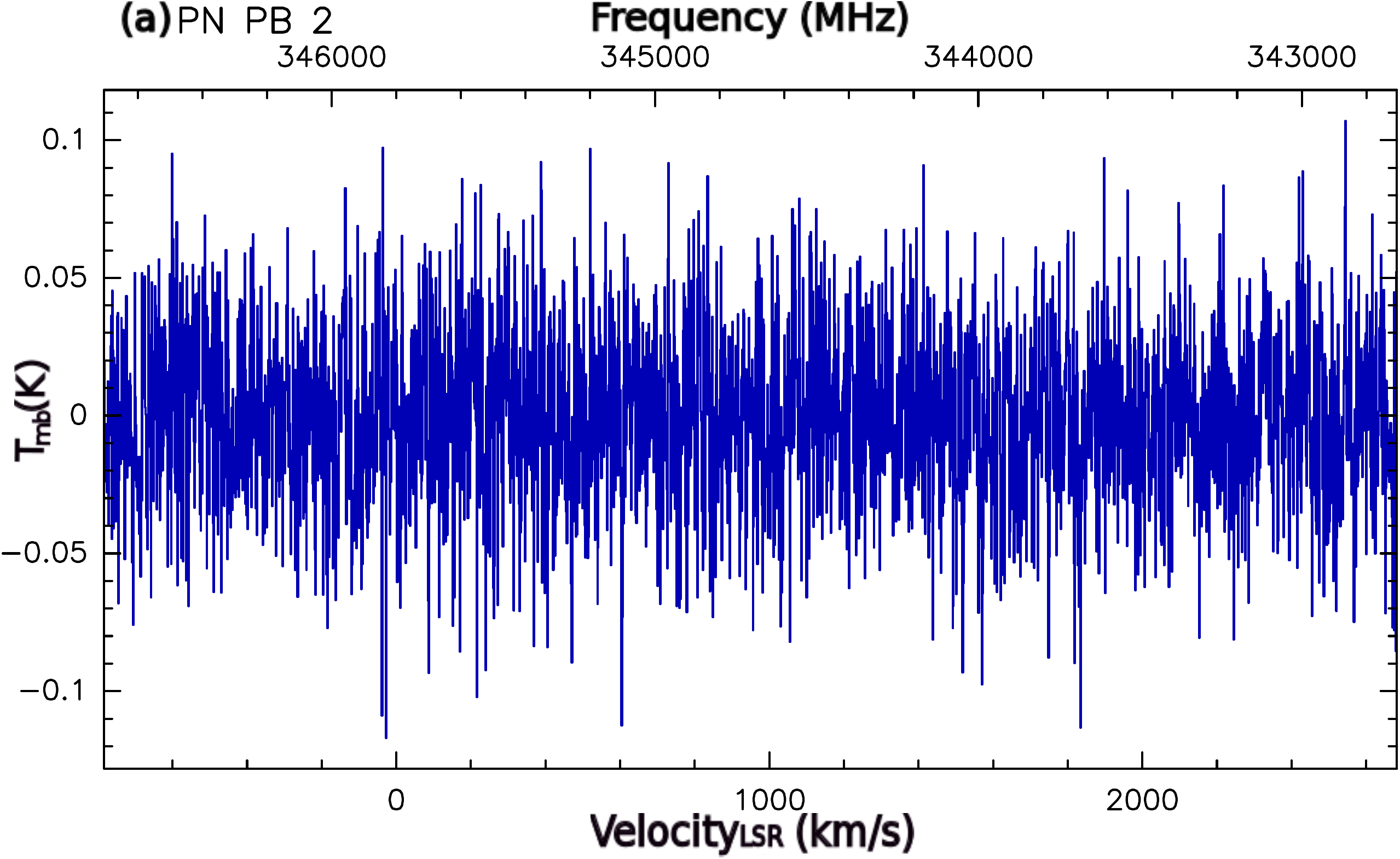}
\includegraphics[width=6cm, height=5.5cm]{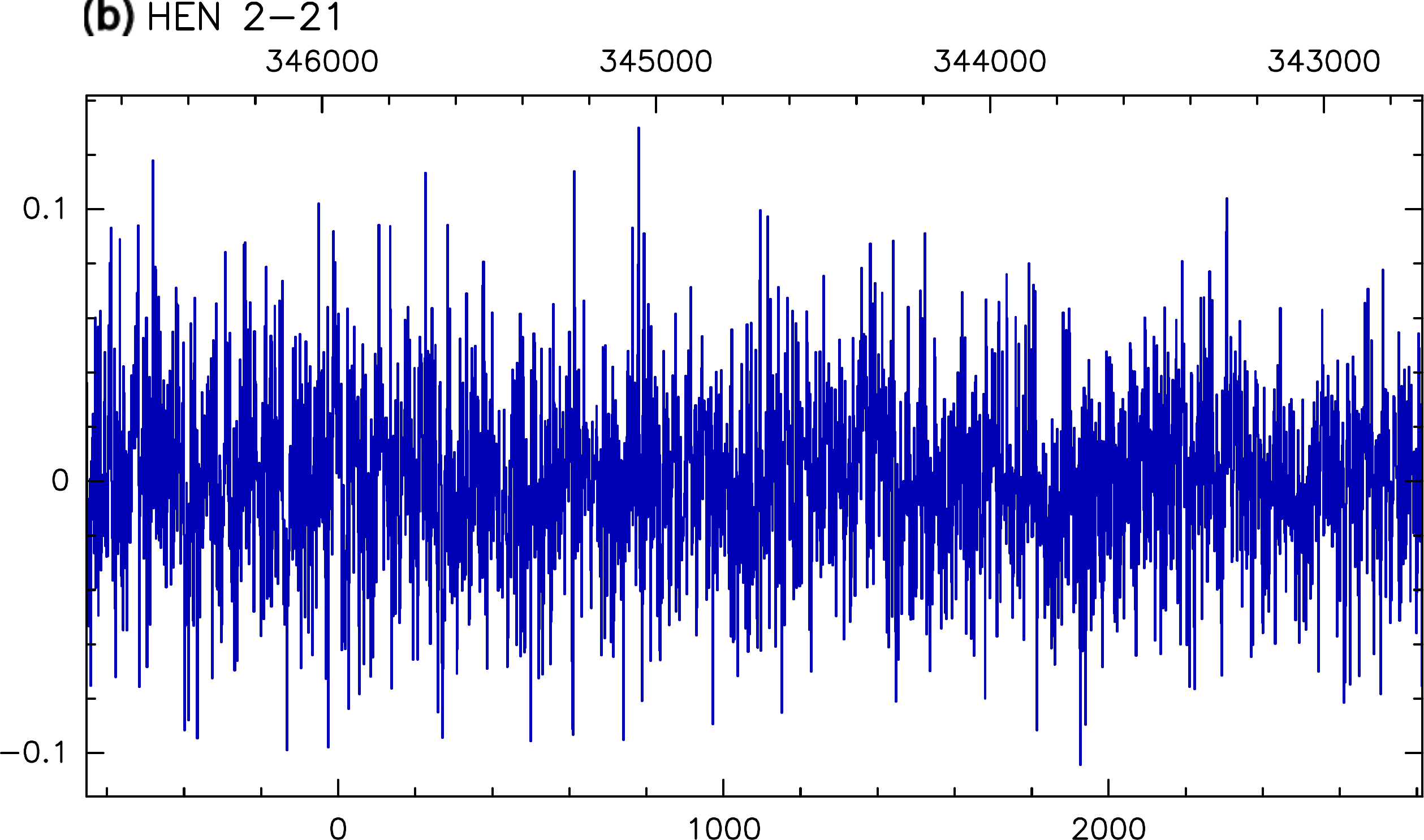}
\includegraphics[width=6cm, height=5.5cm]{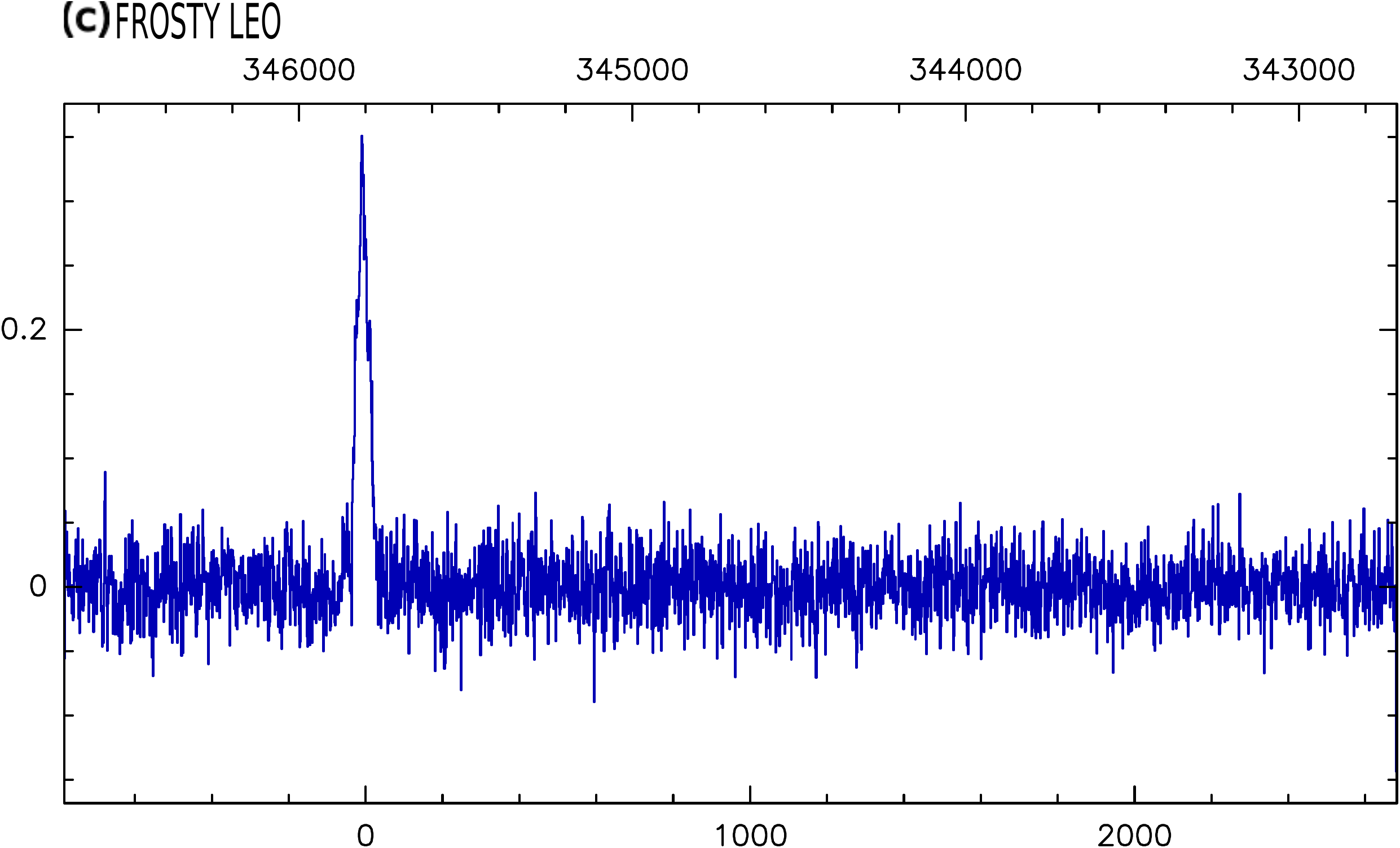}
\vspace{1cm}}
\hbox{
\centering
\vspace{1cm}
\includegraphics[width=6cm, height=5.5cm]{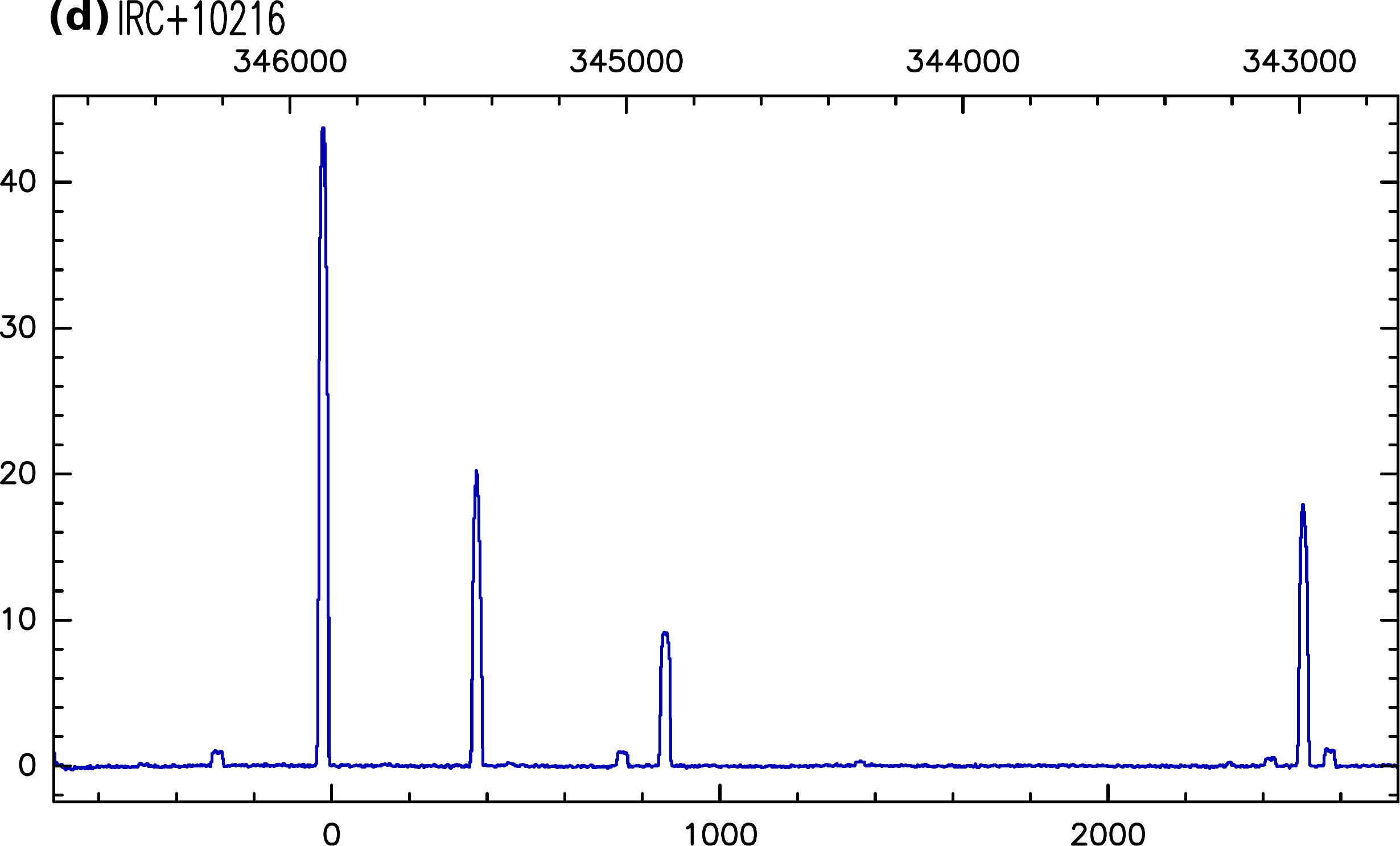}
\includegraphics[width=6cm, height=5.5cm]{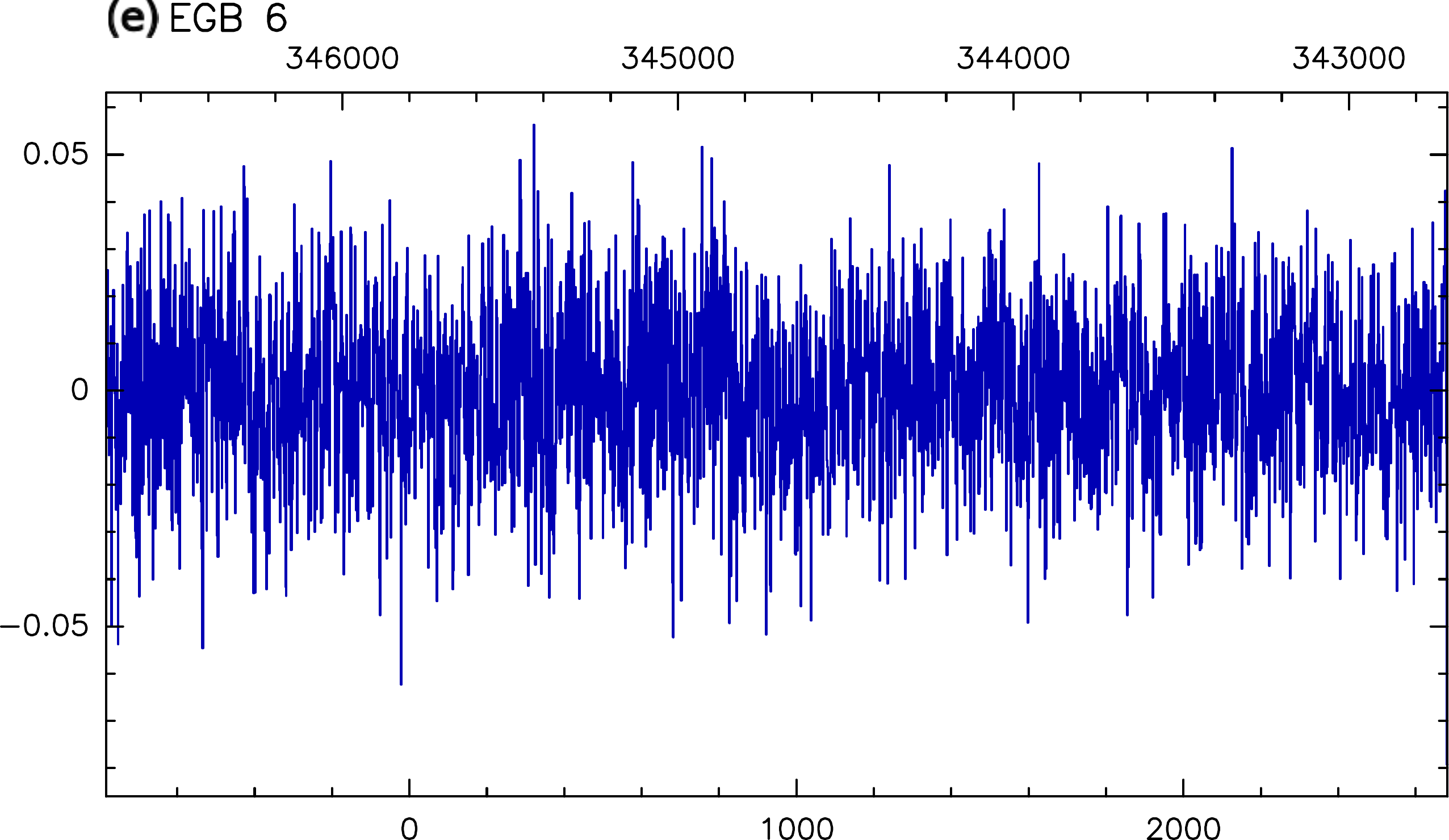}
\includegraphics[width=6cm, height=5.5cm]{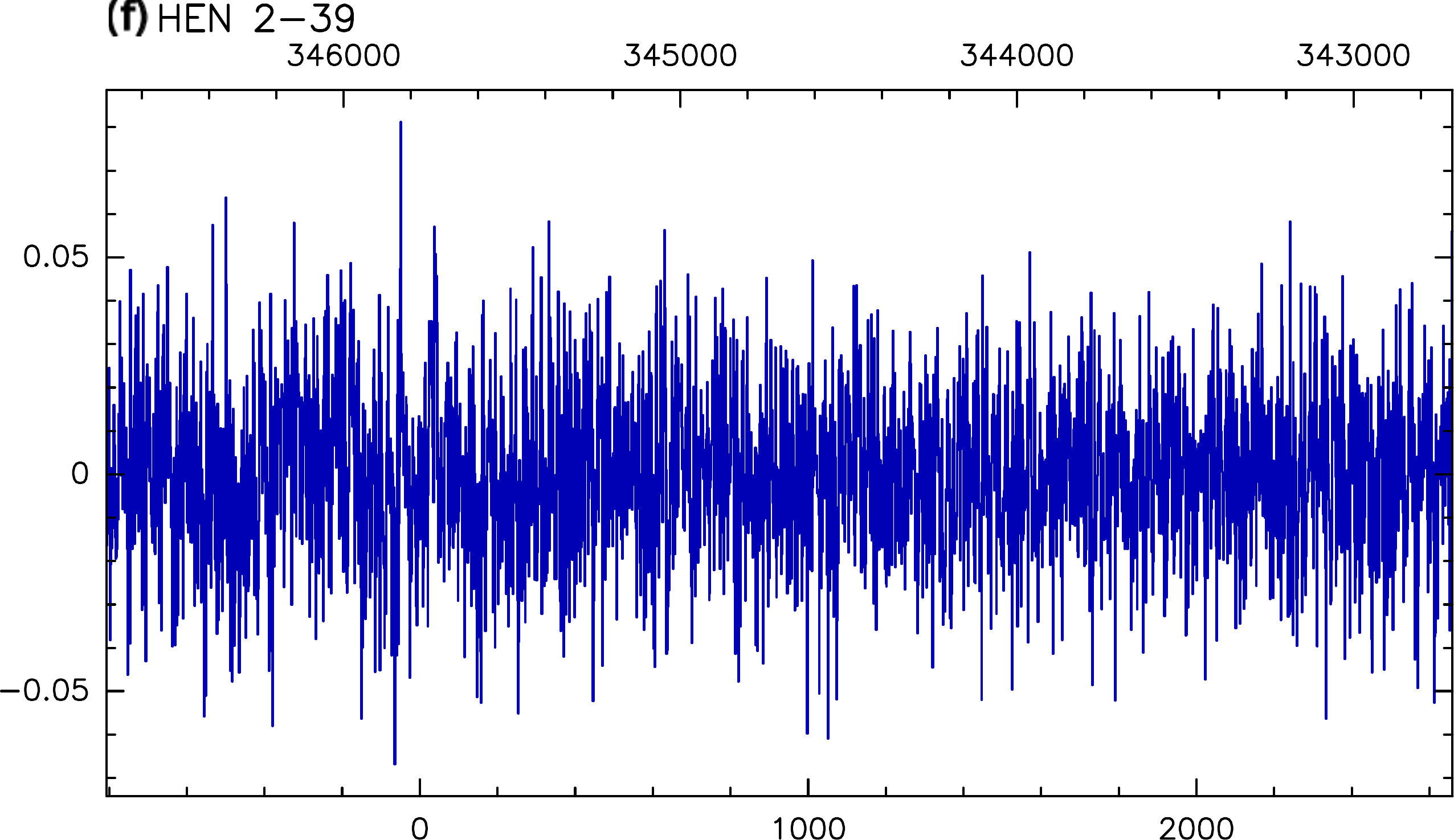}}
\hbox{
\centering
\includegraphics[width=6cm, height=5.5cm]{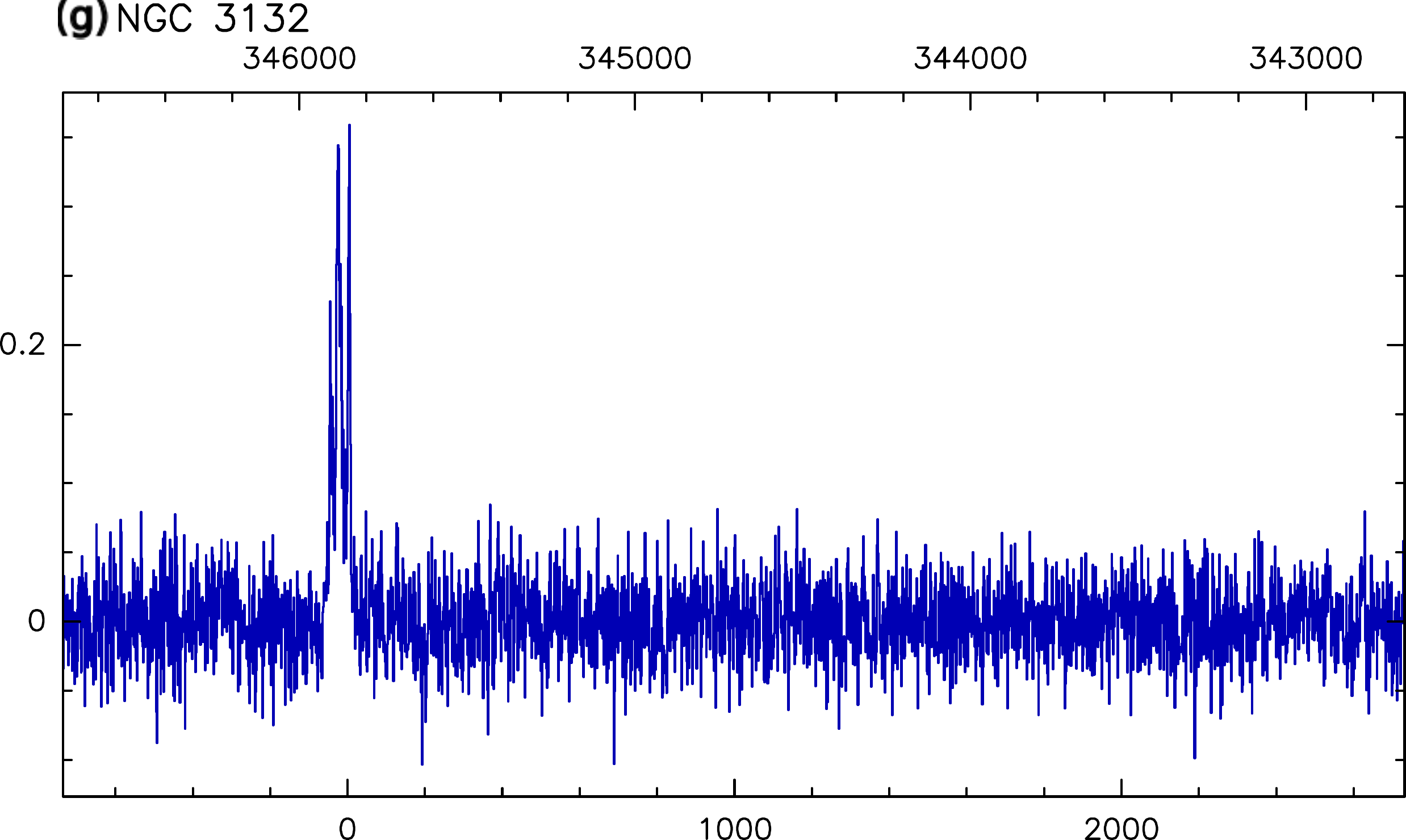}
\includegraphics[width=6cm, height=5.5cm]{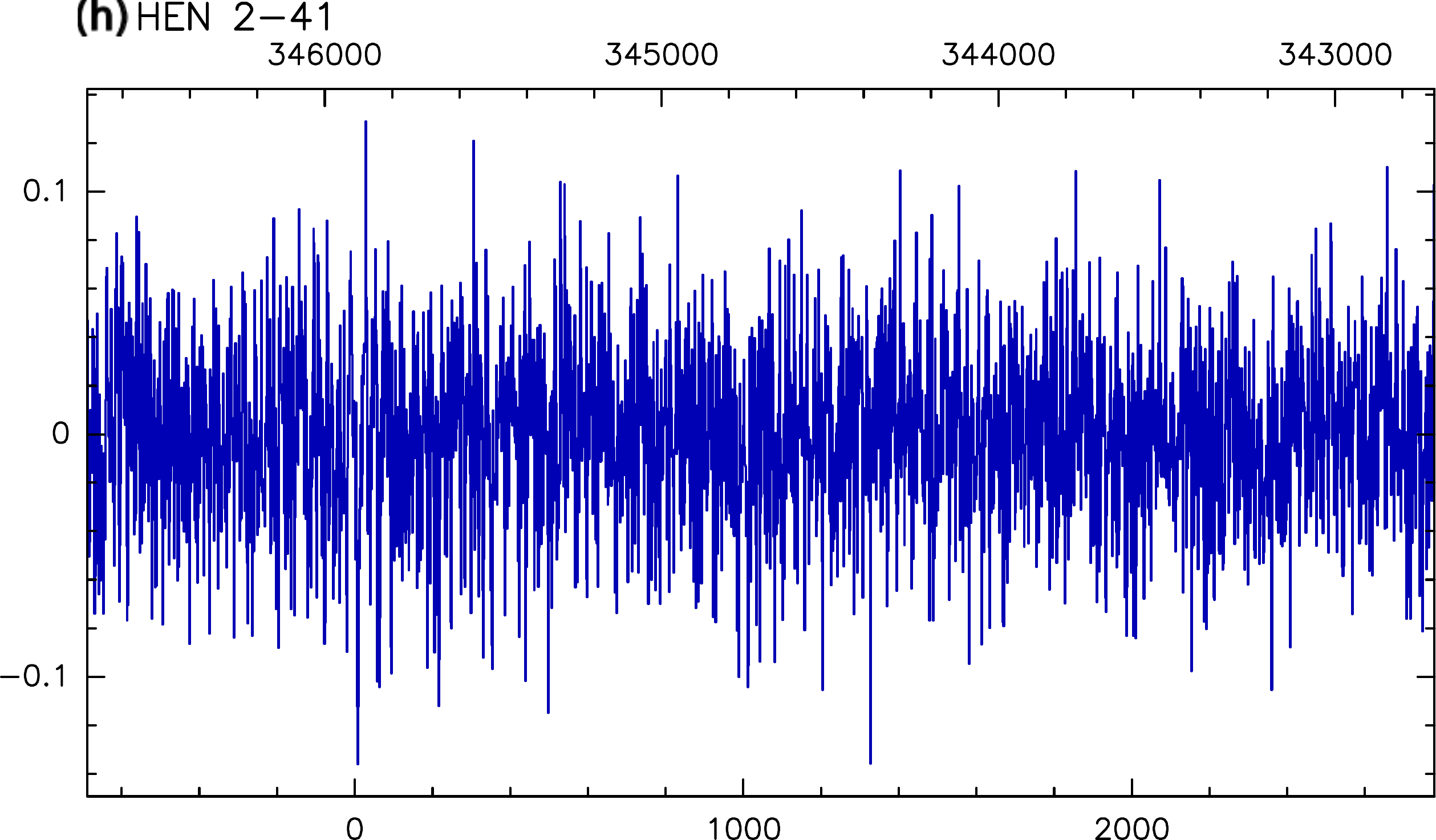}
\includegraphics[width=6cm, height=5.5cm]{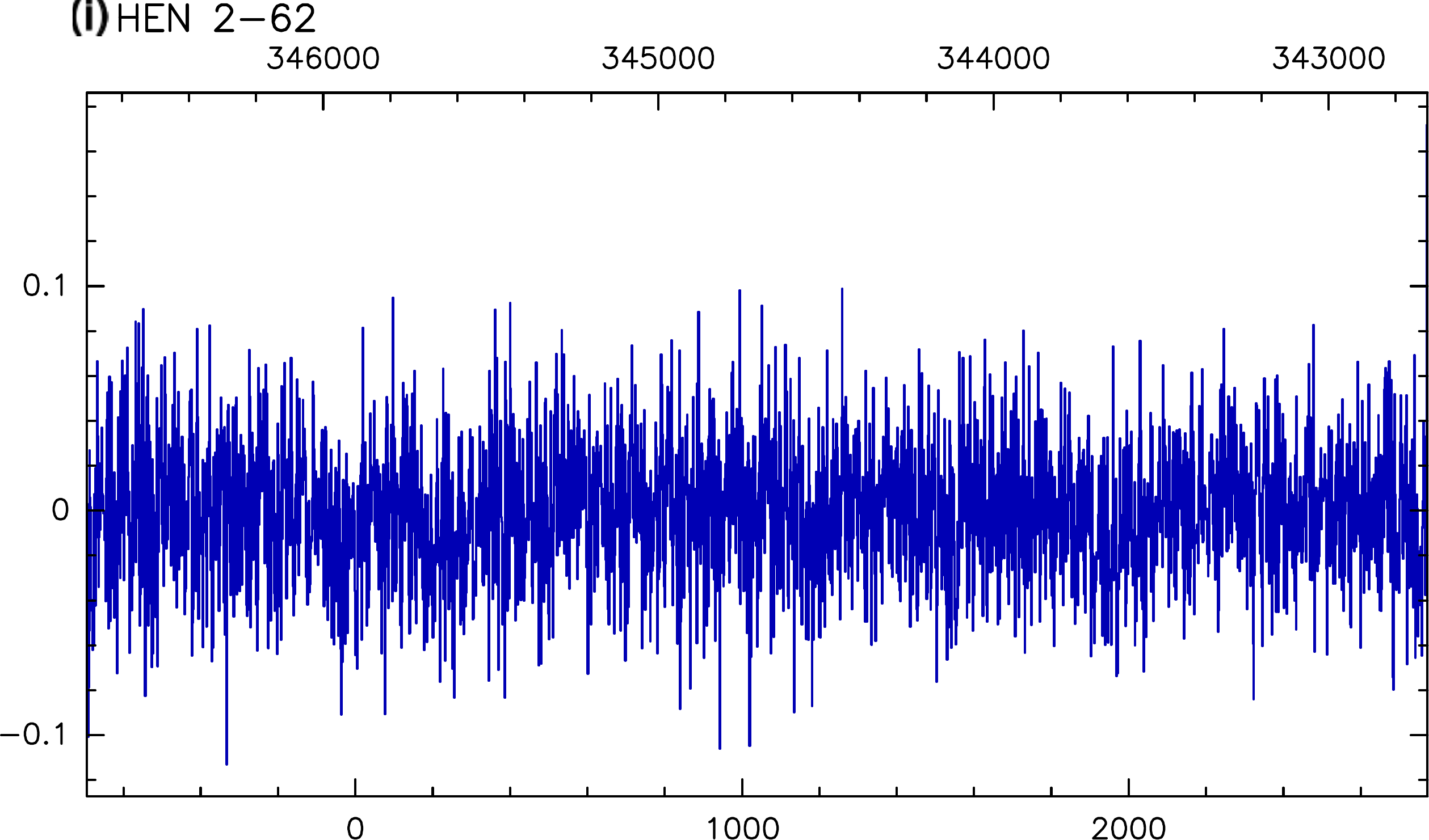}}
\caption[]{pPNe and PNe observations using the APEX telescope.}
\label{Fig2}
\end{figure*} 

\begin{figure*}
\vspace{2cm}
\centering
\hbox{
\centering
\includegraphics[width=6cm, height=5.5cm]{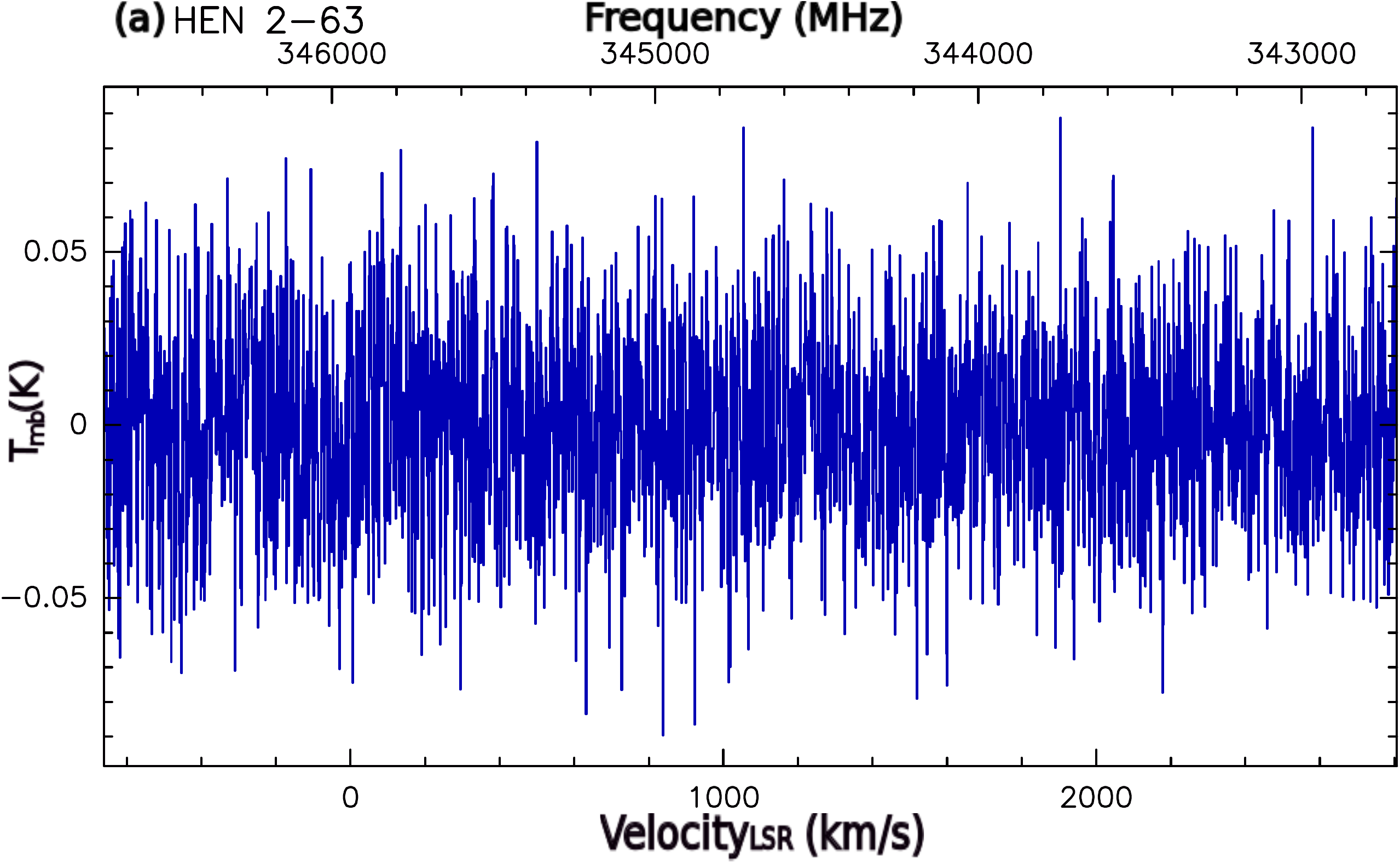}
\includegraphics[width=6cm, height=5.5cm]{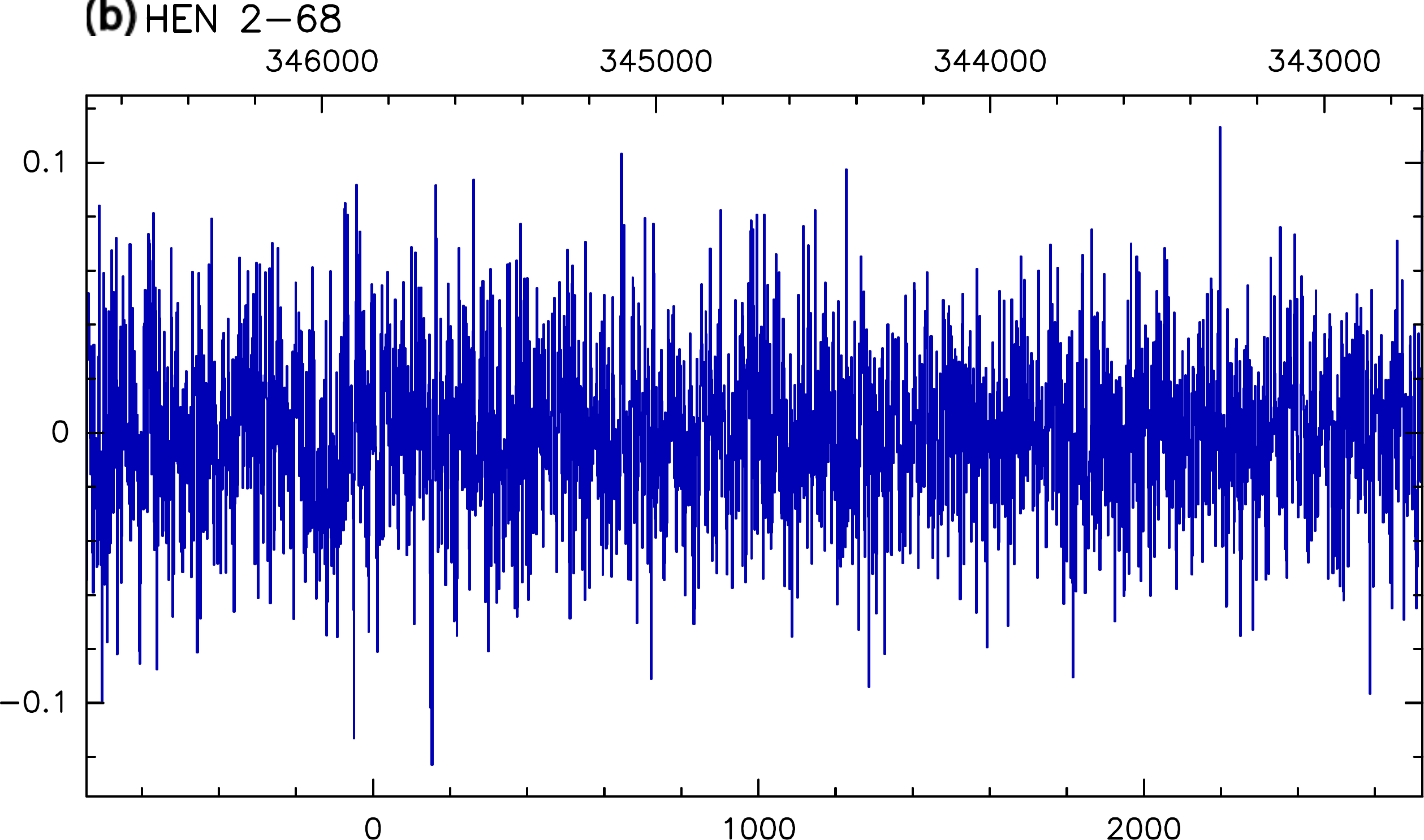}
\includegraphics[width=6cm, height=5.5cm]{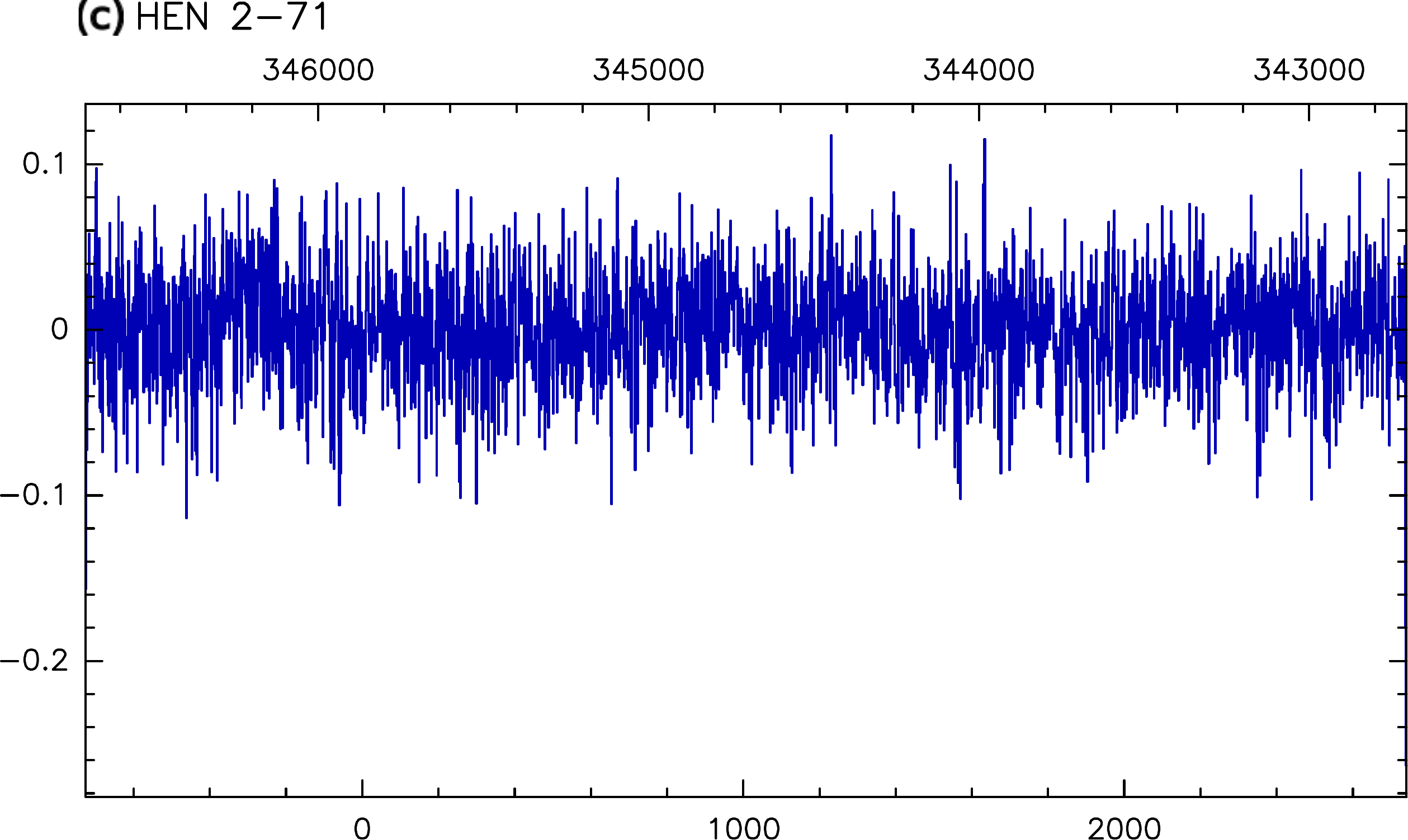}
\vspace{1cm}}
\hbox{
\centering
\vspace{1cm}
\includegraphics[width=6cm, height=5.5cm]{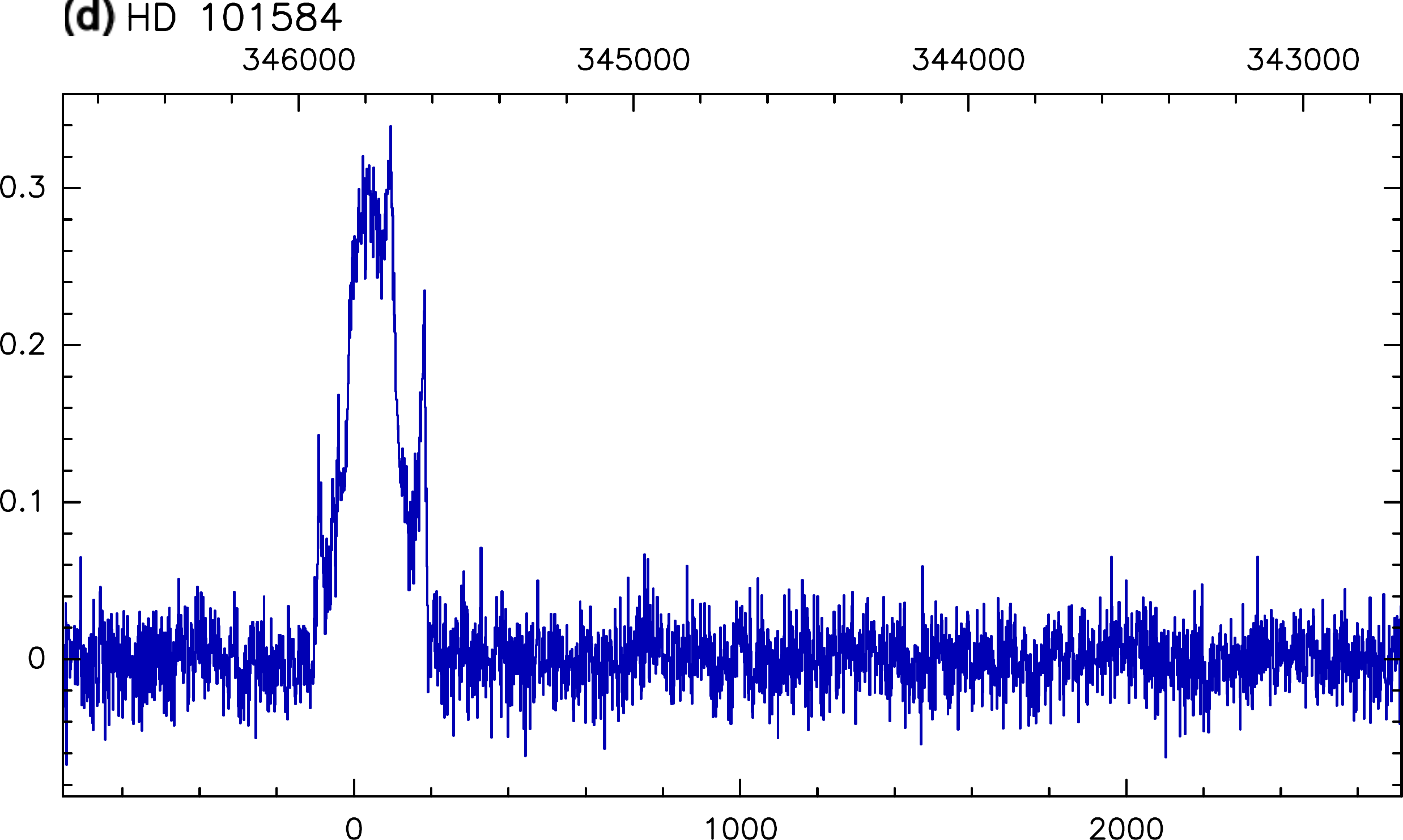}
\includegraphics[width=6cm, height=5.5cm]{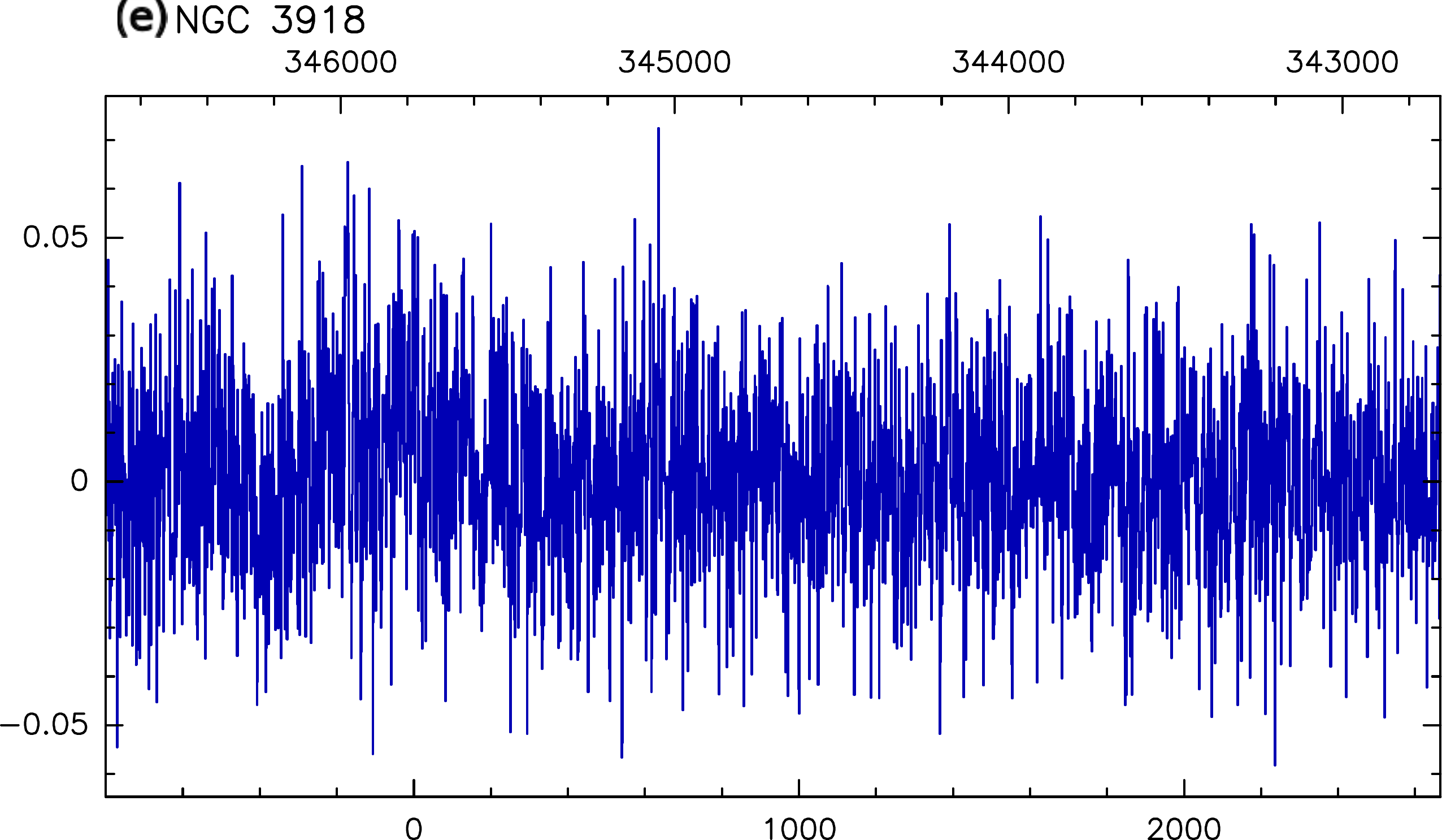}
\includegraphics[width=6cm, height=5.5cm]{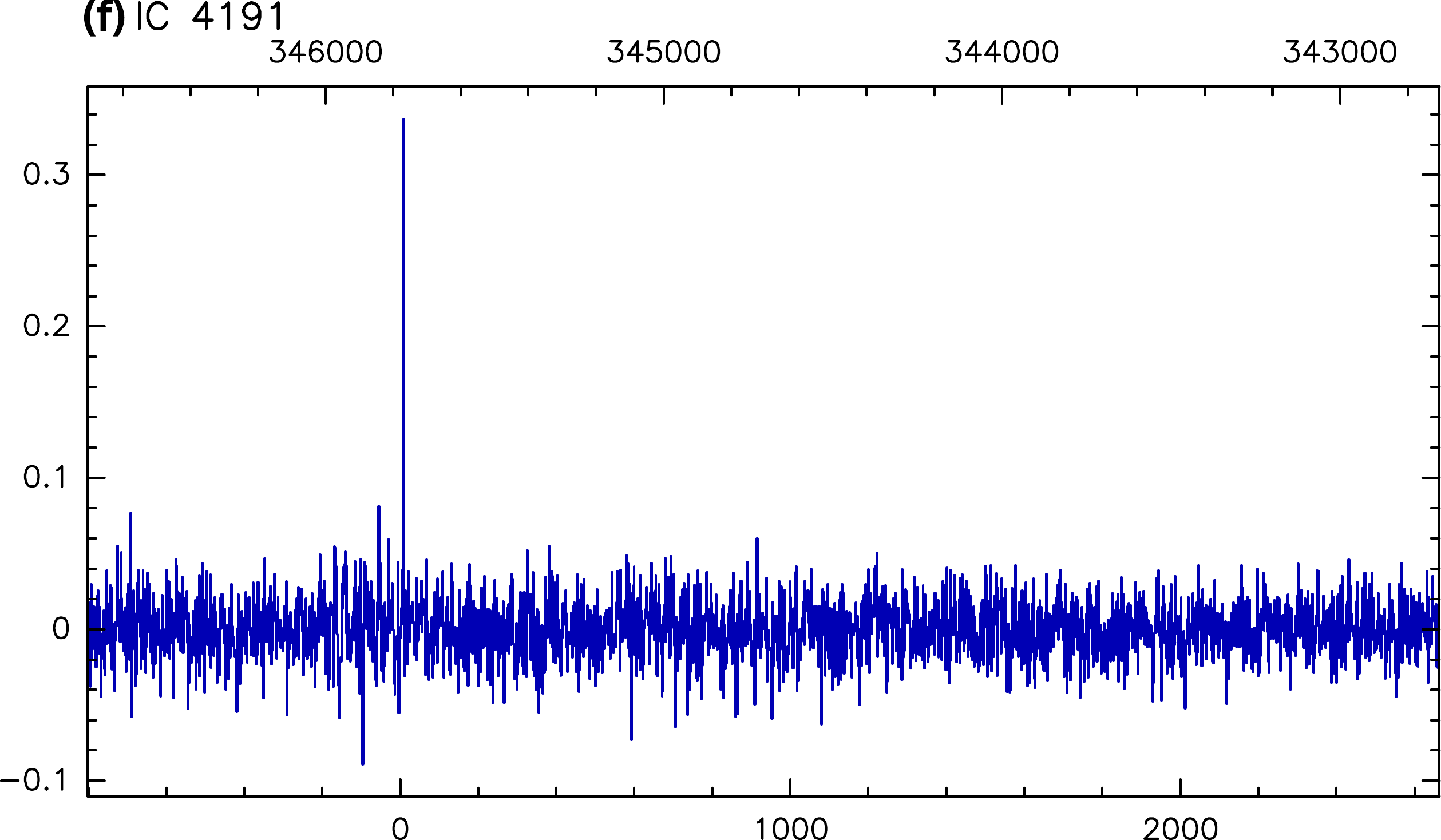}}
\hbox{
\centering
\includegraphics[width=6cm, height=5.5cm]{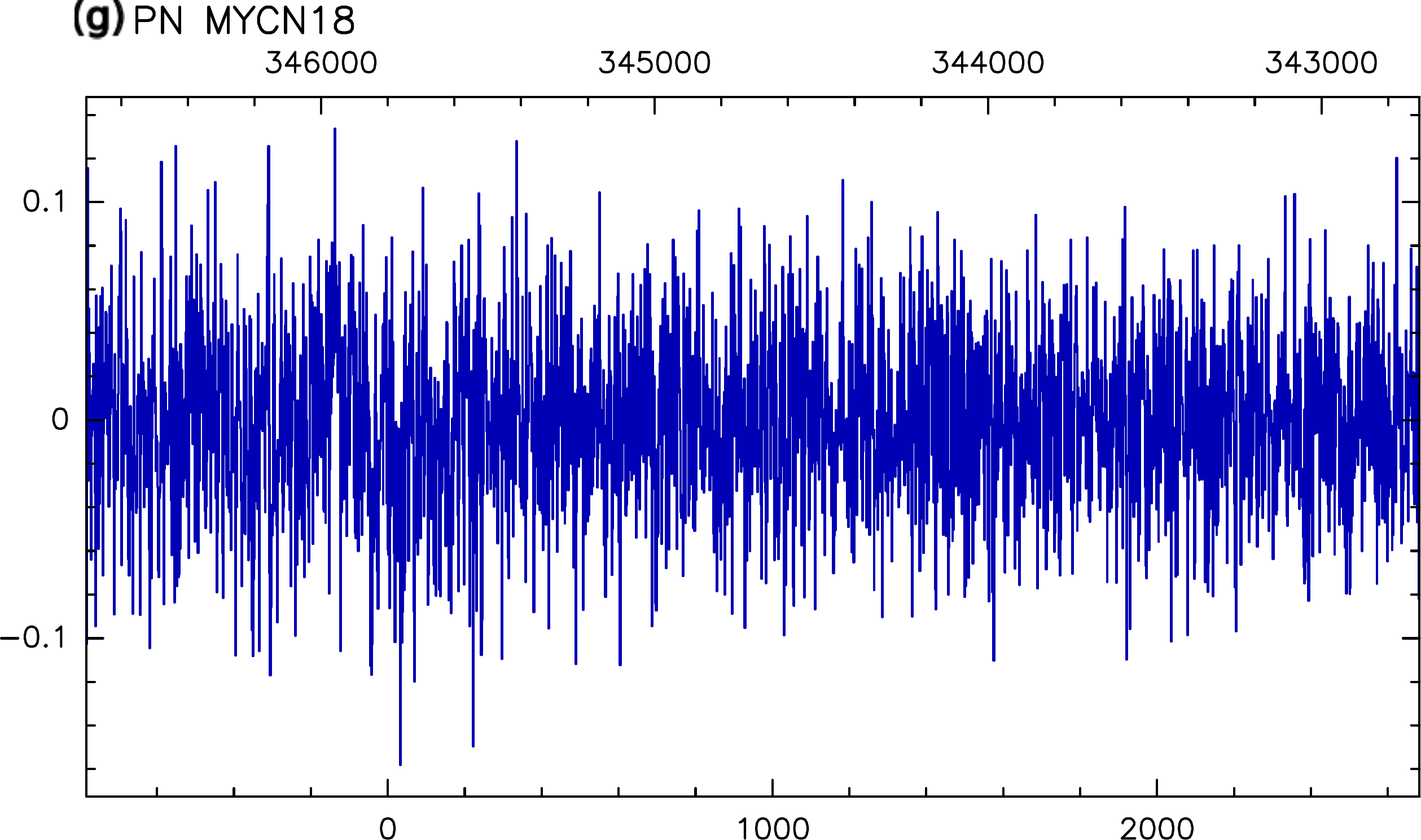}
\includegraphics[width=6cm, height=5.5cm]{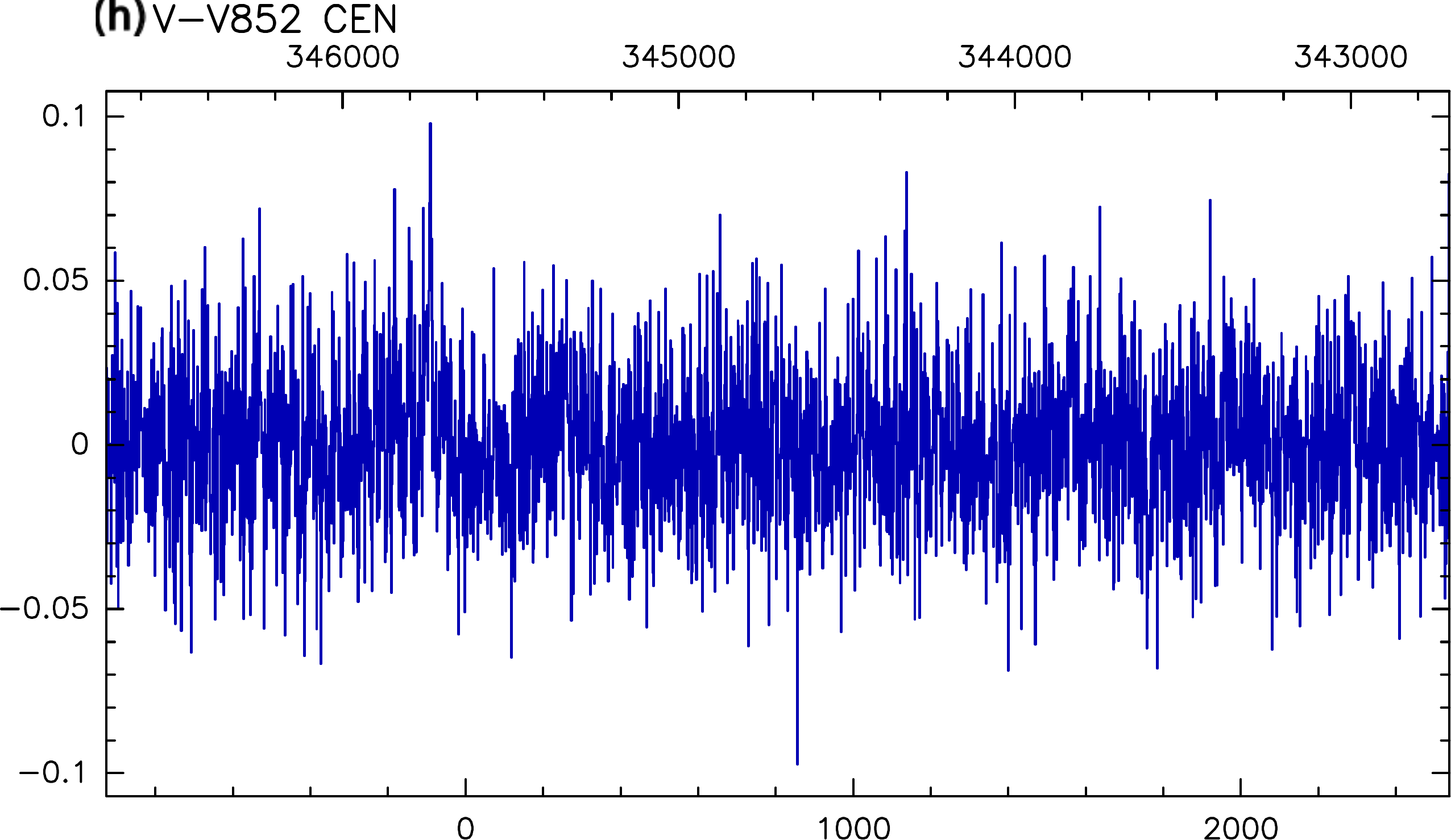}
\includegraphics[width=6cm, height=5.5cm]{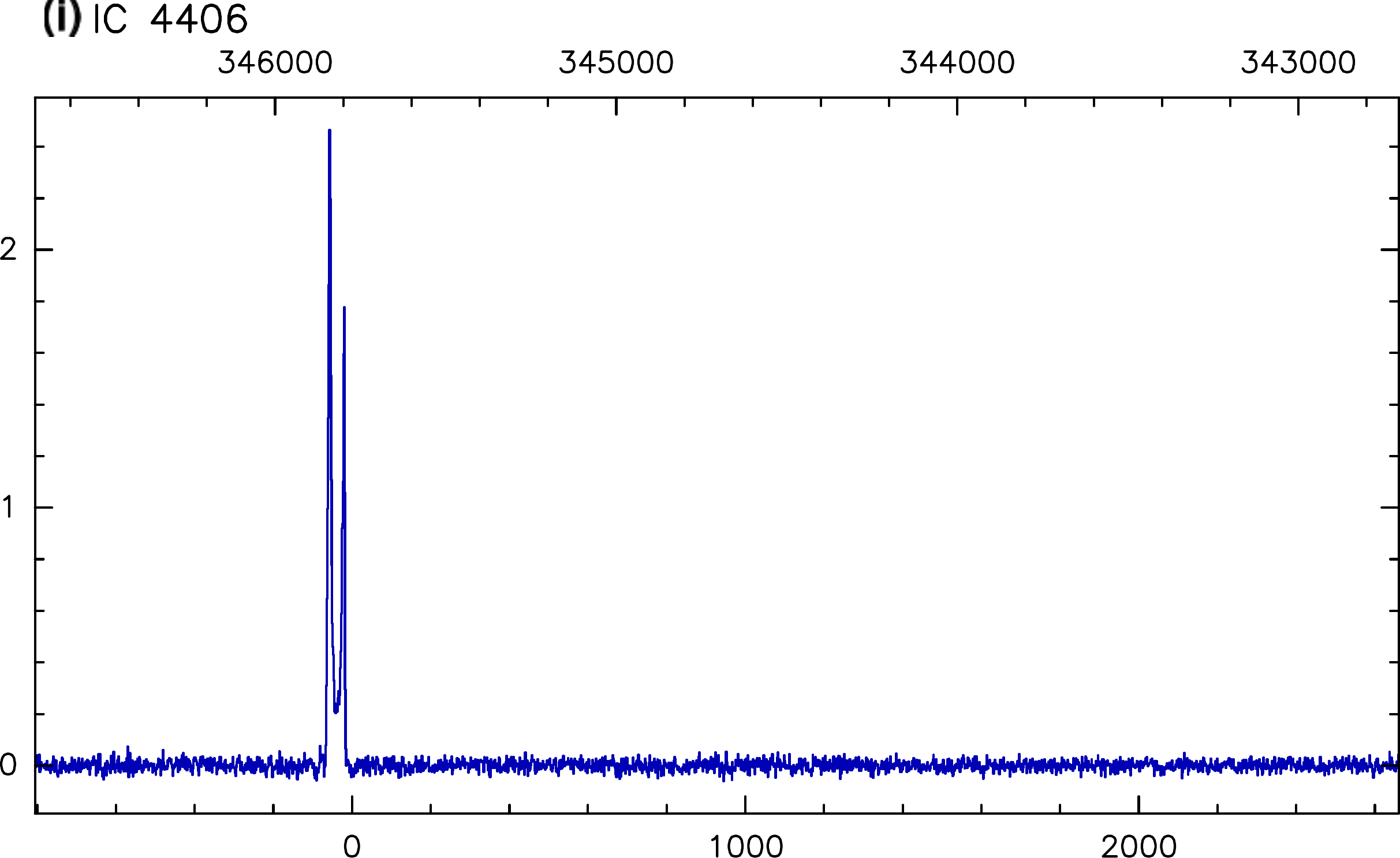}}
\caption[]{pPNe and PNe observations using the APEX telescope.}
\label{Fig3}
\end{figure*} 

\begin{figure*}
\vspace{2cm}
\centering
\hbox{
\centering
\includegraphics[width=6cm, height=5.5cm]{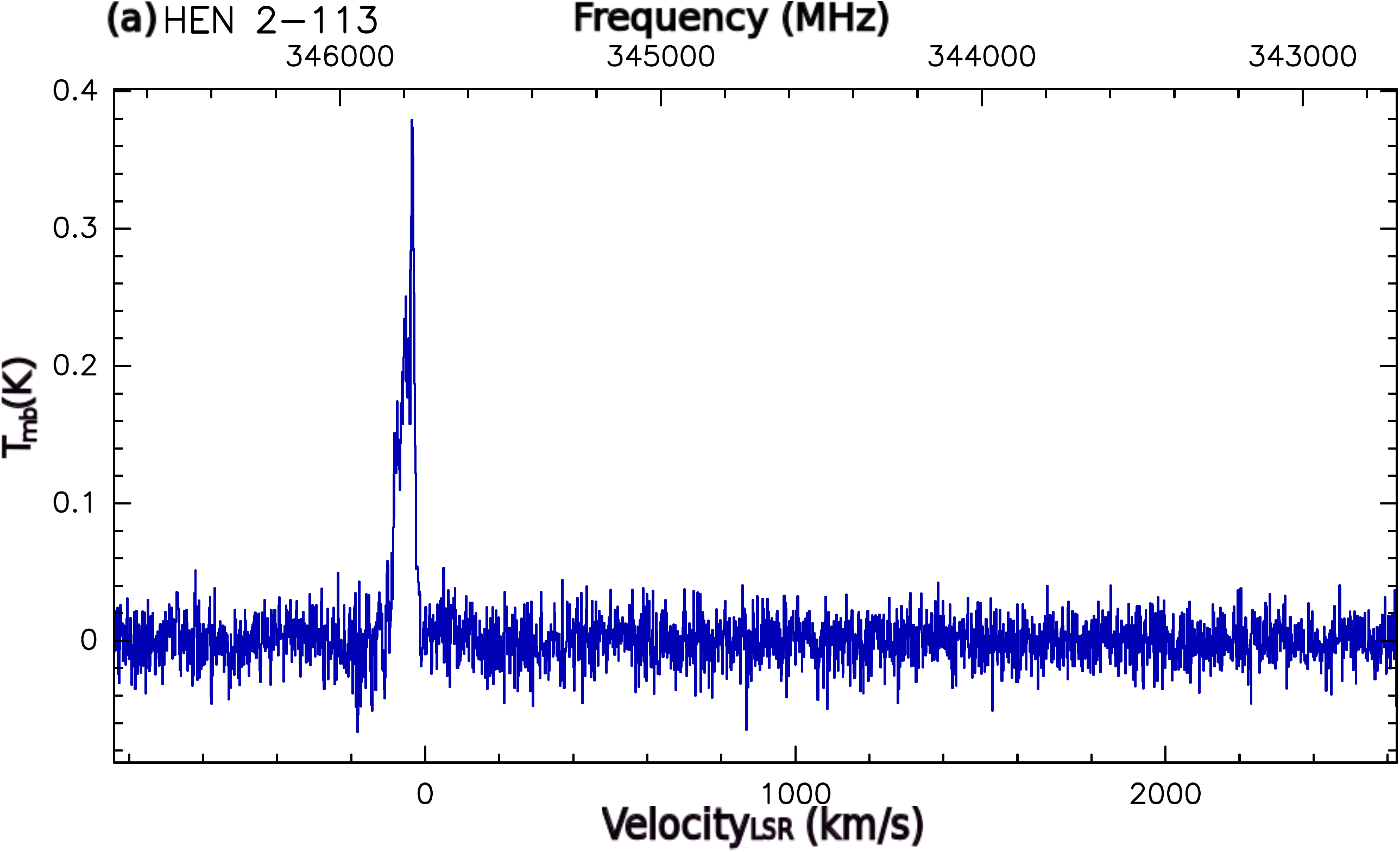}
\includegraphics[width=6cm, height=5.5cm]{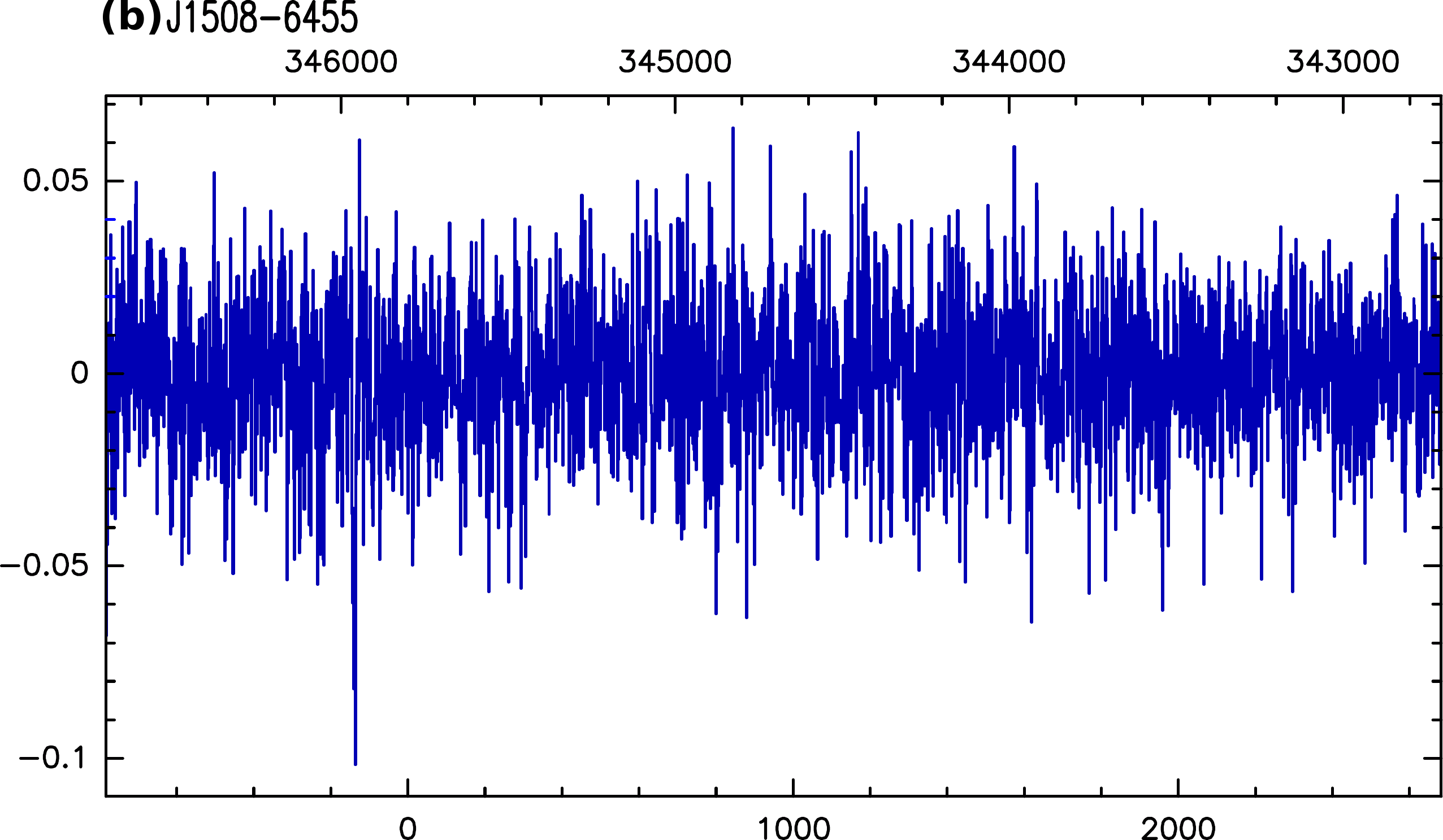}
\includegraphics[width=6cm, height=5.5cm]{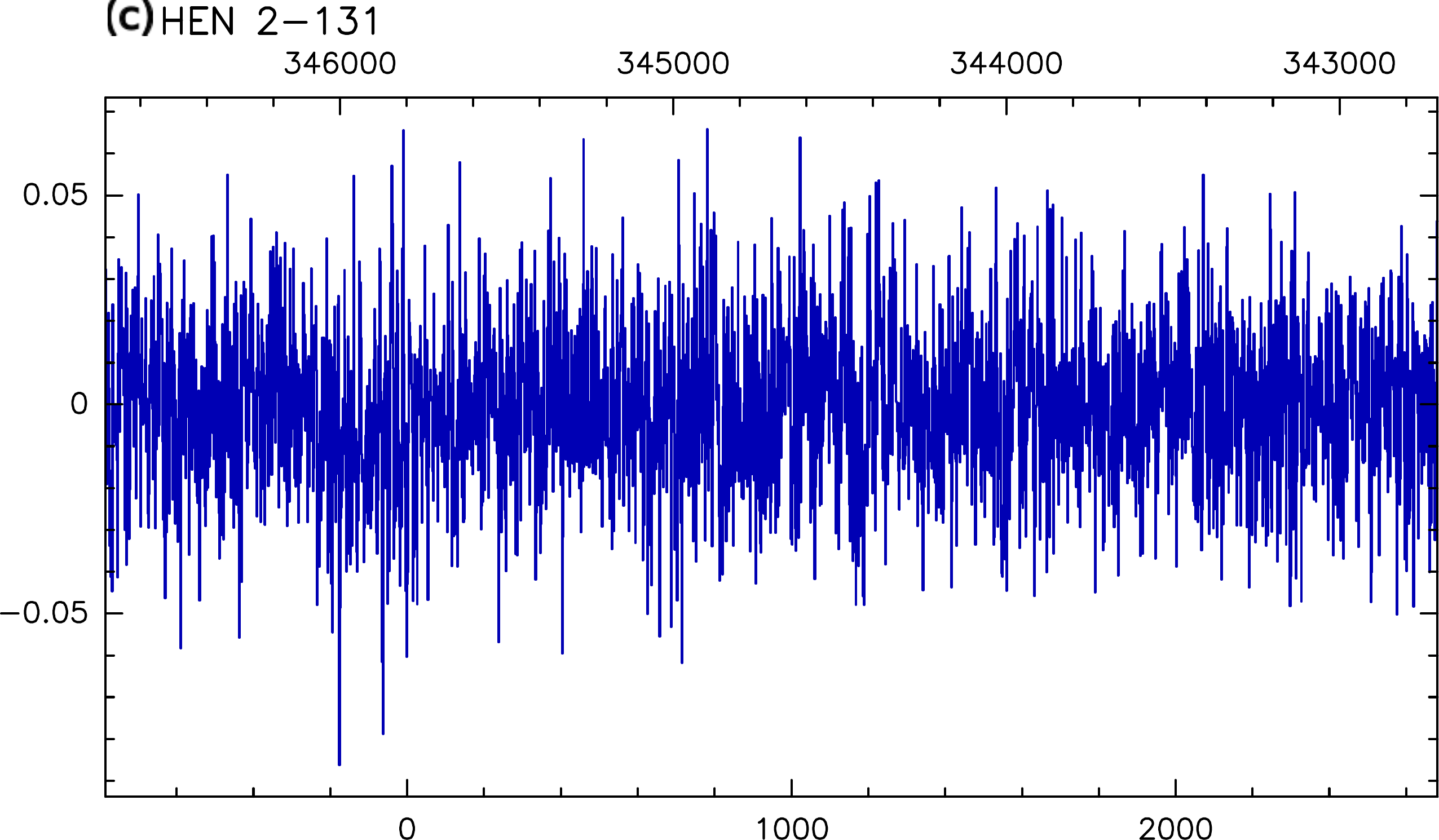}
\vspace{1cm}}
\hbox{
\centering
\vspace{1cm}
\includegraphics[width=6cm, height=5.5cm]{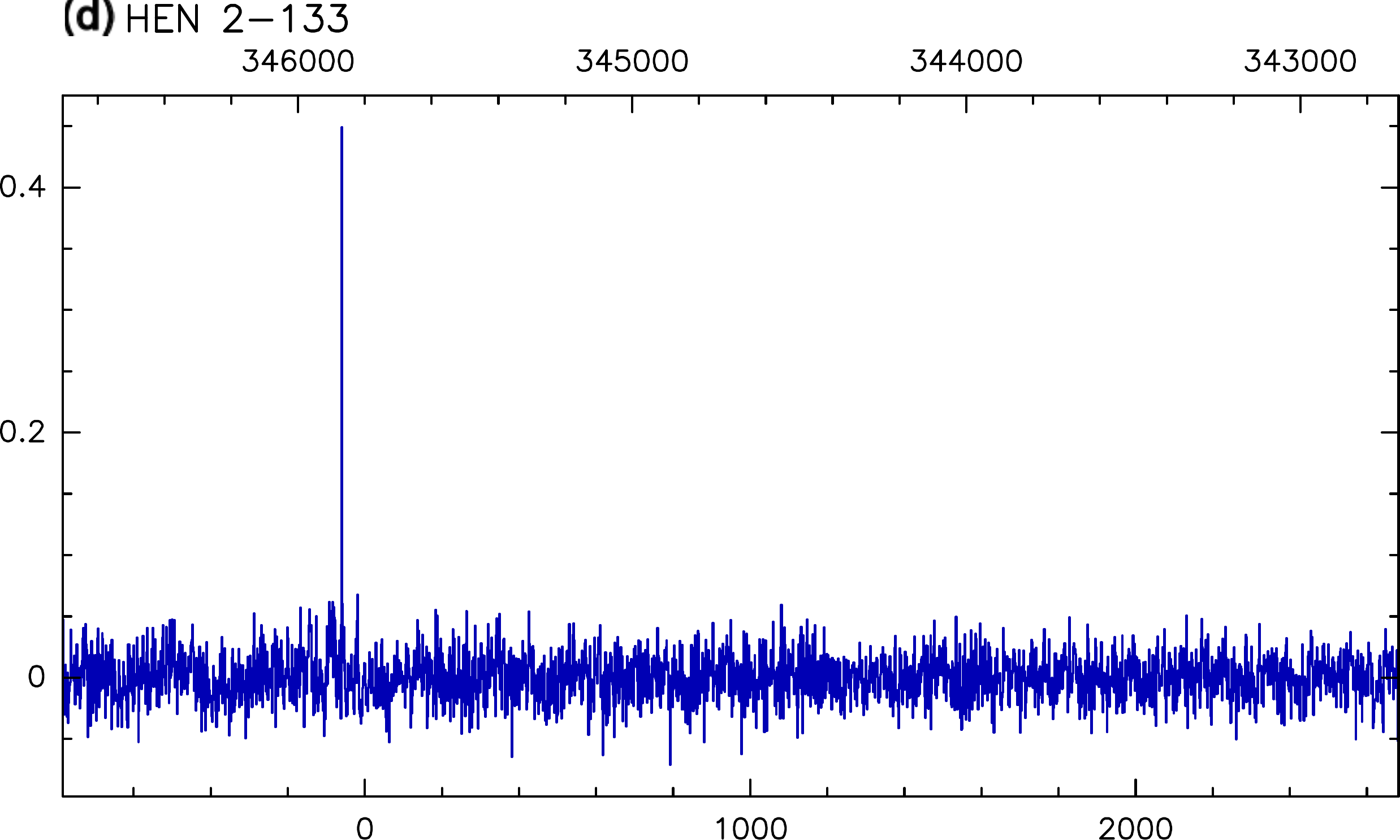}
\includegraphics[width=6cm, height=5.5cm]{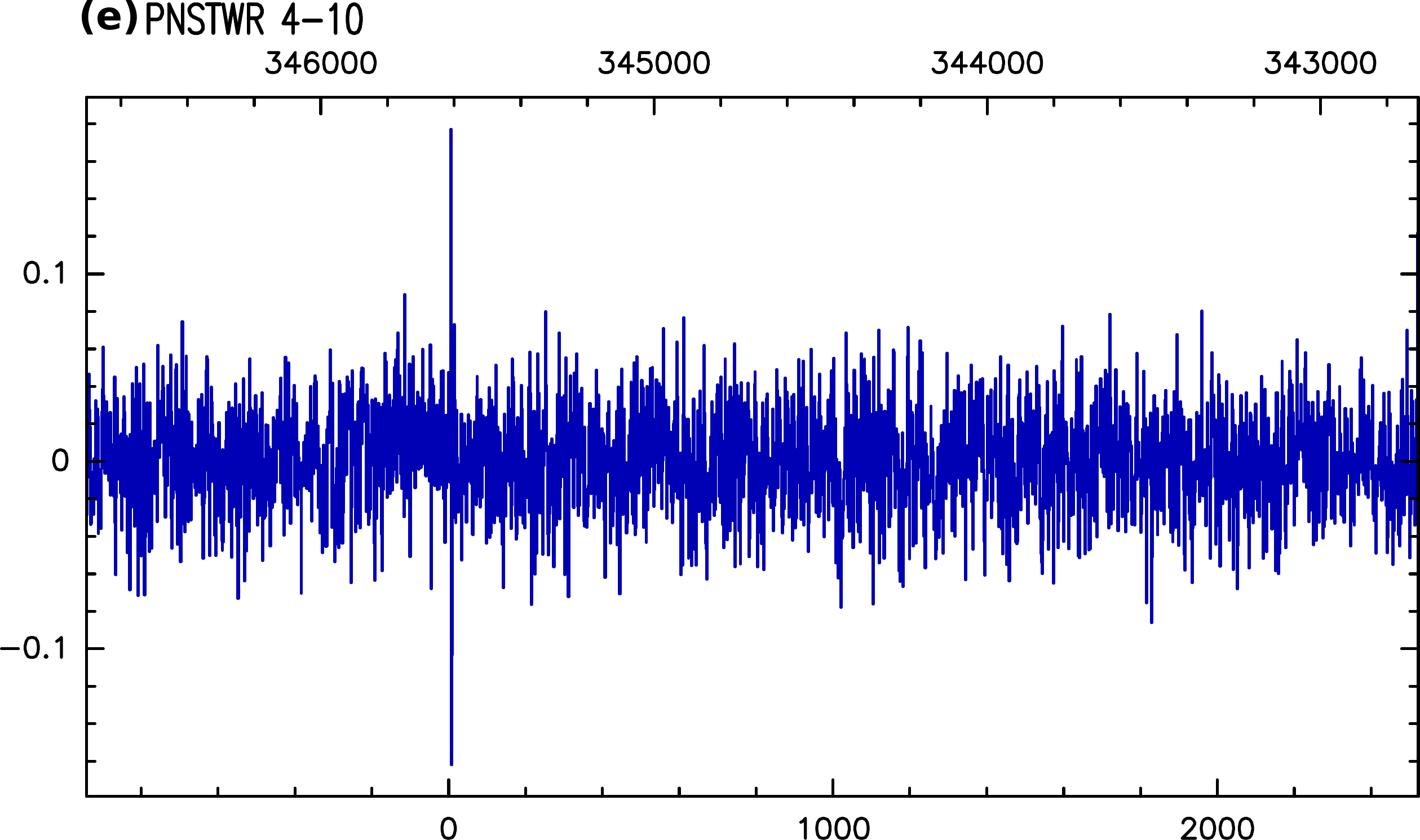}
\includegraphics[width=6cm, height=5.5cm]{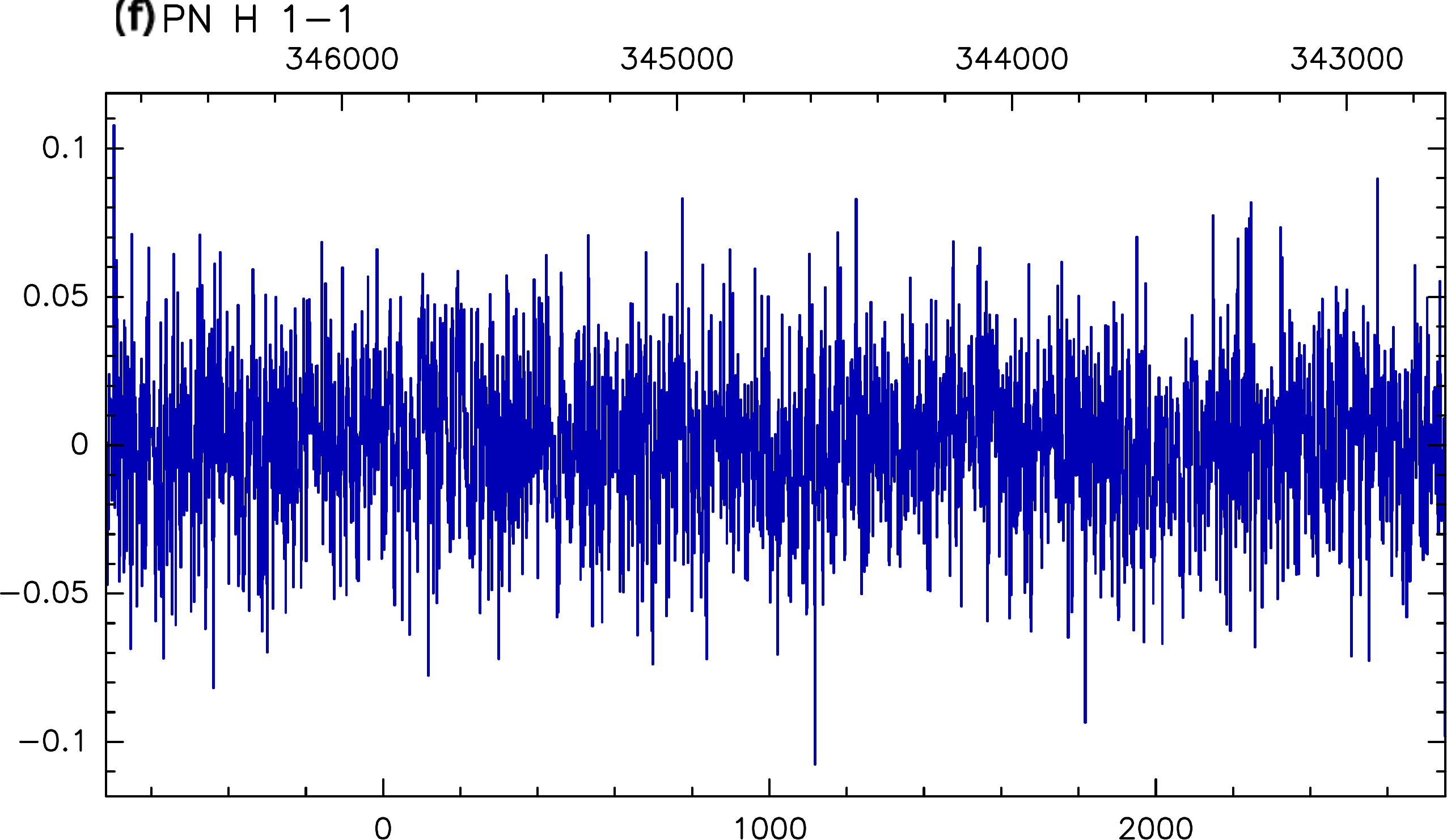}}
\hbox{
\centering
\includegraphics[width=6cm, height=5.5cm]{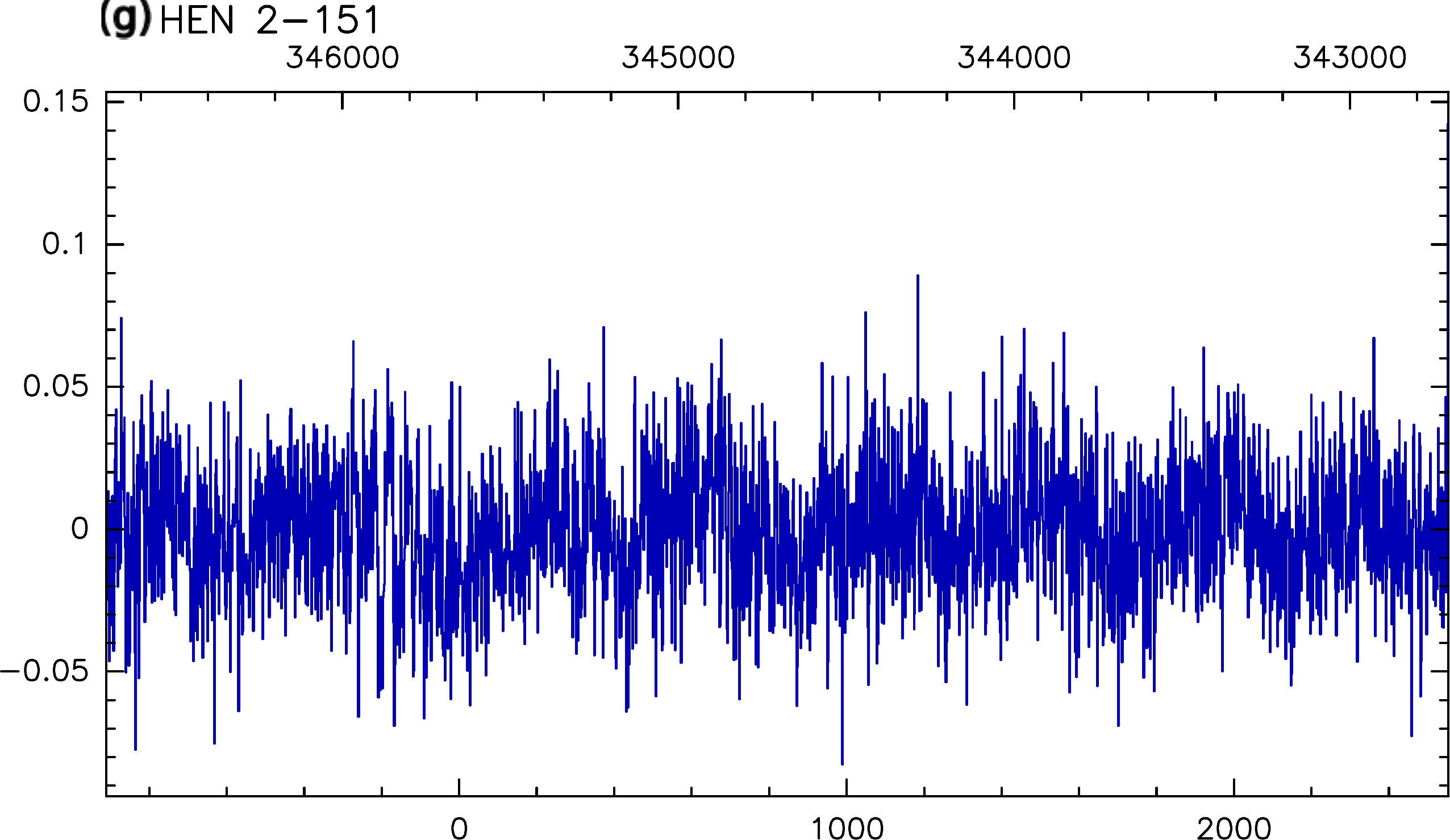}
\includegraphics[width=6cm, height=5.5cm]{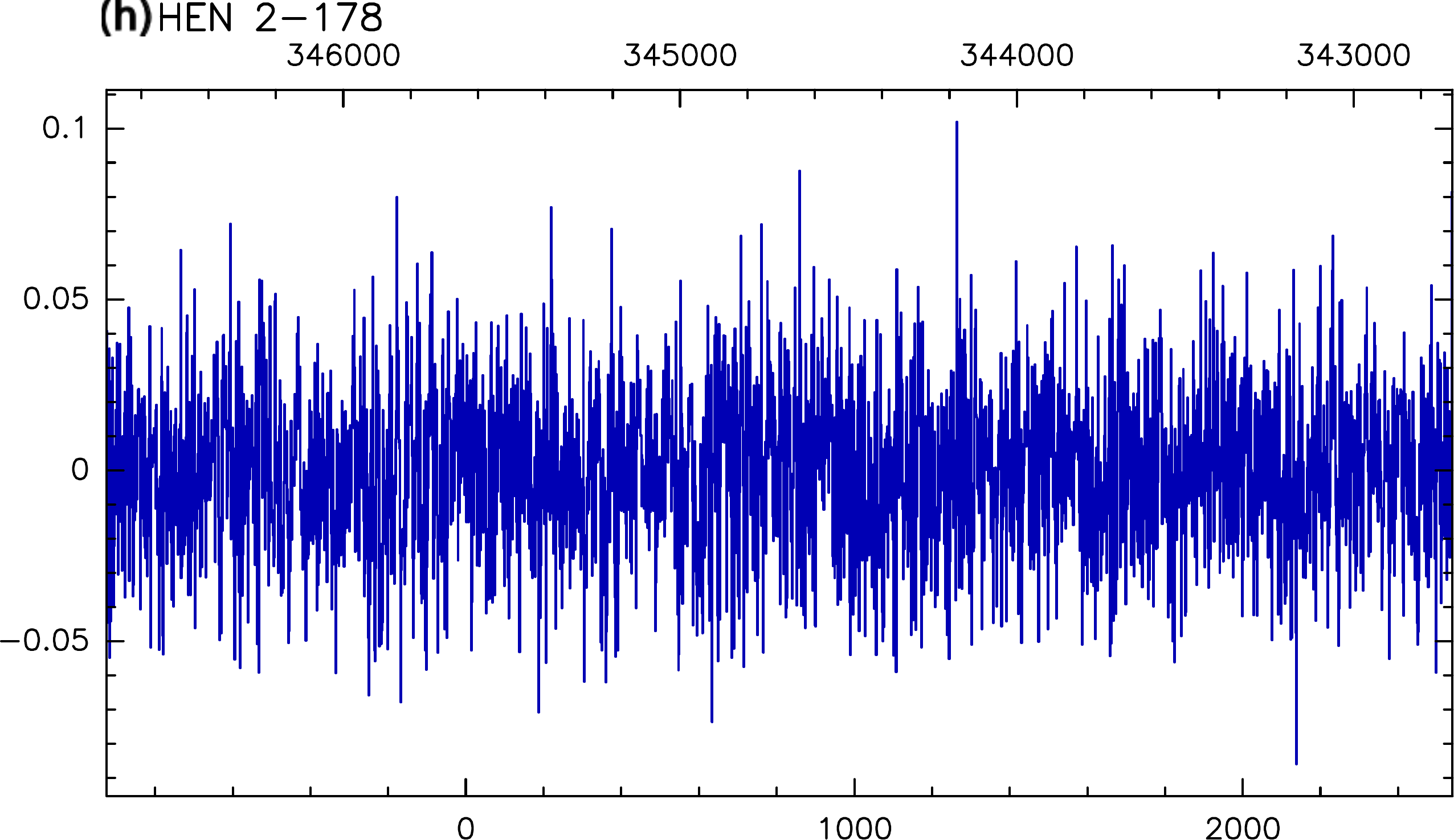}
\includegraphics[width=6cm, height=5.5cm]{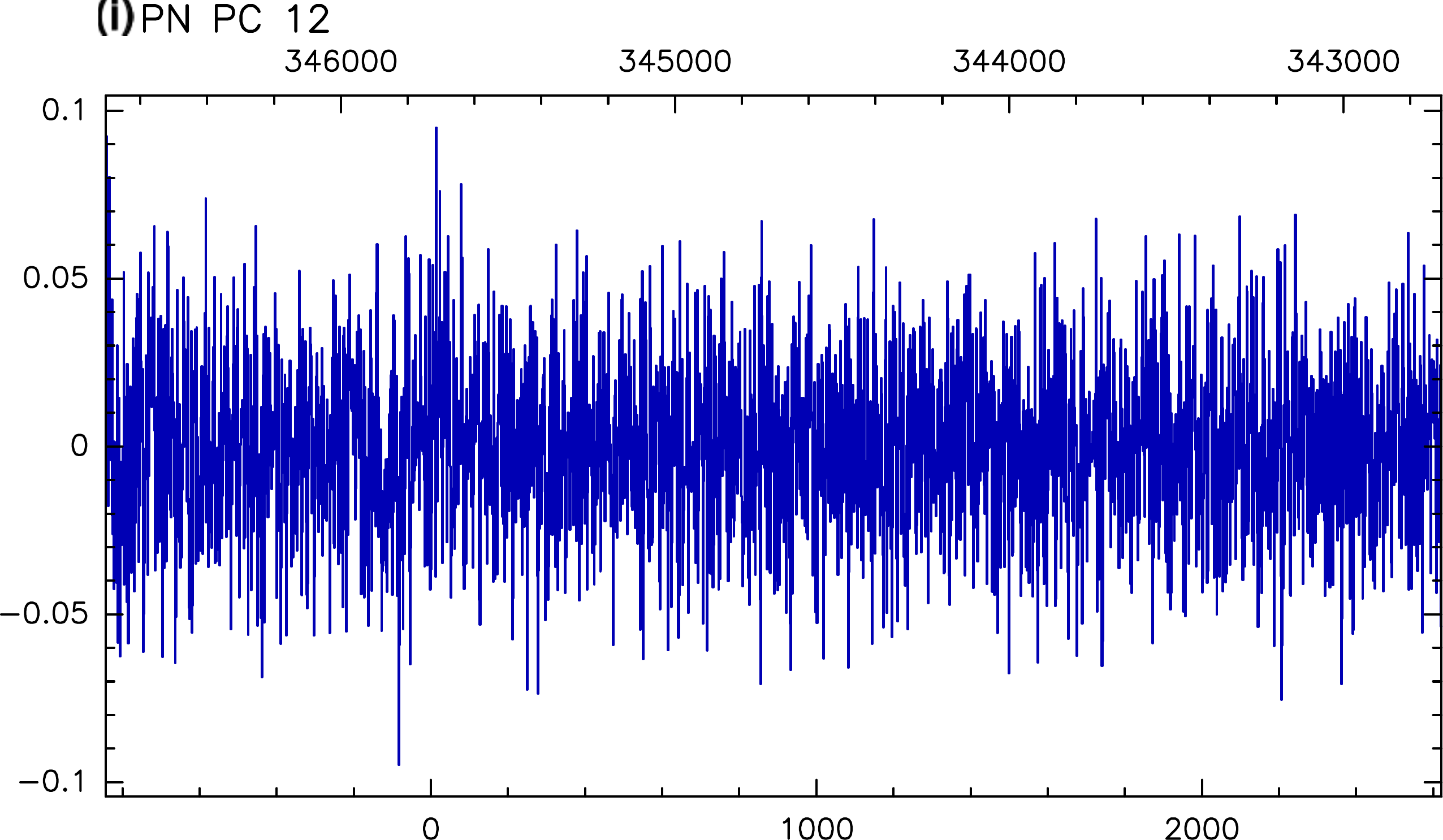}}
\caption[]{pPNe and PNe observations using the APEX telescope.}
\label{Fig4}
\end{figure*} 

\begin{figure*}
\vspace{2cm}
\centering
\hbox{
\centering
\includegraphics[width=6cm, height=5.5cm]{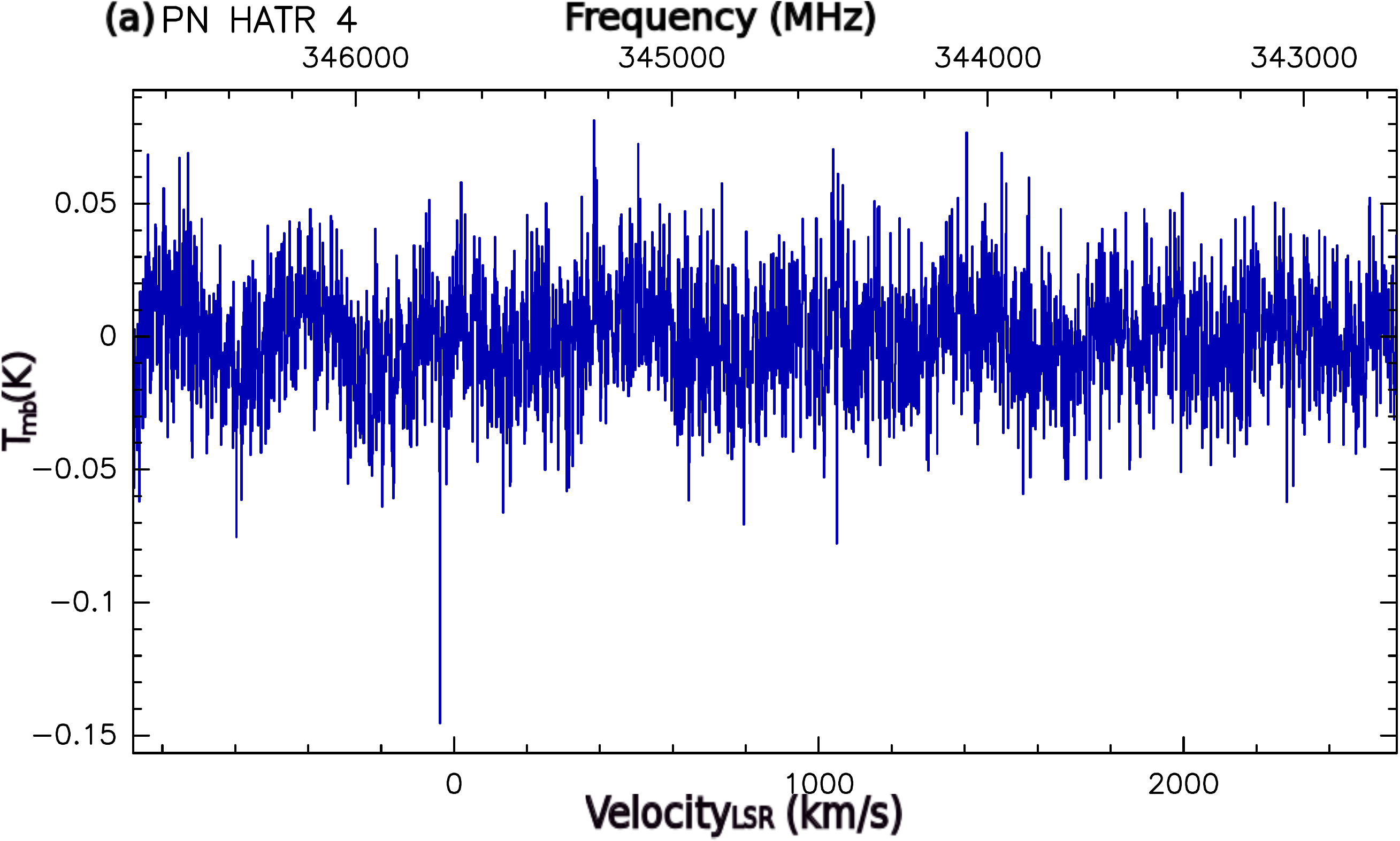}
\includegraphics[width=6cm, height=5.5cm]{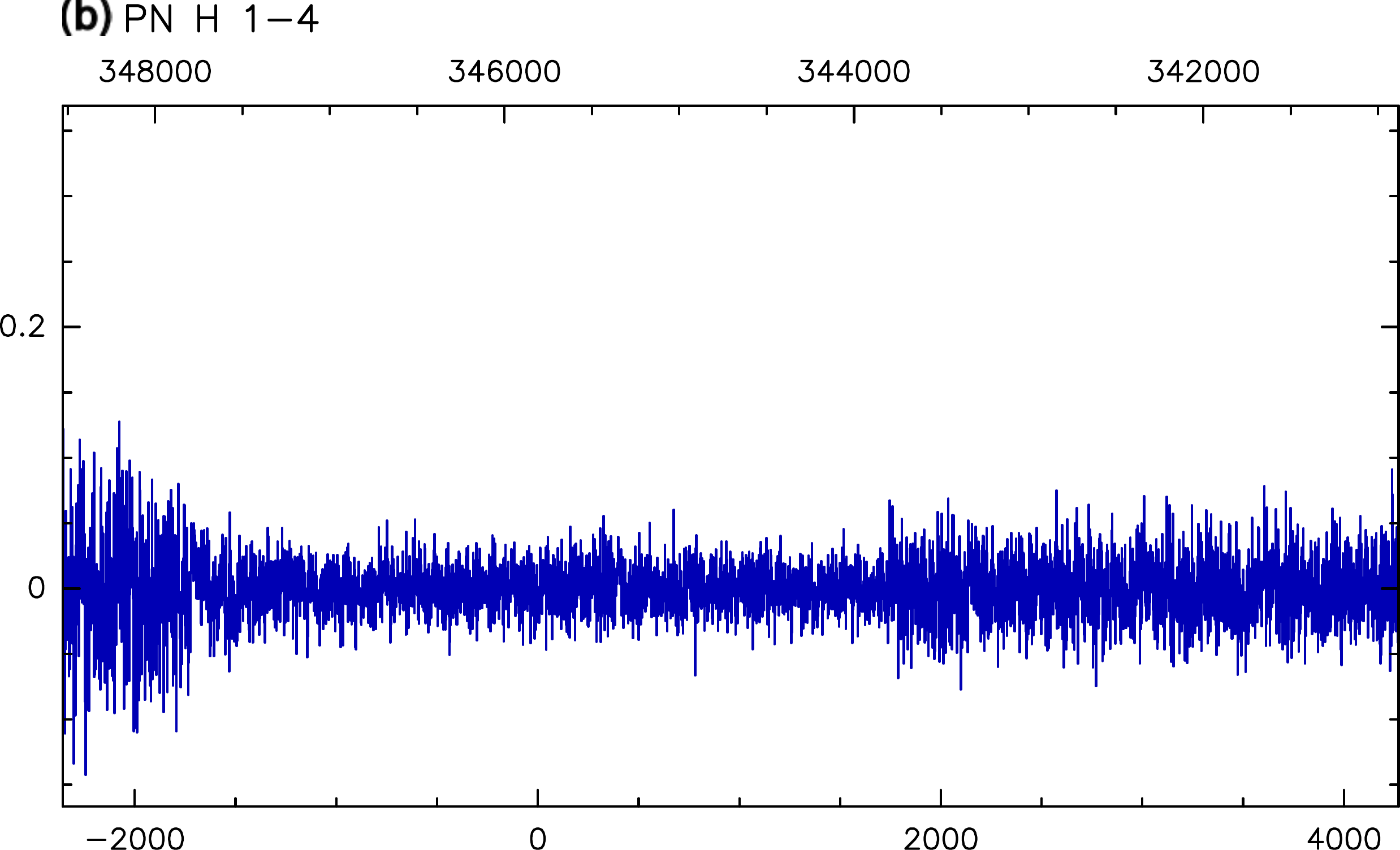}
\includegraphics[width=6cm, height=5.5cm]{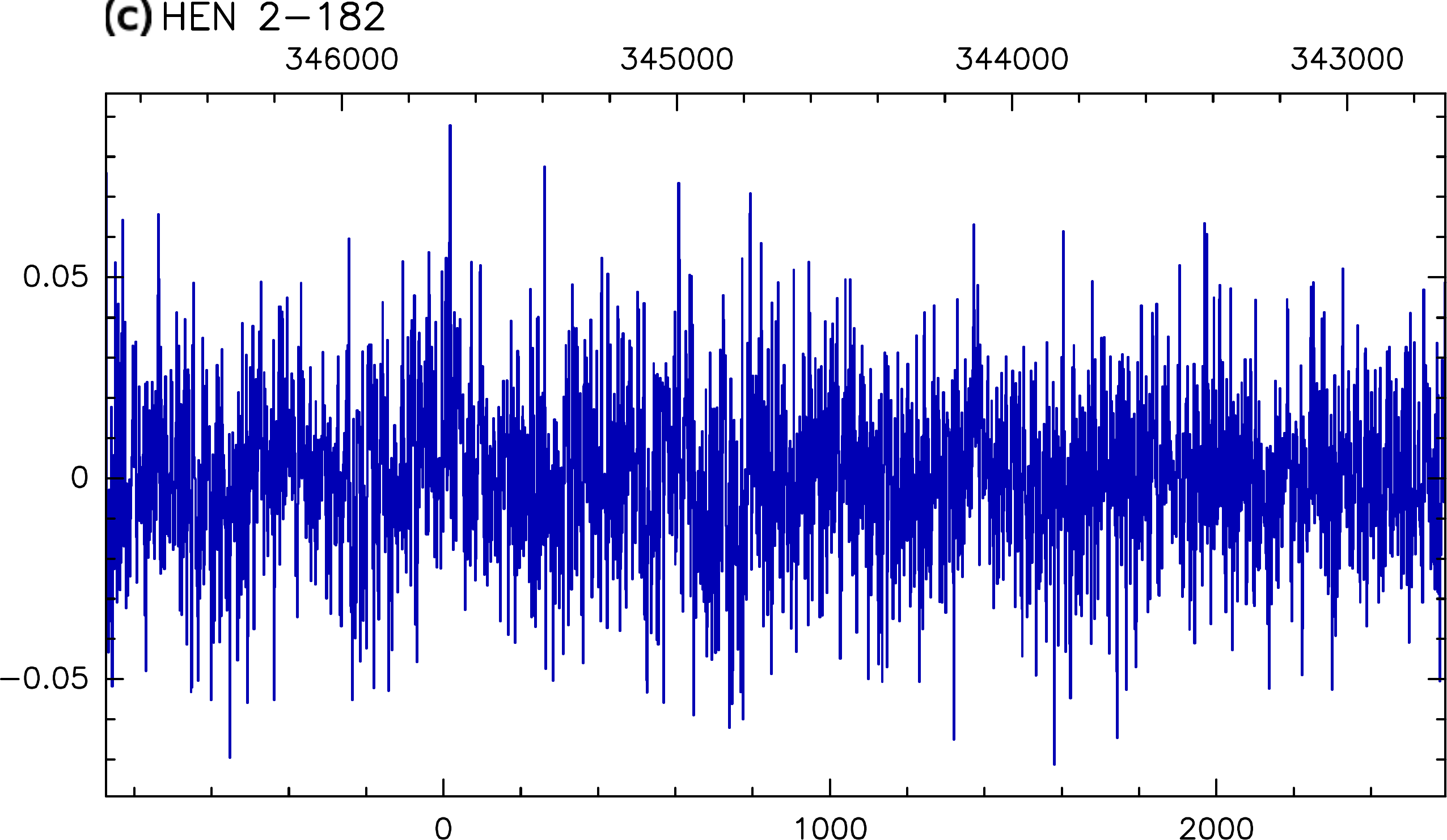}
\vspace{1cm}}
\hbox{
\centering
\vspace{1cm}
\includegraphics[width=6cm, height=5.5cm]{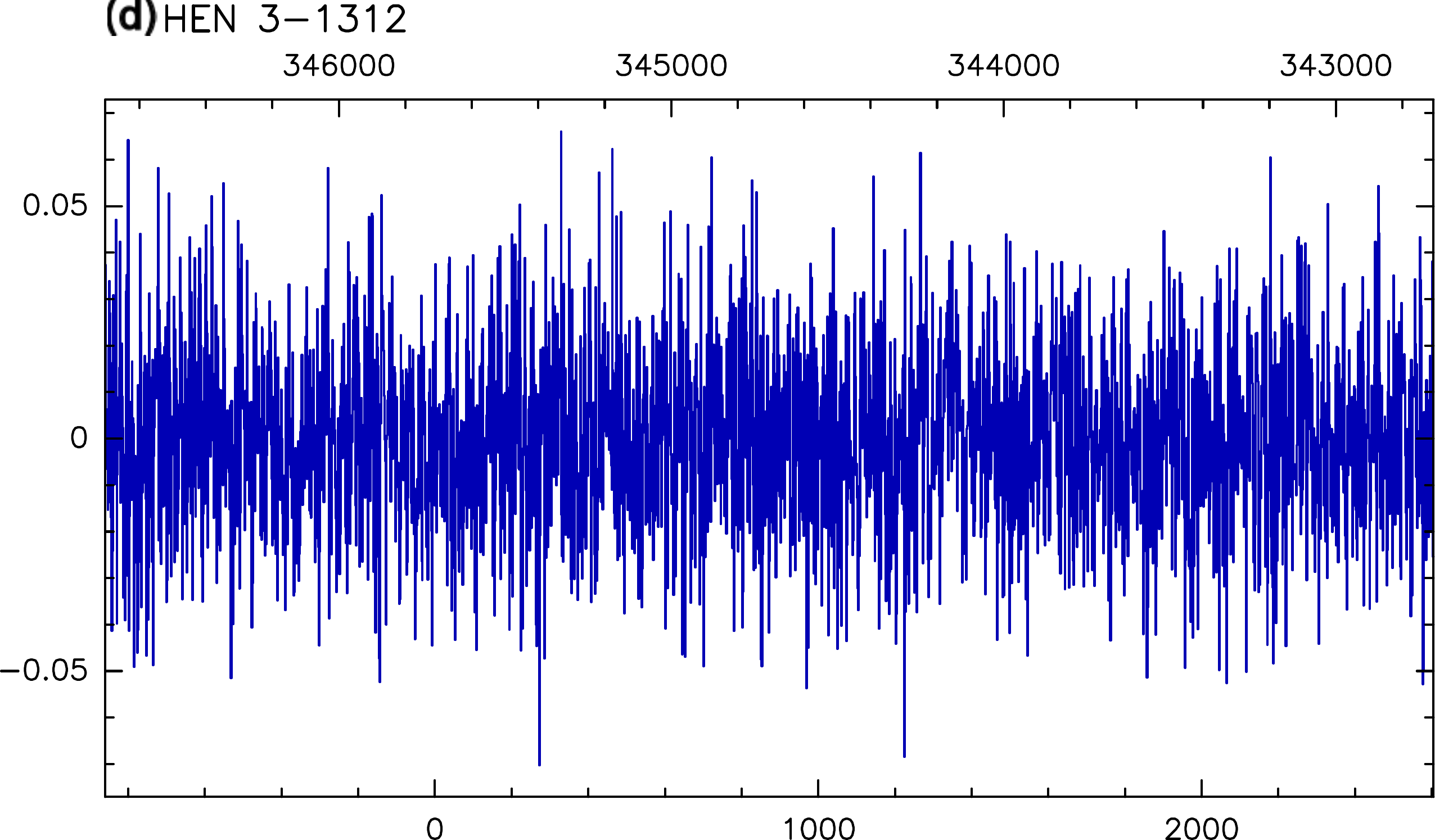}
\includegraphics[width=6cm, height=5.5cm]{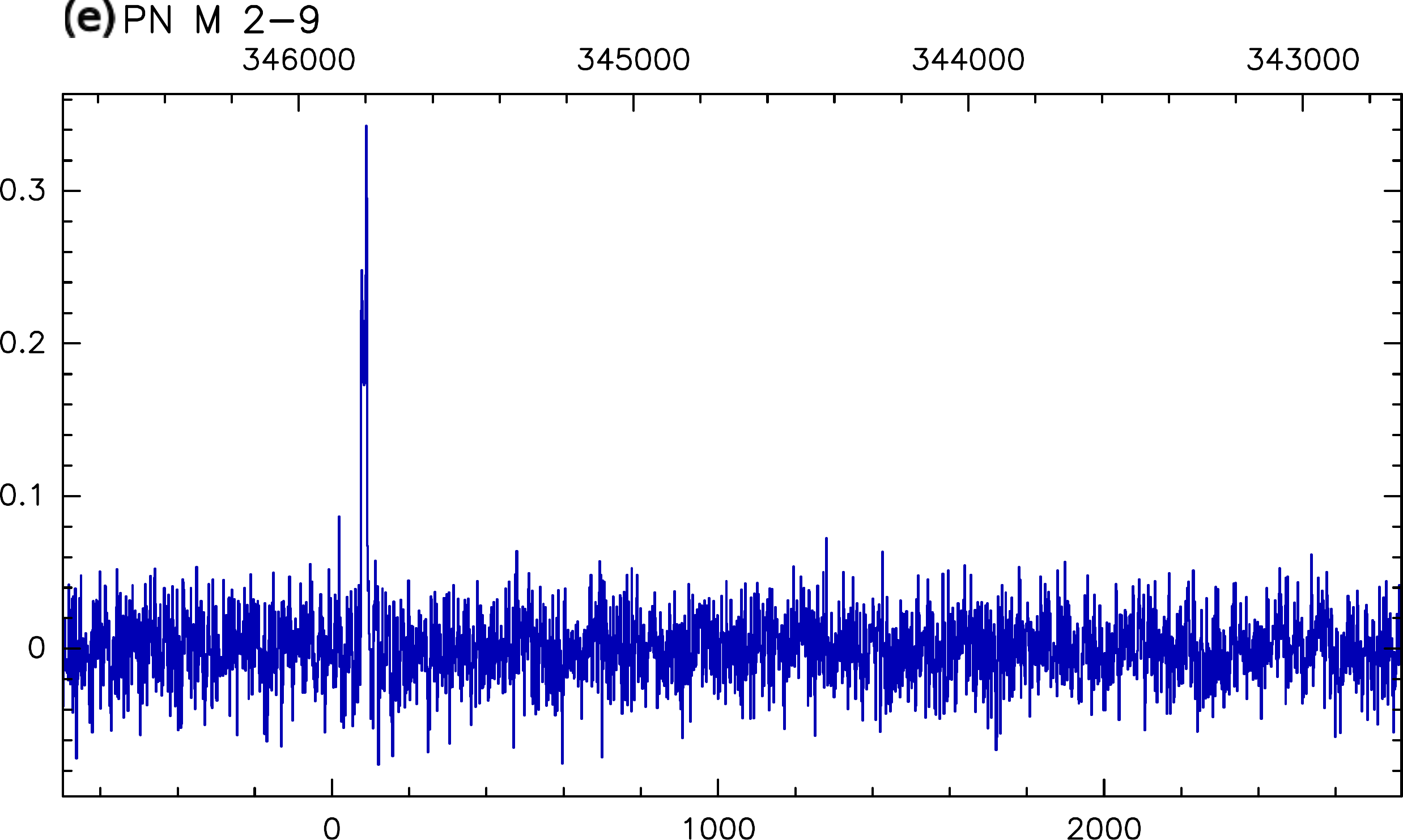}
\includegraphics[width=6cm, height=5.5cm]{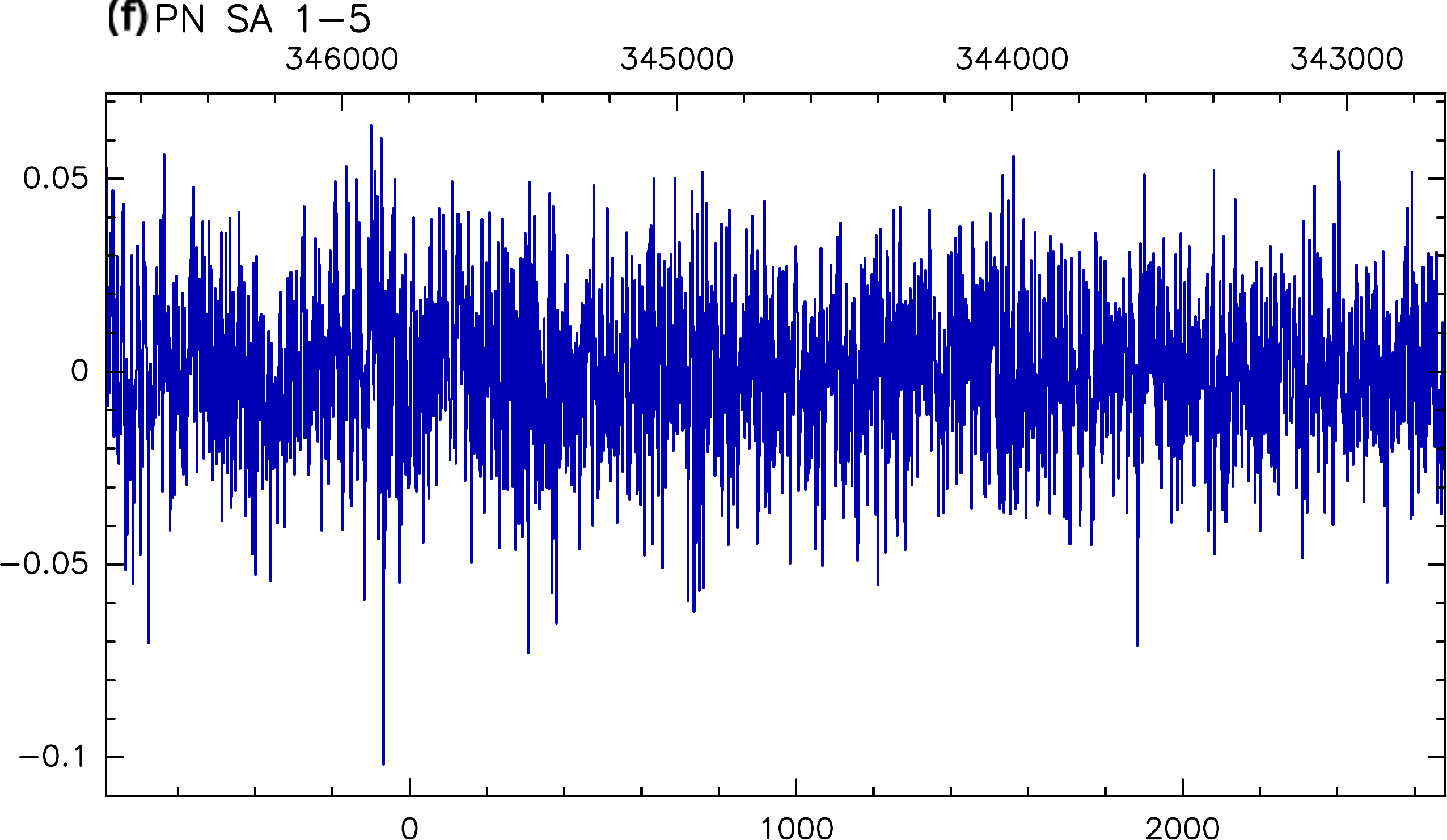}}
\hbox{
\centering
\includegraphics[width=6cm, height=5.5cm]{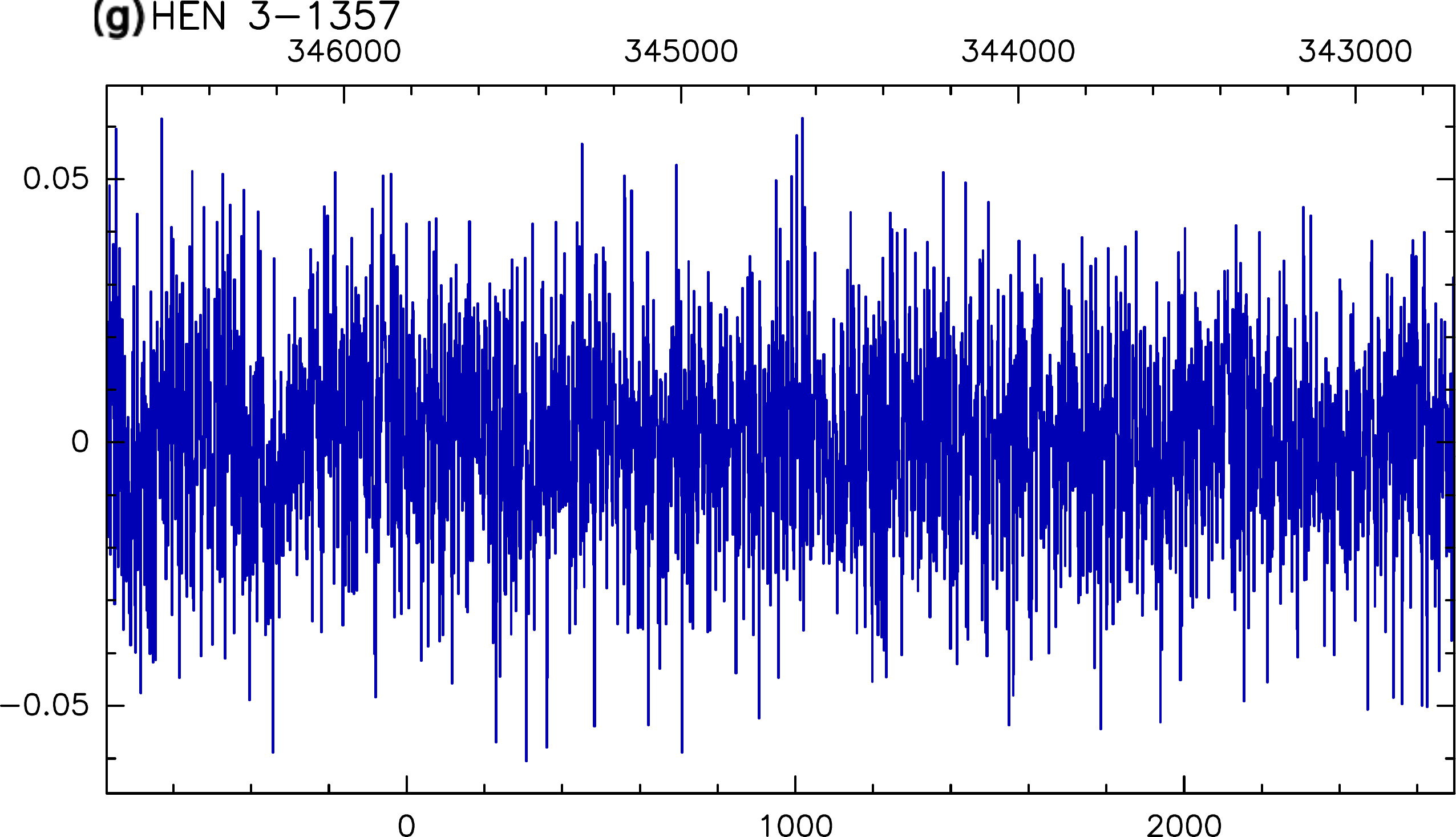}
\includegraphics[width=6cm, height=5.5cm]{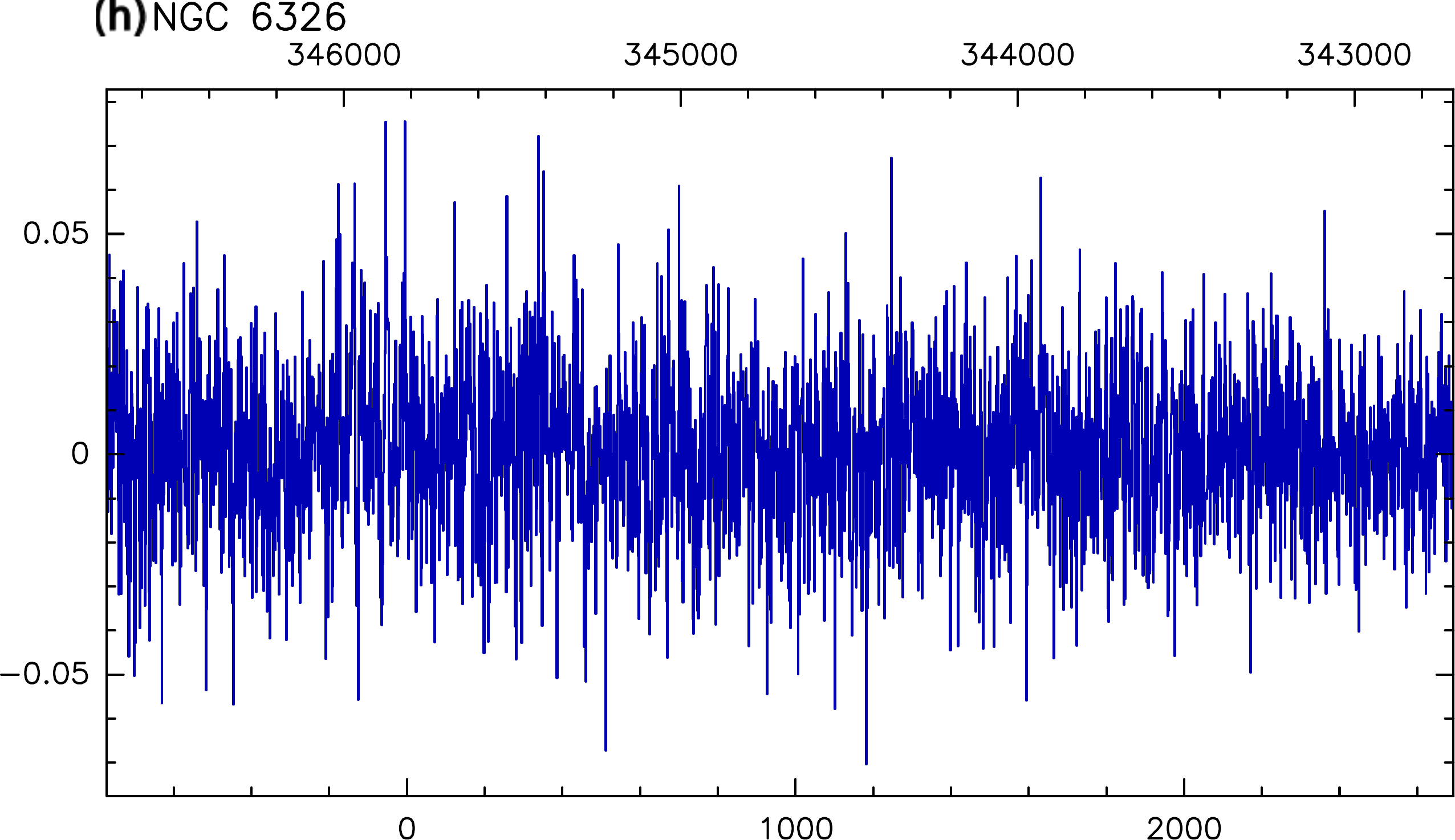}
\includegraphics[width=6cm, height=5.5cm]{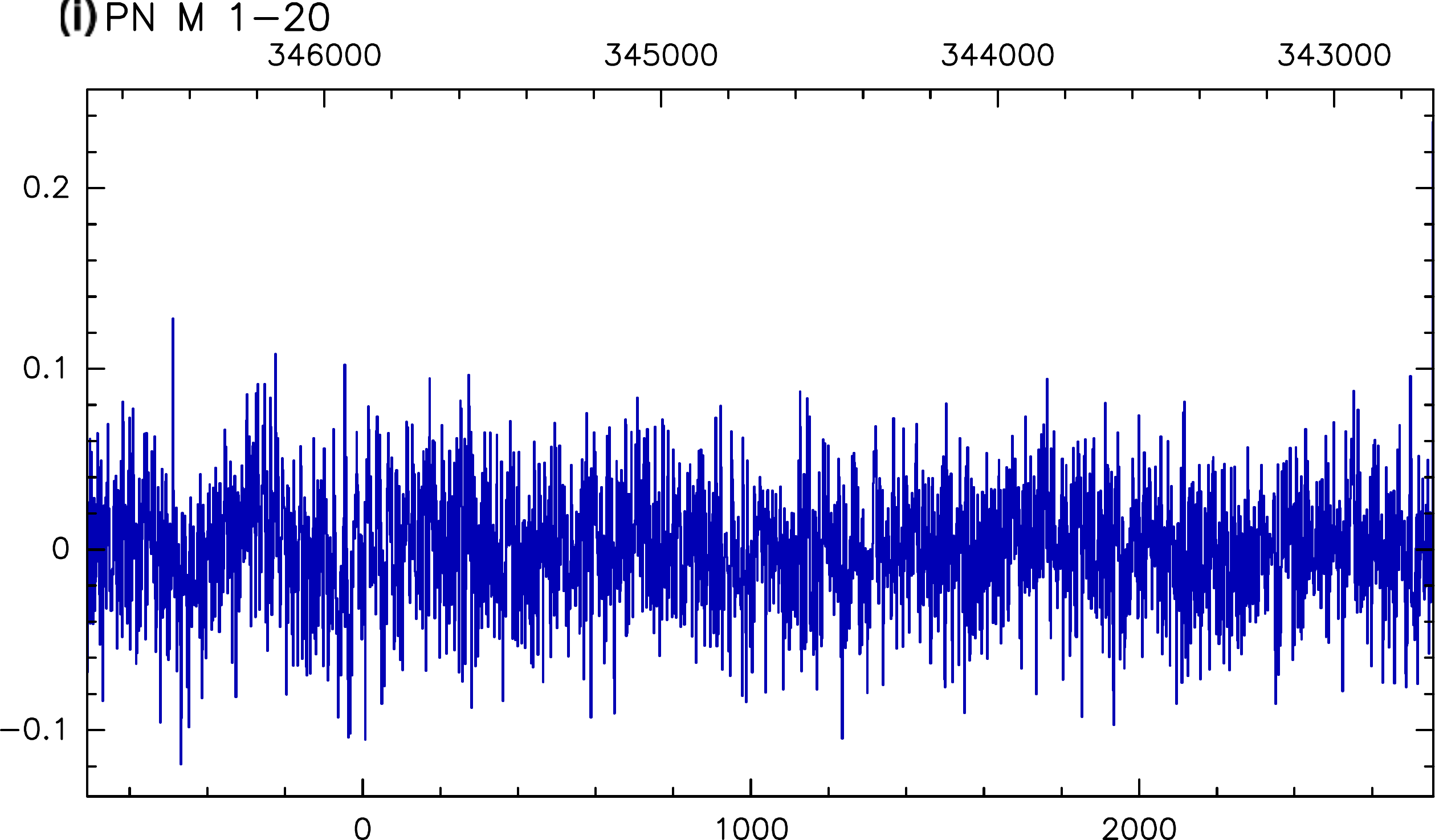}}
\caption[]{pPNe and PNe observations using the APEX telescope.}
\label{Fig5}
\end{figure*} 

\begin{figure*}
\vspace{2cm}
\centering
\hbox{
\centering
\includegraphics[width=6cm, height=5.5cm]{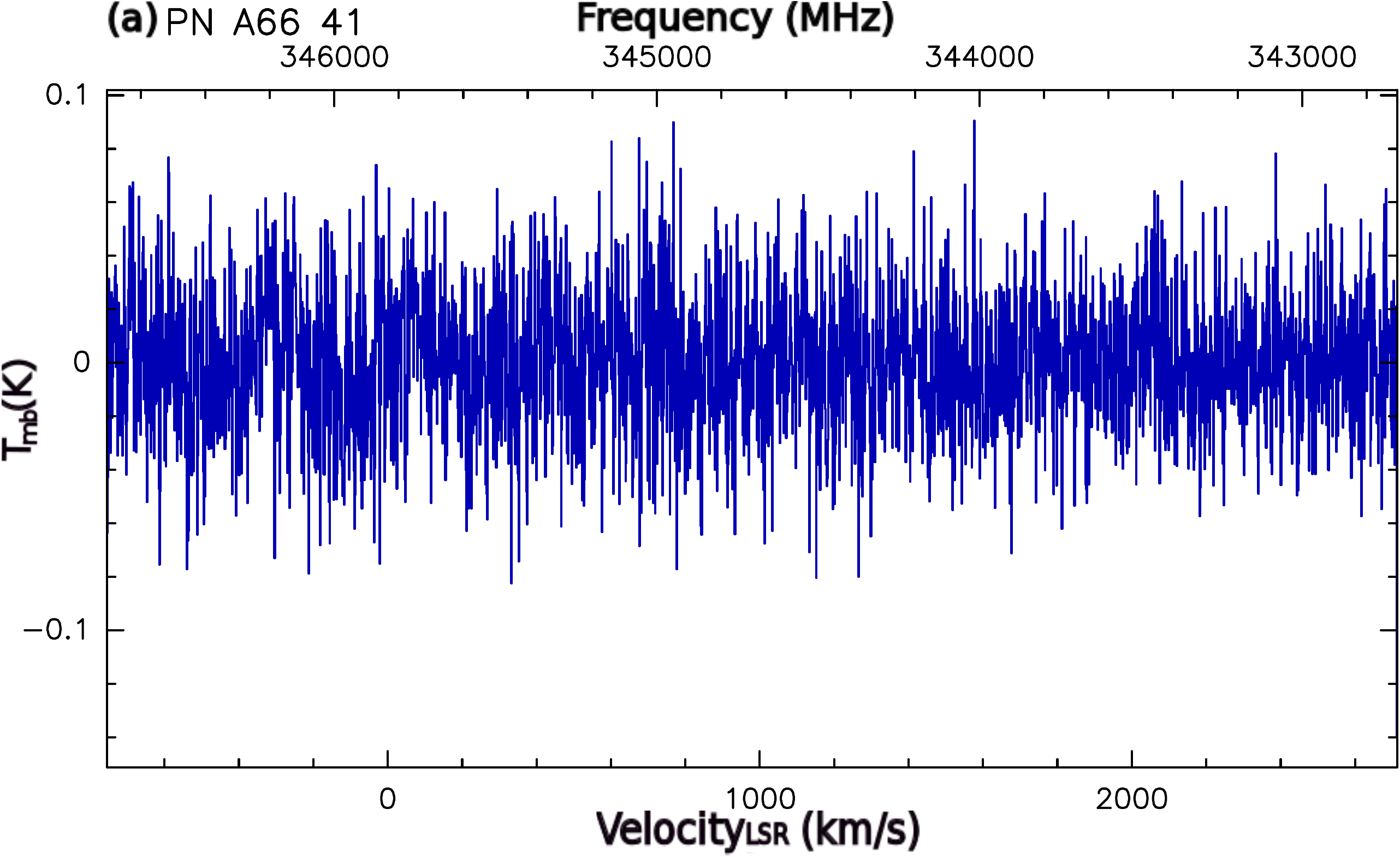}
\includegraphics[width=6cm, height=5.5cm]{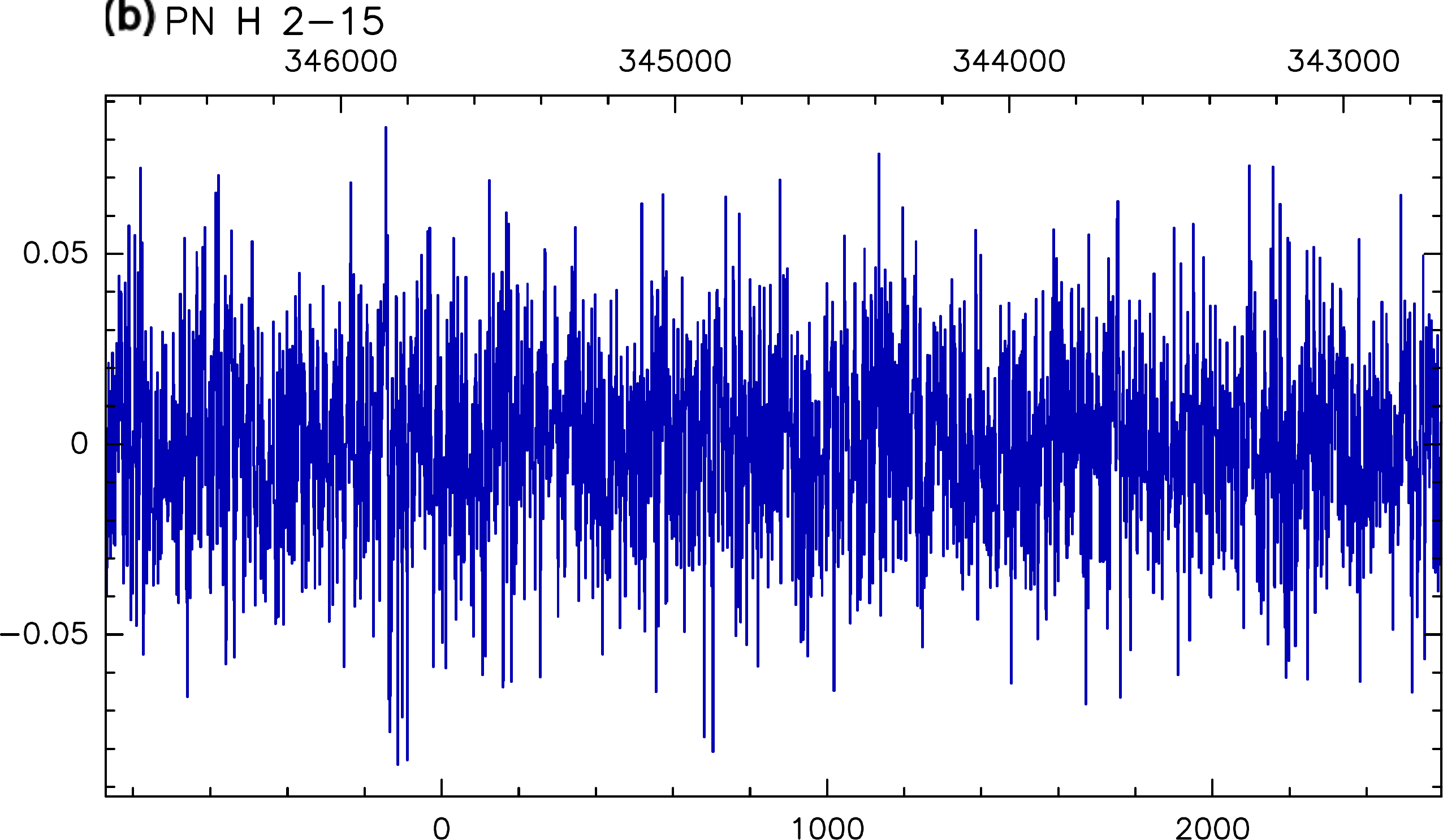}
\includegraphics[width=6cm, height=5.5cm]{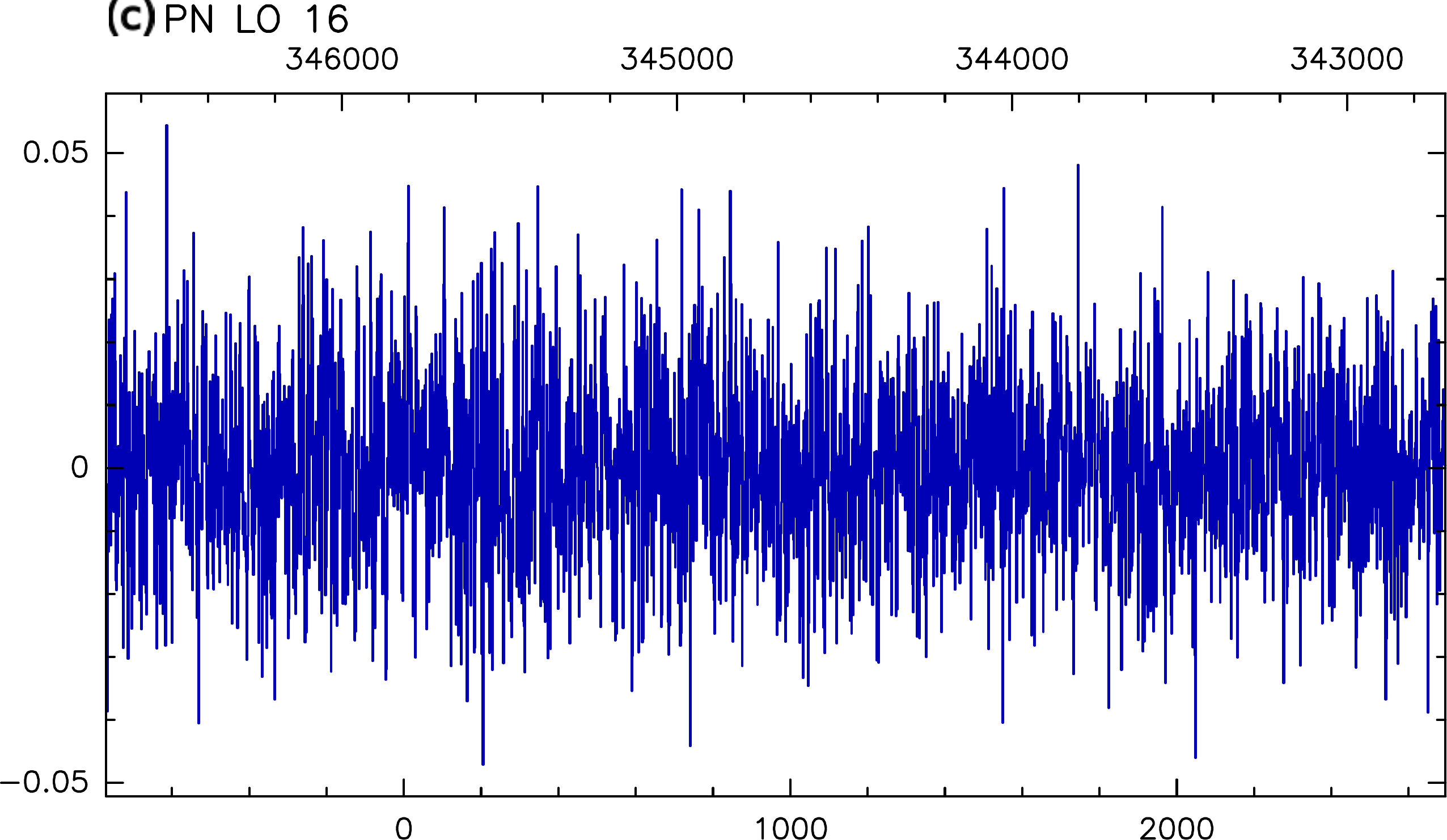}
\vspace{1cm}}
\hbox{
\centering
\vspace{1cm}
\includegraphics[width=6cm, height=5.5cm]{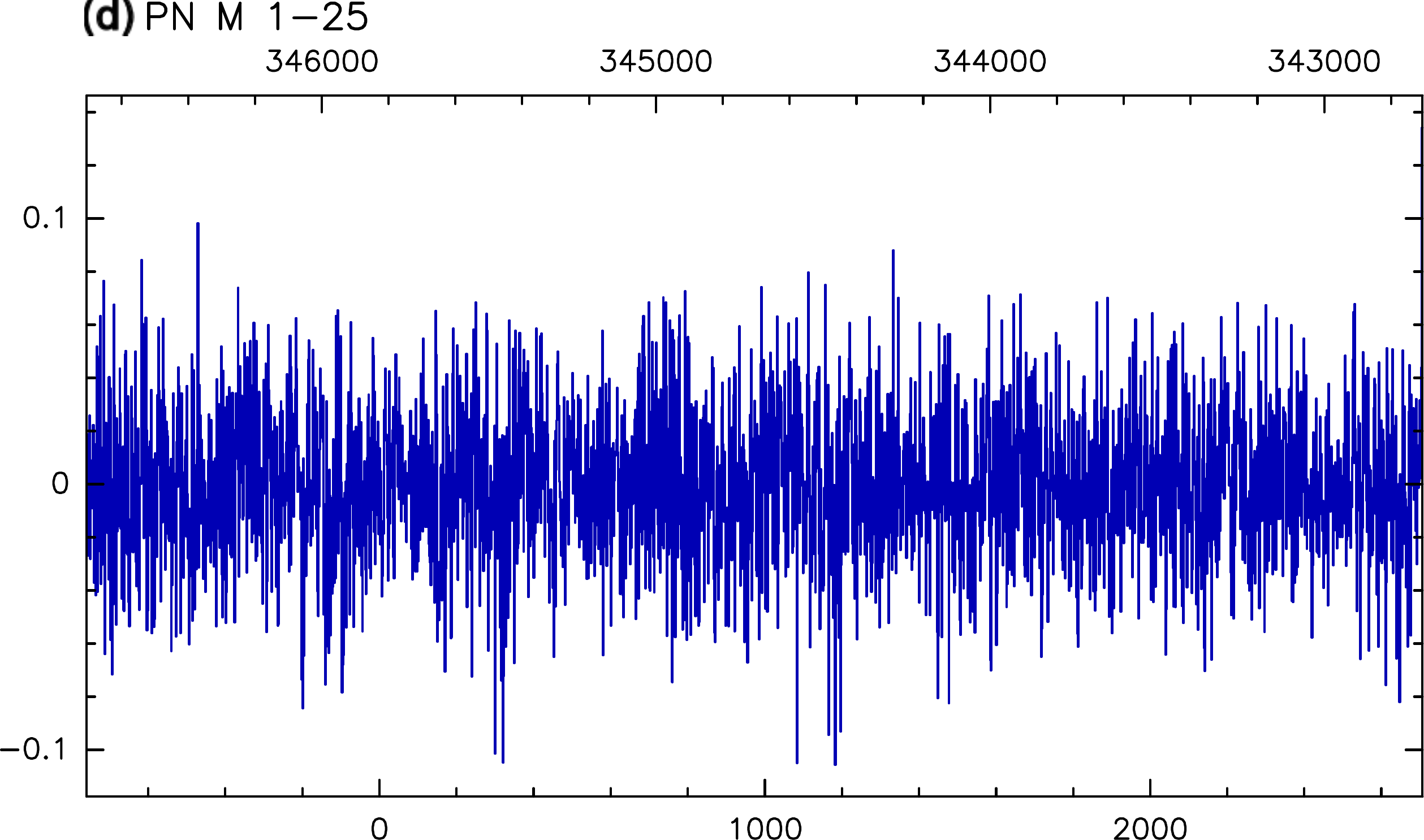}
\includegraphics[width=6cm, height=5.5cm]{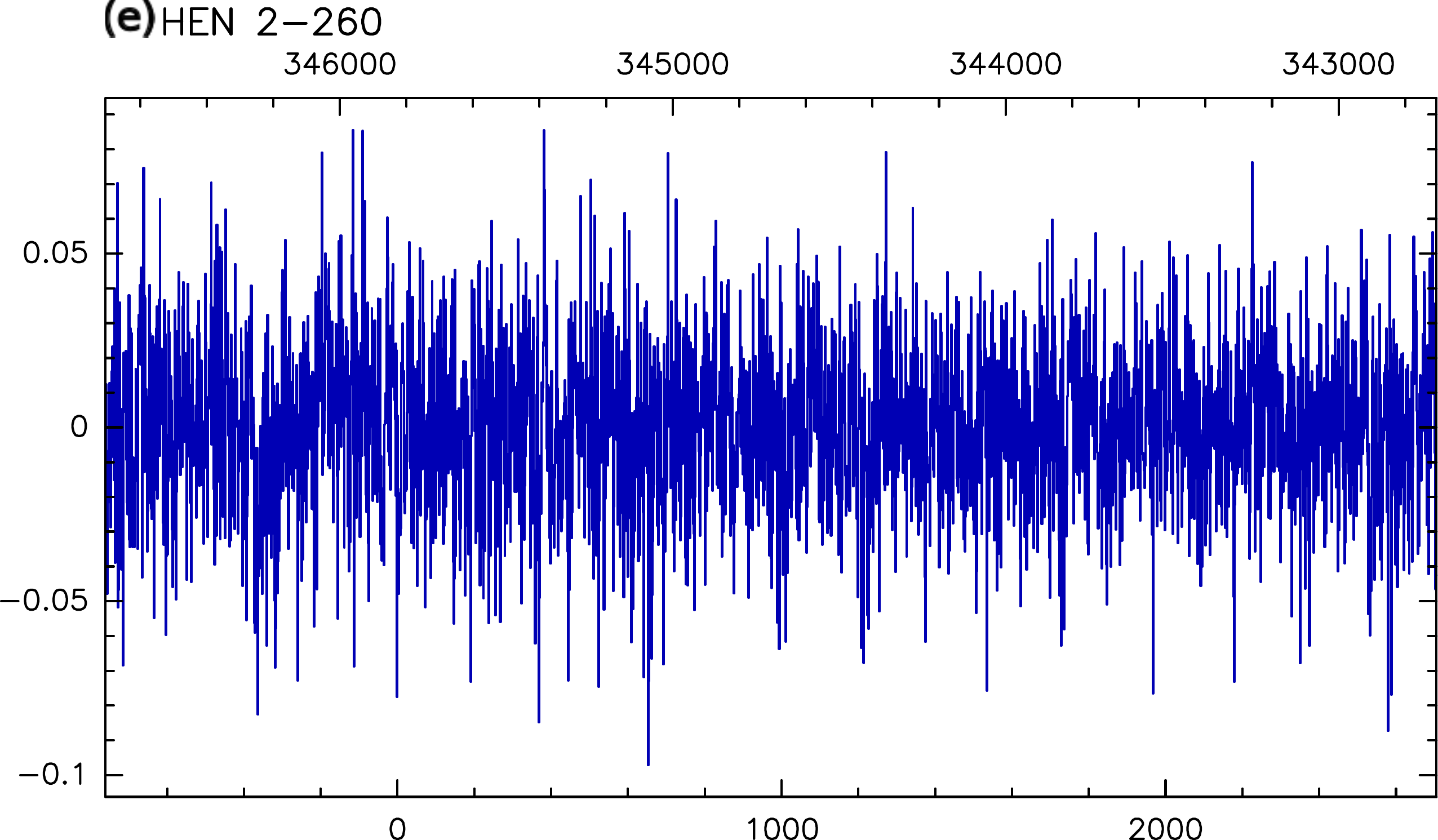}
\includegraphics[width=6cm, height=5.5cm]{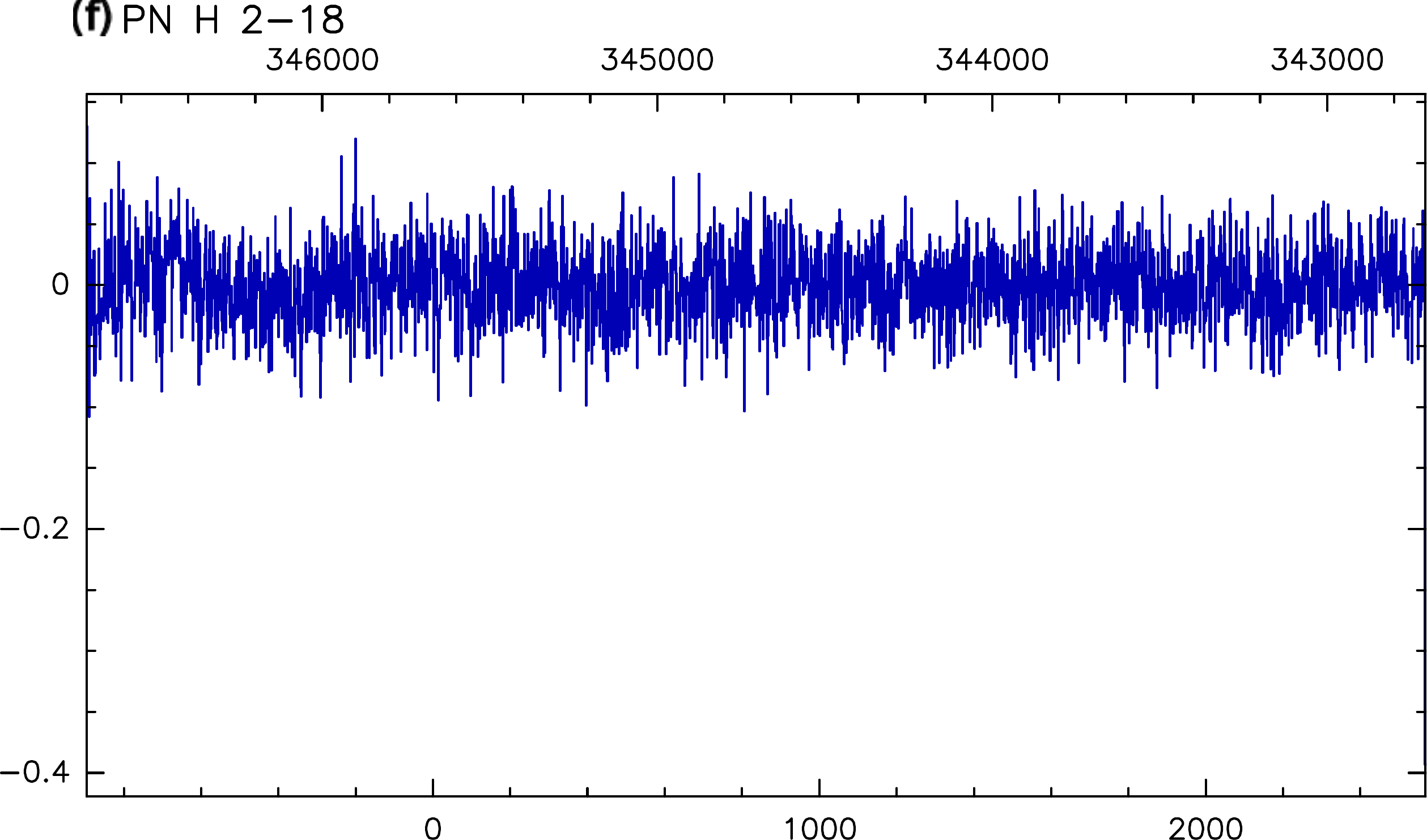}}
\hbox{
\centering
\includegraphics[width=6cm, height=5.5cm]{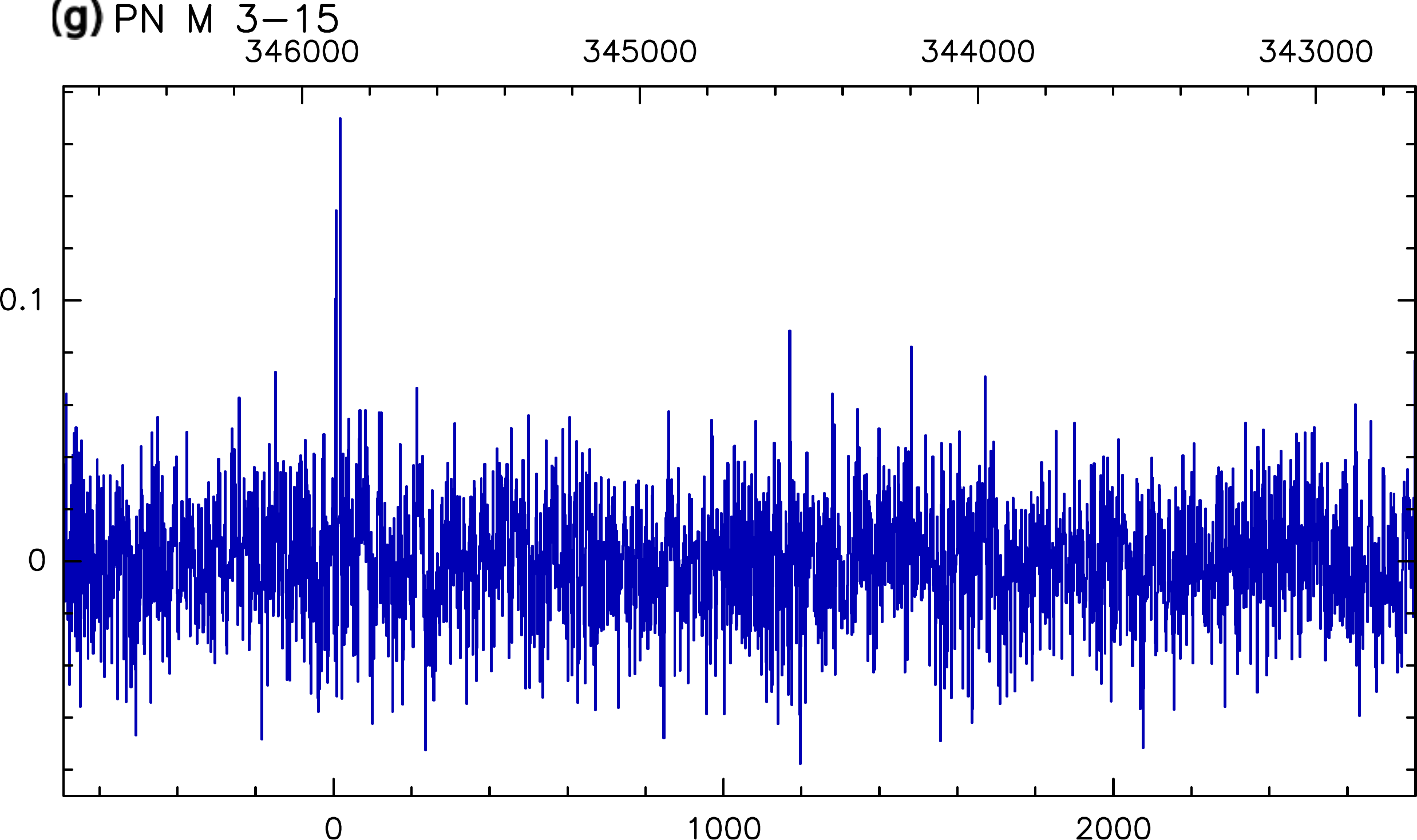}
\includegraphics[width=6cm, height=5.5cm]{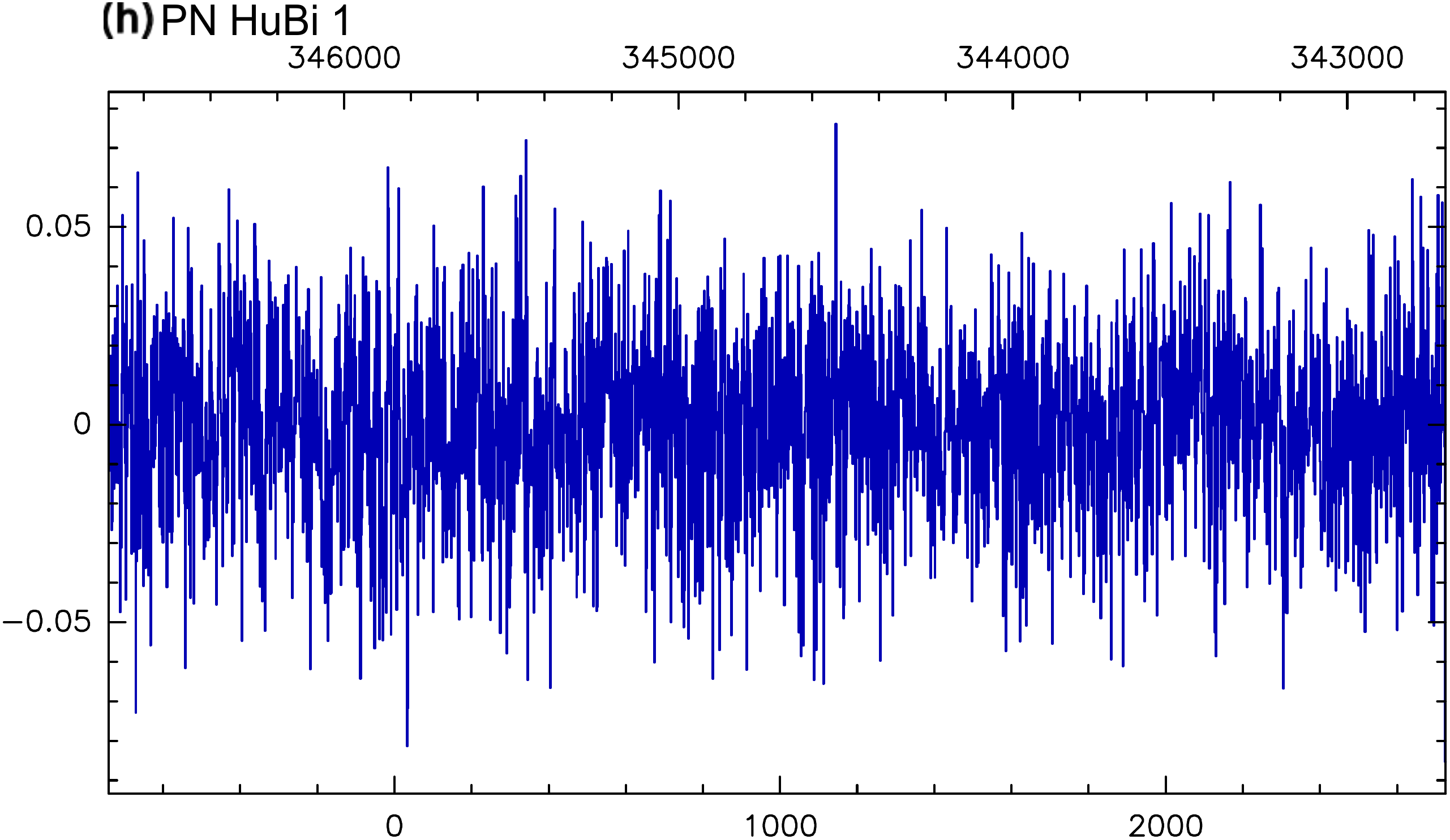}
\includegraphics[width=6cm, height=5.5cm]{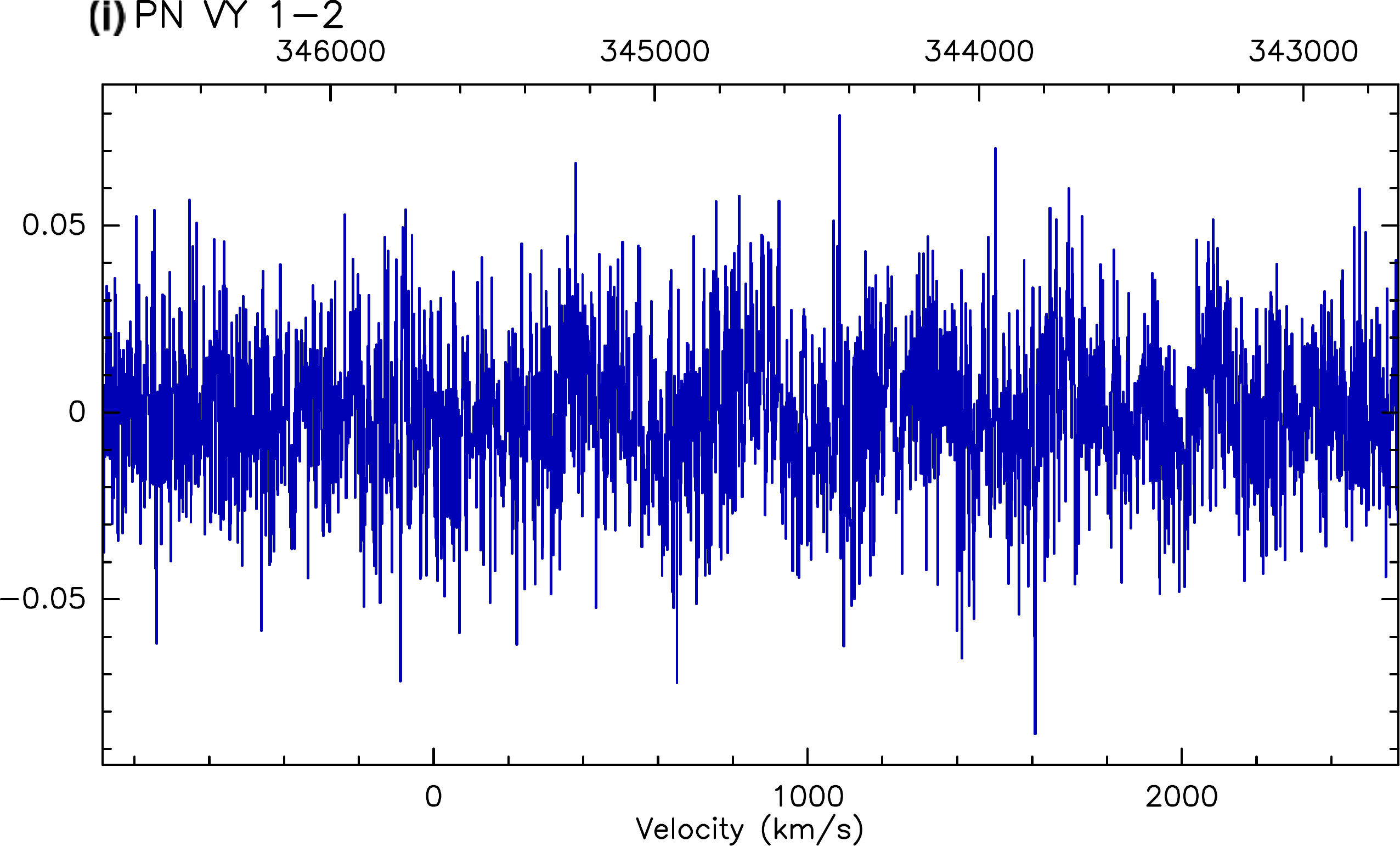}}
\caption[]{pPNe and PNe observations using the APEX telescope.}
\label{Fig6}
\end{figure*} 

\begin{figure*}
\vspace{2cm}
\centering
\hbox{
\centering
\includegraphics[width=6cm, height=5.5cm]{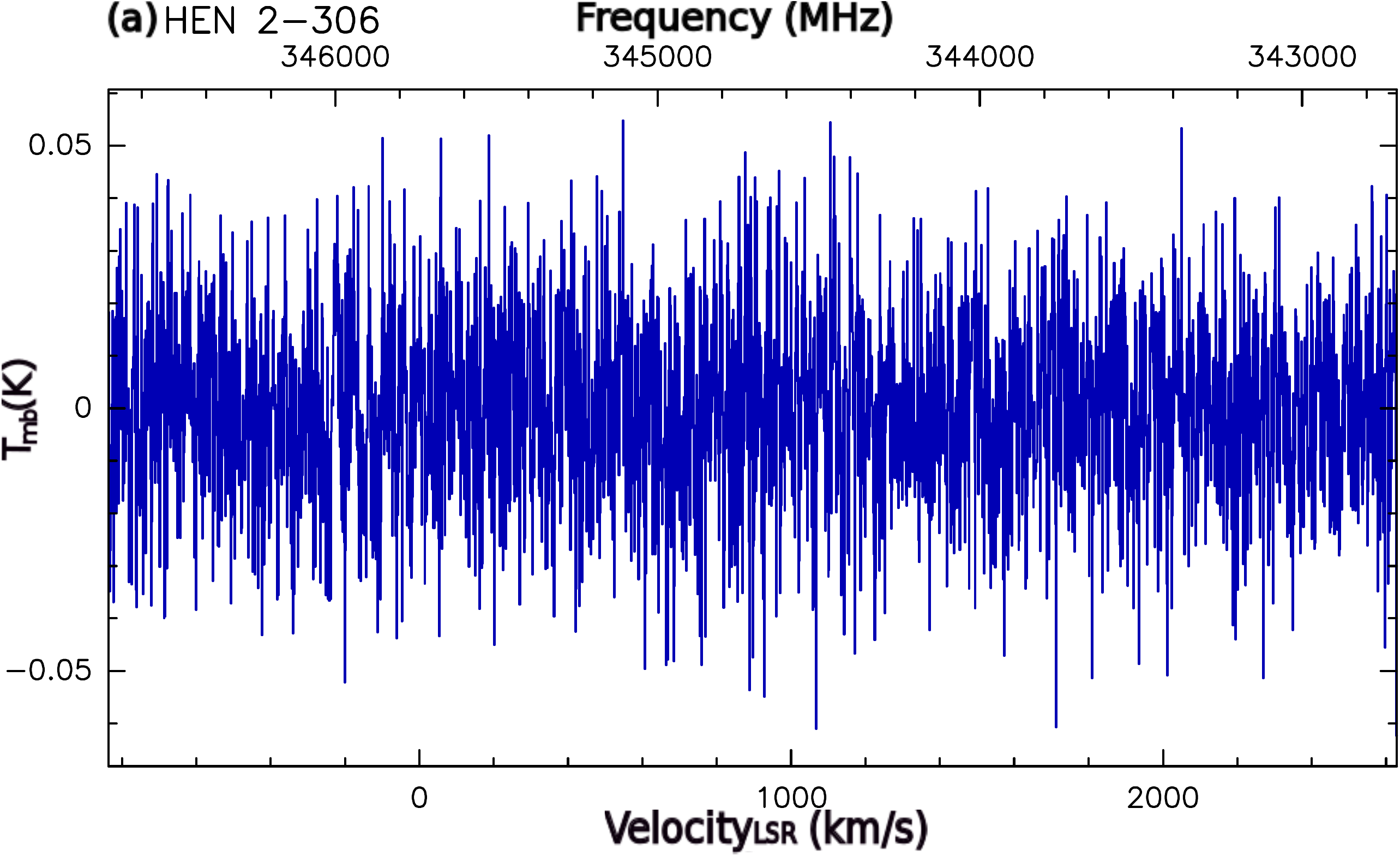}
\includegraphics[width=6cm, height=5.5cm]{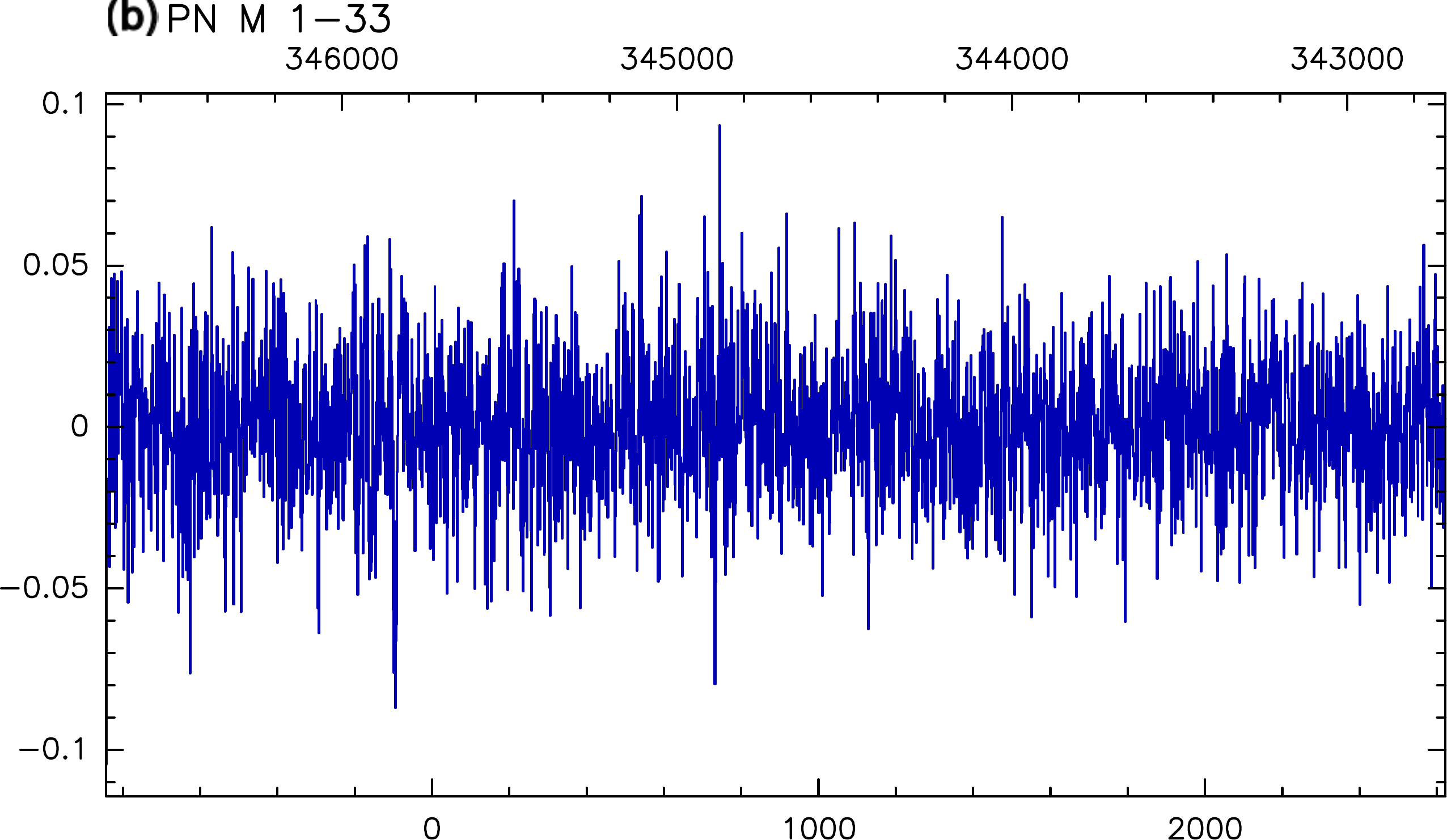}
\includegraphics[width=6cm, height=5.5cm]{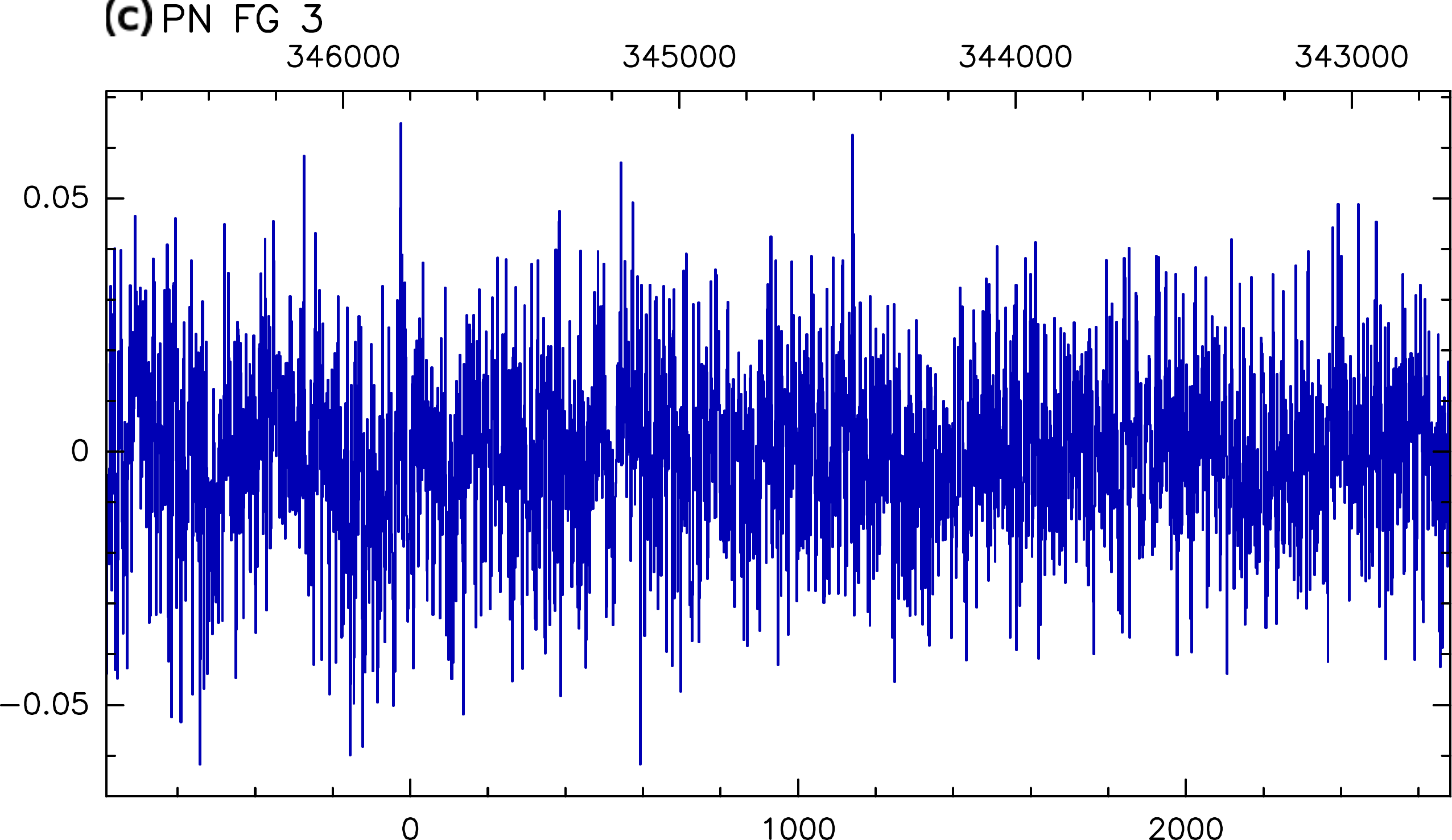}
\vspace{1cm}}
\hbox{
\centering
\vspace{1cm}
\includegraphics[width=6cm, height=5.5cm]{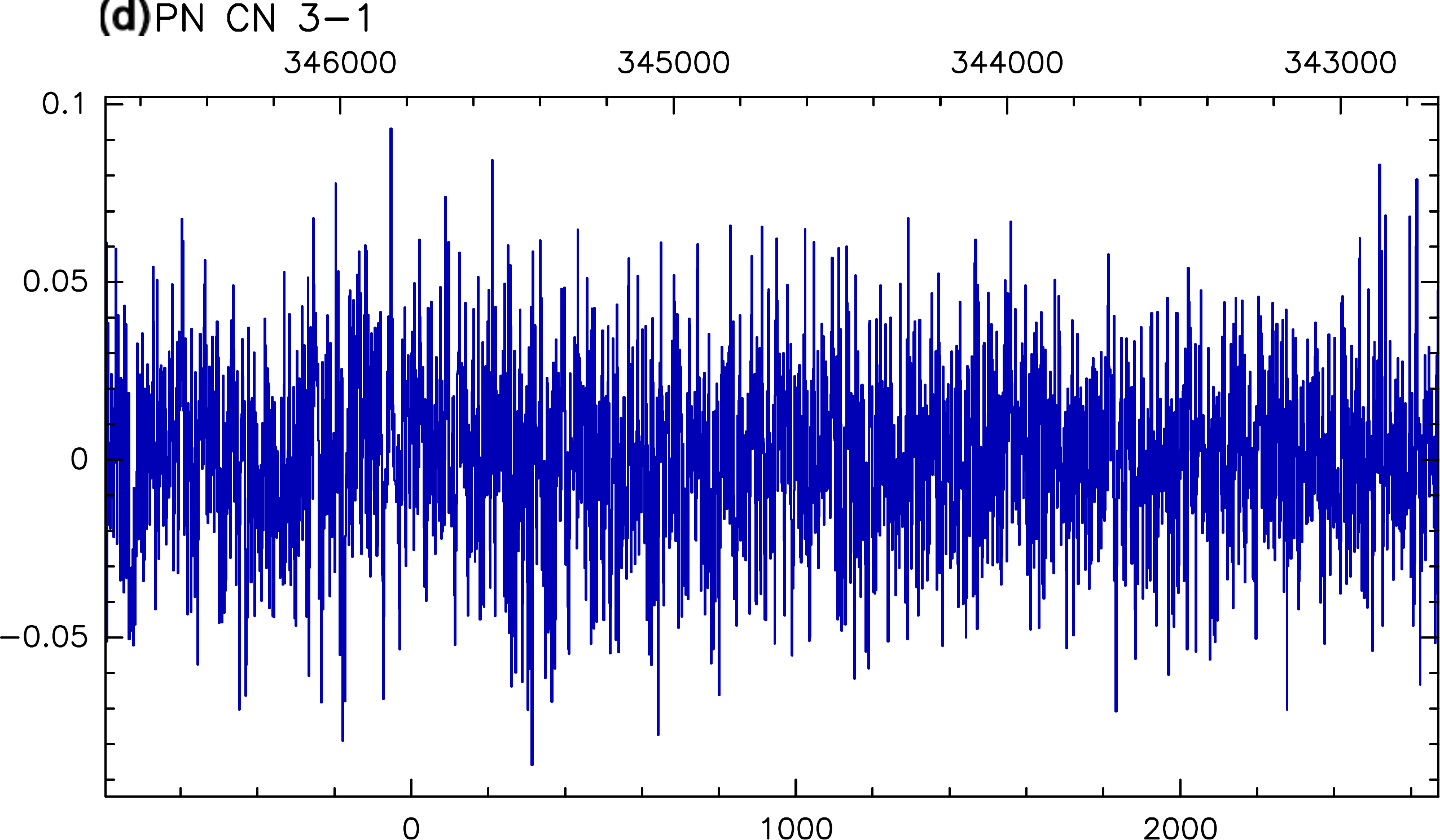}
\includegraphics[width=6cm, height=5.5cm]{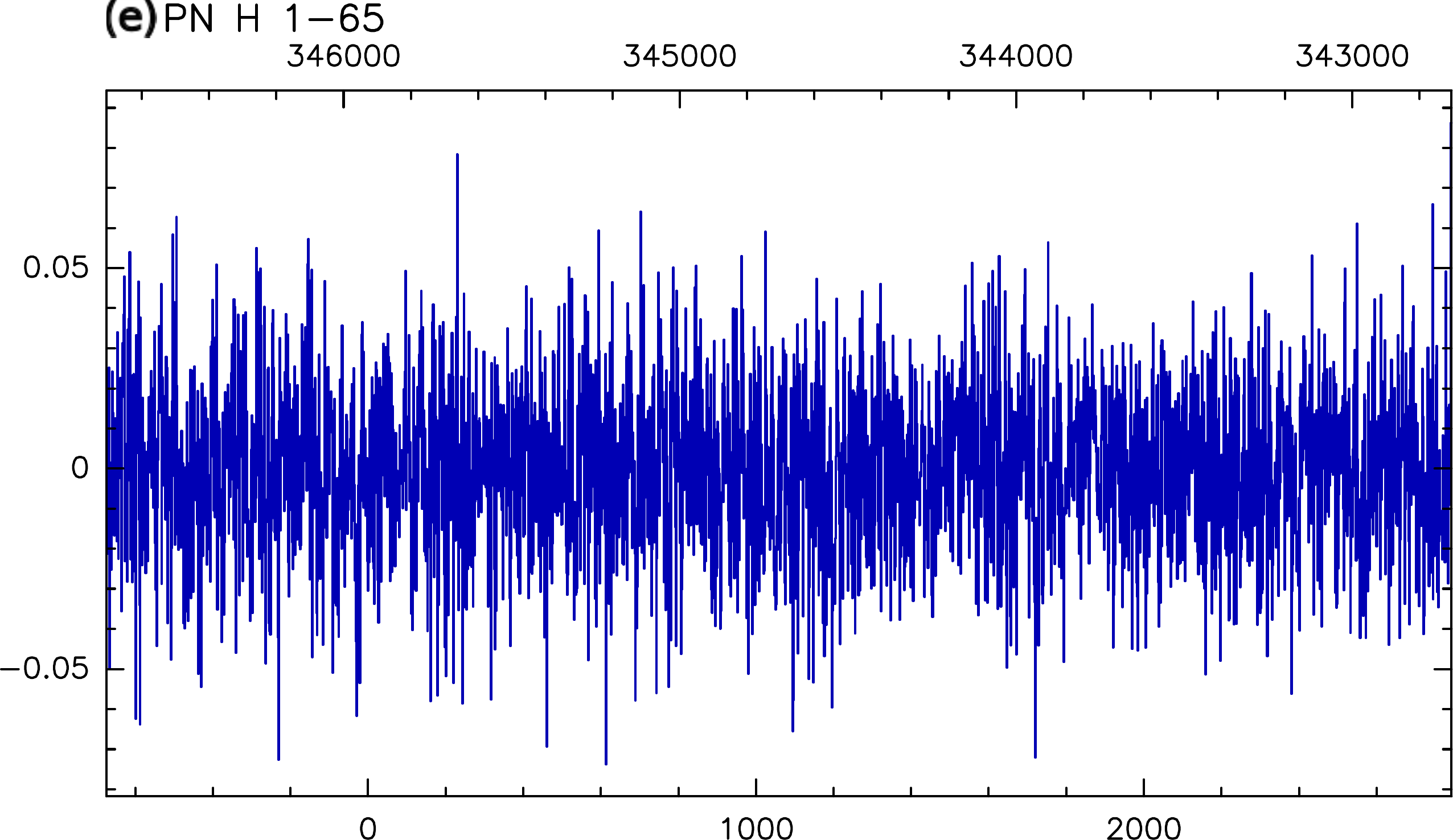}
\includegraphics[width=6cm, height=5.5cm]{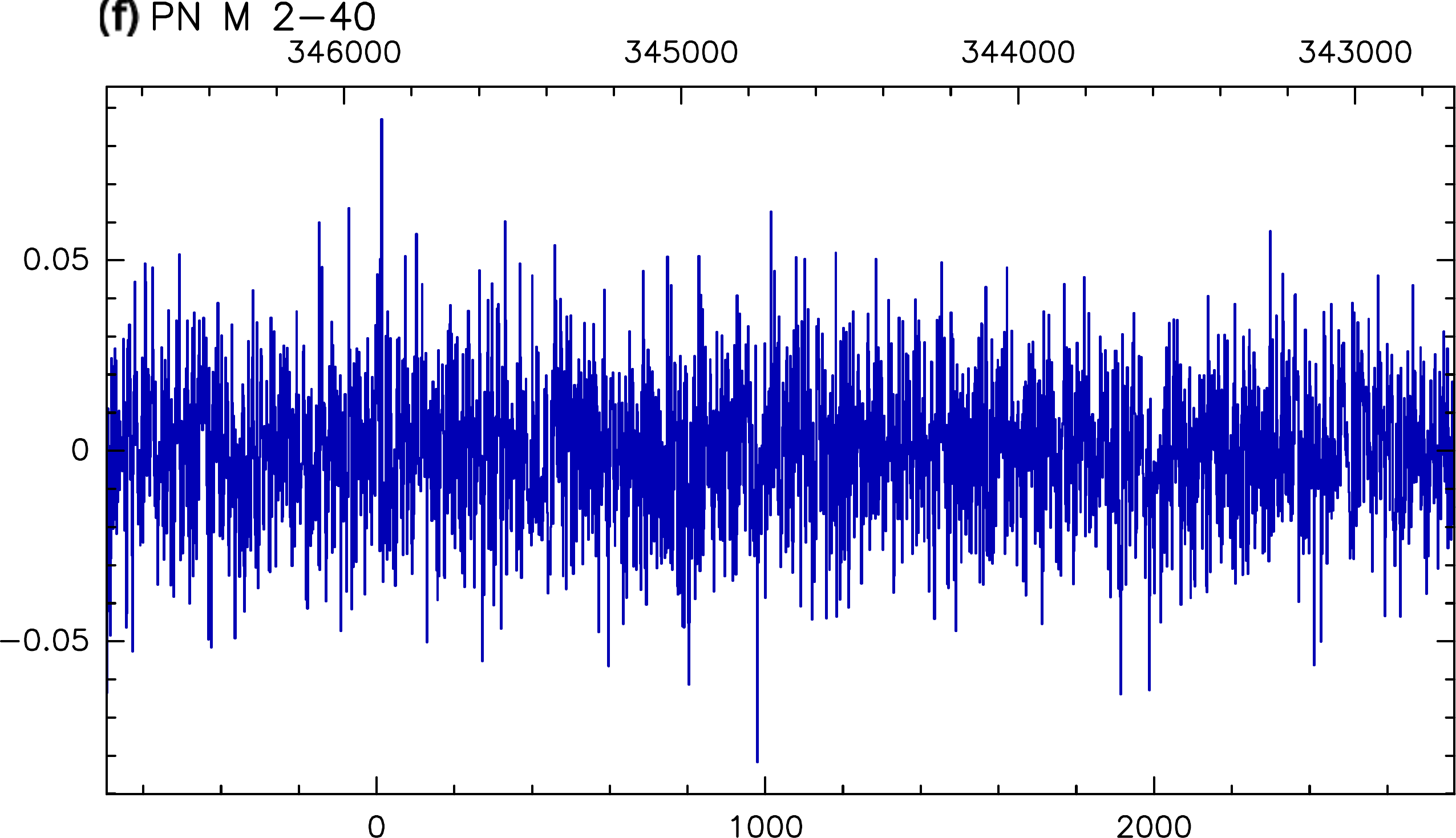}}
\hbox{
\centering
\includegraphics[width=6cm, height=5.5cm]{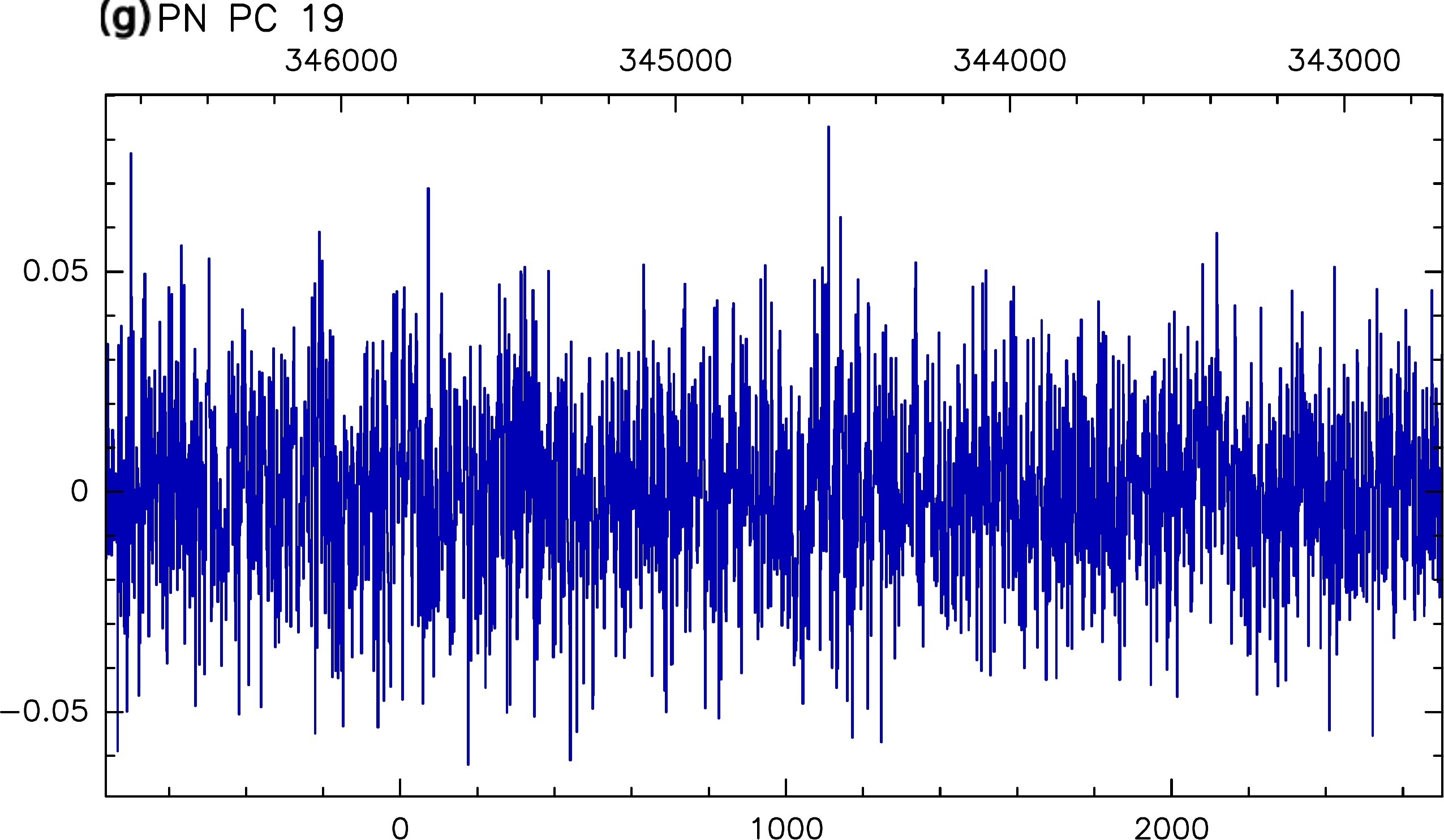}
\includegraphics[width=6cm, height=5.5cm]{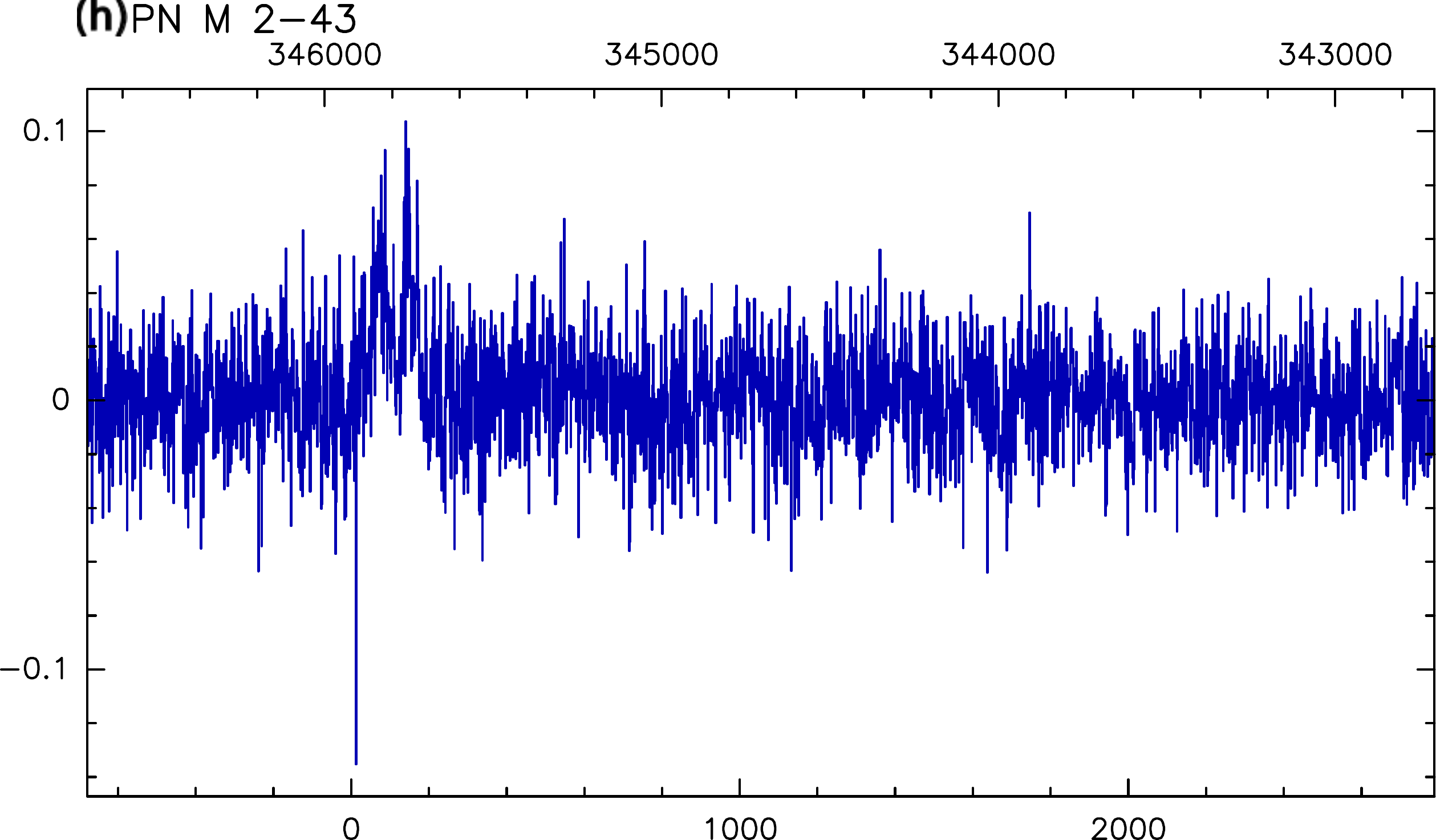}
\includegraphics[width=6cm, height=5.5cm]{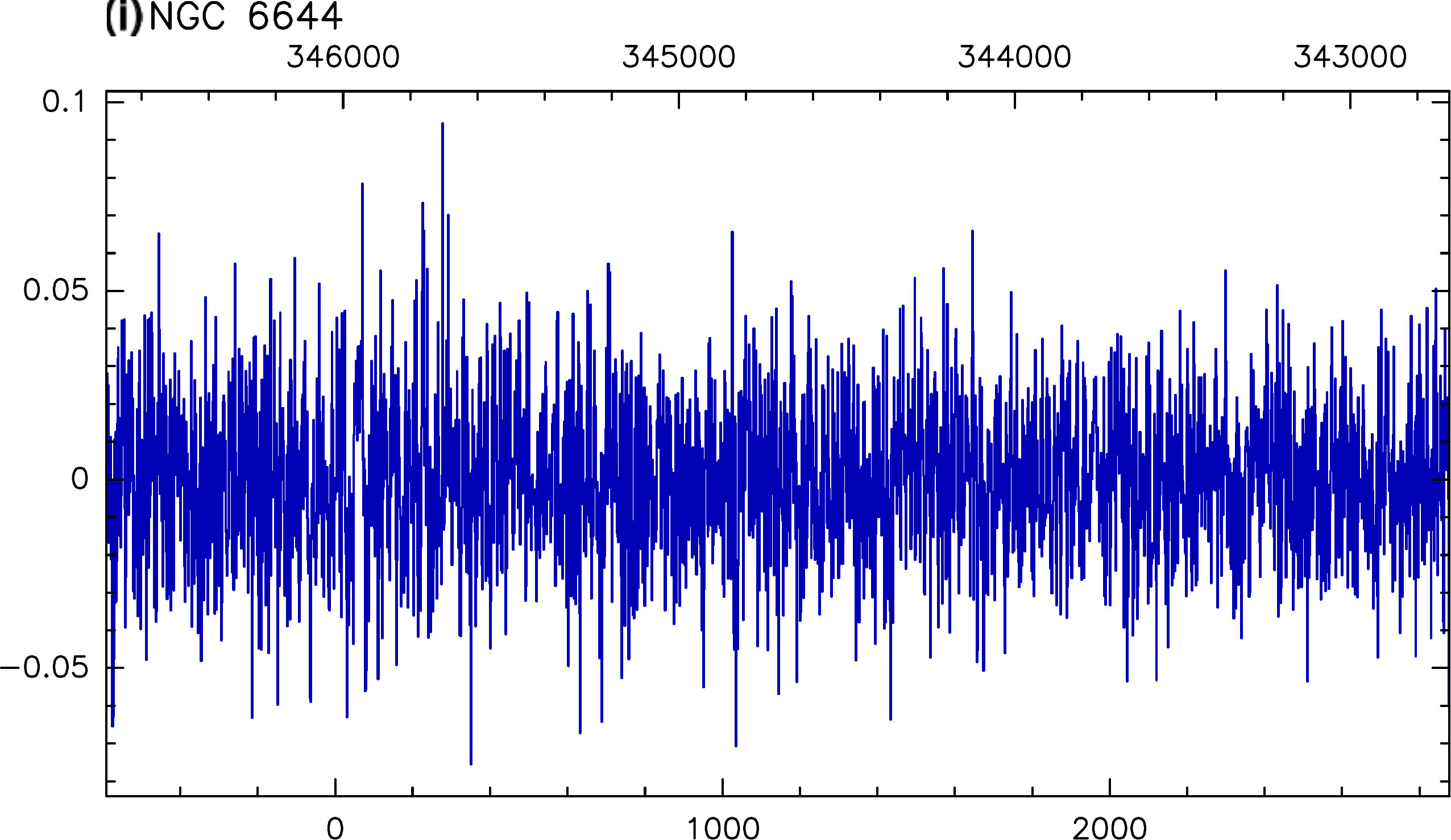}}
\caption[]{pPNe and PNe observations using the APEX telescope.}
\label{Fig7}
\end{figure*} 

\begin{figure*}
\vspace{2cm}
\centering
\hbox{
\centering
\includegraphics[width=6cm, height=5.5cm]{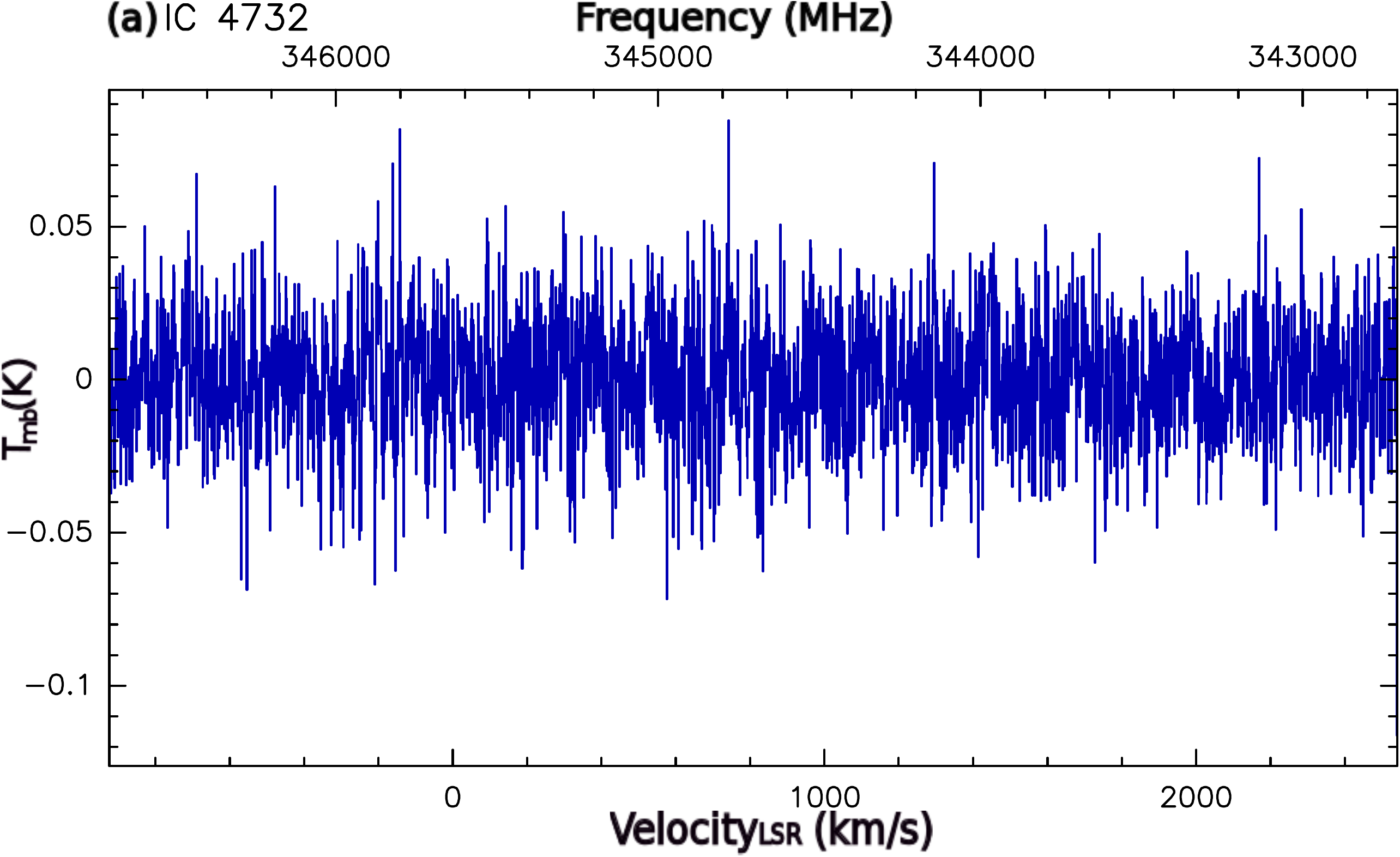}
\includegraphics[width=6cm, height=5.5cm]{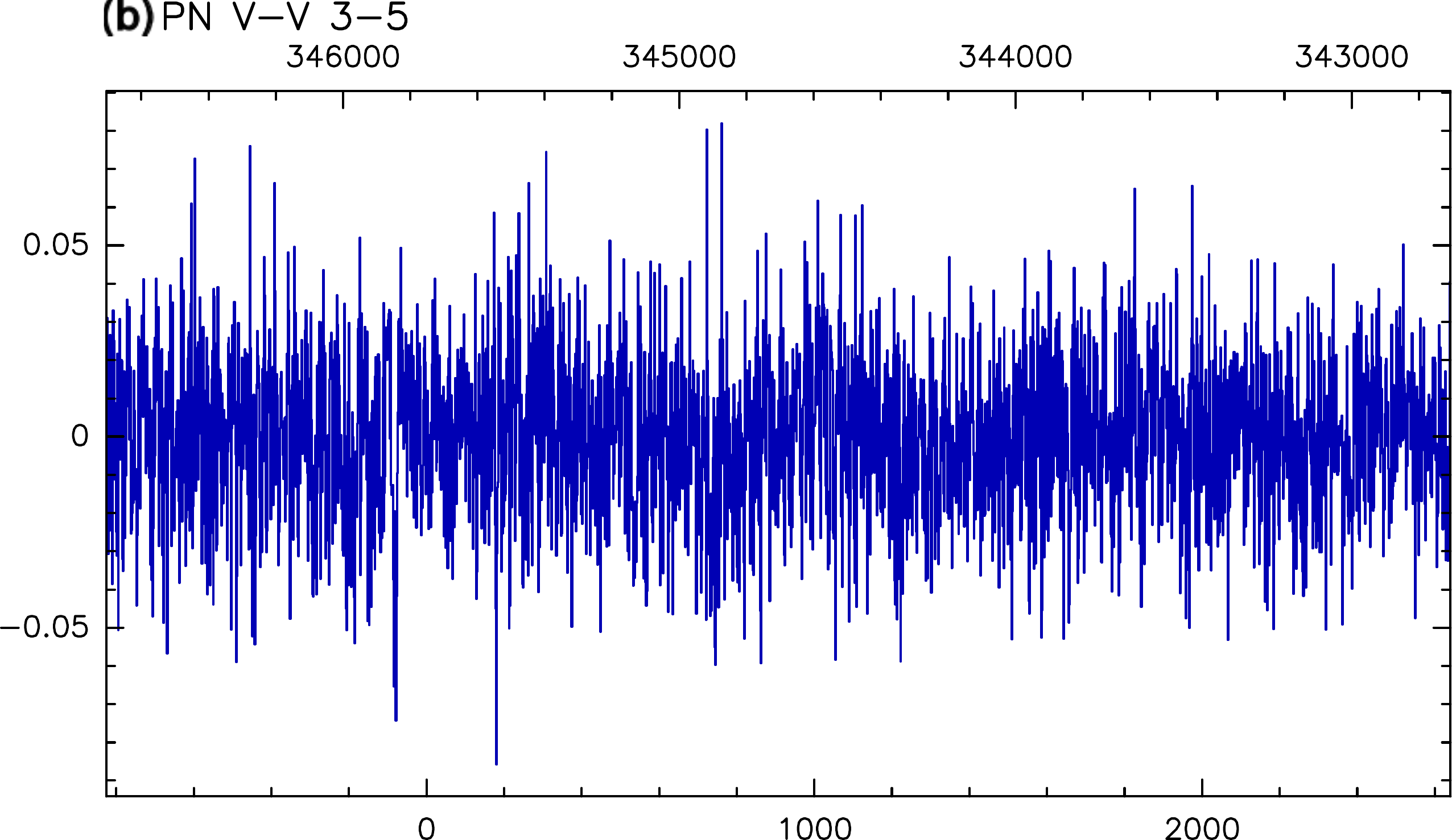}
\includegraphics[width=6cm, height=5.5cm]{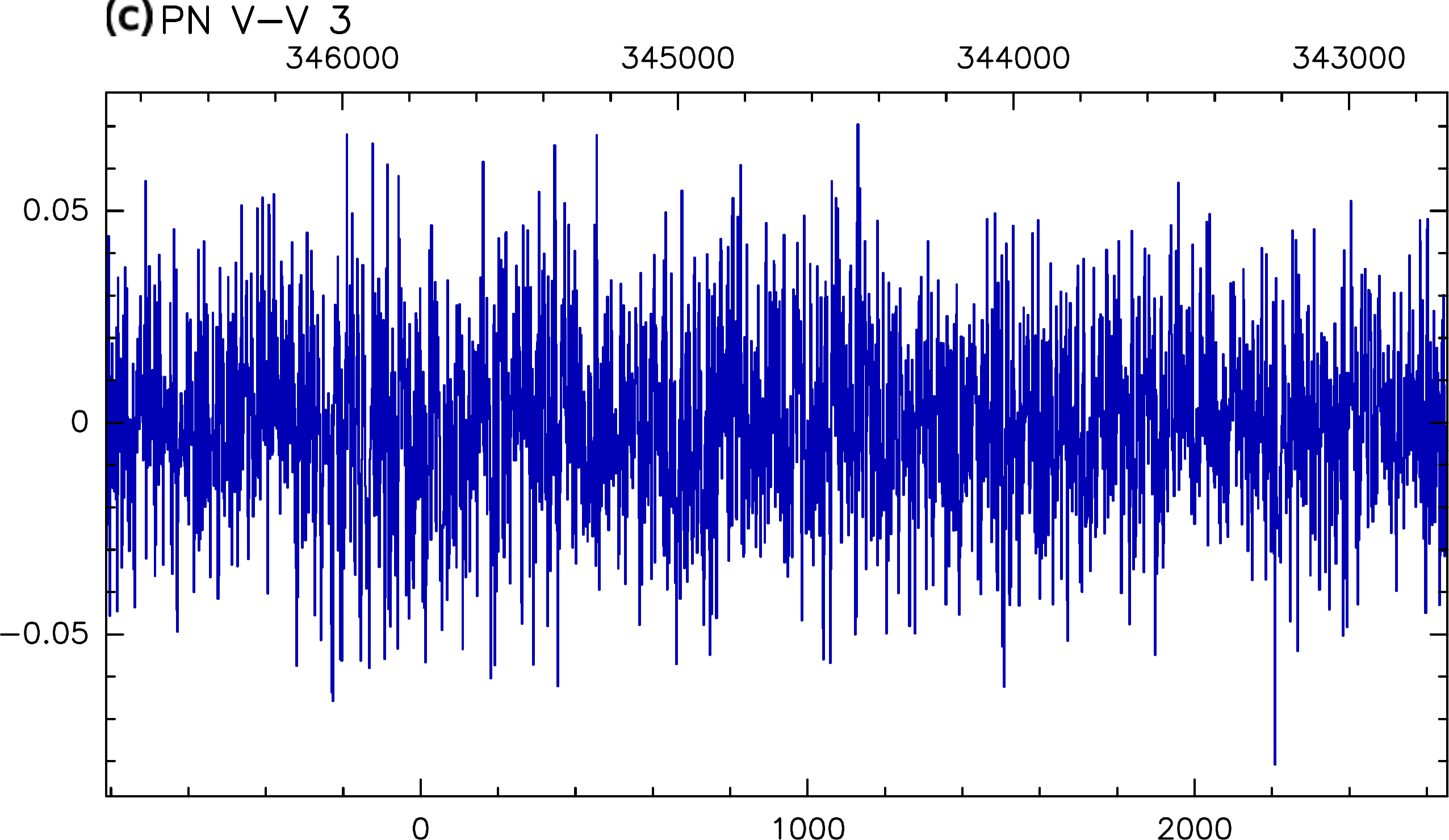}
\vspace{1cm}}
\hbox{
\centering
\vspace{1cm}
\includegraphics[width=6cm, height=5.5cm]{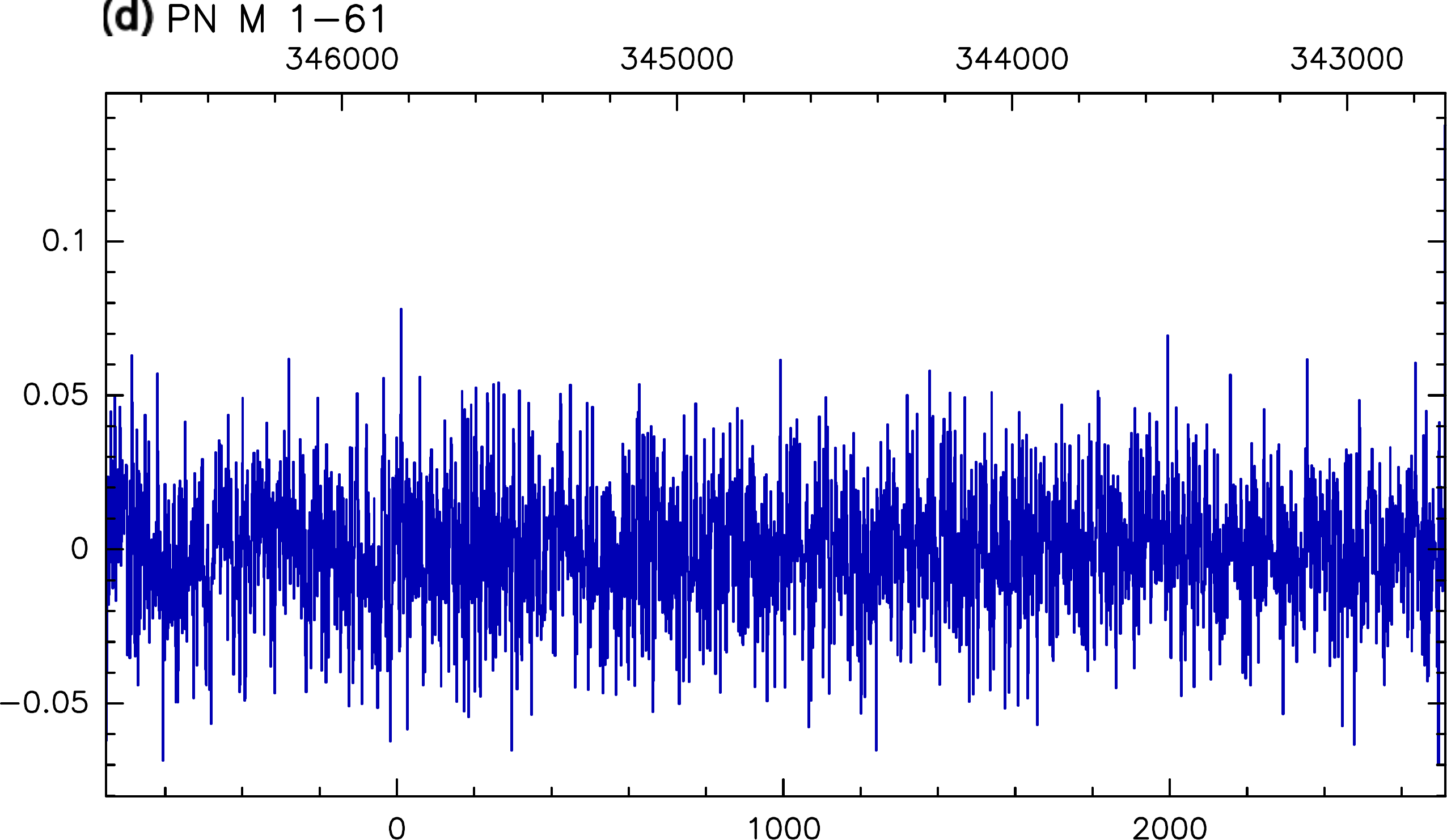}
\includegraphics[width=6cm, height=5.5cm]{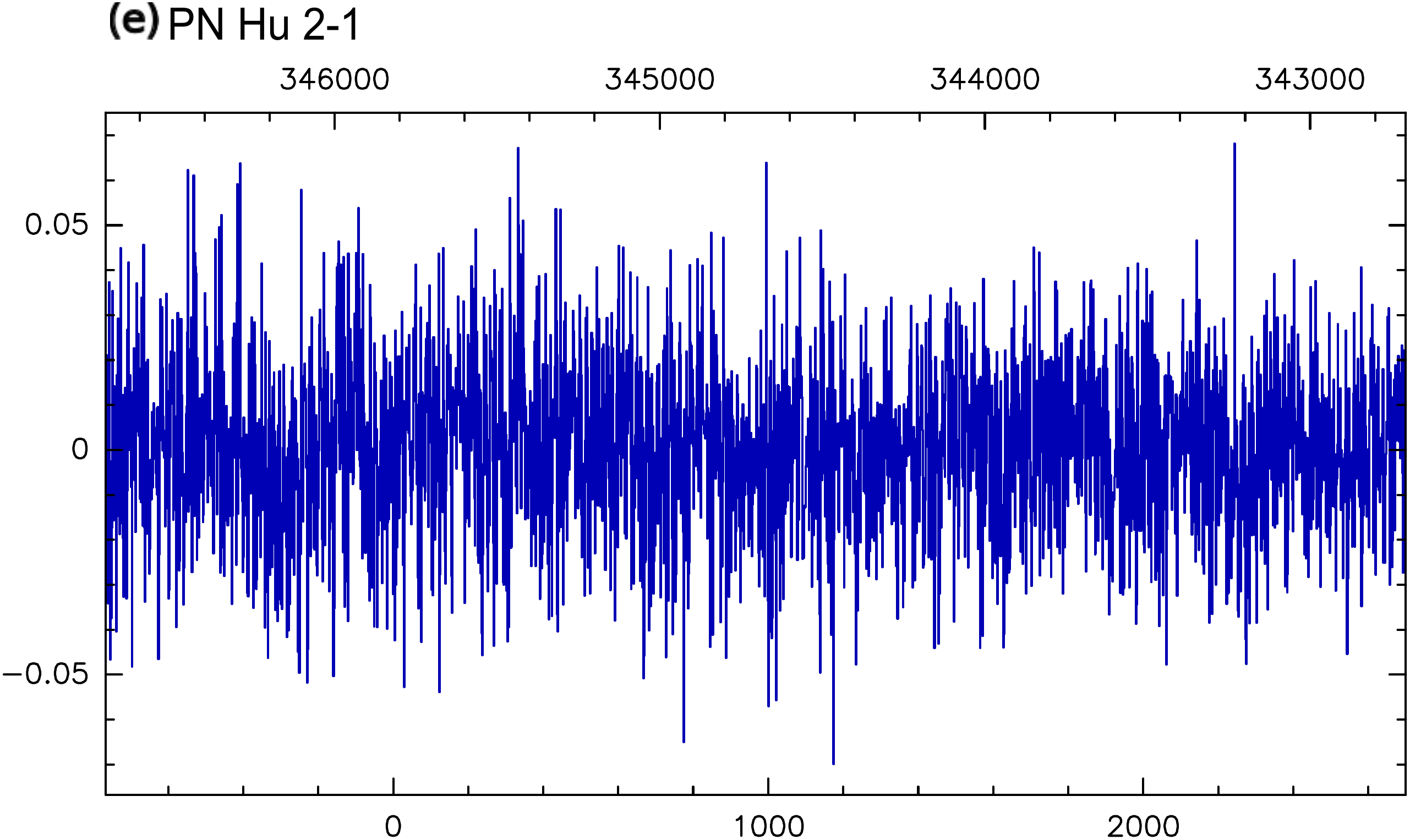}
\includegraphics[width=6cm, height=5.5cm]{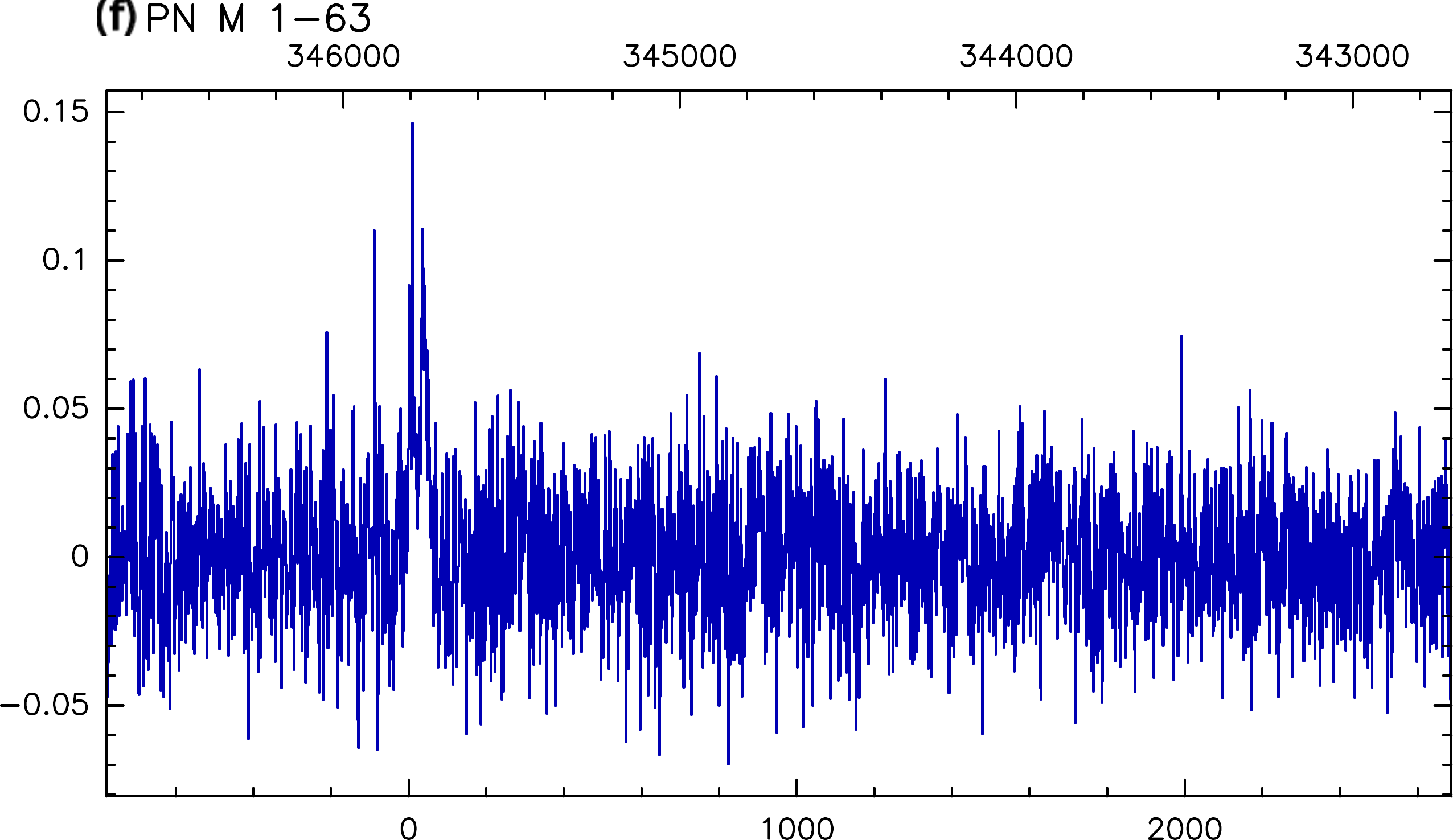}}
\hbox{
\centering
\includegraphics[width=6cm, height=5.5cm]{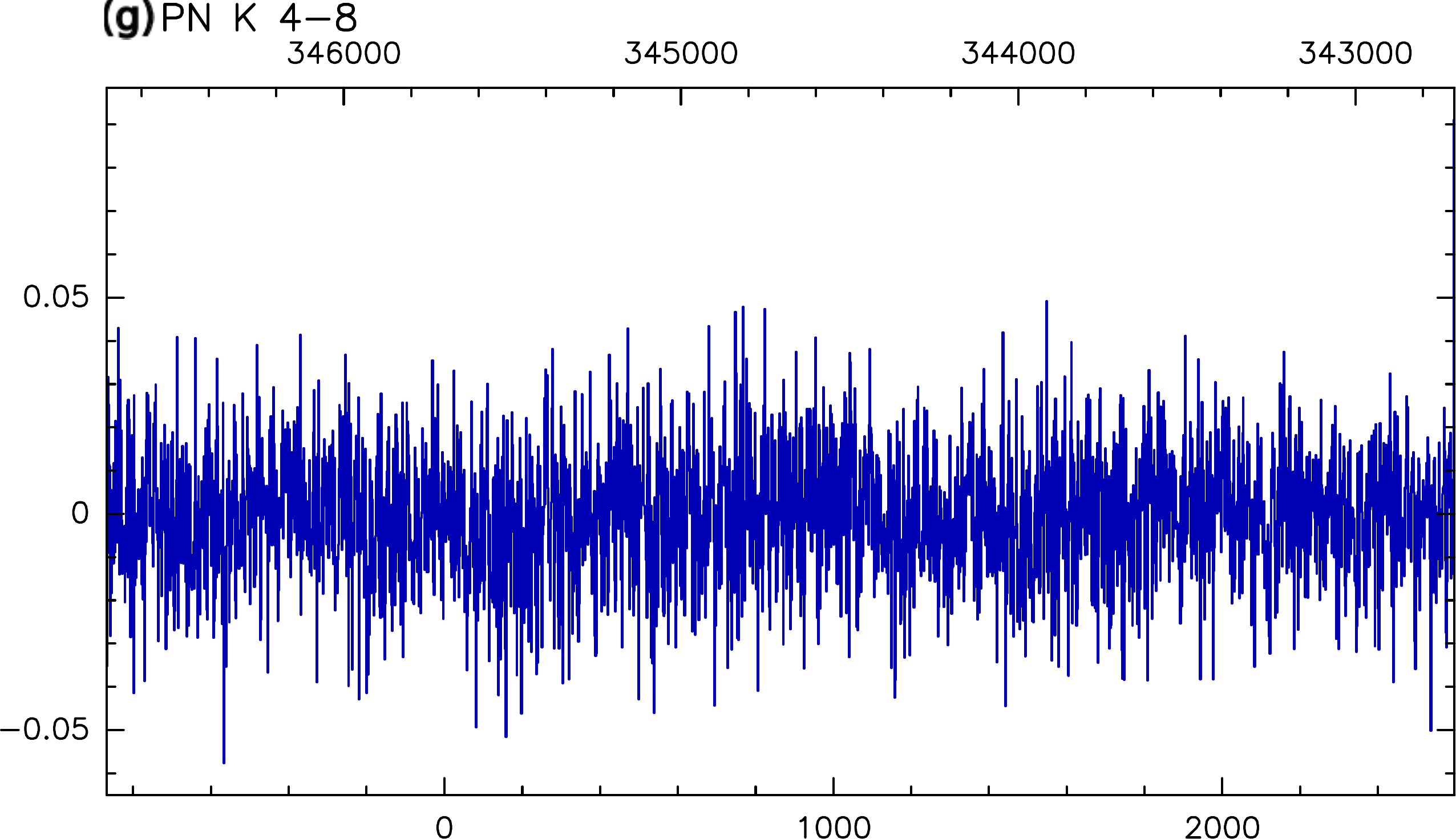}
\includegraphics[width=6cm, height=5.5cm]{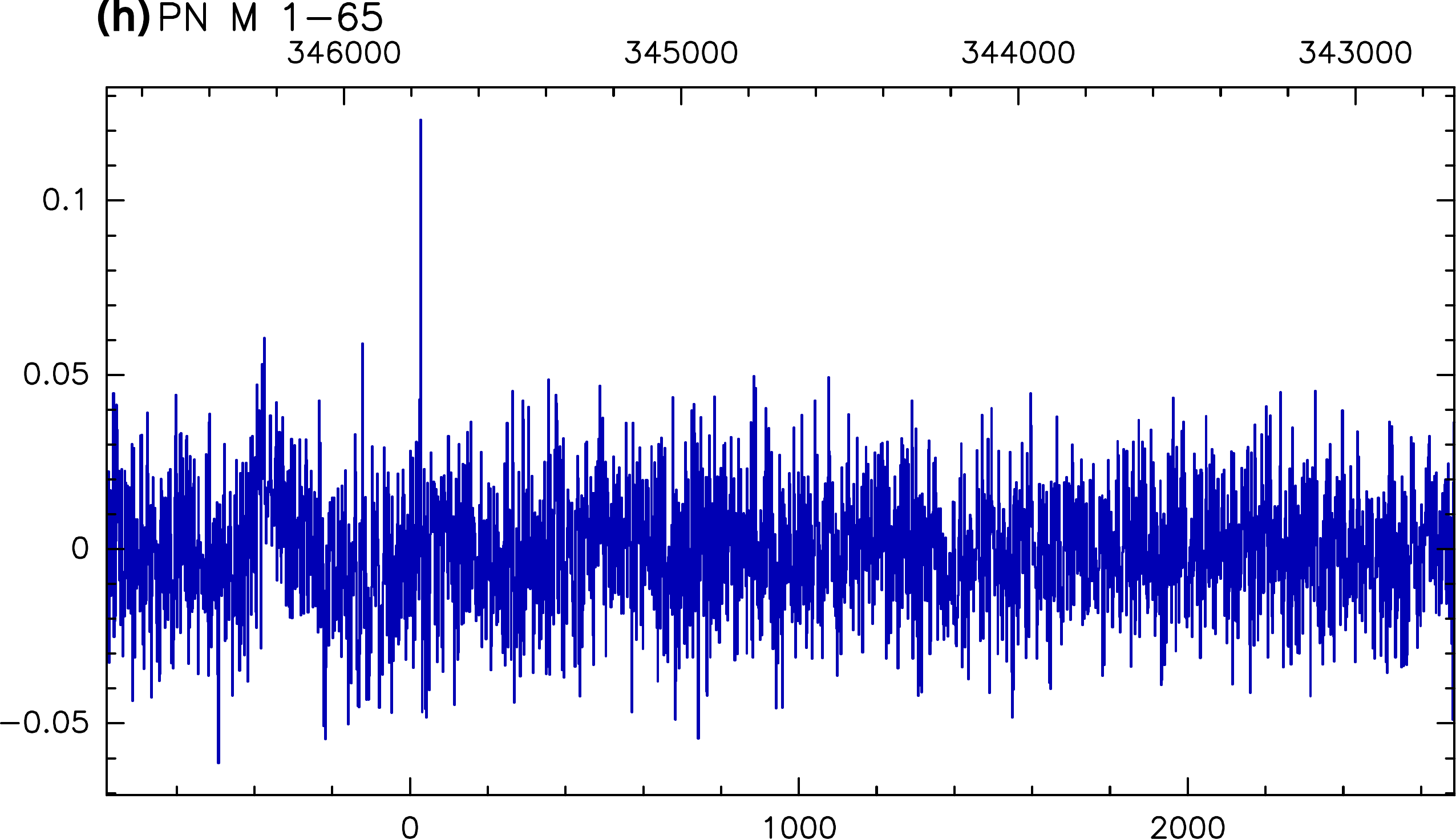}
\includegraphics[width=6cm, height=5.5cm]{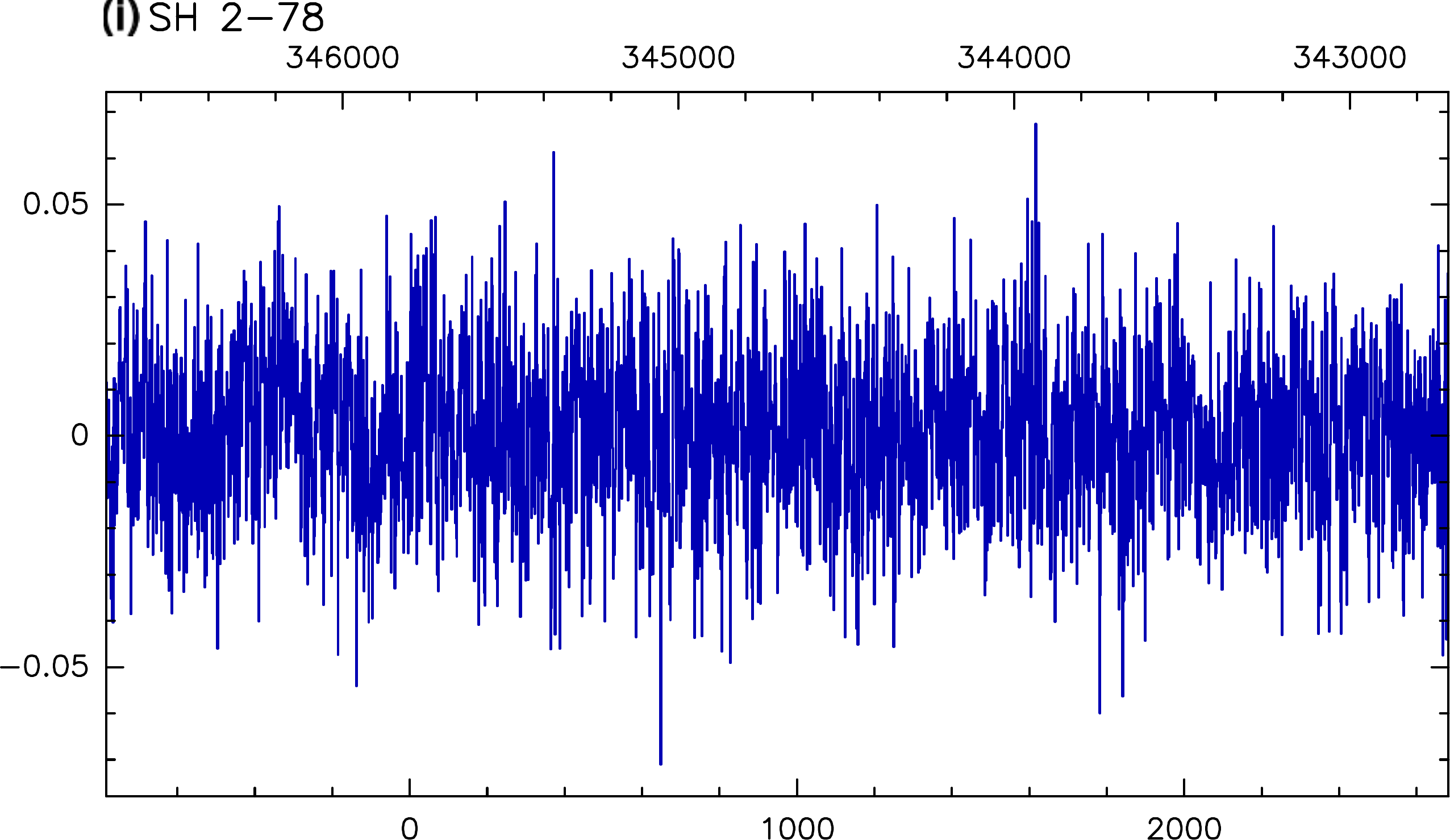}}
\caption[]{pPNe and PNe observations using the APEX telescope.}
\label{Fig8}
\end{figure*} 

\begin{figure*}
\vspace{2cm}
\centering
\hbox{
\centering
\includegraphics[width=6cm, height=5.5cm]{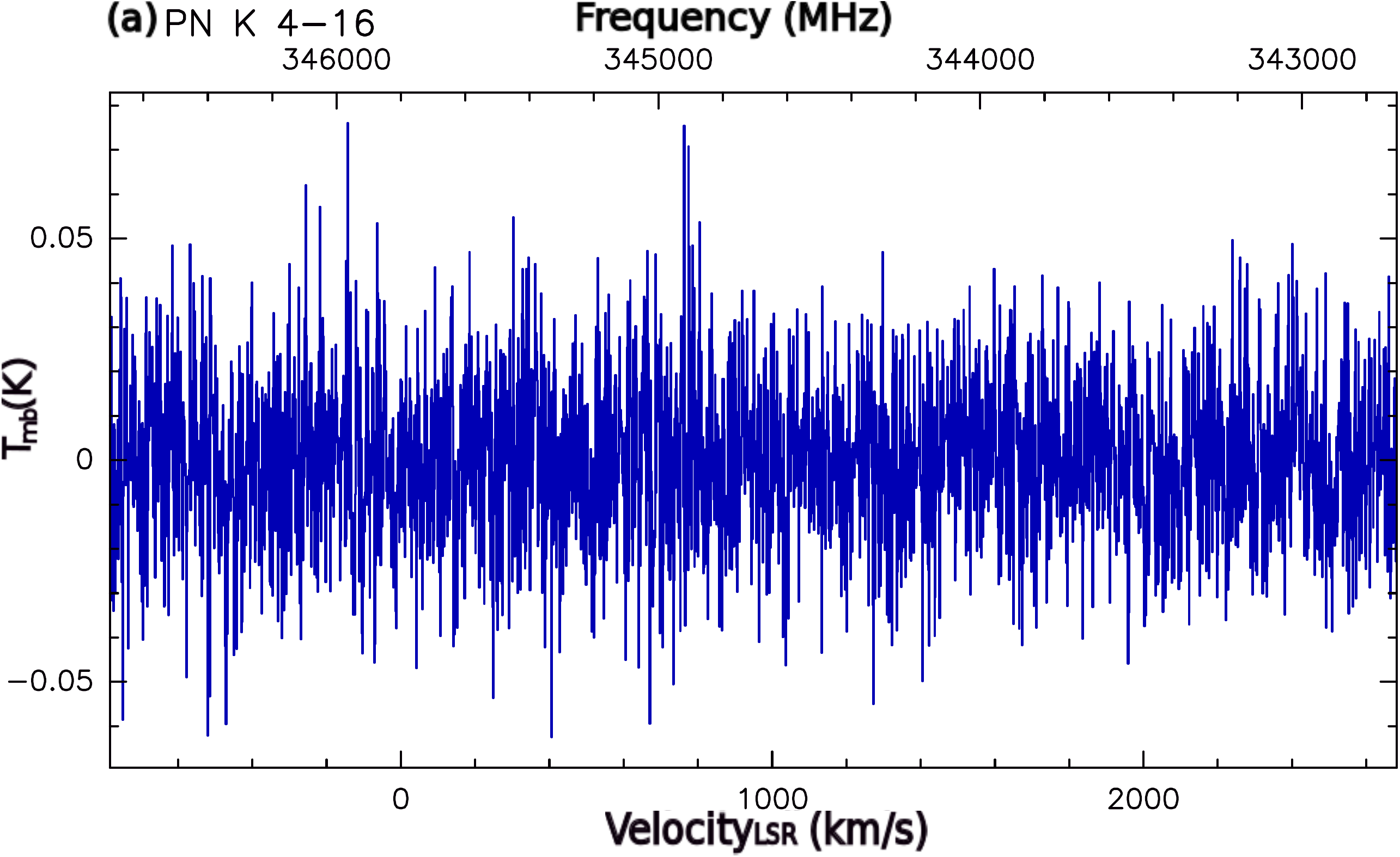}
\includegraphics[width=6cm, height=5.5cm]{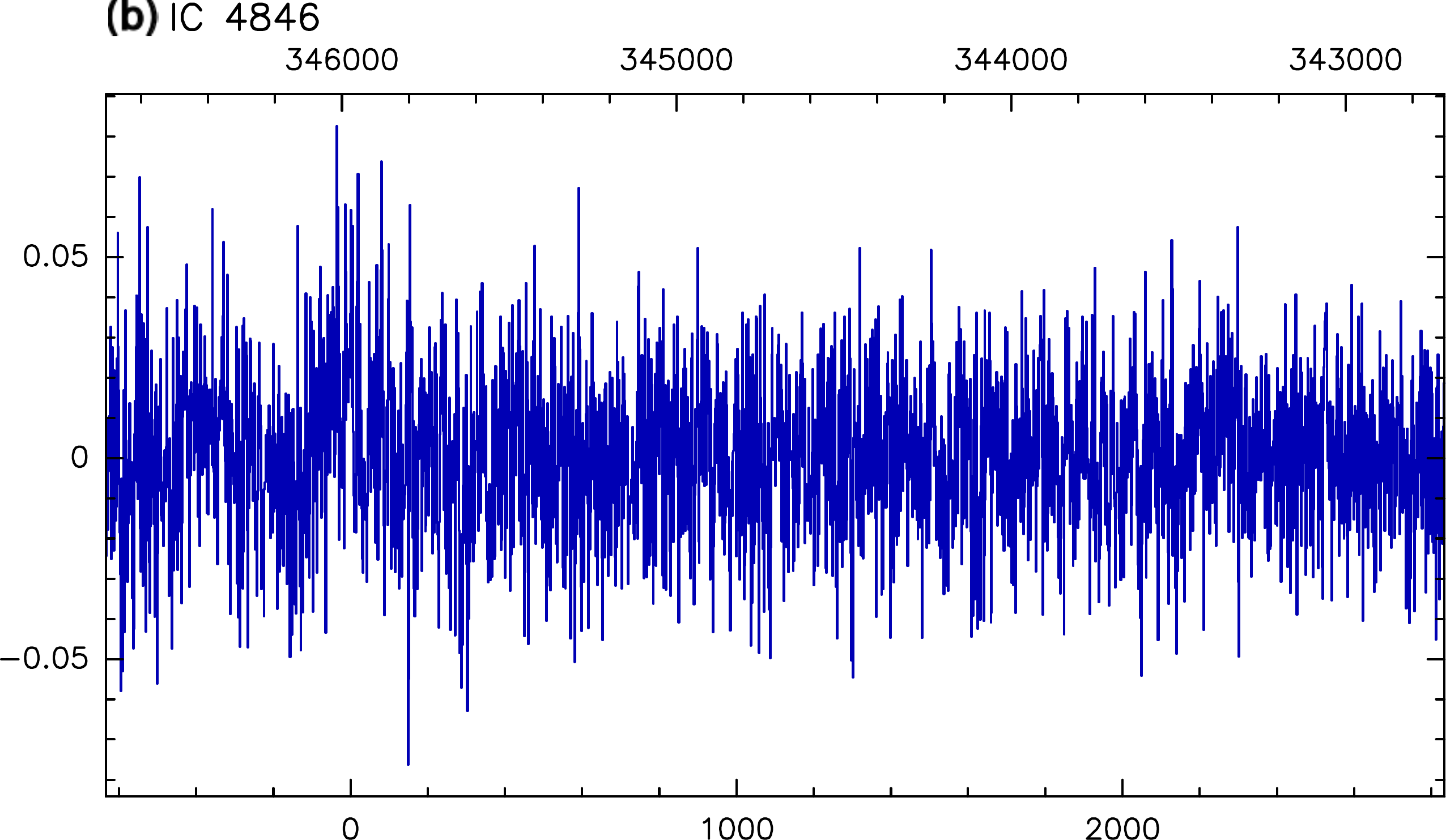}
\includegraphics[width=6cm, height=5.5cm]{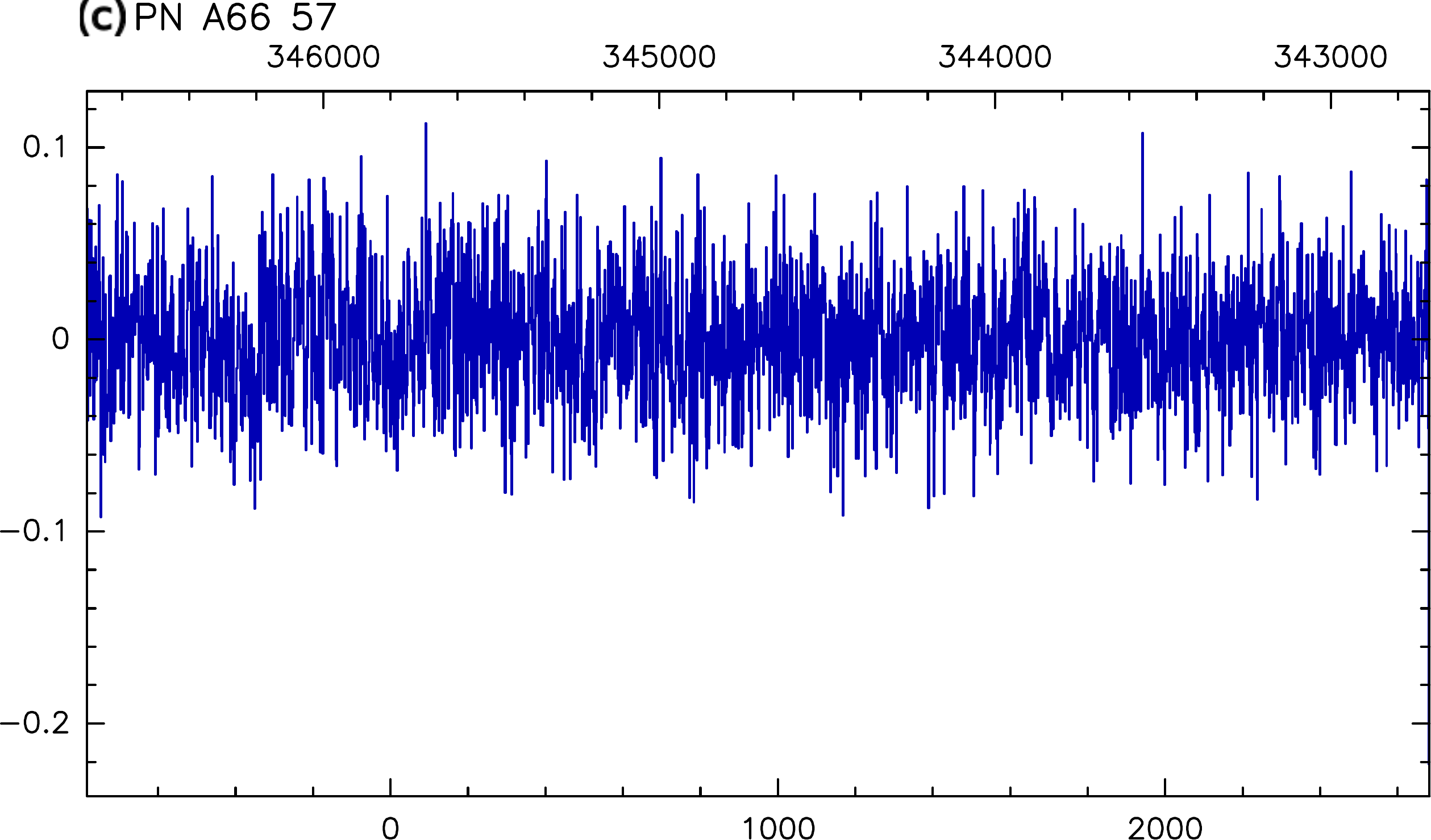}
\vspace{1cm}}
\hbox{
\centering
\vspace{1cm}
\includegraphics[width=6cm, height=5.5cm]{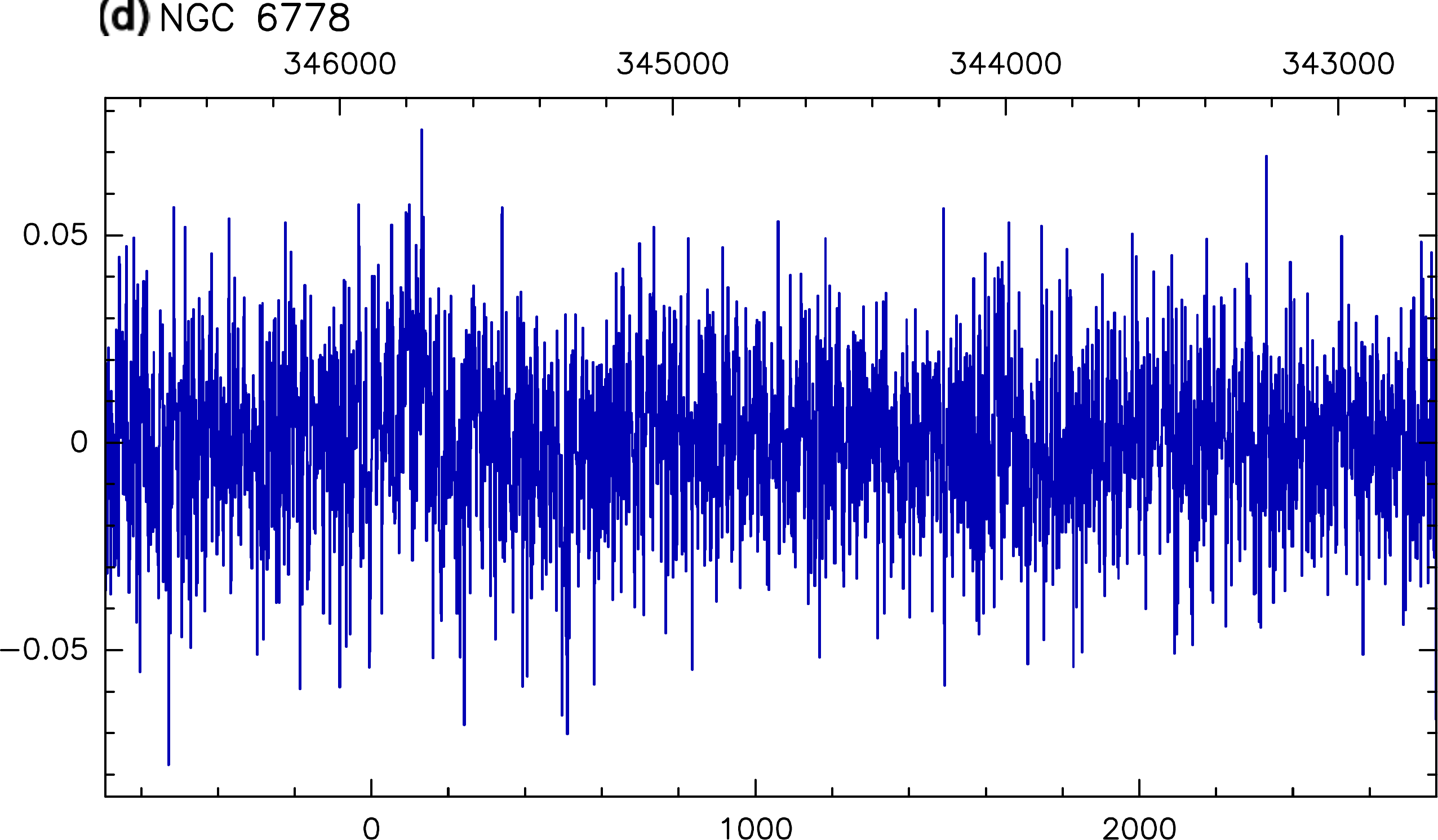}
\includegraphics[width=6cm, height=5.5cm]{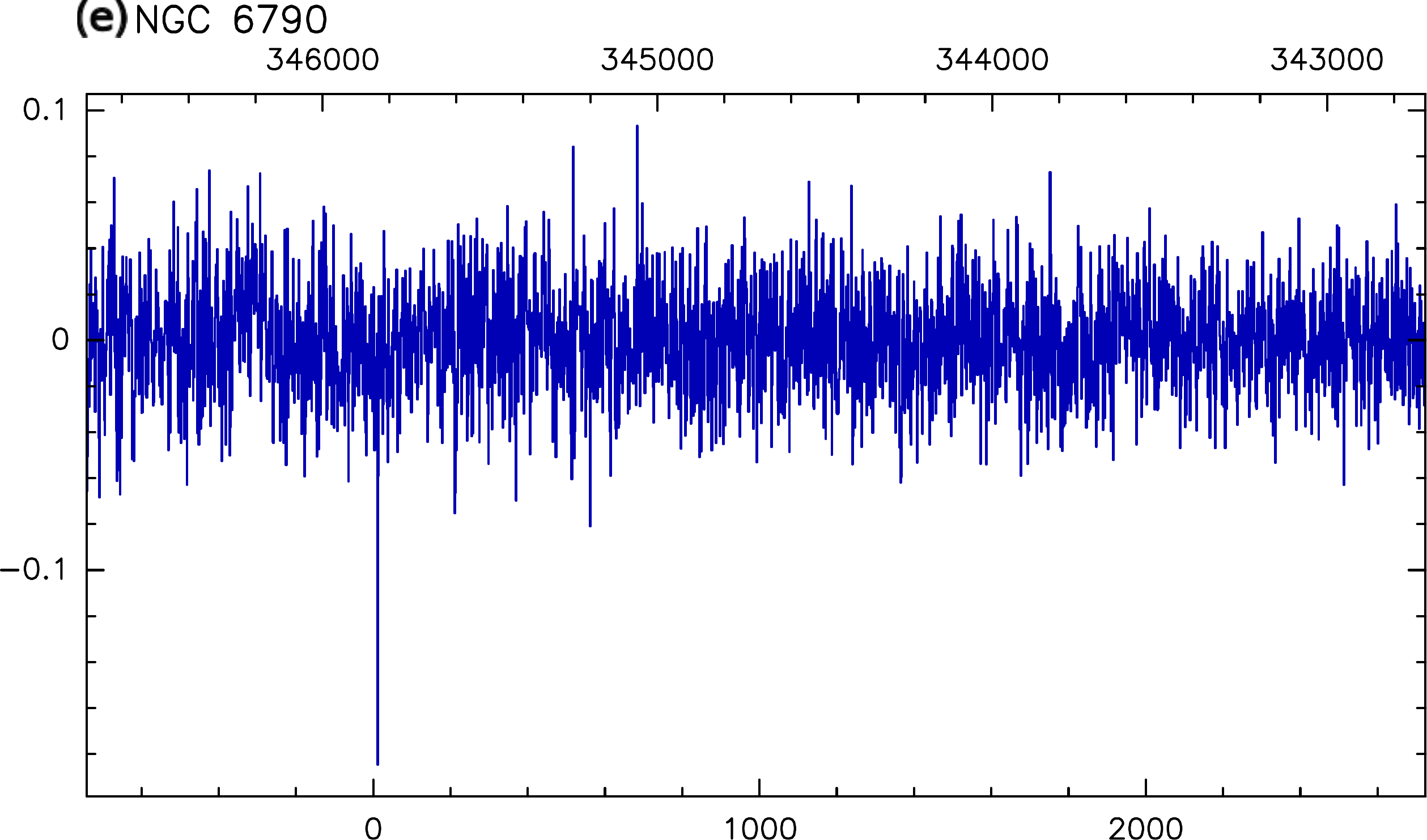}
\includegraphics[width=6cm, height=5.5cm]{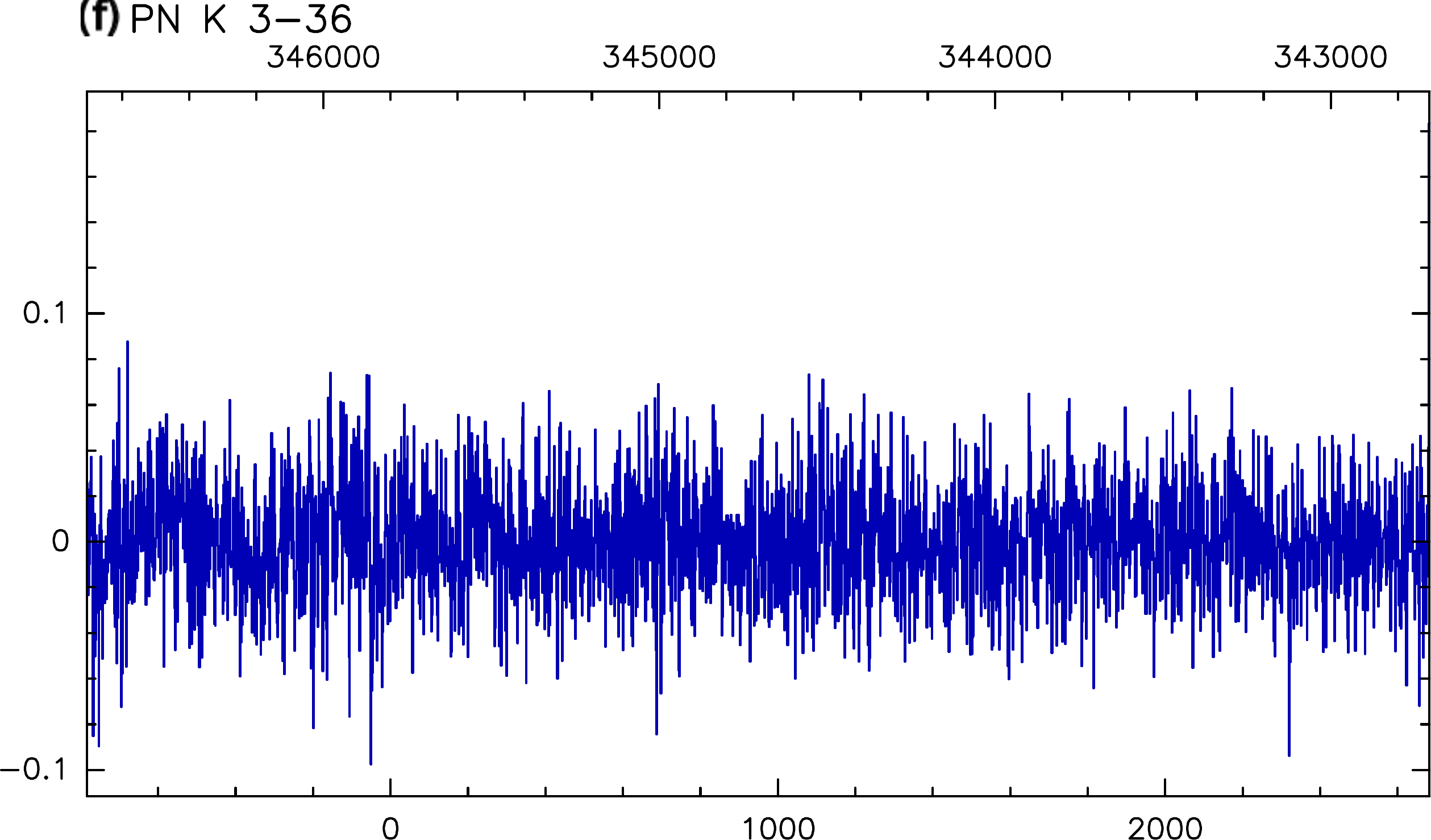}}
\hbox{
\centering
\includegraphics[width=6cm, height=5.5cm]{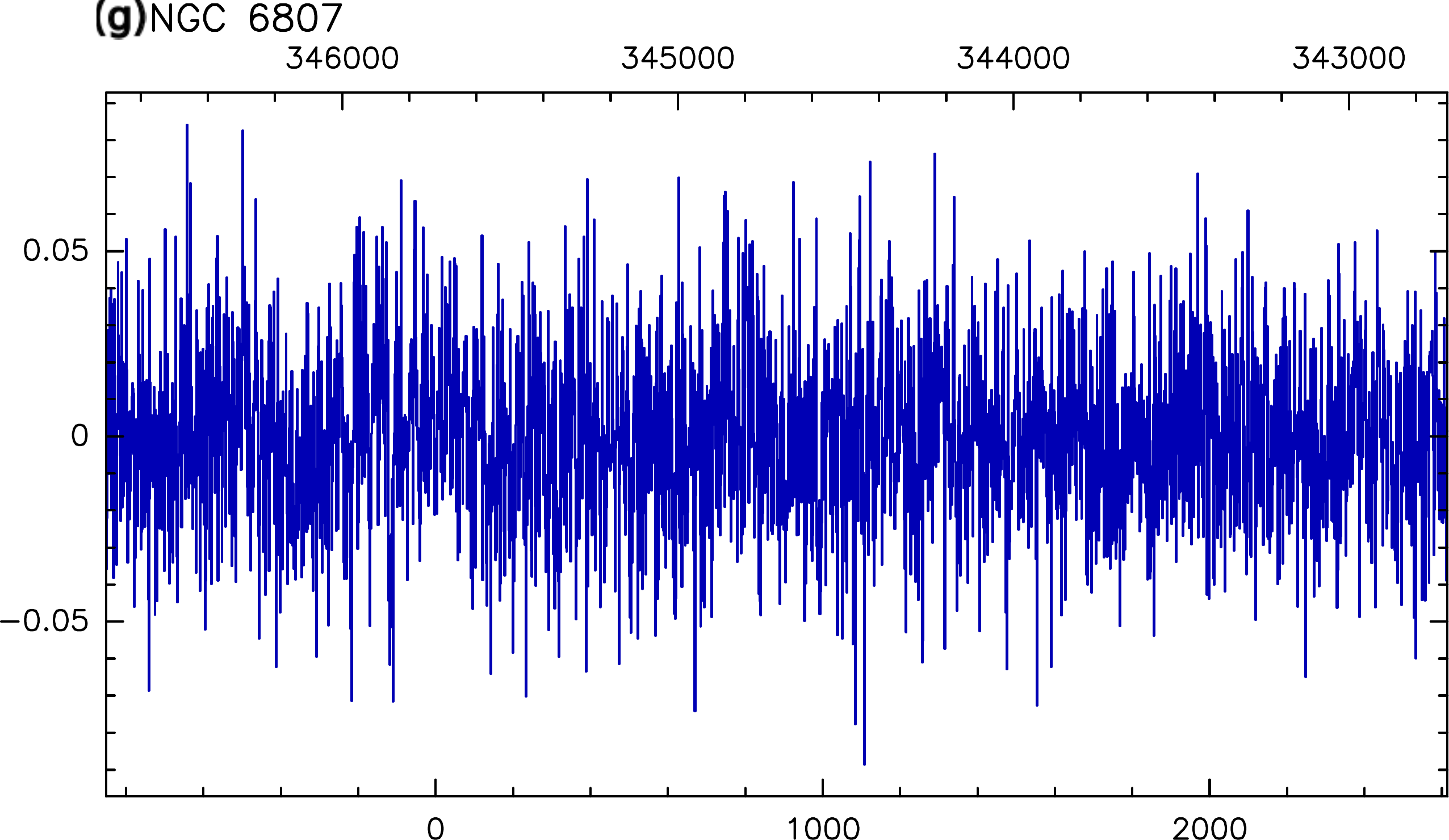}
\includegraphics[width=6cm, height=5.5cm]{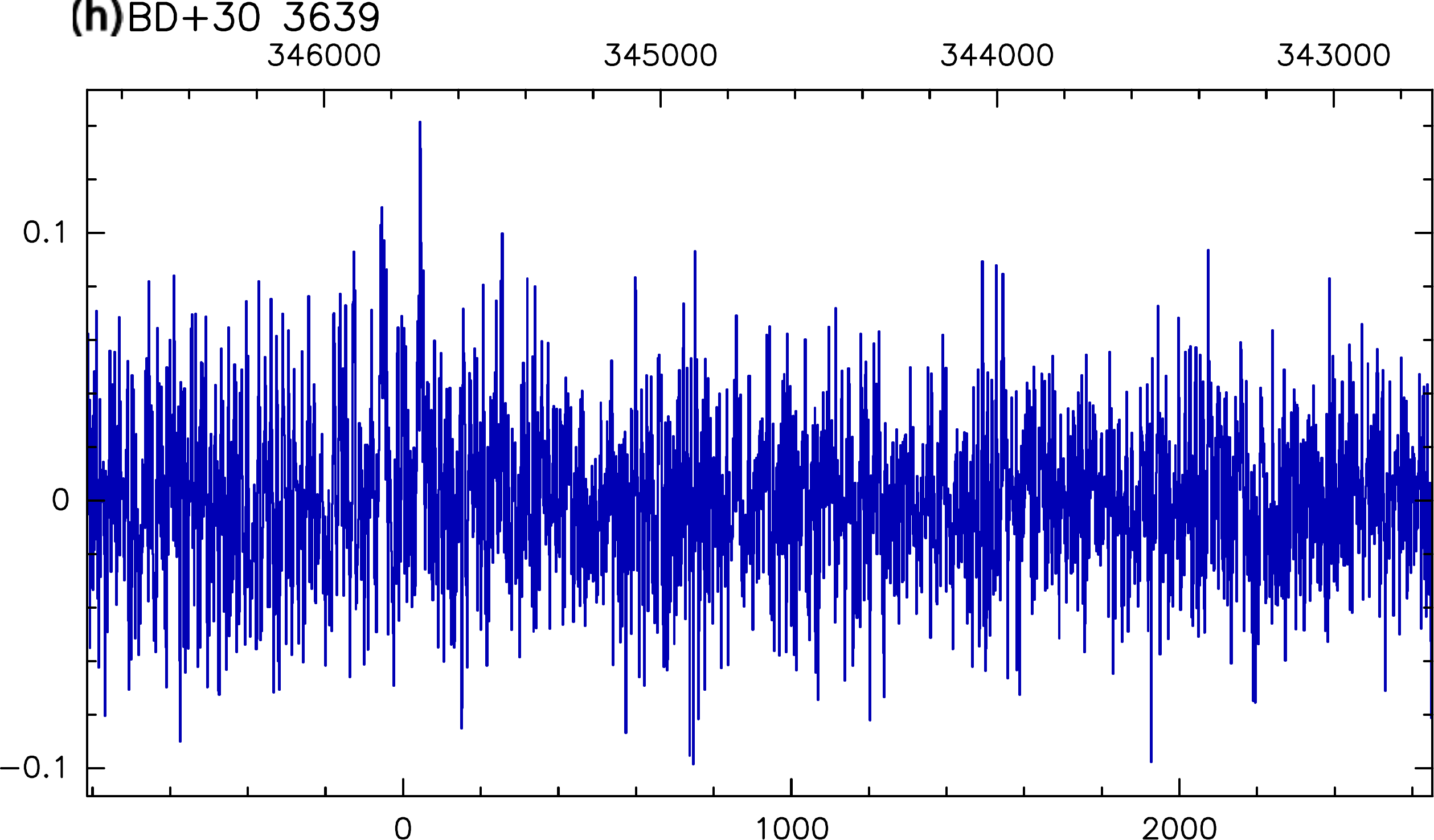}
\includegraphics[width=6cm, height=5.5cm]{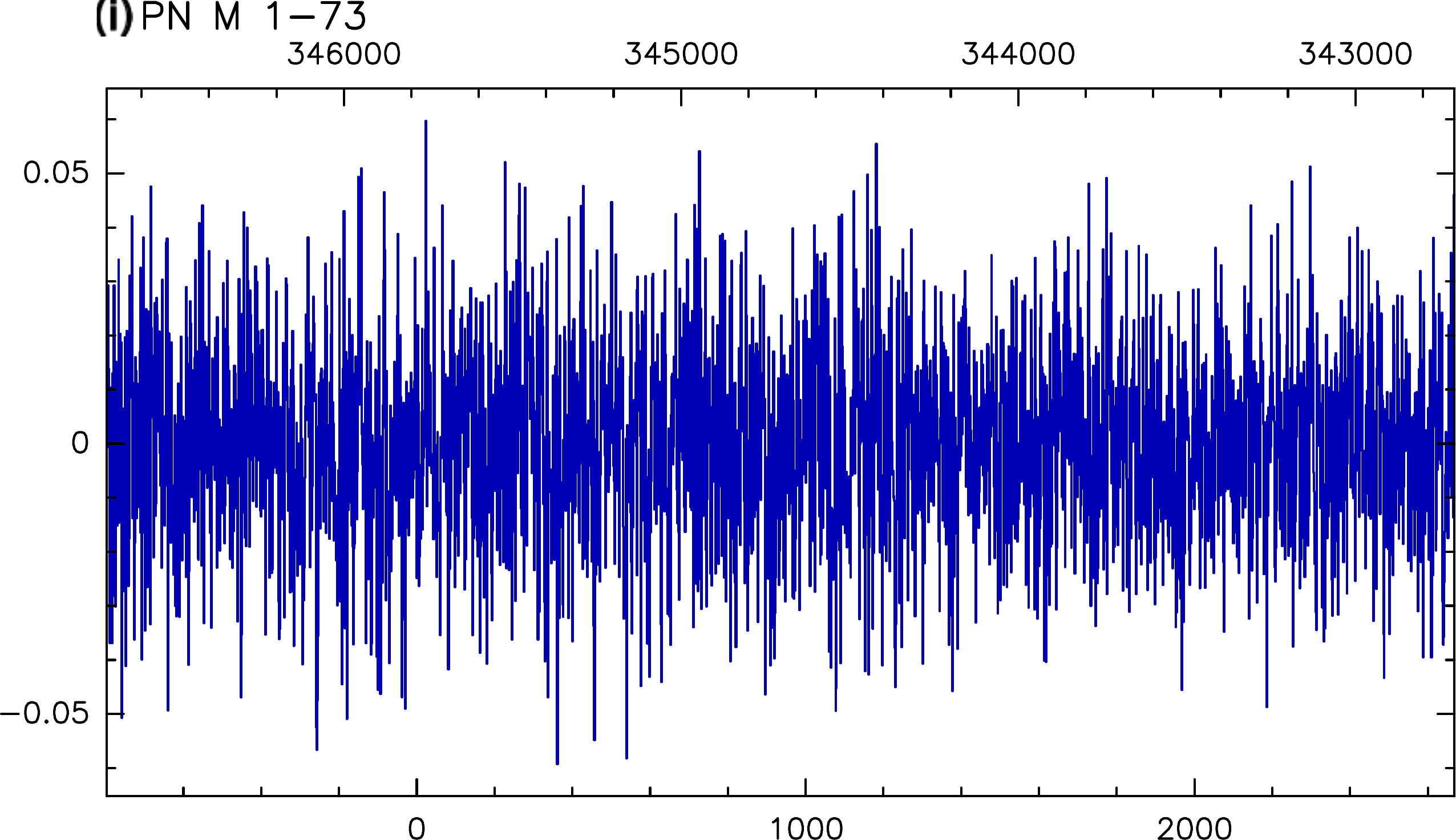}}
\caption[]{pPNe and PNe observations using the APEX telescope.}
\label{Fig9}
\end{figure*} 

\begin{figure*}
\vspace{2cm}
\centering
\hbox{
\centering
\includegraphics[width=6cm, height=5.5cm]{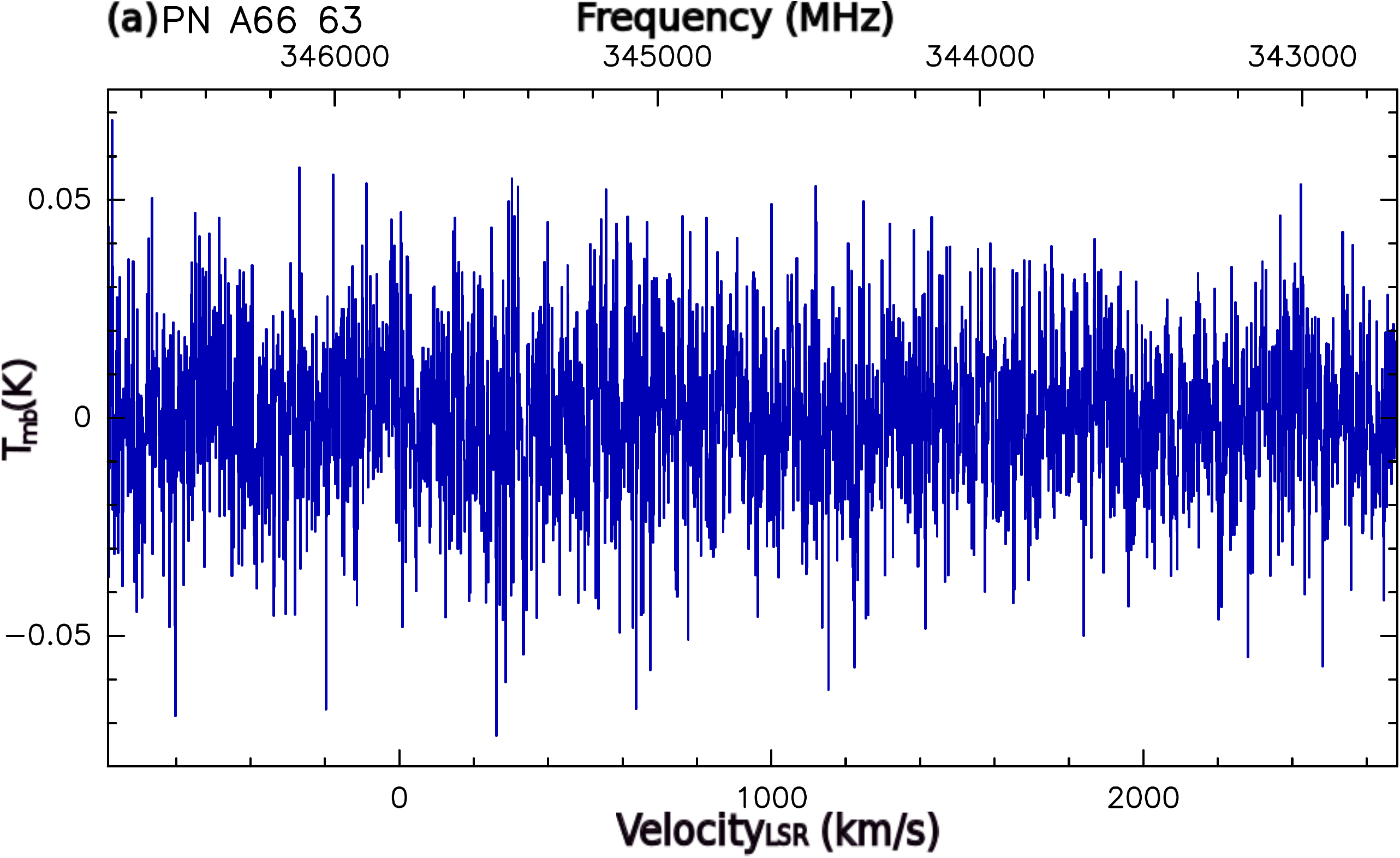}
\includegraphics[width=6cm, height=5.5cm]{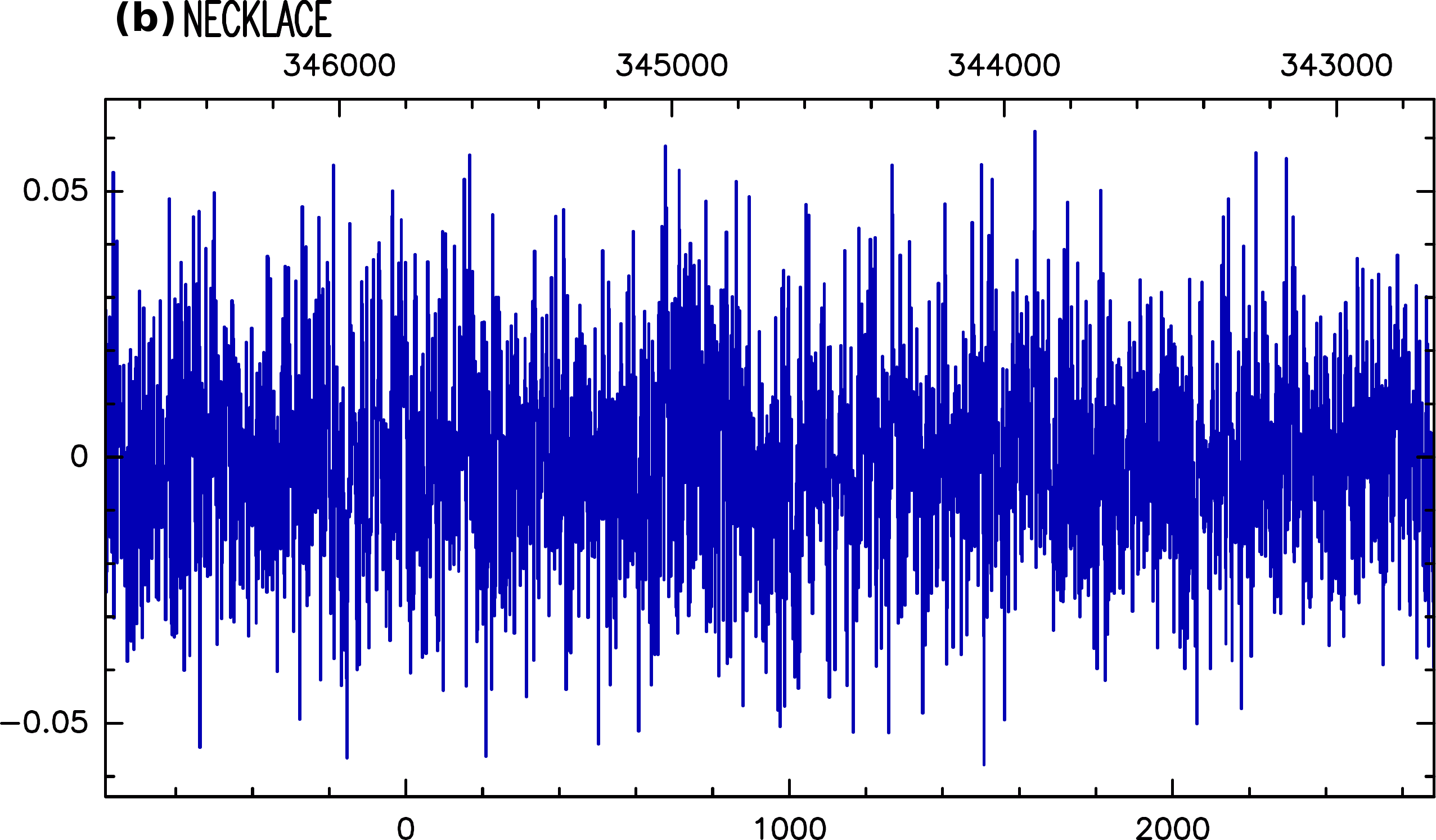}
\includegraphics[width=6cm, height=5.5cm]{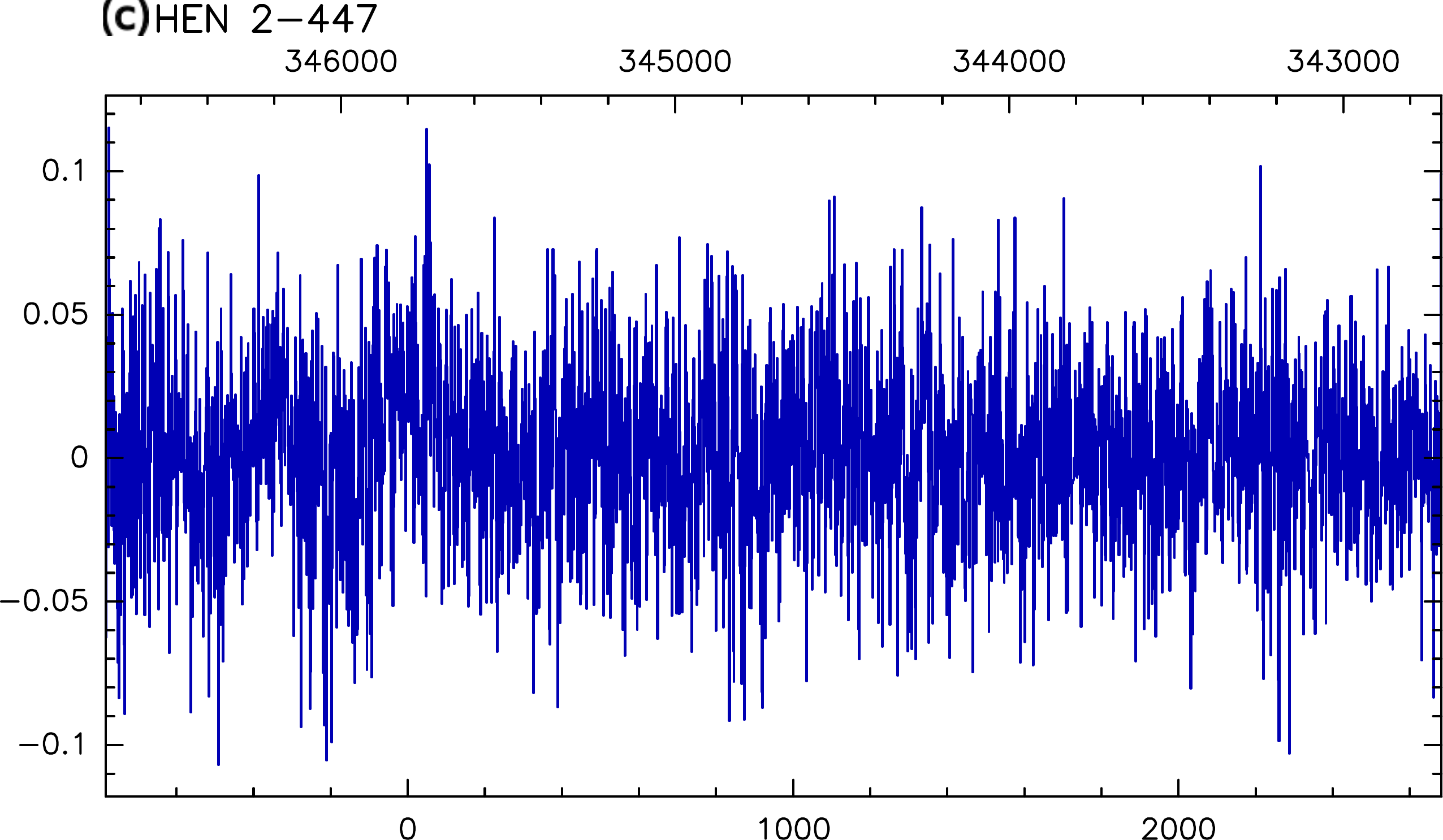}
\vspace{1cm}}
\hbox{
\centering
\vspace{1cm}
\includegraphics[width=6cm, height=5.5cm]{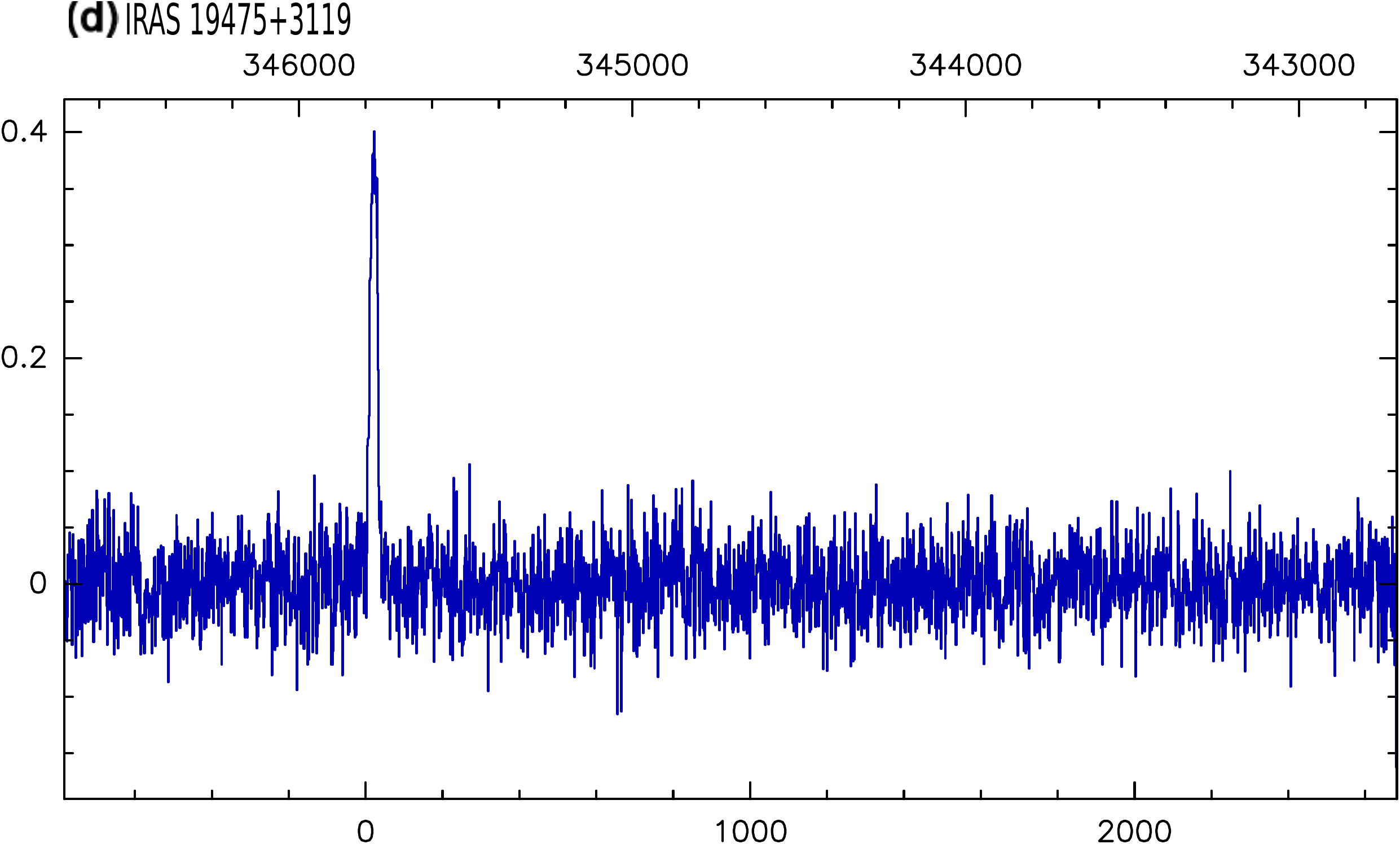}
\includegraphics[width=6cm, height=5.5cm]{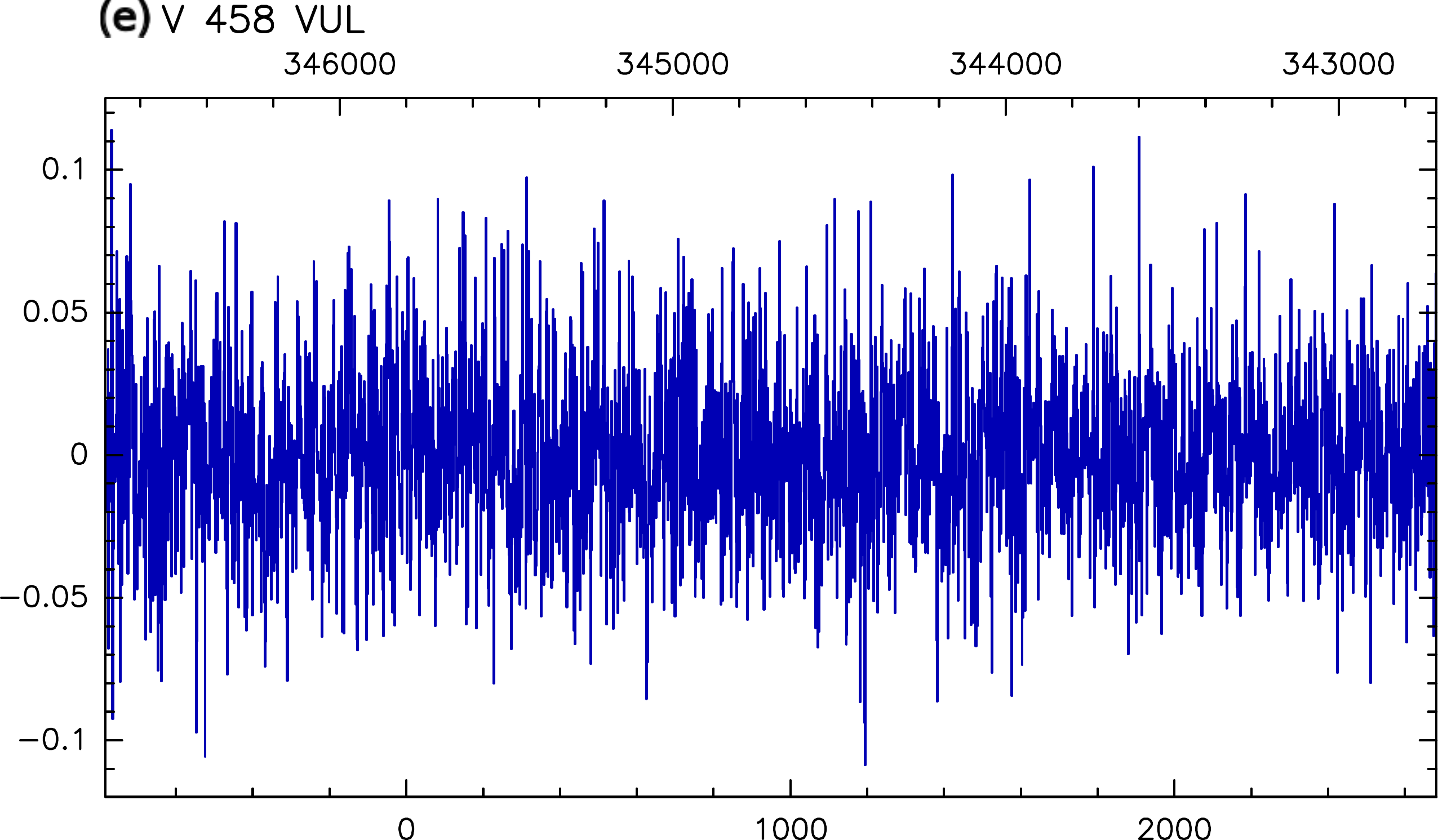}
\includegraphics[width=6cm, height=5.5cm]{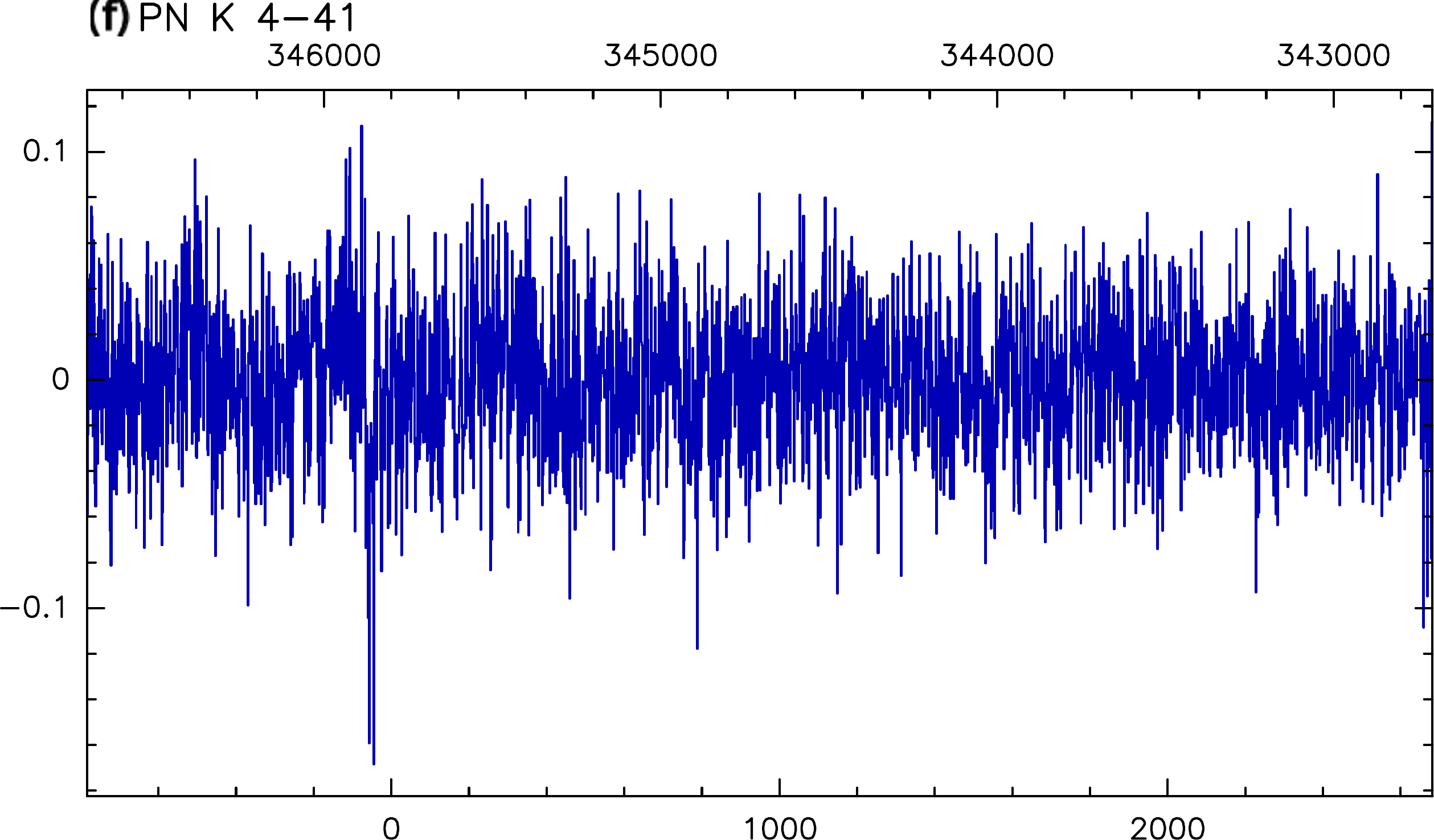}}
\hbox{
\centering
\includegraphics[width=6cm, height=5.5cm]{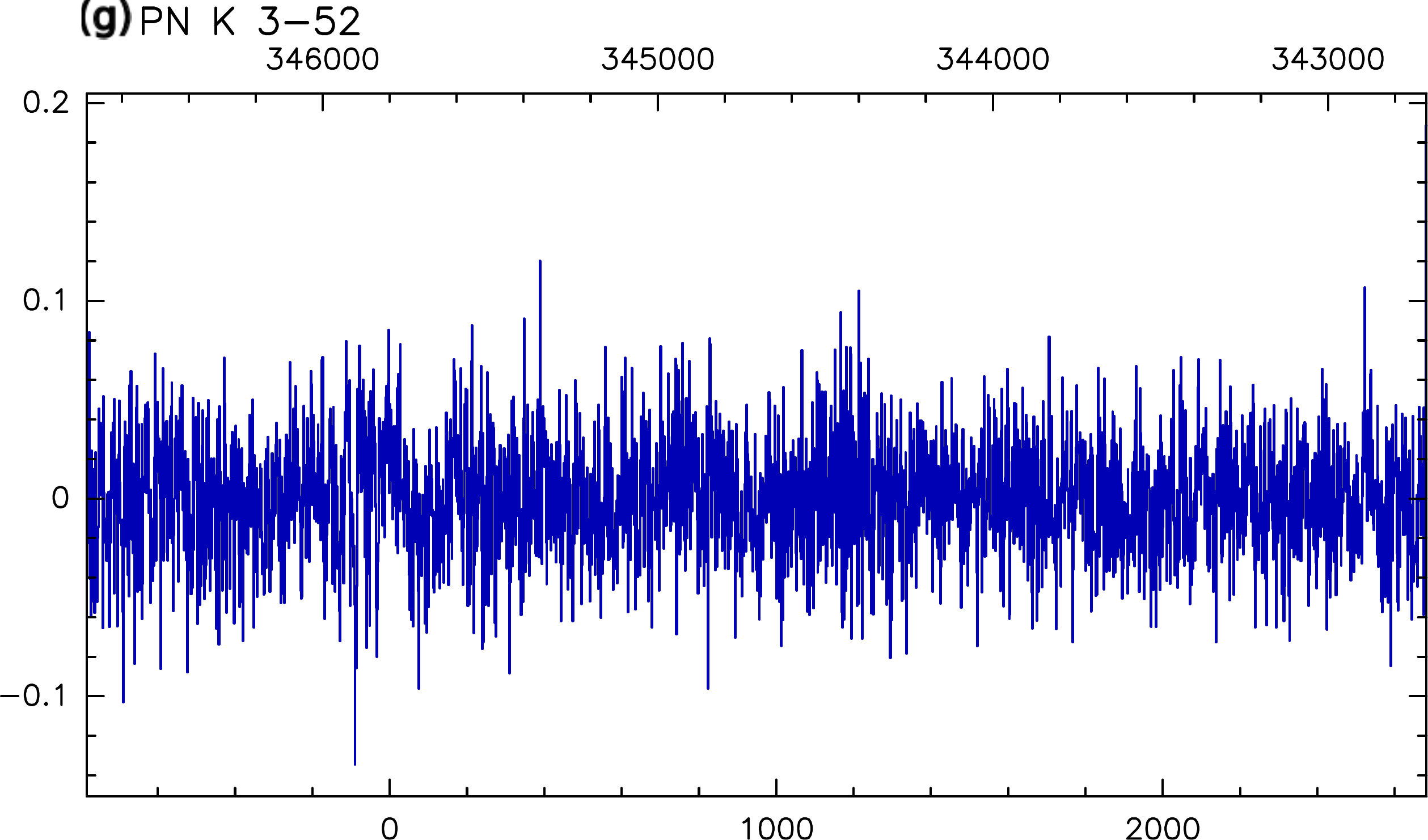}
\includegraphics[width=6cm, height=5.5cm]{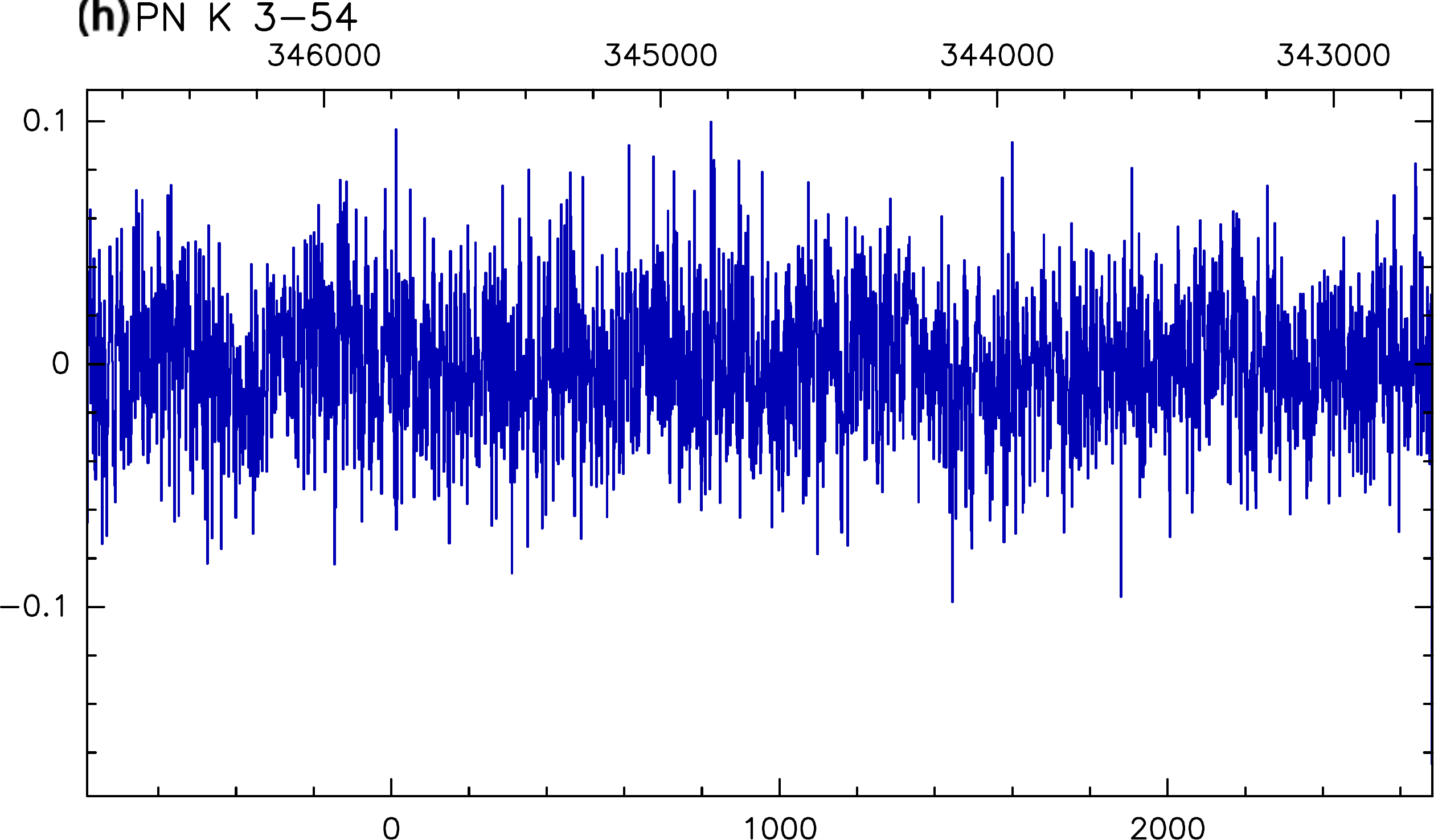}
\includegraphics[width=6cm, height=5.5cm]{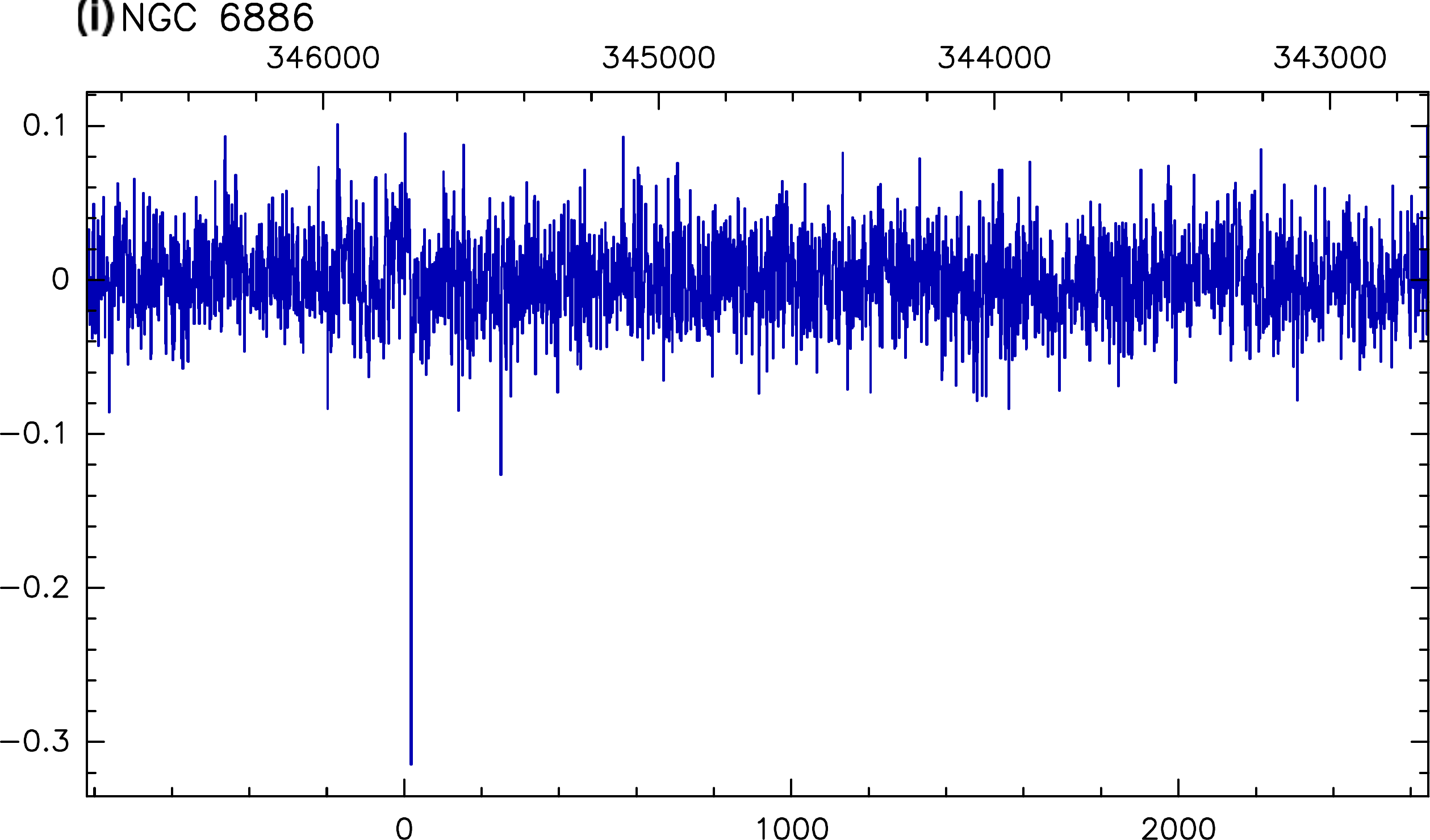}}
\caption[]{pPNe and PNe observations using the APEX telescope.}
\label{Fig10}
\end{figure*} 

\begin{figure*}
\vspace{2cm}
\centering
\hbox{
\centering
\includegraphics[width=6cm, height=5.5cm]{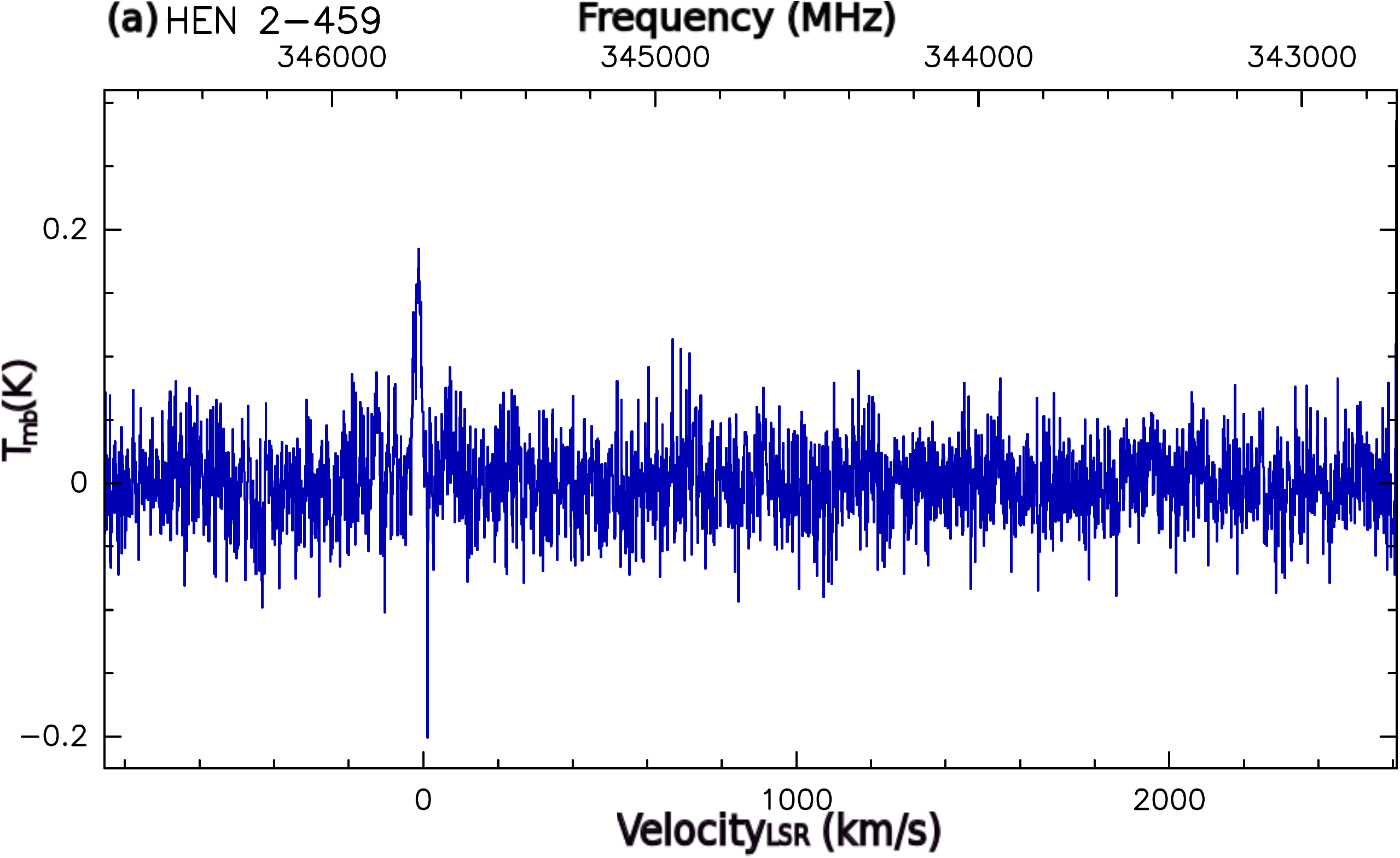}
\includegraphics[width=6cm, height=5.5cm]{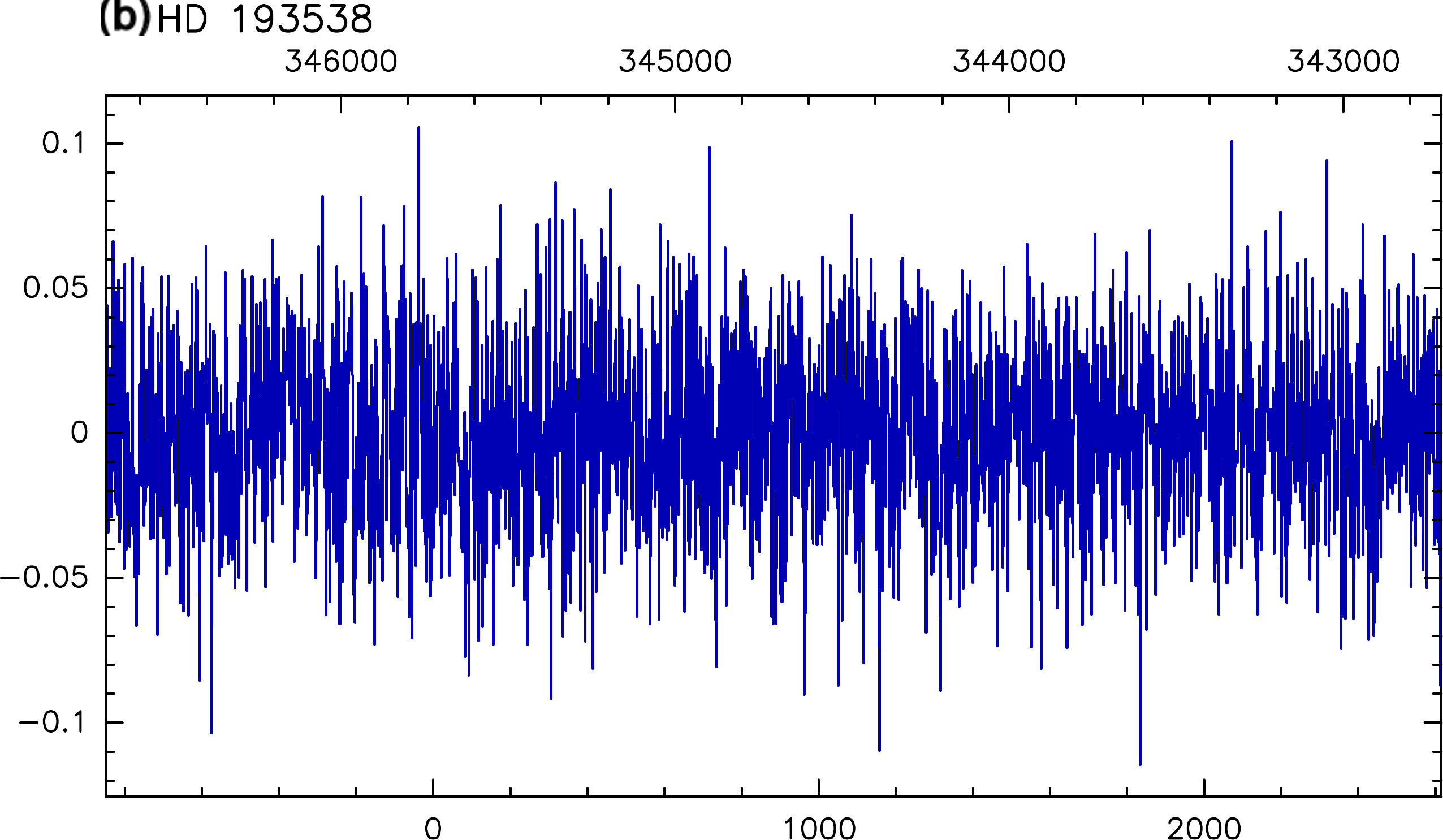}
\includegraphics[width=6cm, height=5.5cm]{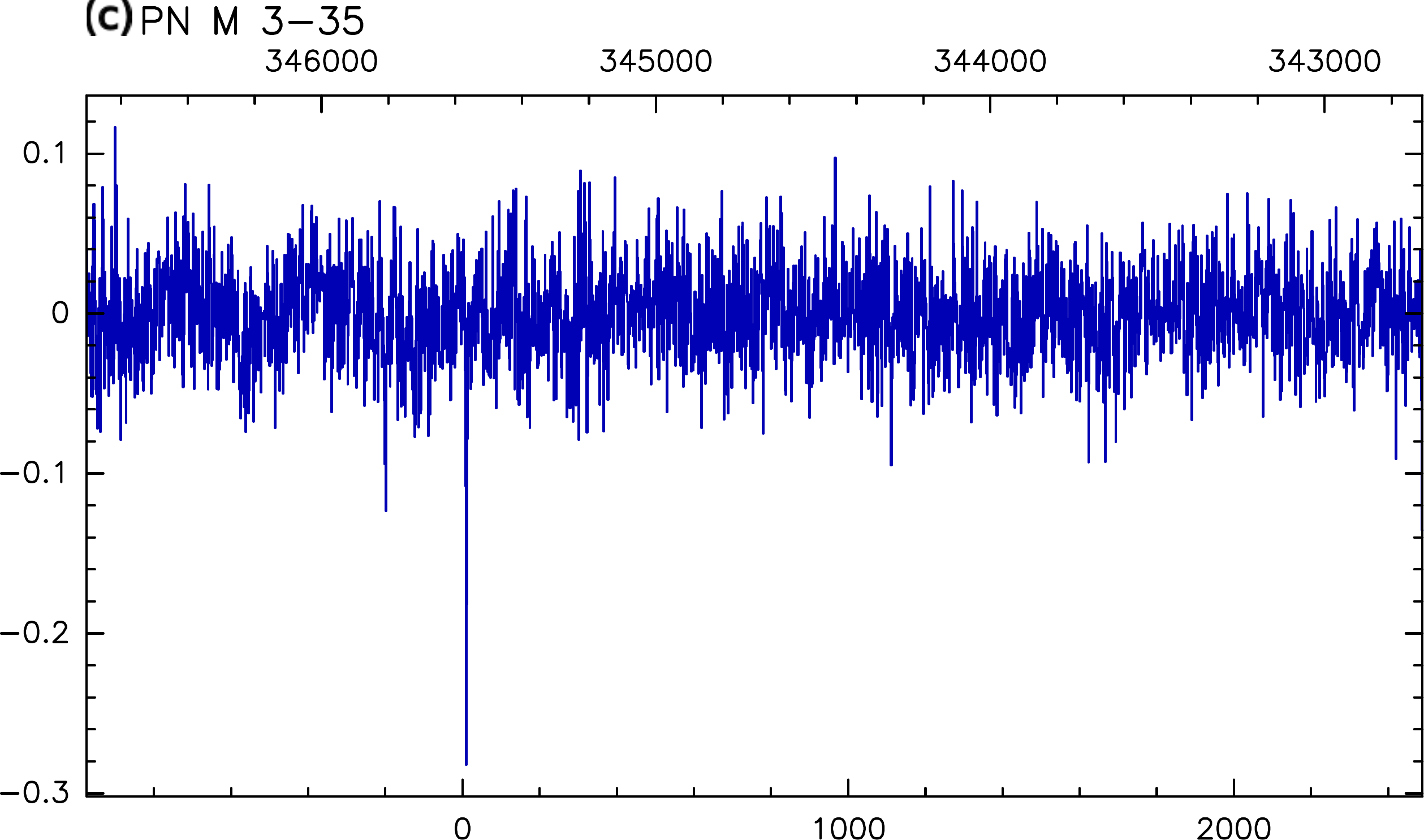}}
\caption[]{pPNe and PNe observations using the APEX telescope.}
\label{Fig11}
\end{figure*} 

\end{appendix}

\end{document}